# Topological Self-Stabilization with Name-Passing Process Calculi

vorgelegt von

**Christina Julia Rickmann**

Matrikelnummer 314392

Technische Universität Berlin

Gutachter:

Prof. Dr.-Ing. Uwe Nestmann, Erstkorrektor

Dr. Stefan Schmid, Zweitkorrektor

Betreuer:

Dipl.-Inf. Christoph Wagner

# Eidesstattliche Erklärung

Hiermit erkläre ich, dass ich die vorliegende Arbeit selbstständig und eigenhändig sowie ohne unerlaubte fremde Hilfe und ausschließlich unter Verwendung der aufgeführten Quellen und Hilfsmittel angefertig habe.

Berlin, 2. Oktober 2015 Christina Julia Rickmann

# Abstract

Topological self-stabilization describes the ability of a distributed system to let the nodes themselves establish a meaningful overlay network. Independent from the initial network topology, the system converges to the desired topology via forwarding, inserting, and deleting links to neighboring nodes.

Name-passing process calculi, like the $\pi$-calculus, are a well-known and widely used method to model concurrent and distributed algorithms. The $\pi$-calculus is designed to naturally express processes with a changing link infrastructure, as the communication between processes may carry information that can be used for a change in the linkage between the processes.

We redesign a simple local linearization algorithm with asynchronous message-passing that was originally designed for a shared memory model. We use an extended localized $\pi$-calculus, a variant of the $\pi$-calculus, to model the algorithm. Subsequently, we formally prove the self-stabilizing properties closure, weak convergence for every arbitrary initial configuration, and strong convergence for two special cases. In our proofs we utilize rather an assertional reasoning than an action-based style. Furthermore, we describe the challenges in proving (strong) convergence in the general case. Additionally, we give strong arguments for strong convergence, supported by further proven lemmata, and discuss different approaches for a formal proof.

# Zusammenfassung

Topologische Selbstabilisierung beschreibt die Fähigkeit eines verteilten Systems, dass die Knoten selbstständing in der Lage sind ein sinnvolles Overlay-Netzwerk zu etablieren. Das System nähert sich unanhängig von der initialen Topologie, durch die Weiterleitung und das Hinzufügen und Löschen von Verbindungen zwischen benachbarten Knoten, der angestrebten Topologie an.

Name-passing Prozesskalküle, wie der $\pi$-Kalkül, sind eine gut erforschte und weitverbreitete Methode zur Modellierung von nebenläufigen und verteilten Algorithmen. Der $\pi$-Kalkül ist konzipiert um in natürlicher Weise Prozesse mit einer sich ändernden Verbindungsinfrastruktur ausdrücken zu können, da die Kommunikation zwischen den Prozessen Informationen übertragen kann, die zu einer Veränderung der Vernetzung genutzt werden können.

Ein einfacher Linearisierungsalgorithmus, ursprünglich für ein System mit Shared Memory entworfen, wird so umgestaltet, dass er mit asynchroner Nachrichtenübertragung funktioniert. Um den Algorithmus zu modellieren, wurde eine Erweitung des Localized $\pi$-Kalküls, eine Variante des $\pi$-Kalküls, verwendet. Darauffolgend werden die selbststabilisierenden Eigenschaften der Abgeschlossenheit, schwache Konvergenz für jede beliebige initiale Konfiguration und starke Konvergenz für zwei Spezialfälle formal bewiesen. In den Beweisen werden Schlussfolgerungen eher basierend auf dem Zustand des Systems gezogen, als einen aktionsbasierten Stil zu verfolgen. Des Weiteren werden die Herrausforderungen des Beweisens von starker Konvergenz für den allgemeinen Fall beschrieben. Zusätzlich werden starke Argumente für starke Konvergenz gegeben, unterstützt von weiteren bewiesenen Lemmata, und verschiedene Ansätze für einen formalen Beweis diskutiert.

# Contents

# Introduction

Undeniably, technology plays an increasing role in our society and everyday life. As hardware and software systems grow in scale and functionality, so does their complexity. An ever increasing complexity comes with an increasing likelihood of errors (Clarke and Wing, 1996). While most of the occurring errors are merely an annoyance, they can also lead to high financial losses (Mars Climate Orbiter Mishap Investigation Board et al., 1999) and in the worst case even to deaths (Leveson and Turner, 1993).

In almost no distributed system all components work correctly the entire time. Therefore fault tolerance is one of the main issues in distributed computing. One interesting and powerful specialization of nonmasking fault tolerance is self-stabilization (Gärtner, 1999). The approach of self-stabilizing systems was first introduced by Dijkstra (1974). According to Dolev (2000), the idea of a self-stabilizing system is that when started in an arbitrary state it always converges to a desired state. Therefore, a self-stabilizing system is able to automatically recover from any transient fault, like process crash with recovery and corrupted random access memory (RAM).

Topological self-stabilization describes a special subclass of self-stabilizing systems. The goal of topological self-stabilization is that the nodes themselves can establish a meaningful overlay network, independent from the initial network topology, via forwarding, inserting, and deleting links to neighboring nodes.

Gall et al. (2014) introduce two variants of a self-stabilizing algorithm for graph linearization. The purpose of linearization is to build an ordered list of nodes according to their unique identifiers (ids). Therefore, whenever a node has two neighbors, both of which have a smaller (respectively a greater) id, it establishes a link between them and deletes its link to the smaller (resp. greater) one. The algorithm works only in a shared memory model as nodes access the variables of all neighbored nodes. However, the motivation for topological self-stabilization is to build robust open distributed systems, like peer-to-peer systems. Therefore, a shared memory model is not the most suitable one.

The goal of this master thesis is to redesign and model the algorithm for an asynchronous message-passing system rather than a shared memory model.

Faults can not only occur in the system itself, but also the design of algorithms. Furthermore, proofs for correctness can be faulty as well. In order to prevent a faulty design of an algorithm and to confirm the correctness of proofs the usage of formal methods is imperative.

The $\pi$-calculus is a well-known and widely used process calculus to model concurrent and distributed systems. According to Milner et al. (1992) it is designed to naturally express processes with a changing link infrastructure, as the communication between processes may carry information that lead to a change in the linkage of processes.

We model the algorithm of Gall et al. (2014) in a self-developed extension of the localized $\pi$-calculus, a variant of the $\pi$-calculus. We extended the calculus in a way similar to Wagner and Nestmann (2014). This allows us to explicitly keep track of the neighbors of each node and therefore of the topology of the system. Each node can receive messages from other nodes via a channel with the same name as its id. To enable a neighboring node to communicate with another neighbor it is sufficient to send it the corresponding id.

In order to assert the correctness of our algorithm we formally prove the self-stabilizing properties closure and weak convergence, utilizing rather assertional reasoning than an action-based style. Furthermore, we give strong arguments for strong convergence.

This thesis is organized as follows. We first introduce technical preliminaries for the better understanding of this work. This includes a short introduction of distributed algorithms and criteria for the differentiation between distributed systems. Afterwards, we describe the different forms of fault tolerance and introduce self-stabilization as a special case of nonmasking fault tolerance. We present the self-stabilizing properties closure and convergence, as well as a weaker variant. We then describe the special subclass of topological self-stabilization and the linearization problem. Subsequently, we introduce process calculi as a formalism for the description and analysis of properties in concurrent systems and the localized $\pi$-calculus as representative.

In Chapter 2, we first briefly describe the original shared memory model and the linearization algorithm of Gall et al. (2014) with the main proof ideas. Thereafter, we introduce an extended version of the localized $\pi$-calculus as basis for an asynchronous message-passing model. We then present our model with basic properties and the redesigned linearization algorithm.

Chapter 3 comprises the formal proofs of closure, strong convergence for two special cases, and weak convergence for arbitrary initial configurations together with all needed lemmata.

Chapter 4 describes the problems for a formal strong convergence proof in the general case. Nevertheless, we give strong arguments for strong convergence for every arbitrary initial configuration and introduce further proven and open lemmata with proof sketches that support this arguments and a formal proof. As a last point, we discuss several approaches for a formal proof of strong convergence.

Subsequently, we conclude with our contributions, a short comparison between the original algorithm and our redesigned one, and a summary of the obtained key findings. Additionally, we give a short overview of the related and the future work.

# 1 Technical Preliminaries

This chapter introduces important fundamentals required for a good understanding of this work. Therefore it serves the classification of the problem and the introduction of principles, concepts and notions.

## 1.1 Distributed Algorithms

In modern computer-based data processing a program is often executed by many different processes. In this context a process can refer to a computer as well as a processor or even a single thread that is executed on a processor (Guerraoui and Rodrigues, 2006). The fundamental problem in developing such a distributed program is that the different processes typically have to work together to solve a shared task.

Distributed algorithms are algorithms, that are designed to be executed on hardware consisting of several such connected processing units (Lynch, 1996). A distributed algorithm is therefore composed of different fragments that are executed concurrently and independent from each other, whereby each of this fragments only has access to limited information of the overall system.

The development of multiprocessor systems and multi-core processors as well as the emergence of global computer networks further increase the omnipresence of distributed algorithms.

## 1.2 Models of Distributed Algorithms

A large variety of concurrent algorithms for a wide range of applications is denoted as distributed algorithms. Originally, the term was only used for processes that are distributed over a large geographical area. Over time the usage of this term broadened, due to many similarities in various settings. Nowadays, it covers also algorithms executed within a local area network or even on a multiprocessor system with shared memory (Lynch, 1996).

Naturally, distributed algorithms differ in the task to be solved by the processes i. e., the problem. The diversity of distributed systems leads to a variety of algorithms solving the same problem.

To evaluate if a distributed algorithm solves a certain problem, it is necessary to possess knowledge about the system in which the algorithm is executed. Therefore hereafter we will present some common classifications for distributed systems and thus distributed algorithms (Guerraoui and Rodrigues, 2006; Lynch, 1996; Attiya and Welch, 2004).

### 1.2.1 Timing Model

One of the most important assumptions about a distributed system is the degree of synchronicity between its processes. Lynch (1996) and others distinguish between synchronous, asynchronous, and partially synchronous systems.

- *Synchronous timing model:* In the synchronous timing model it is assumed that all processes participating in the algorithm take steps simultaneously. This degree of synchronicity is hard to achieve with real world systems. Therefore, it is not considered a realistic assumption for many distributed systems. The advantage of this timing model is that problems can be solved with comparatively simple algorithms. This helps in development and understanding of solutions in more realistic models.

- *Asynchronous timing model:* In contrast in the asynchronous timing model we make no assumptions about the speed of the different processes. Every process executes the steps of the algorithm in its own tempo that may also vary over time. Therefore, the order of steps between process is not determined in general. Since no timing guarantees are given in this model, asynchronous algorithms are applicable to systems with arbitrary timing guarantees. However, in a completely asynchronous model not all problems are solvable at all.

- *Partially synchronous timing model:* The timing assumptions in the partially synchronous model are somewhere in the wide range between the synchronous and asynchronous model. In this model, for example, there could be lower or upper bounds on the speed of executing steps or delivering messages. The partially synchronous timing model is the most realistic. However, the correctness of the algorithm is heavily dependent on the observance of the timing assumptions.

### 1.2.2 Communication Model

Distributed algorithms require an interprocess communication mechanism due to the fact that always multiple processes are involved. Such a mechanism provides the essential possibility for information exchange between processes.

A common abstraction of this interprocess communication is message passing, either via point-to-point or broadcast connections, shared memory, or remote procedure calls (Lynch, 1996).

- *Point-to-Point Communication:* For point-to-point communication it is assumed that two processes are connected via some kind of communication channel. In this model, the topology of connections, assumptions of the transmission time, and the reliability of the channels are of particular interest.
  The topology of the networks describes which process is able to communicate directly with whom. It can be represented as a directed graph, whereby processes are depicted as nodes and point-to-point-connections as edges. An edge $(u, v)$ denotes that a process $u$ can directly send a message to another process $v$ (Lynch, 1996). In this model it is often assumed that between each pair of processes exists a bidirectional channel i. e., every process can send messages to every other process (Guerraoui and Rodrigues, 2006).
  In synchronous message passing, a message is sent and received simultaneously. Therefore, the transmission of the message is instantaneous as sender and receiver have to execute their corresponding steps at the same time. In contrast, during asynchronous message passing, the message transmission can take an arbitrary amount of time and the sender continues with the algorithm directly after sending the message.
  The reliability concerns the kinds of faults that can occur during the message transmission. This will be considered in more detail in Section 1.2.3.

- *Broadcast-Communication*: Broadcast communication is a special variant of message passing. Here it is assumed, that it is possible for a process to send a message to all other processes simultaneously. Just as in in point-to-point communication it can be distinguished between synchronous and asynchronous transmission and there are several degrees of reliability (Attiya and Welch, 2004).

- *Shared Memory*: In this model, processes communicate via so-called registers inside a shared memory. Processes have access through read and write operations to a shared memory cell. If a process writes a value in a specific register, it is possible for another process to directly read this value later.
  This kind of abstraction is suitable for physical shared memory of processes within a multiprocessor system, as well as for virtual shared memory where processes are geographically distributed and do not have actually access to the same hardware (Guerraoui and Rodrigues, 2006).
  Different variants of shared memory can also be distinguished by the number of processes that are allowed to access registers via read or write operations simultaneously.

- *Remote Procedure Call*: In this model, message exchange between processes is hidden. One process is able to make another process to execute a certain procedure. Parameters and return values of these procedures allow for an exchange of further information between the involved processes (Tanenbaum and van Steen, 2008).

### 1.2.3 Models of Faults

Hardly any (non-trivial) system can be considered fault free and components in real-life distributed systems i. e., processes and interprocess communication infrastructure, are often error-prone. Since distributed systems are widely used to assure fault tolerance, it is necessary to take faults into account and to make assumptions about the possible kinds of faults.

In the following, we give a brief overview about various fault models regarding processes and message transmission.

- *Process faults:* A process is seen as faulty if its behavior does not (any longer) correspond to its specification. Guerraoui and Rodrigues (2006) introduce for example four different models of process faults.
  Easiest to treat is the fault assumption that processes can crash i. e., a process fails and afterwards no further actions are exercised. However, until the moment of the fault it behaves according to its specification. This kind of fault is denoted as *crash fault* or *fail-stop*. *Omission faults* describe that a process fails to execute some actions like sending or receiving a messages. Another variant are *crash-recovery faults*. A crashed process recovers eventually and participates once again in the execution of the algorithm. However, this often goes hand in hand with partial or total loss of information. The most severe form of process faults are called *Byzantine faults*. In this model, faulty processes can execute arbitrary actions. Even actions outside of their specification.A process could, for example, send a wrong value to another process. This fault model covers also malicious behavior, since the arbitrary behavior is able to model even an outside attacker on the system.

- *Transmission faults:* Possible faults that can occur during the transmission of messages are *loss* (a message is send, but not delivered), *duplication* (a message is delivered more often than sent), *corruption* (another value is delivered than sent) and *reordering of messages* (messages are delivered out of order).
  Often, assumptions regarding the reliability of message transmission are made. Thereby different levels of reliability can be distinguished. According to Guerraoui and Rodrigues (2006) a *fair-loss*-point-to-point-connection, for example, guarantees that a infinitely often sent message will be delivered, a message that is infinitely often received, was sent infinitely often, and messages are never corrupted.

## 1.3 Fault tolerance

A fault, like the loss of a message, may cause an error, which is an internal system state that does not conform to the specification. An error in turn may further lead to a failure, meaning that the system deviates from its correctness specification (Gärtner, 1999). Fault tolerance describes the ability of a distributed system to continue to provide its service (possibly in a reduced manner) even in case of the appearance of faults in some of its components, instead of leading to a complete system failure (Tanenbaum and van Steen, 2008). When designing for a fault tolerance system, a first prerequisite is to specify the fault class that should be tolerated and afterwards enrich the system with components or concepts that provide protection against faults of the corresponding fault class (Gärtner, 1999). Therefore, the system behaves according to Gärtner (1999) in a well-defined manner once faults, from the considered fault class, occur.

In order to specify any useful behavior of a system Lamport (1977) distinguishes between two different necessary system properties. Namely safety and liveness. Informally, a safety property states that "something bad" never happens in the system. Whereas a liveness property claims that "something good" will eventually happen in every execution of the system. The following example illustrates the meaning of those two properties. A common application for a distributed algorithm is mutual exclusion i. e., an algorithm that ensures the requirement that no two concurrent processes are in their critical section at the same time. In this scenario, the safety property states that there is always at most one process in the critical section, while the liveness property claims that every process that requests to enter a critical section will eventually enter the critical section.

Gärtner (1999) identifies four forms of fault tolerance in dependence of whether safety or liveness properties are ensured by the algorithm in presence of faults.

| | **live** | **not live** |
|---|---|---|
| **safe** | masking | fail safe |
| **not safe** | nonmasking | fault intolerant |

Table 1.1: Four forms of fault tolerance

### 1.3.1 Fault Intolerance

Fault Intolerance means neither liveness nor safety is guaranteed in the presence of faults from a certain fault class. Therefore, the system does not offer any form of fault tolerance for this fault class and is the most trivial, weakest, cheapest, and undesirable form (Gärtner, 1999).

### 1.3.2 Masking

Masking fault tolerance is the other extreme case where both liveness and safety are ensured in case of faults (from a certain fault class). Therefore it is the most strict, and most desirable form of fault tolerance. With masking fault tolerance, the system is able to tolerate faults transparently, i. e., a user of the system cannot determine the occurrence of a fault (Gärtner, 1999). This can be necessary if even a temporary failure of the system would have unacceptable consequences such as the death of humans or unbearably high financial losses. Masking fault tolerance is indeed the most costly form as this can only be achieved through redundancy in time or space (Tanenbaum and van Steen, 2008; Gärtner, 1999). Redundancy in space means multiple instances of components. An example is triple modular redundancy (TMR). In TRM, there are three redundant components that deliver inputs to one voter. If at least two of these inputs are the same, the voter outputs that value. In this way, one faulty component can be masked whether it delivers an illegal input value or no input at all (Tanenbaum and van Steen, 2008). Nevertheless, if the voter or two of the components are erroneous, masking is not possible anymore. More generally speaking, masking fault tolerance is only successful if only a limited part of the components is erroneous. The amount of the erroneous components that can be tolerated depends on the fault model.

### 1.3.3 Fail Safe

Fail safe fault tolerance is one of the two intermediate combinations and guarantees safety but not liveness. The idea is to ensure a safe state in presence of a fault or error. As the system may ceases to show progress, the occurrence of a fault may be observable. (Gärtner, 1999). An example is a traffic light control, where in case of an error all traffic lights are switched to red.

### 1.3.4 Nonmasking

Nonmasking fault tolerance is the last combination. In presence of a fault, liveness is ensured but safety is not guaranteed. In effect, the user may experience a certain amount of incorrect system behavior (i. e., failures) (Gärtner, 1999). For example in a replicated database, a replication variable may not be up to date, but at least liveness is guaranteed in the way that reading and writing requests will always be granted.

Masking fault tolerance is only possible for certain fault classes and it is not realistic to predict all kinds of possible faults and take them into account. Nonmasking fault tolerance is strictly weaker than masking fault and can therefore be used in cases where masking fault tolerance can not be achieved because it is provably impossible or too costly (Gärtner, 1999). Self-stabilization is a specialization of nonmasking fault tolerance, where any kind of transient faults can be tolerated (Gärtner, 1999).

## 1.4 Self-Stabilization

The approach of self-stabilizing systems was first introduced by Dijkstra (1974) in the context of distributed systems. He defined that “we call the system self-stabilizing if and only if, regardless of the initial state [...], the system is guaranteed to find itself in a legitimate state after a finite number of moves.” Since self-stabilization guarantees that a legal state is reached from every starting state, it is not necessary to initialize the system in a certain way. Furthermore, this leads to the ability to tolerate any transient fault since if no new faults occur for a sufficient period of time, the state after the end of the last fault could be considered as a new initial state and the system must recover.

According to Schneider (1993) self-stabilization provides an unified approach for tolerating arbitrary transient faults, including inconsistent initialization, mode change, transmission errors like loss, corruption, or reordering, process failures with recovery and even memory crash. A transient fault is any event that may changes the state of the system, but not its behavior. Therefore, the behavior (i. e., the program code) of the system is not violable. A self-stabilizing algorithm is able to handle any corruption of the random-access memory (RAM) content but not of the read-only memory (ROM) (Dolev, 2000). The property of self-stabilization models the ability of a system to recover from transient failures under the assumption that they cease to occur (Schneider, 1993; Dolev, 2000; Gärtner, 1999). Another characteristic of self-stabilizing algorithms is that they must not terminate (Schneider, 1993). This follows from the fact that the processes must continuously communicate with neighboring nodes. If the participating processes would have a terminated state, the state that all processes are in their terminated state would also be a possible initial configuration. Therefore the task is achieved with no communication at all, which is not possible with a distributed algorithm (Dolev, 2000). As a consequence, the participating processes can not know whether the system is stabilized (i. e., is in a correct configuration).

Dolev (2000) defines a self-stabilizing system as follows:

**Definition 1: Self-Stabilizing System**

A self-stabilizing system can be started in any arbitrary configuration and will eventually exhibit a desired "legal" behavior. We define the desired legal behavior by a set of legal executions denoted *LE*. A set of legal executions is defined for a particular system and a particular task. Every system execution of a self-stabilizing system should have a suffix that appears in *LE*. A configuration *c* is *safe* with regard to a task *LE* and an algorithm if every fair execution of the algorithm that starts from *c* belongs to *LE* (*closure property*). An algorithm is *self-stabilizing* for a task *LE* if every fair execution of the algorithm reaches a safe configuration with relation to *LE* (*convergence property*).

In order to prove convergence a basic proof technique according to Dolev (2000) is a *potential function*. The idea is to define a function over the configuration set and prove that this function monotonically decreases (or increases) with every executed step. Additionally, it has to be shown that after the function reaches a certain threshold, the system is in a safe configuration. Another way to prove convergence are so-called *convergence stairs*. Here the algorithm converges to fulfill $k > 1$ predicates $\mathscr{A}_1, \mathscr{A}_2, \ldots, \mathscr{A}_k$, such that, for every $1 \leq i < k$, $\mathscr{A}_{i+1}$ is a refinement of $\mathscr{A}_i$. Whereby a predicate $\mathscr{A}_{i+1}$ refines the predicate $\mathscr{A}_i$ if $\mathscr{A}_i$ holds whenever $\mathscr{A}_{i+1}$ holds. Then it must be shown whenever $\mathscr{A}_i$ holds every execution reaches a configuration where $\mathscr{A}_{i+1}$ holds for all $i$. The last predicate $\mathscr{A}_k$ defines a safe configuration.
Since the closure property states that every step from a safe or correct configuration leads again to a correct configuration, closure is usually proven through invariants.

An easier to achieve and easier to prove property is weak stabilization. According to Gouda (1995) a system is weakly stabilizing if for every initial configuration there is an execution that reaches a safe or correct configuration. This property is called *weak convergence* and is related to probabilistic convergence.

### 1.4.1 Topological Self-Stabilization

Topological self-stabilization describes a special subclass of self-stabilizing systems. The goal of topological self-stabilization is that the nodes themselves can establish a meaningful overlay network, independent from the initial network topology, via forwarding, inserting and deleting links to neighboring nodes (Gall et al., 2014).

Overlay networks are for example necessary as some algorithms are designed for a certain network topology like a ring or a (spanning) tree. These algorithms may not work properly if the desired network topology is not provided by the system. An overlay network that corresponds to the desired topology will cause the algorithm to behave in the specified way in such a system.

According to Gall et al. (2014) a topological self-stabilizing mechanism guarantees that:

**Definition 2: Topological Self-Stabilization**

1. *Convergence*: By local neighborhood changes (i. e., by creating, forwarding, and deleting links with neighboring nodes), the nodes will eventually form an overlay topology with desirable properties from any initial (and in our case: connected) topology.
2. *Closure*: The system will also stay in a correct configuration provided that no external topological changes occur.

#### Linearization

A desired network topology could be for example a chain. Given a fixed set of nodes $V$ with unique identifiers (ids) and a total order $\leq$ on $V$ the goal is to build an ordered list of the nodes according to their unique ids. Therefore, a topology is desired where only consecutive nodes are connected (whereby $succ(v)$ defines the next to $v$).

**Definition 3: Linear/Chain Graph $G_L$**

Given a set of nodes $V$, the (undirected) *linear/chain graph* $G_L$ is defined as $G_L = (V, E_L)$ such that $\{u, v\} \in E_L$ iff $u = succ(v) \vee v = succ(u)$

Since the successor of any node is (if existent) uniquely defined, the linear graph is also uniquely defined for a given node set $V$. According to Gall et al. (2014) a linearization algorithm is therefore defined as follows:

**Definition 4: Linearization**

A *linearization algorithm* is a distributed self-stabilizing algorithm where

1. an *initial configuration* forms any (undirected) connected graph $G_0 = (V, E_0)$,
2. the only *legal configuration* is the linear topology $G_L = (V, E_L)$ on the nodes $V$, and
3. actions only update the neighborhoods of the nodes.

## 1.5 Name-passing Process Calculi

Name-passing process calculi, like the $\pi$-calculus, are a well-known and widely used method to model concurrent and distributed algorithms.

The $\pi$-calculus according to Milner et al. (1992) is designed to naturally express processes with a changing link infrastructure, as the communication between processes may carry information that can be used to change the linkage between the processes. According to Sangiorgi and Walker (2003) the simplest entities of the $\pi$-calculus are names, which can be thought of as names of communication links. Processes use names to interact, and pass names to one another by mentioning them in interactions. Therefore, all interaction between processes is modeled as message-passing. A process that receives a name, can use it and once again send it in further interactions to another process.

The following definitions characterize basic parts of a process calculus and are close to the definitions in Peters et al. (2013).

**Definition 5: Names**

Let $\mathcal{N}$ be the countably-infinite set of *names*. We use lower case letter $a, b, c, \dots, a_1, \dots$ to range over names. Additionally, let $\tau \notin \mathcal{N}$ and $\overline{\mathcal{N}} = \left\{\overline{a} | a \in \mathcal{N}\right\}$ be the set of *co-names*.

**Definition 6: Process Calculus**

A *process calculus* is a language $\mathcal{L} = \langle \mathcal{P}, \longmapsto \rangle$ with a set of process terms $\mathcal{P}$ (the syntax) and a step relation $\longmapsto : \mathcal{P} \times \mathcal{P}$ (the semantics). The syntax is usually defined by a context-free grammar defining operators. Operators are functions $op : \mathcal{N}^n \times \mathcal{P}^m \to \mathcal{P}$. We call an operator of arity 0 i. e., $m = 0$, a constant.

**Definition 7: Subterms**

Let $\mathscr{L} = \langle \mathscr{P}, \longmapsto \rangle$ be a process calculus and $P \in \mathscr{P}$ a process term. The set of *subterms* of $P = op(x_1, \ldots, x_n, P_1, \ldots, P_m)$ is defined as $\{P\} \cup \{P' | P'$ is subterm of $P_i \wedge 1 \leq i \leq m\}$. Note that every process term is its own *subterm*.

In the following, we will give an intuitive approach in understanding the behavior, i. e., the semantics, of processes respectively process terms in name-passing process calculi. The most important constant and basic process is the *empty process*, usually denoted by 0. This process does not show any behavior and cannot interact with its environment, therefore it can be seen as a deadlocked process. Otherwise processes are capable to perform actions of the forms: output, input, and internal. The (a)synchronous sending of a message with name $y$ over channel $x$ (usually written $\overline{x}\langle y \rangle$) and receiving an arbitrary name over channel $x$ (usually written $x(a)$) allow processes to interact with their environment. Internal actions (usually written $\tau$) are typically caused by interaction, further details are unobservable for the environment. The *input* operator $x(a)$ binds the variable $a$ which is substituted through a corresponding *output* operator $\overline{x}\langle y \rangle$ by the received name $y$ during interaction. The most common and important operator to combine processes $P, Q$ is the *parallel composition* (usually written $P \mid Q$). In parallel composition processes can execute actions independently of each other or interact with each other. Another possibility to bind a name is through a *restriction* operator (usually written $(\nu x) P$). In $(\nu x) P$ the scope of $x$ is restricted to $P$. Whereby a *scope* defines an area in which a particular name is known and can be used. Therefore the components of $P$ can use $x$ to interact with one another but not with other processes outside the scope. However, the sending of $x$ via some other name extends the scope of $x$ by the receiving process. Furthermore, there is typically a way to introduce repetitive behavior, either by a *replication* operator (usually written $!P$) or through *recursion* by process equations.

According to Peters et al. (2013) in process calculi intuitively, the *degree of distributability* corresponds to the number of parallel components that can act independently. Practical experience has shown that it is not possible to implement every $\pi$-calculus term, not even every asynchronous one, in an asynchronous setting while preserving its degree of distributability. One of the limiting factors for how suited a calculus is to model an asynchronous system, are synchronization patterns, especially the so-called synchronization pattern *M*. The synchronization pattern *M* is minimal in the sense that smaller patterns do not limit the distributability in the same way (van Glabbeek et al., 2008). Peters et al. (2013) proved that this pattern is present in the asynchronous $\pi$-calculus. One calculus that does not contain this pattern is according to Brodmann (2014) the localized $\pi$-calculus.

### 1.5.1 Localized Pi-Calculus

The localized $\pi$-calculus is a subcalculus of the asynchronous $\pi$-calculus that restricts how received names can further be used by processes. Intuitively, a process cannot receive messages over a previously received name, but send messages on the corresponding channel and communicate the name to other processes.

The grammar of the *localized $\pi$-calculus* according to Merro and Sangiorgi (1998) has operators of inaction, input prefix, asynchronous output, parallel composition, restriction, and replicated input:

**Definition 8: Syntax of the Localized $\pi$-Calculus**

$$\mathscr{P}_{\mathrm{L}} : P ::= 0 \quad | \quad P \,|\, P \quad | \quad a(x).P \quad | \quad \overline{a}\langle b \rangle \quad | \quad (\nu a)\, P \quad | \quad !a(x).P$$

where in $a(x).P$ name $x$ may not occur free in $P$ in input position.

Figure 1.1: Syntax of the localized $\pi$-calculus

The meaning and consequences of the restriction are described in Sangiorgi and Walker (2003) as follows. In the localized $\pi$-calculus, when a process receives a name, it can only use it for sending or send it to another process, which must itself respect the output-capability constraint. The output-capability constraint arises frequently in applications. For instance, a process is an operating system that is responsible for managing a printer can communicate to another process the capability to send a job to the printer, but not the capability to receive a job intended for the printer. A consequence of the output-capability constraint is that in a process $(\nu x)\, P$ every possible input via $x$ is visible in the syntax of $P$: no input can be created, either inside or outside $P$. In other words, every process that receives via a name is local to the process that created that name. This *locality* property of names gives the localized $\pi$-calculus its name.

# 2 Algorithm

In this chapter, we introduce two variants of a simple linearization algorithm from Gall et al. (2014) designed for a shared memory system. We first describe the characteristics of the system and the model used. Afterwards, we briefly present the algorithm and subsequently illustrate the proof idea to show that the algorithm is a linearization algorithm according to Definition 4.

Gall et al. (2014) claim that the motivation for topological self-stabilization is, to build robust open distributed systems, like peer-to-peer systems. However, a shared memory model is not the most suitable one for this task. Therefore, we redesign the algorithm from Gall et al. (2014) for a system with asynchronous message-passing. We describe an extended version of the localized $\pi$-calculus that is the foundation for our redesign. Thereafter we introduce our model and finally the redesigned algorithm.

## 2.1 Algorithm for shared memory

### 2.1.1 Model for shared memory

First let us introduce the model as presented in Gall et al. (2014) to enable a clear understanding of the original algorithm.

A system consists of a fixed set $V$ of $n$ nodes, whereby each node has a unique integer identifier and nodes are compared by their corresponding id. Each pair of nodes $(u, v)$ shares a boolean variable $e(u, v)$ which specifies an undirected adjacency relation i. e., $u$ and $v$ are neighbors if and only if this shared variable is true i. e., $e(u, v) = 1$. The set of neighbor relations defines an undirected graph, the topology $G = (V, E)$. There is a link between $u$ and $v$ i. e., $\{u, v\} \in E$ if and only if the corresponding shared variable is true, i. e., $e(u, v) = 1$.

A distributed algorithm is executed by each node in the network concurrently. The algorithm for each node is expressed by a set of variables and actions. An action has the form

`< name >  :  < guard >  →  < commands >`

where `< name >` is an action label, `< guard >` is a boolean predicate over the (local and shared) variables of the executing node and `< commands >` is a sequence of commands that may involve any local variables and shared variables of the node itself or its neighbors.

An action is enabled if and only if its guard is true. Every enabled action is passed to a scheduler. An action that is chosen and executed by the scheduler is called a step. The assignments of all local and shared variables define the global state of the system, i. e., a configuration. Each configuration represents a graph since only variables are considered that directly effect the topology. An execution is a sequence of configurations, such that each configuration $c_i + 1$ is the topology resulting from executing step $i$, selected by the scheduler, on the previous configuration $c_i$. A scheduler is treated as a global entity that decides which (independent) actions are executed in parallel to explore the complexity of the algorithms.

### 2.1.2 Linearization Algorithm

Gall et al. (2014) introduce two variants of a linearization algorithm, named $LIN_{all}$ and $LIN_{max}$. Both variants are based on the idea that whenever a node $u$ has two neighbors $v$ and $w$, both of which have a smaller (or both a greater) id, the node can establish a link between them and delete its link to the smaller (respectively greater) one $w$. Therefore there are just two simple linearization rules, left linearization and right linearization. The effects of linearization steps on the topology are depicted in Figure 2.1.

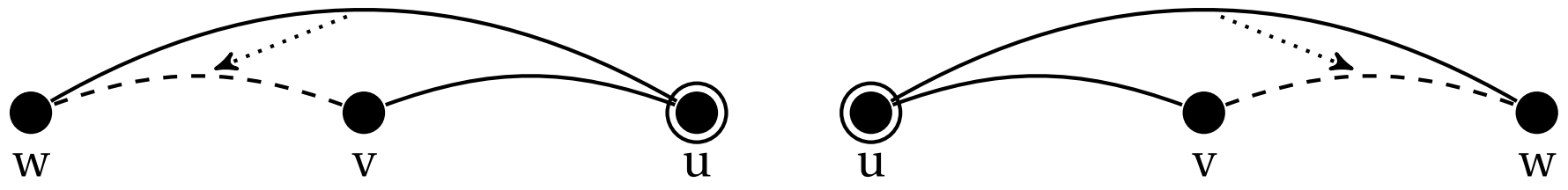

(a) Left Linearization

(b) Right Linearization

Figure 2.1: Linearization steps

The variants of the algorithm only differ in which linearization steps are enabled and therefore proposed to the scheduler. In $LIN_{all}$, every node enables all possible linearization steps on each side. Therefore according to Gall et al. (2014):

**Definition 9: *$LIN_{all}$***

For every node $u$, there are the following rules for every pair of neighbors $v$ and $w$:

**linearize left(v,w)** : $(v, w \in u.L \wedge w < v < u) \rightarrow e(u,w) := 0,\ e(v,w) := 1$

**linearize right(v,w)** : $(v, w \in u.R \wedge u < v < w) \rightarrow e(u,w) := 0,\ e(v,w) := 1$

In $LIN_{max}$, every node $u$ only enables the linearization steps that acquaint the two furthest nodes on the corresponding side and therefore eliminates the longest edges to each side of $u$. Hence according to Gall et al. (2014):

**Definition 10: *LIN*$_{max}$**

Every node $u$ uses the following rules for every pair of neighbors $v$ and $w$:

$$\begin{aligned}
\textbf{linearize left(v,w)} &: (v, w \in u.L) \wedge (w < v < u) \wedge (\nexists x \in u.L \setminus \{w\} : x < v) \\
&\rightarrow e(u,w) := 0,\ e(v,w) := 1 \\
\textbf{linearize right(v,w)} &: (v, w \in u.R) \wedge (u < v < w) \wedge (\nexists x \in u.L \setminus \{w\} : x > v) \\
&\rightarrow e(u,w) := 0,\ e(v,w) := 1
\end{aligned}$$

Whereby $u.L$ denotes the set of all left neighbors of $u$ i. e., all neighbored nodes with a smaller id, and respectively $u.R$ denotes all right neighbors of $u$ i. e., all neighbors with a greater id.

### 2.1.3 Proof Idea

To prove that the algorithm is a linearization algorithm according to Definition 4, it has to be shown with Definition 2 that the closure and convergence properties are fulfilled. In the following, we will shortly picture the idea of the proofs and the main arguments from Gall et al. (2014).

In order to prove the closure property, Gall et al. (2014) show that there is a unique legal configuration. Further they show that any enabled linearization step, regardless of whether left or right linearization, is contradictory to this legal configuration and therefore all actions are disabled. Thus all neighborhoods and, by association, the linearized topology does not change. Accordingly, the system remains in a correct configuration in case no fault occurs.

To prove convergence Gall et al. (2014) introduce a potential function that sums up the length of all edges in the current topology. It holds that the only connected topology with minimal potential is the unique correct configuration. Further, they show that an initial connected topology always stays connected and that in every connected topology, that is not the legal configuration, there is an enabled left or right linearization step. Furthermore, it is shown that with every executed linearization step, the potential function decreases as the new established edge is always strictly shorter than the removed edge. Therefore, the topology converges to the correct configuration in a finite number of steps.

Gall et al. (2014) state in their model that a shared variable $e(u,v)$ can only be changed by $u$ and $v$. However, in the algorithms the shared variables are written not only by the nodes themselves but also by their neighbors. Therefore, the algorithm only works in a setting where all nodes have write access to the whole memory, as the algorithm could be initialized with any topology and hence also with the fully meshed graph. Since this is a very restrictive model, especially in the context of peer-to-peer networks, we are interested in redesigning the algorithm for a setting with asynchronous message-passing.

## 2.2 Model for Asynchronous Message-Passing

In this section, we describe our model for a linearization algorithm in an asynchronous message-passing system. Furthermore, to lower the system assumptions as much as possible, we assume that the processes can only communicate via unidirectional channels. Similar to Gall et al. (2014), we assume that there are $n$ processes in the system and every process has a unique id. To be as general as possible, we only assume that there is a total order on these ids without further specifying the type. Furthermore, we assume that every value in the system can be interpreted as the id of an existing process. Therefore the corruption of RAM and messages is only considered in the way, that the result is an arbitrary value that is still interpretable as an existing id. This could be implemented in a real system through a type check or similar mechanisms. If this is not possible, but illegal values can be detected, we can simply discard such values.

**Assumption 1: Ids** Every process has a unique constant id and every value in the system can be interpreted as the id of an existing process.

**Definition 11: Process identifiers $\mathscr{P}$**

Let $\mathscr{P}$ be the (non-empty) finite *set of unique identifiers* of the processes in the system. Let $\leq$ be a total order on $\mathscr{P}$. Let $|\mathscr{P}| = n \in \mathbb{N}$ then there exists an index function (bijection) $i : \mathscr{P} \to \{1, \ldots, n\}$ and $\forall p \in \mathscr{P}. i(p) = |\{q \in \mathscr{P} | q \leq p\}|$ i. e., $i(p)$ describes the position of $p$ with respect to $\leq$.

We now define the predecessor and the successor of a process as respectively the next smaller and next greater process according to the total order (and $\perp$ if there is none). We call such a pair of processes consecutive.

**Definition 12: Predecessor *pred* and Successor *succ***

The *predecessor* and *successor* of a process $p \in \mathscr{P}$ are then defined as $pred, succ : \mathscr{P} \to (\mathscr{P} \cup \perp)$ with

$$pred(p) = \begin{cases} \perp, & \text{if } i(p) = 1 \\ i^{-1}(m), & \text{if } i(p) = m+1 \end{cases} \quad \text{and} \quad succ(p) = \begin{cases} i^{-1}(m), & \text{if } i(p) = m-1 \\ \perp, & \text{if } i(p) = n \end{cases}$$

**Corollary 1: Predecessor *pred* and Successor *succ***

$$\begin{aligned} &\text{Hence} \quad \forall p \in \mathscr{P}. succ(p) \neq \perp \implies (succ(p) > p \land \forall q \in \mathscr{P}. q > p \implies q \geq succ(p)) \\ &\text{and} \quad \forall p \in \mathscr{P}. pred(p) \neq \perp \implies (pred(p) < p \land \forall q \in \mathscr{P}. q < p \implies q \leq pred(p)) \end{aligned}$$

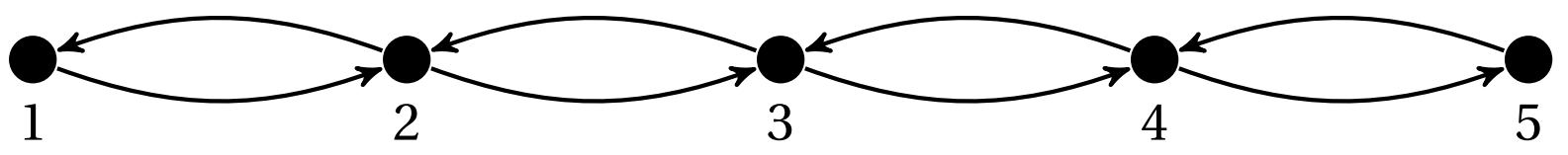

Figure 2.2: Desired network topology, whereby the nodes are ordered according to their id.

In this setting, the desired topology i. e., the overlay network that the processes (represented as nodes) shall establish, is an ordered doubly-linked list according to the total order on the ids (depicted in 2.2 for five nodes and labeled with the position of their id according to the total order). This means that every process has an unidirectional link (represented as edge) to its predecessor (with exception of the smallest process) and its successor (with exception of the greatest) and there are no other links. We call this topology the (directed) linear graph. The undirected linear graph is the undirected variant of this desired topology. Therefore for every pair of consecutive processes at least one of them has a link to the other one.

**Definition 13: Undirected Variant of Graph *undirected***

Let $G = (V, E)$ be a directed graph, then the *undirected variant* $undirected(G) = (V', E')$ is defined as:

$$V = V' \quad \text{and} \quad E' = \{\{u, v\} | (u, v) \in E\}$$

**Definition 14: Desired Topology Graph**

The *(directed) linear graph* $G_{LIN} = (V, E)$ is defined as

$$V = \mathscr{P} \quad \text{and} \quad E = \{(p, q) | p, q \in V \land (p = succ(q) \lor q = succ(p))\}$$

The *undirected linear graph* $UG_{LIN} = (V, E)$ is defined as $UG_{LIN} = undirected(G_{LIN})$ i. e.,

$$V = \mathscr{P} \quad \text{and} \quad E = \{\{p, q\} | p, q \in V \land (p = succ(q) \lor q = succ(p))\}$$

The distance between two processes is the number of nodes between the processes according to the total order. Therefore, a process has a distance of zero to itself and the distance between a pair of consecutive processes is one.

**Definition 15: Distance *dist***

The *distance* between two nodes $p, q \in \mathscr{P}$ is defined as $dist : \mathscr{P} \times \mathscr{P} \to \mathbb{N}$ with

$$dist(p, q) = \begin{cases} |\{r \in \mathscr{P} | p < r \leq q\}|, & \text{if } p < q \\ |\{r \in \mathscr{P} | q < r \leq p\}|, & \text{otherwise} \end{cases}$$

The length of an edge (regardless whether directed or undirected) is defined as the distance of the nodes it is connecting.

**Definition 16: Length of Edges**

Let $G = (\mathscr{P}, E)$ be an arbitrary undirected graph and $G' = (\mathscr{P}, E')$ be an arbitrary directed graph. The *length of an edge* $e = \{p,q\} \in E$ respectively $e = (p,q) \in E'$ is defined as

$$len(e) = dist(\{p,q\})$$

The maximum distance between a pair of processes in the system is the distance of the smallest and the greatest process. This is obviously with the Definitions 15 and 11 $n-1$. This is also the length of the longest possible edge, as the length of edges is defined as the distance between its end-points. We call this maximum distance *maxdist*.

**Definition 17: Maximum Distance**

The maximum distance for the set of process identifiers and therefore the length of the longest possible edge in the network topology is:

$$maxdist = dist(min(\mathscr{P}), max(\mathscr{P})) = n-1$$

### 2.2.1 Extended Localized Pi-Calculus

To model the algorithm, we introduce an extension of the name-passing localized $\pi$-calculus. This extension is based on ideas similar to Wagner and Nestmann (2014). We call the calculus the extended localized $\pi$-calculus and denote it with $\text{eL}\pi = \langle \mathscr{P}_{\text{eL}}, \longmapsto \rangle$. This allows us to define a kind of standard form for a configuration of our algorithm. The local state of all processes and the messages in transit, and therefore the global state of the system, is directly accessible via the parameters of the corresponding process definition. This in turn allows state-based proofs, which is more traditional for distributed algorithms (Lynch, 1996), instead of the action-based style of process calculi.

**Notation: Multisets** Let $a, b, c \in M$ be arbitrary elements of an arbitrary set M. We denote with $Ms = \{\!|a, a, b, c|\!\}$ a multiset and use $\mathbb{N}^M$ as the type of such a set (similar to $2^M$ as notation for the power set). Furthermore, the union $\cup$ of two multisets is the multiset where all appearances of elements in both are added i. e., for example $\{\!|a,a,b,c|\!\} \cup \{\!|a,d|\!\} = \{\!|a,a,a,b,c,d|\!\}$ and the difference $\setminus$ is the multiset where all appearances of elements in the first multiset are decreased by those in the second (but at least zero) i. e., for example $\{\!|a,a,b,c,e,e|\!\} \setminus \{\!|a,d,e,e|\!\} = \{\!|a,b,c|\!\}$. Since sets are only special cases with multiplicity one for all elements, we also use combinations of sets and multisets.

We assume the existence of a countably infinite set **A** containing all channel names, function names, and variables.

**Definition 18: Syntax of the extended Localized $\pi$-Calculus- $\mathscr{P}_{\mathrm{eL}}$**

| | |
|---|---|
| DATA VALUES **V** | $v ::= \bot \mid 0 \mid 1 \mid c \mid (v,v) \mid \{v,\ldots,v\} \mid \{\!|v,\ldots,v|\!\}$, with $c \in \mathbf{A}$ |
| VARIABLE PATTERN | $X ::= x \mid (X,X)$, with $x \in \mathbf{A}$ |
| EXPRESSIONS | $e ::= v \mid X \mid (e,e) \mid f(e)$, with $f \in \mathbf{A}$ |
| PROCESSES **P** | $P ::= 0 \mid P \mid P \mid c(X).P \mid \overline{c}\langle v\rangle \mid (\nu c)P \mid$ **if** $e$ **then** $P$ **else** $P \mid$ **let** $X = e$ **in** $P \mid K(e)$ |
| PROCESS EQUATIONS | $D = \{K_j(X) = P_j\}_{j \in J}$ a finite set of process definitions |

where in $c(X).P$ variable $x$ as part of $X$ may not occur free in $P$ in input position.

Figure 2.3: Syntax of the extended localized $\pi$-calculus

$K(X)$ denotes a parameterized process constant, which is defined with respect to a finite set of process equations $D$ of the form $\{K_j(X) = P_j\}_{j \in J}$. Since we use parameterized process constants, we exclude replication and use instead recursion via process definitions to model repetitive behavior.

Names received as an input and restricted names are *bound names*. The remaining names are *free names*. Accordingly, we assume three sets, the sets of names $\mathsf{n}(P)$ and its subsets of free names $\mathsf{fn}(P)$ and bound names $\mathsf{bn}(P)$, with each term $P$. To avoid name capture or clashes, i. e., to avoid confusion between free and bound names or different bound names, bound names can be mapped to fresh names by *$\alpha$-conversion*. We write $P \equiv_\alpha Q$ if $P$ and $Q$ differ only by $\alpha$-conversion.

The substitution of value $v$ for a variable pattern $X$ in expression $e$ or process $P$ is written $\{^v/_X\}e$ and $\{^v/_X\}P$ respectively. Note that only data values can be substituted for names and that all variables of the pattern $X$ must be free in $P$ (while possibly applying $\alpha$-conversion to avoid capture or name clashes).

Let $[\![e]\!]$ denote the evaluation of expression $e$ which allows results in a data value, defined in the standard way.

**Definition 19: Structural Congruence for the Localized $\pi$-Calculus**

Now we present the structural congruence for the extended localized $\pi$-calculus. The definition is based on the structural congruence for the $\pi$-Calculus.

$$P \equiv Q \text{ if } P \equiv_\alpha Q \qquad P \mid 0 \equiv P \qquad P \mid Q \equiv Q \mid P \qquad P \mid (Q \mid R) \equiv (P \mid Q) \mid R \qquad (\nu n)0 \equiv 0$$
$$P \mid (\nu n)Q \equiv (\nu n)(P \mid Q), \text{if } n \notin \mathsf{fn}(P) \qquad (\nu n)(\nu m)P \equiv (\nu m)(\nu n)P$$
$$\textbf{if } e \textbf{ then } P \textbf{ else } Q \equiv P, \text{ if } [\![e]\!] = 1 \qquad \textbf{if } e \textbf{ then } P \textbf{ else } Q \equiv Q \text{ if } [\![e]\!] = 0$$
$$\textbf{let } X = e \textbf{ in } P \equiv \{^{[\![e]\!]}/_X\}P \qquad K(e) \equiv \{^{[\![e]\!]}/_X\}P \text{ if } (K(X) = P) \in D$$

We are only interested in the interaction between the processes and not with any further environment. Therefore, we only present a reduction semantics for our extended localized $\pi$-calculus, based on the reduction semantics of the $\pi$-calculus.

**Definition 20: Reduction Semantics of the extended Localized $\pi$-Calculus- $\longmapsto$**

$$\texttt{comm:}\ \frac{}{c(X).P \mid \overline{c}\langle v\rangle \longmapsto \{^{v}/_{X}\}P} \qquad \texttt{par:}\ \frac{P \longmapsto P'}{P \mid Q \longmapsto P' \mid Q}$$

$$\texttt{res:}\ \frac{P \longmapsto P'}{(\nu c)\, P \longmapsto (\nu c)\, P'} \qquad \texttt{struct:}\ \frac{P \equiv Q \qquad Q \longmapsto Q' \qquad Q \equiv Q'}{P \longmapsto P'}$$

Figure 2.4: Semantic of the extended localized $\pi$-calculus

**Definition 21: Steps**

We call a single application of this reduction semantics a step. We write $P \longmapsto P'$ and we call $P'$ a derivative of $P$ and denote $P \longmapsto$ if there is such a $P'$ or $P \not\longmapsto$ if not. We write $\Longmapsto$ for the reflexive and transitive closure of $\longmapsto$. We use *execution* to refer to a reduction starting from a particular term.

Every structural extension like function calls, if-then-else-statements, and let-in-statements are evaluated in the structural congruence. Therefore, the evaluation of these constructs is not considered a step on its own. Hence, internal computations are executed as parts of other steps.

As the state after every fault can be seen as a new initial state we assume that there are no faults. An infinite message delay can be seen as message loss. Therefore, every message that is sent is received after finite time. Since the message-passing model is asynchronous, there are no further assumptions regarding the delivering time of messages.

**Assumption 2: No Message Loss** Every message is received after a finite but arbitrary number of steps.

Furthermore, we need an assumption of fairness, as otherwise nodes could starve. A process starves if it never executes a step. Furthermore, a subprocess of a node starves, for example, if the process is only consuming messages, but never tries to find a linearization steps itself. Therefore, without a fairness assumption, it is not possible to show any progress in the system.

**Assumption 3: Fairness** Every continuously enabled subprocess will eventually (after an arbitrary but finite number of steps) execute a step.

The formal definitions are given in Section 2.3.

## 2.3 Linearization Algorithm for Asynchronous Message-Passing

The utilization of a calculus enables us to model the algorithm unambiguously and allows us to formally prove properties of the algorithm. Although in the calculus itself all channels are bidirectional, we only use them as previously stated in an unidirectional manner to lower the system requirements as much as possible. Together with the output-capability restriction of the localized $\pi$-calculus this helps us to ensure that every process can be implemented in an asynchronous setting on a different location. Therefore, the processes really are distributable.

In our algorithm, each node can receive messages from other nodes via a channel with the same name as its id. One could think of the ids as serving as the IP address of the corresponding process. To enable a neighbor to communicate with another process it is sufficient to send it the corresponding id. Therefore, the set **A** contains the ids of all processes in the system i. e., $\mathscr{P} \subset \mathbf{A}$. This simplifies the algorithm and the proof in the sense, that we do not need an additional mapping from the ids to a corresponding channel name of the process and vice versa.

We use function calls, if-then-else-statements, and let-in-statements to model internal or local computations. Since the evaluation of function calls, if-then-else-statements, and let-in-statements are part of the structural congruence, internal computations are part of other steps. A process can receive a message in a step, add a previously received process id to its neighborhood, or perform some local computations and send messages to other processes.

Since in process calculi any kind of behavior is represented by communication, variables must also be modeled through messages. To model local variables we use restricted channels for every process. Each variable is therefore represented through a message in transit that can only be received by the corresponding process. The value of the variable is the value of this message whereby receiving the message corresponds to reading the variable and sending corresponds to writing. In our algorithm, every process $p$ has one local variable $nb_p$, describing the neighborhood of the process i. e., it contains all processes that $p$ knows and therefore can send messages to.

Each process can be in one of two local states. In the state $Alg(p, nb)$ the process $p$ is able to receive a message from another process in the system. In the state $Alg'(p, nb, x)$ the process $p$ can add a previously received process id $x$ to its current neighborhood $nb$. Thereby, a process blocks the reception of messages until the the previously received process id is added to the neighborhood. In both states, the process can additionally try to find a linearization step in its current neighborhood $nb$ based on internal computations. If it finds no possible linearization step in its neighborhood, it sends *keep-alive*-messages to its current neighbors. This can only happen if it knows at most one smaller and one greater process. These *keep-alive*-messages are necessary to prevent deadlocks and also to reach the desired network topology in case only edges are missing in the current topology to form the desired topology.

$$
\begin{aligned}
Alg(p, initNb) = (\nu nb_p)\Big(\ & \overline{nb_p}\langle initNb\rangle \mid \\
& Alg_{rec}(p) \mid \\
& Alg_{match}(p)\Big)
\end{aligned}
$$

$$
\begin{aligned}
Alg'(p, initNb, x) = (\nu nb_p)\Big(\ & \overline{nb_p}\langle initNb\rangle \mid \\
& Alg_{add}(p, x) \mid \\
& Alg_{match}(p)\Big)
\end{aligned}
$$

$$
Alg_{rec}(p) = p(x).Alg_{add}(p, x)
$$

$$
Alg_{add}(p, x) = nb_p(y).\Big(\overline{nb_p}\langle y \cup \{x\}\rangle \mid Alg_{rec}(p)\Big)
$$

$$
\begin{aligned}
Alg_{match}(p) = nb_p(y).(&\textbf{let } x = select(findLin(p, y)) \textbf{ in} \\
&\quad \textbf{if } x = \perp \textbf{ then} \\
&\quad\quad \prod_{j \in y} \overline{j}\langle p\rangle \mid \overline{nb_p}\langle y\rangle \\
&\quad \textbf{else} \\
&\quad\quad \textbf{if } x = (j, k) \textbf{ then} \\
&\quad\quad\quad \textbf{if } j < k \land k < p \textbf{ then} \\
&\quad\quad\quad\quad \overline{j}\langle k\rangle \mid \overline{nb_p}\langle y \setminus \{j\}\rangle \\
&\quad\quad\quad \textbf{else} \\
&\quad\quad\quad\quad \textbf{if } j < k \land p < j \textbf{ then} \\
&\quad\quad\quad\quad\quad \overline{k}\langle j\rangle \mid \overline{nb_p}\langle y \setminus \{k\}\rangle \\
&\quad\quad\quad\quad \textbf{else} \\
&\quad\quad\quad\quad\quad \overline{nb_p}\langle y\rangle \\
&\quad\quad \textbf{else} \\
&\quad\quad\quad \overline{nb_p}\langle y\rangle \\
&\mid Alg_{match}(p))
\end{aligned}
$$

Figure 2.5: Algorithm of one process

In the algorithm, $LeftN\colon \mathscr{P} \times 2^{\mathscr{P}} \to 2^{\mathscr{P}}$ calculates the *left neighborhood* of a process i. e., all neighbors with a smaller id and corresponding $RightN\colon \mathscr{P} \times 2^{\mathscr{P}} \to 2^{\mathscr{P}}$ the *right neighborhood* of a process i. e., all neighbors with a greater id. They are defined as follows:

$$LeftN(p,y) = \{q \in \mathscr{P} \mid q \in y \wedge q < p\}$$
$$RightN(p,y) = \{q \in \mathscr{P} \mid q \in y \wedge q > p\}$$

The function $findLin : \mathscr{P} \times 2^{\mathscr{P}\times\mathscr{P}} \to 2^{\mathscr{P}\times\mathscr{P}}$ calculates all possible linearization steps in the neighborhood of a process, based on the input sets.

$$findLin(p,y) = \{(q,r) \mid q,r \in y \wedge q < r \wedge (q,r \in LeftN(p,y) \vee RightN(p,y))\}$$

The function $select\colon 2^{\mathscr{P}\times\mathscr{P}} \to (\mathscr{P}\times\mathscr{P})$ returns one of these linearization steps

$$select(y) = \begin{cases} \bot & \text{if } y = \emptyset \\ \varepsilon x.x \in y & \text{if } y \neq \emptyset \end{cases}$$

whereby we denote with $\varepsilon$ Hilbert's choice operator returning an arbitrary element. Therefore, this corresponds to the $LIN_{all}$ variant of the algorithm in Gall et al. (2014). With another selection it would equally possible to implement $LIN_{max}$.

**Definition 22: Subprocesses**

For every process $p \in \mathscr{P}$ we denote $Alg_{match}(p)$, $Alg_{rec}(p)$, and $Alg_{add}(p,\cdot)$ as *subprocesses* of $p$. Note that all subprocess are input guarded i. e., they are all waiting to receive a message.

The subprocess $Alg_{rec}(p)$ models the behavior that a process is able to receive a message from another process in the system. When it receives such a message with content $x$, it continues as subprocess $Alg_{add}(p,x)$ and the process itself changes therefore its local state. $Alg_{add}(p,x)$ reads the current value of the neighborhood of $p$ and adds the received process id $x$ to its neighborhood. Afterwards, the process is again able to receive a new message from another process.

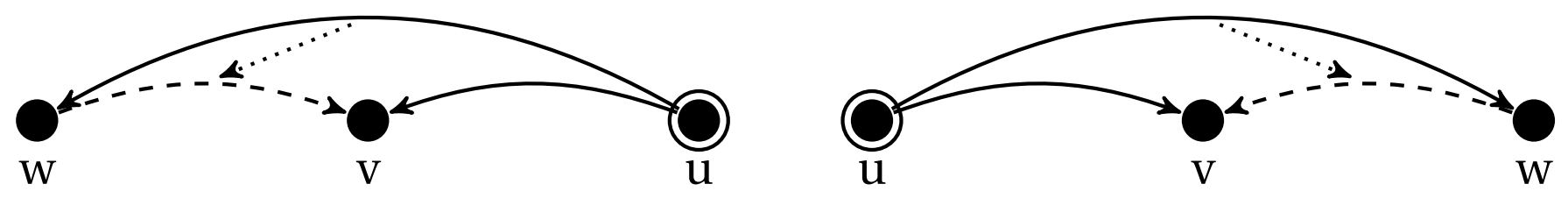

(a) Directed Left Linearization (b) Directed Right Linearization

Figure 2.6: Directed linearization steps

The subprocess $Alg_{match}(p)$ defines the behavior, based on the internal computations of the function $findLin(p,nb)$, in case process $p$ tries to find a linearization step in its neighborhood $nb$. If *select* returns a left linearization step the process behaves according the second case, in case of a right linearization step according to the third case. In both cases, $p$ sends the further away process the id of the other process and deletes it from its neighborhood (as depicted in Figure 2.6). If there is no possible linearization step, *select* returns $\bot$ and the process sends *keep-alive*-messages to its current neighbors. The other cases are only implemented to obtain a complete case distinction but are ruled out by definitions. In all cases, the process is directly able to try to find another linearization step.

So far, we have just defined a single process. As the system is composed of $n$ such processes, we define now the global state of such a system. The global states that serve as starting points for the executions of our algorithm are called initial configurations. Later we show that every global state can serve as such a starting point.

**Definition 23: Initial Configuration**

Let

$\mathscr{P}$ be the *set of unique identifiers* of the processes in the system,

$P, P' \subseteq \mathscr{P}$ with $P \cup P' = \mathscr{P}$ and $P \cap P' = \emptyset$,

$init : \mathscr{P} \to 2^{\mathscr{P}}$ a function that defines for every process $p \in \mathscr{P}$ the neighborhood i. e., which process ids are known by $p$,

$Msgs \in \mathbb{N}^{\mathscr{P} \times \mathscr{P}}$ a multiset that describes the messages in transit and

$add : \mathscr{P} \rightharpoonup \mathscr{P}$ a partial function with $\forall p \in P'.\exists q \in \mathscr{P}.(p,q) \in add$ and $\forall p \in P.\forall q \in \mathscr{P}.(p,q) \notin add$ that describes the adding in progress i. e., where $add(p) = q$ describes that $p$ wants to add $q$ to its neighborhood.

Then an *initial configuration* of the algorithm is defined as the process term:

$$Alg_{ges}\big(P, P', init, Msgs, add\big) = \prod_{j \in P} Alg\big(j, init(j)\big) \mid \prod_{j \in P'} Alg'\big(j, init(j), add(j)\big) \mid \prod_{(j,k) \in Msgs} \overline{j}\langle k \rangle$$

The assumption 1 that only valid values are in the system, i. e., every value can be interpreted as the id of some existing node, is reflected in the types of the parameters. Furthermore, we always assume for every message $(x, y) \in Msgs$ that $x \neq y$ and for every $p \in \mathscr{P}$ that $p \notin nb(p)$ and $add(p) \neq p$. This is not a strong restriction as every process knows its own id and could easily hold its set of neighbors free from itself, and no process ever sends a message on a channel that carries the channel name itself. But active handling would only lead to more cases in the proofs and therefore increase the complexity, without any information gain.

In an initial configuration there is for every process $p \in \mathscr{P}$ exactly one $\overline{nb_p}\langle\cdot\rangle$-message. This message can not be lost or duplicated through a previous fault, since as stated they only serve to model a variable. Since the program code can not get corrupted, a fault can lead to an arbitrary value of a variable but not to its disappearance or duplication. Thus, the value of this message can be an arbitrary set of $\mathscr{P}$ (without $p$ itself), reflecting the arbitrary neighborhood of $p$ as initial state. Therefore, these messages of all processes describe the initial network topology. We will formally define variants of the network topology of a configuration in Chapter 3.2. They differ in whether or not messages still in transit are also taken into account, i. e., whether a message in transit and adding in progress is represented as an edge or not. A configuration describes the global state of the system, consisting of the local state of all processes and the messages in transit.

**Definition 24: Configuration**

Let $\mathscr{I}$ be the set of all initial configurations. We call every process term $C$ that can be reached from any arbitrary initial configuration, i. e., $\exists I \in \mathscr{I}. I \Longmapsto C$ *configuration* (of the algorithm) or process term of the algorithm. We denote the set of all such configurations with $\mathscr{T}$.

**Definition 25: Reachability**

We call a configuration $C'$ *reachable* from a configuration $C$ iff $C \Longmapsto C'$. Further, we say a configuration with a predicate $P$ is *reached* from configuration $C$ iff in every execution there is a configuration $C'$ with $C \Longmapsto C'$ and $P$ holds for $C'$.

**Definition 26: Enabled Subprocess**

We call a subprocesses $S$ of process $p$, with $p \equiv Alg(p,\cdot)$ or $p \equiv Alg'(p,\cdot,\cdot)$, enabled in a configuration $C$ iff $p$ is a subterm of $C$ and there is a subterm $S'$ of $C$ with $S \equiv S'$ and if $S = Alg_{rec}(p)$ a message $m$ as subterm of $C$ else a message $m$ as subterm of $p$ with $S \mid m \longmapsto$.

First of all, we show in Chapter 3 that for every configuration $C$ there are such parameters, so that $C$ is structural equivalent to $Alg_{ges}(\cdot,\cdot,\cdot,\cdot,\cdot)$. We call $Alg_{ges}(\cdot,\cdot,\cdot,\cdot,\cdot)$ the standard form of a configuration and use them as representatives of all structurally equivalent configurations. Furthermore, it then holds that every configuration is structurally equivalent to an initial configuration. This reflects the fact that every configuration can serve as a starting point of our algorithm, as it is represented by a structurally equivalent initial configuration.

We define the topology of a configuration in different variants. Afterwards, we show that if the initial topology is weakly connected, while taking the messages that are still in transit into account, the topology always stays weakly connected.

Subsequently we define correct configurations as configurations with the desired topology and show the uniqueness of such configurations up to structural equivalence and the number of messages in transit.

The algorithm never terminates, as required for self-stabilizing algorithms. If the desired network topology is reached, every process (with exception of the smallest and greatest process) knows exactly one smaller (its predecessor) and one greater process (its successor). The processes continue to send *keep-alive*-messages to their desired neighbors. Since *keep-alive*-messages are the only messages that are send in a correct configuration, the topology does not change anymore and therefore closure holds.

*Keep-alive*-messages are among others necessary to prove convergence, i. e., that a correct configuration is reached, in case that the current topology only lacks edges compared to the desired topology. Without these message the system would deadlock, as no linearization steps are possible in such a configuration. However, unfortunately such message can for example also reestablish edges that where already deleted through linearization steps. This leads to some obstacles and makes proving convergence much more difficult and challenging than in the shared memory model. Therefore, we formally prove convergence for some special cases and weak convergence in the general case i. e., for every arbitrary initial configuration there exist executions that reach a correct configuration.

# 3 Proof

In this chapter, we present the formal proofs of properties of our algorithm. We start with some basic properties and introduce afterwards the topology of a configuration in different variants. A correct configuration is a configuration whose topology forms the desired topology i. e., in our case the (directed) linear graph. Subsequently, we show that whenever such a configuration is reached, the system stays in such a configuration and therefore closure holds. Afterwards, we prove convergence in the cases that the topology either only lacks desired edges or only contains too many edges i. e., every desired edge is already established in the topology. Furthermore, we show weak convergence for the general case.

## 3.1 Basic Properties

In this section, we prove some basic properties for the algorithm to simplify further proofs. First of all, we show that for every configuration $C$ there are parameters with the same properties as in Definition 23, so that $C$ is structurally equivalent to $Alg_{ges}(\cdot,\cdot,\cdot,\cdot,\cdot)$. We call $Alg_{ges}(\cdot,\cdot,\cdot,\cdot,\cdot)$ the standard form of a configuration and use it as representative of all structurally equivalent configurations. Therefore, every configuration is structurally equivalent to an initial configuration.

**Lemma 1: Standardform**

Starting from an arbitrary initial configuration according to Definition 23 the process term of the algorithm has always a form that is structurally equivalent to a term

$$Alg_{ges}\left(P, P', nb, Msgs, add\right)$$

whereby $P, P' \subseteq \mathscr{P}$ with $P \cup P' = \mathscr{P}$ and $P \cap P' = \emptyset$, $nb : \mathscr{P} \to 2^{\mathscr{P}}$ is a function that defines for every process $p \in \mathscr{P}$ the neighborhood i. e., which process ids are known by $p$, $Msgs \in \mathbb{N}^{\mathscr{P} \times \mathscr{P}}$ is the multiset of messages in transit, $add : \mathscr{P} \rightharpoonup \mathscr{P}$ is a partial function with $\forall p \in P'.\exists q \in \mathscr{P}.(p,q) \in add$ and $\forall p \in P.\forall q \in \mathscr{P}.(p,q) \notin add$ i. e., where $add(p) = q$ describes that $p$ is adding $q$ to its neighborhood.

*Proof*:
We prove this by induction over the step semantics. The claim holds for any initial configuration according to Definition 23. Let $Alg_{ges}(P, P', nb, Msgs, add)$ be an arbitrary configuration of the algorithm that is reachable from an initial configuration. Then according to the semantic as defined in Definition 20 there are only the following kinds of steps possible:

First, a process tries to find a linearization step:

Let $p \in P$ be an arbitrary process then

$$\begin{aligned}
& Alg_{ges}(P, P', nb, Msgs, add) \\
\equiv\quad & \prod_{j\in P} Alg(j, nb(j)) \mid \prod_{j\in P'} Alg'(j, nb(j), add(j)) \mid \prod_{(j,k)\in Msgs} \overline{j}\langle k\rangle \\
\equiv\quad & \prod_{j\in P\setminus\{p\}} Alg(j, nb(j)) \mid \prod_{j\in P'} Alg'(j, nb(j), add(j)) \mid \prod_{(j,k)\in Msgs} \overline{j}\langle k\rangle \mid Alg(p, nb(p)) \\
\equiv\quad & \prod_{j\in P\setminus\{p\}} Alg(j, nb(j)) \mid \prod_{j\in P'} Alg'(j, nb(j), add(j)) \mid \prod_{(j,k)\in Msgs} \overline{j}\langle k\rangle \mid \\
& (\nu nb_p)\Big(\overline{nb_p}\langle nb(p)\rangle \mid Alg_{rec}(p) \mid Alg_{match}(p)\Big)
\end{aligned}$$

Now there are four possible cases dependent on $select(findLin(p, nb(p)))$:

- if there is no possible linearization step in the neighborhood of $p$ i. e., $select(findLin(p, nb(p))) = \bot$:

$$\begin{aligned}
\longmapsto\quad & \prod_{j\in P\setminus\{p\}} Alg(j, nb(j)) \mid \prod_{j\in P'} Alg'(j, nb(j), add(j)) \mid \prod_{(j,k)\in Msgs} \overline{j}\langle k\rangle \mid \\
& (\nu nb_p)\Big(Alg_{rec}(p) \mid \prod_{j\in nb(p)} \overline{j}\langle p\rangle \mid \overline{nb_p}\langle nb(p)\rangle \mid Alg_{match}(p)\Big) \\
\equiv\quad & \prod_{j\in P\setminus\{p\}} Alg(j, nb(j)) \mid \prod_{j\in P'} Alg'(j, nb(j), add(j)) \mid \prod_{(j,k)\in Msgs} \overline{j}\langle k\rangle \mid \\
& (\nu nb_p)\Big(\overline{nb_p}\langle nb(p)\rangle \mid Alg_{rec}(p) \mid Alg_{match}(p)\Big) \mid \prod_{j\in nb(p)} \overline{j}\langle p\rangle \\
\equiv\quad & \prod_{j\in P\setminus\{p\}} Alg(j, nb(j)) \mid \prod_{j\in P'} Alg'(j, nb(j), add(j)) \mid \prod_{(j,k)\in Msgs} \overline{j}\langle k\rangle \mid \\
& Alg(p, nb(p)) \mid \prod_{j\in nb(p)} \overline{j}\langle p\rangle \\
\equiv\quad & \prod_{j\in P} Alg(j, nb(j)) \mid \prod_{j\in P'} Alg'(j, nb(j), add(j)) \mid \prod_{(j,k)\in Msgs\cup\{(j,p)\mid j\in nb(p)\}} \overline{j}\langle k\rangle \\
\equiv\quad & Alg_{ges}(P, P', nb, Msgs \cup \{(j,p) \mid j \in nb(p)\}, add)
\end{aligned}$$

- if *select* returns a left linearization step i. e.,

$select\big(findLin\big(p, nb(p)\big)\big) = (q, r) \wedge (q < r \wedge r < p)$ :

$$\begin{aligned}
\longmapsto \quad & \prod_{j\in P\setminus\{p\}} Alg\big(j, nb(j)\big) \mid \prod_{j\in P'} Alg'\big(j, nb(j), add(j)\big) \mid \prod_{(j,k)\in Msgs} \overline{j}\langle k\rangle \mid \\
& (\nu nb_p)\Big(Alg_{rec}(p) \mid \overline{q}\langle r\rangle \mid \overline{nb_p}\big\langle nb(p)\setminus\{q\}\big\rangle \mid Alg_{match}(p)\Big) \\
\equiv \quad & \prod_{j\in P\setminus\{p\}} Alg\big(j, nb(j)\big) \mid \prod_{j\in P'} Alg'\big(j, nb(j), add(j)\big) \mid \prod_{(j,k)\in Msgs} \overline{j}\langle k\rangle \mid \\
& (\nu nb_p)\Big(\overline{nb_p}\big\langle nb(p)\setminus\{q\}\big\rangle \mid Alg_{rec}(p) \mid Alg_{match}(p)\Big) \mid \overline{q}\langle r\rangle \\
\equiv \quad & \prod_{j\in P\setminus\{p\}} Alg\big(j, nb(j)\big) \mid \prod_{j\in P'} Alg'\big(j, nb(j), add(j)\big) \mid \prod_{(j,k)\in Msgs} \overline{j}\langle k\rangle \mid \\
& Alg\big(p, nb(p)\setminus\{j\}\big) \mid \overline{q}\langle r\rangle \\
\equiv \quad & \prod_{j\in P} Alg\big(j, nb'(j)\big) \mid \prod_{j\in P'} Alg'\big(j, nb'(j), add(j)\big) \mid \prod_{(j,k)\in Msgs\cup\{(q,r)\}} \overline{j}\langle k\rangle \\
\equiv \quad & Alg_{ges}\big(P, P', nb', Msgs\cup\{(q,r)\}, add\big) \\
& \text{where } nb'(x) = \begin{cases} nb(x), & \text{if } x \neq p \\ nb(p)\setminus\{q\}, & \text{if } x = p \end{cases}
\end{aligned}$$

- if *select* returns a right linearization step i. e., $select\big(findLin\big(p, nb(p)\big)\big) = (q, r) \wedge (q < r \wedge p < q)$ :

$$\begin{aligned}
\longmapsto \quad & \prod_{j\in P\setminus\{p\}} Alg\big(j, nb(j)\big) \mid \prod_{j\in P'} Alg'\big(j, nb(j), add(j)\big) \mid \prod_{(j,k)\in Msgs} \overline{j}\langle k\rangle \mid \\
& (\nu nb_p)\Big(Alg_{rec}(p) \mid \overline{r}\langle q\rangle \mid \overline{nb_p}\big\langle nb(p)\setminus\{r\}\big\rangle \mid Alg_{match}(p)\Big) \\
\equiv \quad & \prod_{j\in P\setminus\{p\}} Alg\big(j, nb(j)\big) \mid \prod_{j\in P'} Alg'\big(j, nb(j), add(j)\big) \mid \prod_{(j,k)\in Msgs} \overline{j}\langle k\rangle \mid \\
& (\nu nb_p)\Big(\overline{nb_p}\big\langle nb(p)\setminus\{r\}\big\rangle \mid Alg_{rec}(p) \mid Alg_{match}(p)\Big) \mid \overline{r}\langle q\rangle \\
\equiv \quad & \prod_{j\in P\setminus\{p\}} Alg\big(j, nb(j)\big) \mid \prod_{j\in P'} Alg'\big(j, nb(j), add(j)\big) \mid \prod_{(j,k)\in Msgs} \overline{j}\langle k\rangle \\
& \mid Alg\big(p, nb(p)\setminus\{r\}\big) \mid \overline{r}\langle q\rangle \\
\equiv \quad & \prod_{j\in P} Alg\big(j, nb'(j)\big) \mid \prod_{j\in P'} Alg'\big(j, nb'(j), add(j)\big) \mid \prod_{(j,k)\in Msgs\cup\{(r,q)\}} \overline{j}\langle k\rangle \\
\equiv \quad & Alg_{ges}\big(P, P', nb', Msgs\cup\{(r,q)\}, add\big) \\
& \text{where } nb'(x) = \begin{cases} nb(x), & \text{if } x \neq p \\ nb(p)\setminus\{r\}, & \text{if } x = p \end{cases}
\end{aligned}$$

- else:

$$\longmapsto \quad \prod_{j\in P\setminus\{p\}} Alg\big(j, nb(j)\big) \mid \prod_{j\in P'} Alg'\big(j, nb(j), add(j)\big) \mid \prod_{(j,k)\in Msgs} \overline{j}\langle k\rangle \mid$$

$$
\begin{aligned}
& (\nu nb_p)\Big(Alg_{rec}(p) \mid \overline{nb_p}\langle nb(p)\rangle \mid Alg_{match}(p)\Big) \\
\equiv \quad & \prod_{j\in P\setminus\{p\}} Alg\big(j, nb(j)\big) \mid \prod_{j\in P'} Alg'\big(j, nb(j), add(j)\big) \mid \prod_{(j,k)\in Msgs} \overline{j}\langle k\rangle \mid \\
& (\nu nb_p)\Big(\overline{nb_p}\langle nbp\rangle \mid Alg_{rec}(p) \mid Alg_{match}(p)\Big) \\
\equiv \quad & \prod_{j\in P\setminus\{p\}} Alg\big(j, nb(j)\big) \mid \prod_{j\in P'} Alg'\big(j, nb(j), add(j)\big) \mid \prod_{(j,k)\in Msgs} \overline{j}\langle k\rangle \mid \\
& Alg\big(p, nb(p)\big) \\
\equiv \quad & \prod_{j\in P} Alg\big(j, nb(j)\big) \mid \prod_{j\in P'} Alg'\big(j, nb(j), add(j)\big) \mid \prod_{(j,k)\in Msgs} \overline{j}\langle k\rangle \\
\equiv \quad & Alg_{ges}\big(P, P', nb, Msgs, add\big)
\end{aligned}
$$

Let $p \in P'$ be an arbitrary process that tries to find a linearization step, then similar to the last case it holds that exactly one of the following four steps is possible:

- if $select\big(findLin(p, nb(p))\big) = \bot$ :

$$Alg_{ges}\big(P, P', nb, Msgs, add\big) \longmapsto Alg_{ges}\big(P, P', nb, Msgs \cup \{(x,p) | x \in nb(p)\}, add\big)$$

- if $select\big(findLin(p, nb(p))\big) = (j,k) \wedge (j < k \wedge k < p)$ :

$$Alg_{ges}\big(P, P', nb, Msgs, add\big) \longmapsto Alg_{ges}\big(P, P', nb', Msgs \cup \{(j,k)\}, add\big)$$

$$\text{and } \; nb'(x) = \begin{cases} nb(x), & \text{if } x \neq p \\ nb(p) \setminus \{j\}, & \text{if } x = p \end{cases}$$

- if $select\big(findLin(p, nb(p))\big) = (j,k) \wedge (j < k \wedge p < j)$ :

$$Alg_{ges}\big(P, P', nb, Msgs, add\big) \longmapsto Alg_{ges}\big(P, P', nb', Msgs \cup \{(k,j)\}, add\big)$$

$$\text{and } \; nb'(x) = \begin{cases} nb(x), & \text{if } x \neq p \\ nb(p) \setminus \{k\}, & \text{if } x = p \end{cases}$$

- else:

$$Alg_{ges}\big(P, P', nb, Msgs, add\big) \longmapsto Alg_{ges}\big(P, P', nb, Msgs, add\big)$$

Second, a process receives a message:

Let $p \in P$ be an arbitrary process and $\exists q \in \mathscr{P}.(p,q) \in Msgs$ then

$$
\begin{aligned}
& Alg_{ges}\big(P, P', nb, Msgs, add\big) \\
\equiv \quad & \prod_{j\in P} Alg\big(j, nb(j)\big) \mid \prod_{j\in P'} Alg'\big(j, nb(j), add(j)\big) \mid \prod_{(j,k)\in Msgs} \overline{j}\langle k\rangle
\end{aligned}
$$

$$\begin{aligned}
\equiv\;& \prod_{j\in P\setminus\{p\}} Alg\big(j,nb(j)\big) \mid \prod_{j\in P'} Alg'\big(j,nb(j),add(j)\big) \mid \prod_{(j,k)\in Msgs\setminus\{(p,q)\}} \overline{j}\langle k\rangle \mid \\
& Alg\big(p,nb(p)\big) \mid \overline{p}\langle q\rangle \\
\equiv\;& \prod_{j\in P\setminus\{p\}} Alg\big(j,nb(j)\big) \mid \prod_{j\in P'\cup\{p\}} Alg'\big(j,nb(j),add(j)\big) \mid \prod_{(j,k)\in Msgs\setminus\{(p,q)\}} \overline{j}\langle k\rangle \mid \\
& (\nu nb_p)\Big(\overline{nb_p}\langle nb(p)\rangle \mid Alg_{rec}(p) \mid Alg_{match}(p)\Big) \mid \overline{p}\langle q\rangle \\
\equiv\;& \prod_{j\in P\setminus\{p\}} Alg\big(j,nb(j)\big) \mid \prod_{j\in P'} Alg'\big(j,nb(j),add(j)\big) \mid \prod_{(j,k)\in Msgs\setminus\{(p,q)\}} \overline{j}\langle k\rangle \mid \\
& (\nu nb_p)\Big(\overline{nb_p}\langle nb(p)\rangle \mid p(x).Alg_{add}(p,x) \mid Alg_{match}(p)\Big) \mid \overline{p}\langle q\rangle \\
\longmapsto\;& \prod_{j\in P\setminus\{p\}} Alg\big(j,nb(j)\big) \mid \prod_{j\in P'} Alg'\big(j,nb(j),add(j)\big) \mid \prod_{(j,k)\in Msgs\setminus\{(p,q)\}} \overline{j}\langle k\rangle \mid \\
& (\nu nb_p)\Big(\overline{nb_p}\langle nb(p)\rangle \mid \{q/x\}Alg_{add}(p,x) \mid Alg_{match}(p)\Big) \\
\equiv\;& \prod_{j\in P\setminus\{p\}} Alg\big(j,nb(j)\big) \mid \prod_{j\in P'} Alg'\big(j,nb(j),add(j)\big) \mid \prod_{(j,k)\in Msgs\setminus\{(p,q)\}} \overline{j}\langle k\rangle \mid \\
& Alg'\big(p,nb(p),q\big) \\
\equiv\;& \prod_{j\in P\setminus\{p\}} Alg\big(j,nb(j)\big) \mid \prod_{j\in P'\cup\{p\}} Alg'\big(j,nb(j),add\cup\{(p,q)\}(j)\big) \mid \prod_{(j,k)\in Msgs\setminus\{(p,q)\}} \overline{j}\langle k\rangle \\
\equiv\;& Alg_{ges}\big(P\setminus\{p\},P'\cup\{p\},nb,Msgs\setminus\{(p,q)\},add\cup\{(p,q)\}\big)
\end{aligned}$$

Third, a process is adding another process to its neighborhood:

Let $p \in P'$ be an arbitrary process then $\exists q \in \mathscr{P}.(p,q) \in add$ and

$$\begin{aligned}
& Alg_{ges}\big(P,P',nb,Msgs,add\big) \\
\equiv\;& \prod_{j\in P} Alg\big(j,nb(j)\big) \mid \prod_{j\in P'} Alg'\big(j,nb(j),add(j)\big) \mid \prod_{(j,k)\in Msgs} \overline{j}\langle k\rangle \\
\equiv\;& \prod_{j\in P} Alg\big(j,nb(j)\big) \mid \prod_{j\in P'\setminus\{p\}} Alg'\big(j,nb(j),add(j)\big) \mid \prod_{(j,k)\in Msgs} \overline{j}\langle k\rangle \mid \\
& Alg'\big(p,nb(p),q\big) \\
\equiv\;& \prod_{j\in P} Alg\big(j,nb(j)\big) \mid \prod_{j\in P'\setminus\{p\}} Alg'\big(j,nb(j),add(j)\big) \mid \prod_{(j,k)\in Msgs} \overline{j}\langle k\rangle \mid \\
& (\nu nb_p)\Big(\overline{nb_p}\langle nb(p)\rangle \mid Alg_{add}(p,q) \mid Alg_{match}(p)\Big) \\
\equiv\;& \prod_{j\in P} Alg\big(j,nb(j)\big) \mid \prod_{j\in P'\setminus\{p\}} Alg'\big(j,nb(j),add(j)\big) \mid \prod_{(j,k)\in Msgs} \overline{j}\langle k\rangle \mid \\
& (\nu nb_p)\Big(\overline{nb_p}\langle nb(p)\rangle \mid nb_p(y).\Big(\overline{nb_p}\langle y\cup\{q\}\rangle \mid Alg_{rec}(p)\Big) \mid Alg_{match}(p)\Big) \\
\longmapsto\;& \prod_{j\in P} Alg\big(j,nb(j)\big) \mid \prod_{j\in P'\setminus\{p\}} Alg'\big(j,nb(j),add(j)\big) \mid \prod_{(j,k)\in Msgs} \overline{j}\langle k\rangle \mid
\end{aligned}$$

$$
\begin{aligned}
& (\nu nb_p)\Big((\{^{nb(p)}/_y\}\overline{nb_p}\langle y \cup \{q\}\rangle \mid Alg_{rec}(p)) \mid Alg_{match}(p)\Big) \\
\equiv \quad & \prod_{j\in P} Alg(j, nb(j)) \mid \prod_{j\in P'\setminus\{p\}} Alg'(j, nb(j), add(j)) \mid \prod_{(j,k)\in Msgs} \overline{j}\langle k\rangle \mid \\
& Alg(p, nb(p) \cup \{q\}) \\
\equiv \quad & \prod_{j\in P\cup\{p\}} Alg(j, nb'(j)) \mid \prod_{j\in P'\setminus\{p\}} Alg'(j, nb'(j), add(j)) \mid \prod_{(j,k)\in Msgs} \overline{j}\langle k\rangle \\
\equiv \quad & Alg_{ges}(P \cup \{p\}, P' \setminus \{p\}, nb', Msgs, add \setminus \{(p,q)\}) \\
& \text{where } nb'(x) = \begin{cases} nb(x), & \text{if } x \neq p \\ nb(p) \cup \{q\}, & \text{if } x = p \end{cases}
\end{aligned}
$$

□

**Notation: Configuration Components** Let $A$ be an arbitrary configuration of the algorithm then there are parameters $P$, $P'$, $nb$, $Msgs$ and $add$ such that $A \equiv Alg_{ges}(P, P', nb, Msgs, add)$ and we denote in the following :

$$P_A = P \quad , \quad P'_A = P' \quad , \quad nb_A = nb \quad , \quad Msgs_A = Msgs \quad \text{and} \quad add_A = add$$

**Corollary 2: Steps**

For every process term $Alg_{ges}(P, P', nb, Msgs, add)$ there are always exactly the following steps (up to structural congruence) possible:

- $\forall p \in \mathscr{P}.select(findLin(p, nb(p))) = \bot \implies$

  $$Alg_{ges}(P, P', nb, Msgs, add) \longmapsto Alg_{ges}(P, P', nb, Msgs \cup \{(j,p) | j \in nb(p)\}, add)$$

- $\forall p \in \mathscr{P}.select(findLin(p, nb(p))) = (j,k) \wedge (j < k \wedge k < p) \implies$

  $$Alg_{ges}(P, P', nb, Msgs, add) \longmapsto Alg_{ges}(P, P', nb', Msgs \cup \{(j,k)\}, add)$$

  $$\text{with } nb'(x) = \begin{cases} nb(x), & \text{if } x \neq p \\ nb(p) \setminus \{j\}, & \text{if } x = p \end{cases}$$

- $\forall p \in \mathscr{P}.select(findLin(p, nb(p))) = (j,k) \wedge (j < k \wedge p < j) \implies$

  $$Alg_{ges}(P, P', nb, Msgs, add) \longmapsto Alg_{ges}(P, P', nb', Msgs \cup \{(k,j)\}, add)$$

  $$\text{and } nb'(x) = \begin{cases} nb(x), & \text{if } x \neq p \\ nb(p) \setminus \{k\}, & \text{if } x = p \end{cases}$$

- $\forall p \in \mathscr{P}.\text{if } select(findLin(p, nb(p)))$ is something else $\Longrightarrow$

$$Alg_{ges}(P, P', nb, Msgs, add) \longmapsto Alg_{ges}(P, P', nb, Msgs, add)$$

- $\forall p \in P.\exists q \in \mathscr{P}.(p, q) \in Msgs \Longrightarrow$

$$\begin{aligned} &Alg_{ges}(P, P', nb, Msgs, add) \\ &\longmapsto Alg_{ges}(P \setminus \{p\}, P' \cup \{p\}, nb, Msgs \setminus \{(p, q)\}, add \cup \{(p, q)\}) \end{aligned}$$

- $\forall p \in P'.\exists q \in \mathscr{P}.(p, q) \in add \wedge$

$$Alg_{ges}(P, P', nb, Msgs, add) \longmapsto Alg_{ges}(P \cup \{p\}, P' \setminus \{p\}, nb', Msgs, add \setminus \{(p, q)\})$$

$$\text{with } nb'(x) = \begin{cases} nb(x), & \text{if } x \neq p \\ nb(p) \cup \{q\}, & \text{if } x = p \end{cases}$$

**Corollary 3: Unique Neighborhood-messages**

For every process $p \in \mathscr{P}$ there is always exactly one $\overline{nb_p}\langle\cdot\rangle$ message in the system.

Since every process only receives messages from other processes in the system via a channel name that corresponds to its unique id and the $nb_p$ channels are restricted, the receiver of every message is uniquely determined.

**Lemma 2: Unique Receiver**

There are no two processes that could receive the same message and therefore the receiver of a message is determinable. Let $Alg_{ges}(P, P', nb, Msgs, add)$ be a process term of the algorithm according to Lemma 1 then it holds that for all $m \in Msgs$ there is exactly one $p \in \mathscr{P}$ that could receive $m$.
*Proof:*
According to the definition of the process term

$$Alg_{ges}(P, P', nb, Msgs, add) = \prod_{j \in P} Alg(j, nb(j)) \mid \prod_{j \in P'} Alg'(j, nb(j), add(j)) \mid \prod_{(j,k) \in Msgs} \overline{j}\langle k\rangle$$

and therefore every message $m \in Msgs$ has always the form $\overline{j}\langle k\rangle$ with $j, k \in \mathscr{P}$. As the process ids $\mathscr{P}$ are according to Assumption 1 unique and every process $p \in \mathscr{P}$ only receives messages on the channel name $p$ (and $nb_p$), the claim holds. □

Every subprocess is input guarded and gets enabled by the existence of a corresponding message. Since there is for every process $p$ always one $\overline{nb_p}\langle\cdot\rangle$ and other messages cannot be received by other processes and messages not get lost, every enabled subprocess stays enabled until it executes a step.

**Lemma 3: Subprocesses stay Enabled**

Every subprocess that is enabled at some point stays enabled until it executed a step.
*Proof*:
We prove this by induction over the execution.
Let $A \equiv Alg_{ges}\left(P, P', nb, Msgs, add\right)$ be an arbitrary configuration of the algorithm according to Lemma 1. Let $A'$ be an arbitrary configuration with $A \longmapsto A'$. We have to show that every subprocess that was enabled in $A$, with exception of the subprocess that executed the step, is still enabled in $A'$. According to Corollary 2 there are the following cases:

- Every process $p \in \mathscr{P}$ can execute a step to find a linearization step in every configuration as $P \cup P' = \mathscr{P}$, $Alg_{match}\left(p\right)$ is a subterm of both $Alg\left(p, nb(p)\right)$ and $Alg'\left(p, nb(p), add(p)\right)$, and according to Corollary 3 there is always exactly one $nb_p$-message. Therefore, the subprocess $Alg_{match}\left(p\right)$ is always enabled.

- Let $p \in P$ be an arbitrary process and there is $q \in \mathscr{P}$ with $(p, q) \in Msgs$ i. e., $p$ could receive a message in $A$ and therefore the subprocess $Alg_{rec}\left(p\right)$ is enabled.
If the executed step was a step where any process $r \in \mathscr{P}$ tried to find a match, then it holds according to Corollary 2 that $P_{A'} = P$, $P'_{A'} = P'$ and $Msgs_{A'} = Msgs \cup X$ with $X \subset \mathscr{P} \times \mathscr{P}$. Therefore, it holds $p \in P_{A'}$ and $(p, q) \in Msgs_{A'} = Msgs \cup X$ and the step that $p$ can receive $\overline{p}\langle q \rangle$ and therefore the subprocess $Alg_{rec}\left(p\right)$ is still enabled in $A'$.
If the executed step was a step where a process $r \in P$ received a message $\overline{r}\langle s \rangle$ with $s \in \mathscr{P}$, then there are two cases. If $r = p$ then the subprocess $Alg_{rec}\left(p\right)$ executed the step. Otherwise it holds that $r \neq p$ and according to Corollary 2 that $P_{A'} = P \setminus \{r\}$, $P'_{A'} = P' \cup \{r\}$ and $Msgs_{A'} = Msgs \setminus \{(r, s\}$. Therefore, it holds $p \in P'_{A'} = P' \cup \{r\}$ and $(r, s) \in Msgs_{A'} = Msgs \setminus \{(r, s\}$ and the step that $p$ can receive $\overline{p}\langle q \rangle$ and therefore the subprocess $Alg_{rec}\left(p\right)$ is still enabled in $A'$.
If the executed step was a step where a process $r \in P'$ added a process $s$ to its neighborhood i. e., $add(r) = s$, then it holds that $r \neq p$ as $P \cap P' = \emptyset$ and according to Corollary 2 that $P_{A'} = P \cup \{r\}$, $P'_{A'} = P' \setminus \{r\}$ and $Msgs_{A'} = Msgs$. Therefore, it holds $p \in P'_{A'} = P' \setminus \{r\}$ and $(p, q) \in Msgs_{A'} = Msgs$ and the step that $p$ can receive $\overline{p}\langle q \rangle$ and therefore the subprocess $Alg_{rec}\left(p\right)$ is still enabled in $A'$.

- Let $p \in P'$ be an arbitrary process and $q \in \mathscr{P}$ with $add(p) = q$ i. e., $p$ could add $q$ to its neighborhood in $A$ and therefore the subprocess $Alg_{add}\left(p, q\right)$ is enabled.
If the executed step was a step where any process $r \in \mathscr{P}$ tried to find a linearization step, then it holds according to Corollary 2 that $P_{A'} = P$, $P'_{A'} = P'$ and $add_{A'} = add$. Therefore, it holds $p \in P'_{A'}$ and $add_{A'}(p) = q$ and the step that $p$ can add $q$ to its neighborhood and therefore the subprocess $Alg_{add}\left(p, q\right)$ is still enabled in $A'$.
If the executed step was a step where a process $r \in P$ received a message $\overline{r}\langle s \rangle$ with $s \in \mathscr{P}$, then it holds that $r \neq p$ as $P \cap P' = \emptyset$ and according to Corollary 2 that $P_{A'} = P \setminus \{r\}$, $P'_{A'} = P' \cup \{r\}$ and $add_{A'} = add \cup \{(r, s)\}$. Therefore, it holds $p \in P'_{A'} = P' \cup \{r\}$ and $add_{A'}(p) = q$ and the step that $p$ can add $q$ to its neighborhood and therefore the subprocess $Alg_{add}(p,q)$ is still enabled in $A'$.
If the executed step was a step where a process $r \in P'$ added a process $s$ to its neighborhood i. e., $add(r) = s$, then there are two cases. If $r = p$ then it holds that $s = q$ and the subprocess $Alg_{add}(p,q)$ executed the step. Otherwise $r \neq p$ and it holds according to Corollary 2 that $P_{A'} = P \cup \{r\}$, $P'_{A'} = P' \setminus \{r\}$ and $add_{A'} = add \setminus \{(r,s\}$. Therefore, it holds $p \in P'_{A'} = P' \setminus \{r\}$ and $add_{A'}(p) = q$ and the step that $p$ can add $q$ to its neighborhood and therefore the subprocess $Alg_{add}(p,q)$ is still enabled in $A'$.

□

Since every message in transit is received eventually and the addition of an previously received process id is executed eventually, there is always a configuration reached such that the corresponding process id is in the neighborhood of the receiving process (nevertheless it can be deleted later).

**Lemma 4: Progression of adding a Process to the Neighborhood**

Let $A$ be an arbitrary configuration of the algorithm according to Lemma 1. Let $p \in \mathscr{P}$ be an arbitrary process with $p \in P'_A$ and $q \in \mathscr{P}$ with $add_A(p) = q$. Then there will be a step where the process $p$ adds $q$ to its neighborhood i. e., there will be configurations $R, R'$ reached with $A \Longmapsto R$ and $R \longmapsto R'$ and it holds that:

$$R \longmapsto R' \equiv Alg_{ges}\left(P_R \cup \{p\}, P'_R \setminus \{p\}, nb_{R'}, Msgs_R, add_R \setminus \{(p,q)\}\right)$$

$$\text{with } nb_{R'}(x) = \begin{cases} nb_R(x), & \text{if } x \neq p \\ nb_R(p) \cup \{q\}, & \text{if } x = p \end{cases}$$

and therefore $q \in nb_{R'}(p)$.
*Proof:*
From $p \in P'_A$ and $add_A(p) = q$ it follows that the subprocess $Alg_{add}(p,q)$ is enabled and according to Lemma 3 it stays enabled until it executes a step. According to the fairness assumption 3 $Alg_{add}(p,q) = nb_p(y).\left(\overline{nb_p}\langle y \cup \{q\}\rangle \mid Alg_{rec}(p)\right)$ will execute a step after a finite number of steps. Let $R$ be the configuration before this step. Then it holds according to Corollary 2

$$R \longmapsto R' \equiv Alg_{ges}\left(P_R \cup \{p\}, P'_R \setminus \{p\}, nb_{R'}, Msgs_R, add_R \setminus \{(p,q)\}\right)$$

$$\text{with } nb_{R'}(x) = \begin{cases} nb_R(x), & \text{if } x \neq p \\ nb_R(p) \cup \{q\}, & \text{if } x = p \end{cases}$$

and therefore $q \in nb_{R'}(p)$. □

**Lemma 5: Progression of receiving a Message**

Let $A$ be an arbitrary configuration of the algorithm according to Lemma 1. Let $p \in \mathscr{P}$ be an arbitrary process with $q \in \mathscr{P}$ and $(p,q) \in Msgs_A$. Then there will be a step where process $p$ receives the message i. e., there will be configurations $R, R'$ reached with $A \Longmapsto R$ and $R \longmapsto R'$ and it holds that:

$$R \longmapsto R' \equiv Alg_{ges}\big(P_R \setminus \{p\}, P'_R \cup \{p\}, nb_R, Msgs_R \setminus \{(p,q)\}, add_R \cup \{(p,q)\}\big)$$

and therefore $p \in P'_{R'}$ and $add_{R'}(p) = q$.
*Proof*:
According to the Assumption 2 that there is no message loss, the message will be eventually (after a finite number of steps) received. According to Lemma 2, $p$ is the only process that can receive this message. Let $R$ be the configuration before this step. Then it holds according to Corollary 2

$$R \longmapsto R' \equiv Alg_{ges}\big(P_R \setminus \{p\}, P'_R \cup \{p\}, nb_R, Msgs_R \setminus \{(p,q)\}, add_R \cup \{(p,q)\}\big)$$

and therefore $p \in P'_{R'}$ and $add_{R'}(p) = q$ . □

**Corollary 4: Progression of adding a Process from a Message to the Neighborhood**

Let $A$ be an arbitrary configuration of the algorithm according to Lemma 1. Let $p \in \mathscr{P}$ be an arbitrary process with $q \in \mathscr{P}$ and $(p,q) \in Msgs_A$. Then there will be a step where the process $p$ receives the message and later a step where it adds $q$ to its neighborhood i. e., there will be configurations $R, R', R'', R'''$ reached with $A \Longmapsto R$ , $R \longmapsto R'$ , $R \Longmapsto R''$ and $R'' \longmapsto R'''$ and it holds that:

$$R \longmapsto R' \equiv Alg_{ges}\big(P_R \setminus \{p\}, P'_R \cup \{p\}, nb_R, Msgs_R \setminus \{(p,q)\}, add_R \cup \{(p,q)\}\big)$$

and therefore $p \in P'_{R'}$ and $add_{R'}(p) = q$ and

$$R'' \longmapsto R''' \equiv Alg_{ges}\big(P_{R''} \cup \{p\}, P'_{R''} \setminus \{p\}, nb_{R'''}, Msgs_{R''}, add_{R''} \setminus \{(p,q)\}\big)$$
$$\text{with } nb_{R'''}(x) = \begin{cases} nb_{R''}(x), & \text{if } x \neq p \\ nb_{R''}(p) \cup \{q\}, & \text{if } x = p \end{cases}$$

and therefore $q \in nb_{R'''}(p)$.
This follows directly as the conclusion of Lemma 5 is the precondition of Lemma 4.

Whether a process finds a linearization step or not, depends on the number of neighbors a process has. If a process has more than two neighbors, at least two of them must be on the same side (i. e., both have smaller or greater process ids) and therefore there must be a possible linearization step. If a process has less than two neighbors, then there cannot be a possible linearization step and the process sends therefore *keep-alive*-messages . If a process has exactly two neighbors, there is a possible linearization step if both are on the same side and otherwise it sends *keep-alive*-messages . Hence, sending *keep-alive*-messages and finding a linearization step are mutually exclusive, but either one or the other happens if a process tries to find a linearization step.

**Lemma 6: Number of Neighbors $|nb(\cdot)|$**

Let $A$ be an arbitrary configuration of the algorithm. Let $p \in \mathcal{P}$ be an arbitrary process that tries to find a linearization step i. e., the subprocess $Alg_{match}(p)$ executes a step in $A$. Directly from the definition of $select(findLin(p, nb_A(p)))$ and the property that $\leq$ is a total order (Definition 11) it follows:

- $|nb_A(p)| < 2 \implies p$ sends *keep-alive*-messages , i. e.,

  $$select(findLin(p, nb_A(p))) = \perp$$

- $|nb_A(p)| = 2 \implies$ if one of the processes is smaller and the other one greater i. e.,

  $$|LeftN(nb_A(p), p)| = 1 \quad \wedge \quad |RightN(nb_A(p), p)| = 1,$$
  $p$ sends *keep-alive*-messages i. e.,
  $$select(findLin(p, nb_A(p))) = \perp$$

  otherwise there is an enabled linearization step i. e.,

  $$select(findLin(p, nb_A(p))) = (j, k) \wedge j < k \wedge ((k < p) \vee (p < j))$$

- $|nb_A(p)| > 2 \implies$ there is an enabled linearization step i. e.,

  $$select(findLin(p, nb_A(p))) = (j, k) \wedge j < k \wedge ((k < p) \vee (p < j))$$

- if $p$ can send *keep-alive*-messages $\implies |nb_A(p)| \leq 2$
- there is an enabled linearization step for $p \iff p$ can not send *keep-alive*-messages
- $p$ can send *keep-alive*-messages $\iff$ there is no enabled linearization step for $p$

## 3.2 Topology of Configuration

In this section, we define the network topology of a configuration. We introduce variants, which differ in whether we regard the direction of the edges, and if we take the messages that are in transit into account or not. The network topology graphs with messages describe how the neighborhood sets would be if all current messages in transit were received and processed (depicted in Figure 3.1). Therefore, the network topology graphs without messages of a configuration are always subgraphs of the topology graphs with messages.

**Notation: Topology (without messages)** A solid line represents a process in the neighborhood, a dashed line an adding in progress, and a dotted line a message in transit.

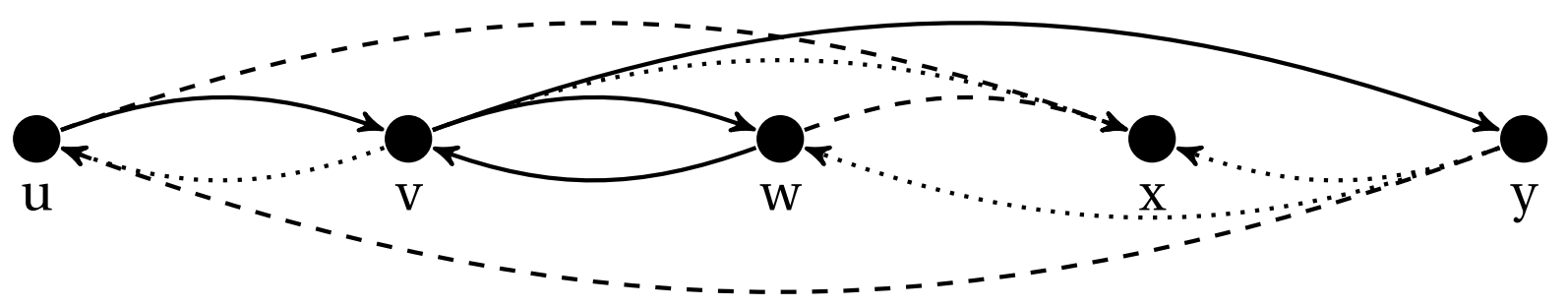

(a) Topology without Messages

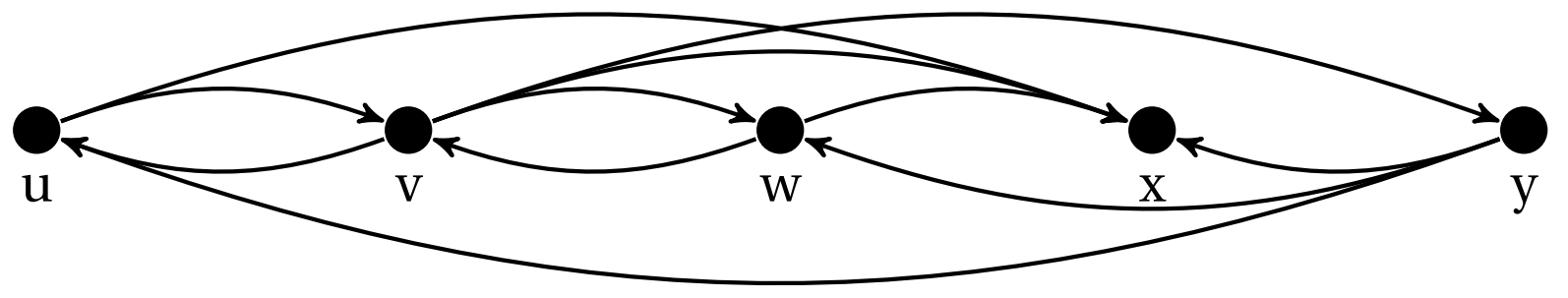

(b) Topology with Messages

Figure 3.1: Topology with and without messages whereby solid lines represent the edges

**Notation: Subgraph** We denote $G' = (V', E')$ a subgraph of $G = (V, E)$, written $G' \subseteq G$ iff $V' = V$ and $E' \subseteq E$.

**Definition 27: Network Topology Graph**

Let $A \equiv Alg_{ges}(P, P', nb, Msgs, add)$ be an arbitrary configuration of the algorithm according to Lemma 1. Then the *(directed) network topology graph* $NT(A) = (V, E)$ is defined as follows:

$$V = P \cup P' = \mathscr{P} \quad \text{and} \quad E = \{(p,q) | p, q \in V \wedge q \in nb(p)\}$$

**Definition 28: Network Topology Graph with Messages**

Let $A \equiv Alg_{ges}(P, P', nb, Msgs, add)$ be an arbitrary configuration of the algorithm according to Lemma 1. Then the *(directed) network topology graph with messages* $NTM(A) = (V, E)$ is defined as follows:

$$V = P \cup P' = \mathscr{P} \quad \text{and} \quad E = \{(p,q) | p, q \in V \wedge (q \in nb(p) \vee (p,q) \in Msgs \vee add(p) = q)\}$$

**Corollary 5: *NT* ⊆ *NTM***

Let $A$ be an arbitrary configuration of the algorithm. Then the directed network topology graph is always a subgraph of the directed network topology graph with messages i. e.,

$$NT(A) \subseteq NTM(A)$$

**Definition 29: Undirected Topology Graph**

Let $A \equiv Alg_{ges}\left(P, P', nb, Msgs, add\right)$ be an arbitrary configuration of the algorithm according to Lemma 1. Then the *undirected network topology graph* $UNT(A) = (V, E)$ is defined as follows:

$$V = P \cup P' = \mathscr{P} \quad \text{and} \quad E = \{\{p, q\} | p, q \in V \wedge q \in nb(p)\}$$

**Corollary 6: *undirected*(*NT*) = *UNT***

Directly from the Definitions 27 and 29 it follows that for an arbitrary configuration $A$ it holds that

$$undirected(NT(A)) = UNT(A)$$

**Definition 30: Undirected Topology Graph with Messages**

Let $A \equiv Alg_{ges}\left(P, P', nb, Msgs, add\right)$ be an arbitrary configuration of the algorithm according to Lemma 1. Then the *undirected network topology graph with messages* $UNTM(A) = (V, E)$ is defined as follows:

$$V = P \cup P' = \mathscr{P} \quad \text{and} \quad E = \{\{p, q\} | p, q \in V \wedge (q \in nb(p) \vee (p, q) \in Msgs \vee add(p) = q)\}$$

**Corollary 7: *UNT* ⊆ *UNTM***

Let $A$ be an arbitrary configuration of the algorithm according to Lemma 1. Then the undirected network topology graph is always a subgraph of the undirected network topology graph with messages i. e.,

$$UNT(A) \subseteq UNTM(A)$$

**Corollary 8: *undirected*(*NTM*) = *UNTM***

Directly from the Definitions 28 and 30 it follows that for an arbitrary configuration $A$ it holds that

$$undirected(NTM(A)) = UNTM(A)$$

**Notation: Topology Graph Components** Let $A$ be an arbitrary configuration of the algorithm, we denote for $NT(A) = (\mathscr{P}, E)$ , $NTM(A) = (\mathscr{P}, E')$ , $UNT(A) = (\mathscr{P}, E'')$ , $UNTM(A) = (\mathscr{P}, E''')$ in the following :

$$E_{NT(A)} = E \quad , \quad E_{NTM(A)} = E' \quad , \quad E_{UNT(A)} = E'' \quad \text{and} \quad E_{UNTM(A)} = E'''$$

We show that if the topology graph with messages of the initial configuration is weakly connected, than the topology graph with messages of all reachable configurations is weakly connected. The only steps that lead to a removal of edges are linearization steps. Nevertheless, if a process executes a linearization step, the removed edge can always be simulated by the new introduced edge and the edge to the other neighbor of the executing process. Therefore, linearization can not result in partitioning the network topology.

**Lemma 7: Connectivity**

Let $A_0 \equiv Alg_{ges}\left(P_0, P'_0, init, Msgs_0, add_0\right)$ be an initial configuration according to Definition 23 and let $A \equiv Alg_{ges}\left(P, P', nb, Msgs, add\right)$ be an arbitrary reachable configuration according to Lemma 1. Then, it holds that if the initial undirected topology graph with messages i. e., $UNTM(A_0)$ is connected, also $UNTM(A)$ is connected.
*Proof*:
We will prove the lemma by induction over the execution.
$UNTM(A_0)$ is connected by assumption. Let $A_i \equiv Alg_{ges}\left(P_i, P'_i, nb_i, Msgs_i, add_i\right)$ be an arbitrary reachable configuration with $UNTM(A_i)$ is connected. Let $A_{i+1}$ be an arbitrary configuration with $A_i \longmapsto A_{i+1}$, according to Corollary 2 there are only the following cases possible:

- If a process $p \in \mathscr{P}$ tries to find match but $select\big(findLin\left(p, nb_i(p)\right)\big) = \bot$, it sends *keep-alive*-messages to any neighbor i. e., to any process $j \in nb_i(p)$. It holds that

  $$A_{i+1} \equiv Alg_{ges}\left(P_i, P'_i, nb_i, Msgs_i \cup \{(j,p) | j \in nb_i(p)\}, add_i\right)$$

  and therefore $UNTM(A_{i+1}) = UNTM(A_i)$ and connected by assumption.

- If a process $p \in \mathscr{P}$ finds a match with a left-linearization step i. e., it holds that $select\big(findLin\left(p, nb_i(p)\right)\big) = (j,k) \wedge (j < k \wedge k < p)$, it sends the smaller process (and therefore the process with the greater distance) $j \in nb_i(p)$ a message and deletes it from its neighborhood. It holds that

  $$A_{i+1} \equiv Alg_{ges}\left(P_i, P'_i, nb_{i+1}, Msgs_i \cup \{(j,k)\}, add_i\right)$$
  $$\text{with } nb_{i+1}(x) = \begin{cases} nb_i(x), & \text{if } x \neq p \\ nb_i(p) \setminus \{j\}, & \text{if } x = p \end{cases}$$

  Let $UNTM(A_i) = (V_i, E_i)$, then it holds that the edge is removed or not (dependent on whether there is still a message in transit with same edge or not) and therefore $UNTM(A_{i+1}) = (V_i, E_i \cup \{\{j,k\}\})$ or $UNTM(A_{i+1}) = (V_i, (E_i \cup \{\{j,k\}\}) \setminus \{\{p,j\}\})$. The first case is true if any of the following holds $(p,j) \in Msgs_i, (j,p) \in Msgs_i, (j,p) \in add_i$ or $p \in nb_i(j)$ but the (potentially) addition of a new edge cannot disconnect the topology graph and therefore $UNTM(A_{i+1})$ is still connected. In the second case similarly only the removal of $\{p,j\}$ could lead to a disconnection. It holds obviously that $\{j,k\} \in ((E_i \cup \{\{j,k\}\}) \setminus \{\{p,j\}\})$ since $k < p$. Since $k \in nb_{i+1}(p)$, it holds that $\{p,k\} \in ((E_i \cup \{\{j,k\}\}) \setminus \{\{p,j\}\})$. Therefore, every path that used the edge $\{p,j\}$ can be simulated by using the edges $\{p,k\}$ and $\{k,j\}$ and it holds that $UNTM(A_{i+1})$ is still connected.

- If a process $p \in \mathscr{P}$ finds a match with a right-linearization step i. e., it holds that $select\big(findLin(p, nb_i(p))\big) = (j,k) \wedge (j < k \wedge p < j)$, it sends the greater process (and therefore the process with the greater distance) $k \in nb_i(p)$ a message and deletes it from its neighborhood. It holds that

$$A_{i+1} \equiv Alg_{ges}\big(P_i, P'_i, nb_{i+1}, Msgs_i \cup \{(k,j)\}, add_i\big)$$
$$\text{with } nb_{i+1}(x) = \begin{cases} nb_i(x), & \text{if } x \neq p \\ nb_i(p) \setminus \{k\}, & \text{if } x = p \end{cases}$$

  and $p \in \mathscr{P}$ and $j,k \in nb_i(p)$. The proof that $UNTM(A_{i+1})$ is connected is similar to the left-linearization step.

- If $select\big(findLin(p, nb_i(p))\big)$ could have any other return value, it holds that

$$A_{i+1} \equiv Alg_{ges}\big(P_i, P'_i, nb_i, Msgs_i, add_i\big) = A$$

  and therefore $UNTM(A_{i+1}) = UNTM(A_i)$ and connected by assumption.

- If a process $p \in P_i$ receives a message from another process $q \in \mathscr{P}$ i. e., $(p,q) \in Msgs_i$, the message is consumed but $p$ will later add $q$ to its neighborhood. It holds that

$$A_{i+1} \equiv Alg_{ges}\big(P_i \setminus \{p\}, P'_i \cup \{p\}, nb_i, Msgs_i \setminus \{(p,q)\}, add_i \cup \{(p,q)\}\big)$$

  and therefore $UNTM(A_{i+1}) = UNTM(A_i)$ and connected by assumption.

- If a process $p \in P'_i$ adds another process $q \in \mathscr{P}$ to its neighborhood i. e., $(p,q) \in add_i$, it holds that

$$A_{i+1} \equiv Alg_{ges}\big(P_i \cup \{p\}, P'_i \setminus \{p\}, nb_{i+1}, Msgs_i, add_i \setminus \{(p,q)\}\big)$$
$$\text{with } nb_{i+1}(x) = \begin{cases} nb_i(x), & \text{if } x \neq p \\ nb_i(p) \cup \{q\}, & \text{if } x = p \end{cases}$$

  and hence $UNTM(A_{i+1}) = UNTM(A_i)$ and connected by assumption.

□

## 3.3 Correct Configuration

A configuration is correct if every process exactly knows its consecutive processes. To ensure that also no other connections will be established through messages, it must additionally hold that every message in transit contains the id of a consecutive process of the receiver. Therefore, the network topology with and without messages must be the desired topology i. e., the linear graph. With exception of the number of these messages and the local state of the processes, the correct configuration is uniquely defined.

**Definition 31: Correct Configuration**

Let $A \equiv Alg_{ges}(P, P', nb, Msgs, add)$ be a process term of the algorithm according to Lemma 1. Then, $A$ is a *correct configuration* iff the directed topology network graph without and with messages is the linear graph i. e.,

$$NTM(A) = G_{LIN} \quad \wedge \quad NT(A) = G_{LIN}$$

**Notation: ⊥-free Set** Let $A$ be an arbitrary set, in the following $A_\perp$ is an abbreviation defined as

$$A_\perp = A \setminus \{\perp\}$$

**Lemma 8: Up to - Uniqueness of Correct Configuration**

The correct configuration is unique up to structural congruence, the number of messages in the system and the state of the processes.
*Proof*:
Let $A \equiv Alg_{ges}(P, P', nb, Msgs, add)$ be an arbitrary configuration of the algorithm according to Lemma 1 and $A$ is a correct configuration i. e., $NTM(A) = G_{LIN} \wedge NT(A) = G_{LIN}$. Then it holds according to the Definitions 14, 27 and 28 that $A$ is a correct configuration iff:

- $P \cup P' = \mathscr{P}$ and $P \cap P' = \emptyset$
- $\forall p \in \mathscr{P}.nb(p) = \{succ(p), pred(p)\}_\perp$
- $\forall (p, q) \in Msgs.(q = succ(p)) \vee (q = pred(p))$
- $\forall p \in P'.(add(p) = succ(p)) \vee (add(p) = pred(p))$

□

A weaker property is described by an undirected correct configuration. Here, we only demand that the undirected topology graph with message must be the undirected linear graph (as depicted in Figure 3.2). Therefore, the neighborhood of each process is a subset of the consecutive processes and the messages in transit must satisfy the same requirement as in the case of a correct configuration. In order to ensure connectivity, between each pair of consecutive processes there must be at least one connection while taking the messages in transit into account. Similarly to a correct configuration, an undirected configuration is uniquely defined with exception of the number of messages, the state of a process, and the type of the connection, i. e., a message, addition in progress, or neighborhood relation, between a consecutive pair of processes.

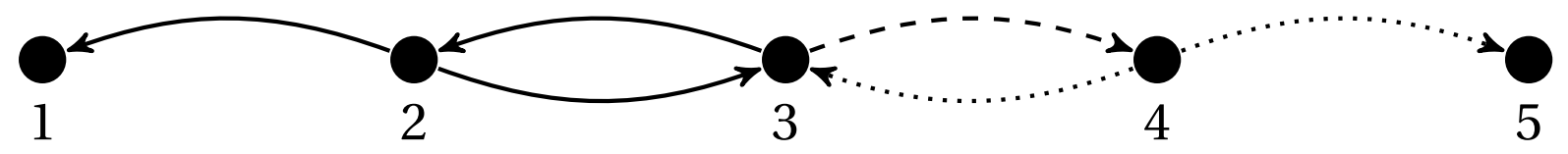

Figure 3.2: Topology of an undirected correct configuration, whereby the nodes are ordered according to their ids

**Definition 32: Undirected Correct Configuration**

Let $A \equiv Alg_{ges}\left(P, P', nb, Msgs, add\right)$ be a process term of the algorithm according to Lemma 1. Then $A$ is an *undirected correct configuration* iff the undirected topology network graph with messages is the undirected linear graph i. e.,

$$UNTM(A) = UG_{LIN}$$

**Lemma 9: Up to - Uniqueness of undirected Correct Configuration**

The undirected correct configuration is unique up to structural congruence, the number of messages in the system, the state of the processes, and the type of connections.
*Proof.*
Let $A \equiv Alg_{ges}\left(P, P', nb, Msgs, add\right)$ be a process term of the algorithm according to Lemma 1 and $A$ is an undirected correct configuration i. e., $UNTM(A) = UG_{LIN}$. Then it holds according to the Definitions 14 and 30 that $A$ is an undirected correct configuration iff:

- $P \cup P' = \mathscr{P}$ and $P \cap P' = \emptyset$
- $\forall p \in \mathscr{P}.nb(p) \subseteq \{succ(p), pred(p)\}_\perp$
- $\forall (p,q) \in Msgs.(q = succ(p)) \vee (q = pred(p))$
- $\forall p \in P'.(add(p) = succ(p)) \vee (add(p) = pred(p))$
- $\forall p,q \in \mathscr{P}.succ(p) = q \implies q \in nb(p) \vee p \in nb(q) \vee (p,q) \in Msgs \vee (q,p) \in Msgs \vee (p \in P' \wedge add(p) = q) \vee (q \in P' \wedge add(q) = p)$ (connectivity)

□

## 3.4 Closure

In this section, we introduce several closure properties that are based on the properties that only a process itself can remove processes from its neighborhood and a process never removes the process id from a desired neighbor i. e., its predecessor and successor. Furthermore, if every process knows (a subset) of desired neighbors but no other processes, then there are no possible linearization steps anymore in the system. All process can only send *keep-alive*-messages to their known desired neighbors and receive such messages.

As already stated, the only steps that remove edges i. e., process ids from the neighborhood of a process, are linearization steps. Therefore, whenever a process does not execute any linearization steps, its neighborhood can be expanded by reception of messages from other processes but it cannot shrink.

**Lemma 10: Growing Neighborhood**

If a process not executes a linearization step, its neighborhood can only grow.
Let $A$ be the process term of the algorithm according to Lemma 1. Let $R$ be an arbitrary reachable configuration $A \Longmapsto R$. If the subprocess $Alg_{match}(p)$ does not executes a linearization step in an intermediated configuration $C$ i. e., $A \Longmapsto C \longmapsto C' \Longmapsto R$ and it holds either $select(findLin(p, nb_C(p))) = (j,k) \wedge (j < k \wedge k < p)$ or $select(findLin(p, nb_C(p))) = (j,k) \wedge (j < k \wedge p < j)$, then it follows that $nb_A(p) \subseteq nb_R(p)$ .
*Proof*:
We proof this by induction over the step semantics.
Obviously it holds $nb_A \subseteq nb_A$. Let $p' \in \mathscr{P}$ be an arbitrary process. Let $C$ be an arbitrary configuration with $A \Longmapsto C \longmapsto R$ and $nb_A(p) \subseteq nb_C(p)$. We have to show that if the subprocess $Alg_{match}(p')$ does not executes the step or $select(findLin(p', nb_C(p'))) \neq (j,k) \wedge (j < k \wedge k < p')$ and $select(findLin(p', nb_C(p'))) \neq (j,k) \wedge (j < k \wedge p' < j)$, it holds $nb_A(p') \subseteq nb_R(p')$. According to Corollary 2 there are only the following cases:

- $\forall p \in \mathscr{P}.select(findLin(p, nb_C(p))) = \bot \implies Alg_{ges}(P_C, P'_C, nb_C, Msgs_C, add_C) \longmapsto R$

  with $R \equiv Alg_{ges}(P_C, P'_C, nb_C, Msgs_C \cup \{(j,p) | j \in nb_C(p)\}, add_C)$

  Since $nb_R = nb_C$, it holds obviously that $\forall p \in \mathscr{P}.nb_R(p) = nb_C(p)$. With $nb_A(p') \subseteq nb_C(p')$ it holds that $nb_A(p') \subseteq nb_C(p') = nb_R(p')$.

- $\forall p \in \mathscr{P}.select(findLin(p, nb_C(p))) = (j,k) \wedge (j < k \wedge k < p) \implies$

  $Alg_{ges}(P_C, P'_C, nb_C, Msgs_C, add_C) \longmapsto R$
  with $R \equiv Alg_{ges}(P_C, P'_C, nb_R, Msgs_C \cup \{(j,k)\}, add_C)$

  $$\text{and } nb_R(x) = \begin{cases} nb_C(x), & \text{if } x \neq p \\ nb_C(p) \setminus \{j\}, & \text{if } x = p \end{cases}$$

Let $p \in \mathscr{P}$ be the executing process. Since $Alg_{match}(p)$ is the executing subprocess and $select\big(findLin\big(p, nb_C(p)\big)\big) = (j,k) \wedge (j < k \wedge k < p)$, it holds that $p' \neq p$ and therefore with $nb_A(p') \subseteq nb_C(p')$ and $nb_R(p') = nb_C(p')$ that $nb_R(p') = nb_C(p') \supseteq nb_A(p')$.

- $\forall p \in \mathscr{P}.select\big(findLin\big(p, nb_C(p)\big)\big) = (j,k) \wedge (j < k \wedge p < j) \implies$

$$Alg_{ges}\big(P_C, P'_C, nb_C, Msgs_C, add_C\big) \longmapsto R$$
$$\text{with } R \equiv Alg_{ges}\big(P_C, P'_C, nb_R, Msgs_C \cup \{(k,j)\}, add_C\big)$$
$$\text{and } nb_R(x) = \begin{cases} nb_C(x), & \text{if } x \neq p \\ nb_C(p) \setminus \{k\}, & \text{if } x = p \end{cases}$$

Let $p \in \mathscr{P}$ be the executing process. Since $Alg_{match}(p)$ is the executing subprocess and $select\big(findLin\big(p, nb_C(p)\big)\big) = (j,k) \wedge (j < k \wedge p < j)$, it holds that $p' \neq p$ and therefore with $nb_A(p') \subseteq nb_C(p')$ and $nb_R(p') = nb_C(p')$ that $nb_R(p') = nb_C(p') \supseteq nb_A(p')$.

- $\forall p \in \mathscr{P}.$if $select\big(findLin\big(p, nb_C(p)\big)\big)$ is something else $\implies$

$$Alg_{ges}\big(P_C, P'_C, nb_C, Msgs_C, add_C\big) \longmapsto R \equiv Alg_{ges}\big(P_C, P'_C, nb_C, Msgs_C, add_C\big)$$

Since $nb_R = nb_C$, it holds obviously that $\forall p \in \mathscr{P}.nb_R(p) = nb_C(p)$. With $nb_A(p') \subseteq nb_C(p')$ it holds that $nb_A(p') \subseteq nb_C(p') = nb_R(p')$.

- $\forall p \in P_C.\exists q \in \mathscr{P}.(p,q) \in Msgs_C \implies Alg_{ges}\big(P_C, P'_C, nb_C, Msgs_C, add_C\big) \longmapsto R$

$$\text{with } R \equiv Alg_{ges}\big(P_C \setminus \{p\}, P'_C \cup \{p\}, nb_C, Msgs_C \setminus \{(p,q)\}, add_C \cup \{(p,q)\}\big)$$

Since $nb_R = nb_C$, it holds obviously that $\forall p \in \mathscr{P}.nb_R(p) = nb_C(p)$. With $nb_A(p') \subseteq nb_C(p')$ it holds that $nb_A(p') \subseteq nb_C(p') = nb_R(p')$.

- $\forall p \in P'_C.\exists q \in \mathscr{P}.(p,q) \in add_C \wedge Alg_{ges}\big(P_C, P'_C, nb_C, Msgs_C, add_C\big) \longmapsto R$

$$\text{with } R \equiv Alg_{ges}\big(P_C \cup \{p\}, P'_C \setminus \{p\}, nb_R, Msgs_C, add_C \setminus \{(p,q)\}\big)$$
$$\text{and } nb_R(x) = \begin{cases} nb_C(x), & \text{if } x \neq p \\ nb_C(p) \cup \{q\}, & \text{if } x = p \end{cases}$$

Let $p \in \mathscr{P}$ be the executing process. If $p \neq p'$, it holds with $nb_A(p') \subseteq nb_C(p')$ and $nb_R(p') = nb_C(p')$ that $nb_R(p') = nb_C(p') \supseteq nb_A(p')$. If $p = p'$, it holds that there is $q \in \mathscr{P}$ and with $nb_A(p') \subseteq nb_C(p')$ that $nb_R(p') = nb_C(p') \cup \{q\} \supseteq nb_C(p') \supseteq nb_A(p')$.

□

Therefore, only in linearization steps a process removes processes from its neighborhood. However, if a process executes a linearization step it never removes a correct neighbor i. e., its successor or predecessor, as it always removes the process that is further away. Therefore, if a process knows a correct neighbor, this correct neighbor remains in the neighborhood for every reachable configuration.

**Lemma 11: Preservation of Correct Neighbors**

Let $A$ be an arbitrary configuration of the algorithm according to Lemma 1. Whenever a process $p \in \mathscr{P}$ knows a correct neighbor i. e., $succ(p)$ or $pred(p)$, it will not remove this neighbor anymore. Let $R$ be an arbitrary reachable configuration i. e., $A \Longmapsto R$, then it holds that

$$\forall p \in \mathscr{P}.(succ(p) \in nb_A(p) \implies succ(p) \in nb_R(p)) \quad \wedge$$
$$(pred(p) \in nb_A(p) \implies pred(p) \in nb_R(p))$$

*Proof*:
We prove this by induction over the step semantics.
Let $p \in \mathscr{P}$ be an arbitrary process with $succ(p) \in nb_A(p)$. Obviously it holds $nb_A = nb_A$. Let $A'$ be an arbitrary configuration with $A \Longmapsto A' \longmapsto R$ and $succ(p) \in nb_{A'}(p)$. We have to show that $succ(p) \in nb_R(p)$. According to Lemma 10 it holds that if the executed step is not performed by the subprocess $Alg_{match}(p)$ or $select\big(findLin\big(p, nb_{A'}(p)\big)\big) \neq (j,k) \wedge (j < k \wedge k < p)$ and $select\big(findLin\big(p, nb_{A'}(p)\big)\big) \neq (j,k) \wedge (j < k \wedge p < j)$, it holds that $nb_{A'}(p) \subseteq nb_R(p)$ and therefore with $succ(p) \in nb_{A'}(p)$ it follows $succ(p) \in nb_R(p)$. If the subprocess $Alg_{match}(p)$ executed the step, there are only the following cases left:

- $select\big(findLin\big(p, nb_{A'}(p)\big)\big) = (j,k) \wedge (j < k \wedge k < p)$:

$$Alg_{ges}\big(P_{A'}, P'_{A'}, nb_{A'}, Msgs_{A'}, add_{A'}\big) \longmapsto R$$
$$\text{with } R \equiv Alg_{ges}\big(P_{A'}, P'_{A'}, nb_R, Msgs_{A'} \cup \{(j,k)\}, add_{A'}\big)$$
$$\text{and } nb_R(x) = \begin{cases} nb_{A'}(x), & \text{if } x \neq p \\ nb_{A'}(p) \setminus \{j\}, & \text{if } x = p \end{cases}$$

  According to Corollary 1 it holds that $succ(p) > p$ and therefore $succ(p) \neq j$. Hence $succ(p) \in nb_R(p) = nb_{A'}(p) \setminus \{j\}$.

- $select\big(findLin\big(p, nb_{A'}(p)\big)\big) = (j,k) \wedge (j < k \wedge p < j)$:

$$Alg_{ges}\big(P_{A'}, P'_{A'}, nb_{A'}, Msgs_{A'}, add_{A'}\big) \longmapsto R$$
$$\text{with } R \equiv Alg_{ges}\big(P_{A'}, P'_{A'}, nb_R, Msgs_{A'} \cup \{(k,j)\}, add_{A'}\big)$$
$$\text{and } nb_R(x) = \begin{cases} nb_{A'}(x), & \text{if } x \neq p \\ nb_{A'}(p) \setminus \{k\}, & \text{if } x = p \end{cases}$$

  According to Corollary 1 it holds that $succ(p) \leq j < k$ and therefore $succ(p) \neq k$. Hence $succ(p) \in nb_R(p) = nb_{A'}(p) \setminus \{k\}$.

The proof for $pred(p)$ is similar. □

Therefore, it follows directly that in the directed and undirected topology without messages edges between consecutive neighbors are preserved as the edges depict the neighborhood sets.

**Corollary 9: Preservation of Correct Edges**

Let $A \equiv Alg_{ges}\left(P, P', nb, Msgs, add\right)$ be an arbitrary configuration of the algorithm according to Lemma 1. For every reachable configuration $R$ i. e., $A \Longmapsto R$ it holds with Definitions 27 and 29 that:

$$\forall p \in \mathscr{P}.(p, succ(p)) \in E_{NT(A)} \implies (p, succ(p)) \in E_{NT(R)},$$
$$\forall p \in \mathscr{P}.(p, pred(p)) \in E_{NT(A)} \implies (p, pred(p)) \in E_{NT(R)} \text{ and}$$
$$\forall p \in \mathscr{P}.\{p, succ(p)\} \in E_{UNT(A)} \implies \{p, succ(p)\} \in E_{UNT(R)}$$

This preservation holds furthermore also for correct edges in the topologies with messages. Here, an edge can also represent an adding of a process to be performed or a message in transit. But every id carried by a messages in transit cannot get lost and is therefore received and processed eventually. Therefore, also edges between desired neighbors that represent adding or messages are preserved.

**Lemma 12: Preservation of Correct Edges in Topology Graphs with Messages**

Let $A \equiv Alg_{ges}\left(P, P', nb, Msgs, add\right)$ be an arbitrary configuration of the algorithm according to Lemma 1. For every reachable configuration $R$ i. e., $A \Longmapsto R$ it holds that:

$$\forall p \in \mathscr{P}.(p, succ(p)) \in E_{NTM(A)} \implies (p, succ(p)) \in E_{NTM(R)},$$
$$\forall p \in \mathscr{P}.(p, pred(p)) \in E_{NTM(A)} \implies (p, pred(p)) \in E_{NTM(R)} \text{ and}$$
$$\forall p \in \mathscr{P}.\{p, succ(p)\} \in E_{UNTM(A)} \implies \{p, succ(p)\} \in E_{UNTM(R)}$$

*Proof.*
Let $p \in \mathscr{P}$ be an arbitrary process with $(p, succ(p)) \in E_{NTM(A)}$, then there are according to Definition 28 the following cases:

- $succ(p) \in nb(p)$: According to Definition 27 and Corollaries 9 and 5 it holds for every reachable configuration $R$ that $(p, succ(p)) \in E_{NTM(R)}$
- $add(p) = q$: Then it holds that the subprocess $Alg_{add}\left(p, succ(p)\right)$ is enabled and according to Lemma 3 it stays enabled until it executes a step. According to the fairness assumption 3 $Alg_{add}\left(p, succ(p)\right) = nb_p(y).\left(\overline{nb_p}\langle y \cup \{succ(p)\}\rangle \mid Alg_{rec}\left(p\right)\right)$ will execute a step after a finite number of steps. Up to this point it holds for every intermediate configuration $C$ that $add_C(p) = succ(p)$ and therefore $(p, succ(p)) \in E_{NTM(C)}$. Let $C'$ be the configuration before this step. Then it holds according to Corollary 2

$$C' \longmapsto R' \equiv Alg_{ges}\left(P_{C'} \cup \{p\}, P'_{C'} \setminus \{p\}, nb', Msgs_{C'}, add_{C'} \setminus \{(p, succ(p))\}\right)$$

$$\text{with } nb'(x) = \begin{cases} nb_{C'}x, & \text{if } x \neq p \\ nb_{C'}p \cup \{succ(p)\}, & \text{if } x = p \end{cases}$$

and therefore $succ(p) \in nb'(p)$ and with Definition 28 $(p, succ(p)) \in E_{NTM(R')}$. According to the previous case, for every further reachable configuration $R$ it holds also that $(p, succ(p)) \in E_{NTM(R)}$.

- $(p, succ(p)) \in Msgs$: According to the Assumption 2 that there is no message loss, the message will be eventually (after a finite number of steps) received and for every configuration $C$ up to this point it holds that $(p, succ(p)) \in Msgs_C$ and therefore $(p, succ(p)) \in E_{NTM(C)}$. According to Lemma 2 $p$ is the only process that can receive this message. Let $C'$ be the configuration before this step. Then it holds according to Corollary 2 $C' \longmapsto R'$ with

$$R' \equiv Alg_{ges}\left(P_{C'} \setminus \{p\}, P'_{C'} \cup \{p\}, nb_{C'}, Msgs_{C'} \setminus \{(p, succ(p))\}, add_{C'} \cup \{(p, succ(p))\}\right)$$

  Therefore $add_{R'}(p) = succ(p)$ and with Definition 28 $(p, succ(p)) \in E_{NTM(R')}$. We have now reduced the problem to the previous case and for every further reachable configuration $R$ it holds also that $(p, succ(p)) \in E_{NTM(R)}$.

The proofs for $pred(p)$ and $UNTM$ are similar. □

The topology of a configuration contains the desired configuration i. e., the linear graph is a subgraph of the topology, if every correct edge is already established but possibly undesired edges are still existent in addition. Since every correct edge is always preserved, it follows directly that whenever the linear graph is a subgraph of a configurations topology, this property applies also to any reachable configuration.

**Corollary 10: Closure for $NTM \subseteq G_{LIN} \wedge NT \subseteq G_{LIN}$**

It follows directly from Definition 14, Corollary 9 and Lemma 12 that if $A$ is a configuration with $G_{LIN} \subseteq NTM(A) \wedge G_{LIN} \subseteq NT(A)$ it holds for every reachable configuration $R$ i. e., $A \Longmapsto R$, that

$$G_{LIN} \subseteq NTM(R) \quad \wedge \quad G_{LIN} \subseteq NT(R)$$

In the topology of an undirected correct configuration every edge is a correct edge but there are possibly correct edges missing. Through preservation of correct edges also this property is invariant and therefore carries over to every reachable configuration. Hence, every configuration that is reachable from an undirected correct configuration is itself an undirected correct configuration. Further, for every reachable configuration it holds that there are possibly still correct edges missing but none of the already established correct edges was removed.

**Lemma 13: Closure for Undirected Correct Configuration**

Let $A \equiv Alg_{ges}\big(P, P', nb, Msgs, add\big)$ be an arbitrary configuration of the algorithm and $A$ is an undirected correct configuration, then it holds for every reachable configuration $C$ that $C$ is also an undirected correct configuration.
*Proof*:
We prove this by induction over the step semantic i. e., every possible step leads again to an undirected correct configuration.
Since $A$ is an undirected correct configuration according to Definition 32, it holds that $UNTM(A) = UG_{LIN}$. According to Lemma 9 it holds that

$$\begin{aligned}
&P \cup P' = \mathscr{P} \quad \text{and} \quad P \cap P' = \emptyset,\\
&\forall p \in \mathscr{P}.nb(p) \subseteq \{succ(p), pred(p)\}_{\perp},\\
&\forall (p,q) \in Msgs.(q = succ(p)) \vee (q = pred(p)),\\
&\forall p \in P'.(add(p) = succ(p)) \vee (add(p) = pred(p)) \text{ and}\\
&\forall p, q \in \mathscr{P}. \quad \big(succ(p) = q \implies \big(q \in nb(p) \vee p \in nb(q) \vee (p,q) \in Msgs \vee (q,p) \in Msgs\\
&\qquad\qquad \vee (p \in P' \wedge add(p) = q) \vee (q \in P' \wedge add(q) = p)\big)\big) \text{ (i. e., is connected).}
\end{aligned}$$

Since $\forall p \in \mathscr{P}.nb(p) \subseteq \{succ(p), pred(p)\}_{\perp}$ and the Definition of $select(findLin())$, it holds that $\forall p \in \mathscr{P}.select\big(findLin\big(p, nb(p)\big)\big) = \perp$. According to Corollary 2 there are only the following steps possible:

- $\forall p \in \mathscr{P}.select\big(findLin\big(p, nb(p)\big)\big) = \perp \implies Alg_{ges}\big(P, P', nb, Msgs, add\big) \longmapsto R$ with

  $$R \equiv Alg_{ges}\big(P, P', nb, Msgs \cup \{(j,p) | j \in nb(p)\}, add\big)$$

  Let $p \in \mathscr{P}$ be an arbitrary process with $select\big(findLin\big(p, nb(p)\big)\big) = \perp$. Since the only change in the configuration concerns the multiset of messages and $nb(p) \subseteq \{succ(p), pred(p)\}_{\perp}$, it still holds that

  $$\forall (p', q) \in Msgs \cup \{(j,p) | j \in nb(p)\}.(q = succ(p')) \vee (q = pred(p'))$$

  and therefore $R \equiv Alg_{ges}\big(P, P', nb, Msgs \cup \{(j,p) | j \in nb(p)\}, add\big)$ is an undirected correct configuration according to Lemma 9.

- $\forall p \in P.\exists q \in \mathscr{P}.(p,q) \in Msgs \implies Alg_{ges}\big(P, P', nb, Msgs, add\big) \longmapsto R$ with

  $$R \equiv Alg_{ges}\big(P \setminus \{p\}, P' \cup \{p\}, nb, Msgs \setminus \{(p,q)\}, add \cup \{(p,q)\}\big)$$

  Let $p \in P$ be an arbitrary process and $q \in \mathscr{P}$ with $(p,q) \in Msgs$. According to the assumptions it holds that $q = succ(p) \vee q = pred(p)$ and therefore

$$\forall p' \in P' \cup \{p\}.(add(p') = succ(p')) \vee (add(p') = pred(p'))$$

Additionally, with the assumptions it obviously holds that:

$$\big(P \setminus \{p\}\big) \cup \big(P' \cup \{p\}\big) = \mathscr{P} \quad \text{and} \quad \big(P \setminus \{p\}\big) \cap \big(P' \cup \{p\}\big) = \emptyset \quad \text{and}$$
$$\forall (p', q') \in Msgs \setminus \{(p, q)\}.(q' = succ(p')) \vee (q' = pred(p'))$$

Since $UNTM(A)$ is connected, it also holds according to Lemma 7 that the topology is still connected. Therefore $R \equiv Alg_{ges}\big(P \setminus \{p\}, P' \cup \{p\}, nb, Msgs \setminus \{(p, q)\}, add \cup \{(p, q)\}\big)$ is an undirected correct configuration according to Lemma 9.

- $\forall p \in P'.\exists q \in \mathscr{P}.(p, q) \in add \wedge Alg_{ges}\big(P, P', nb, Msgs, add\big) \longmapsto R$

$$R \equiv Alg_{ges}\big(P \cup \{p\}, P' \setminus \{p\}, nb', Msgs, add \setminus \{(p, q)\}\big)$$
$$\text{with } nb'(x) = \begin{cases} nb(x), & \text{if } x \neq p \\ nb(p) \cup \{q\}, & \text{if } x = p \end{cases}$$

Let $p \in P'$ be an arbitrary process and $q \in \mathscr{P}$ with $(p, q) \in add$. According to the assumptions it holds that $q = succ(p) \vee q = pred(p)$ and therefore

$$nb(p) \cup \{q\} \subseteq \{succ(p), pred(p)\}_{\perp}$$

Additionally, with the assumptions it obviously holds that:

$$\big(P \cup \{p\}\big) \cup \big(P' \setminus \{p\}\big) = \mathscr{P} \quad \text{and} \quad \big(P \cup \{p\}\big) \cap \big(P' \setminus \{p\}\big) = \emptyset \quad \text{and}$$
$$\forall p' \in P' \setminus \{p\}.\big((p', succ(p')) \in add \setminus \{(p, q)\}\big) \vee \big((p', pred(p')) \in add \setminus \{(p, q)\}\big)$$

Since $UNTM(A)$ is connected, it also holds according to Lemma 7 that the topology is still connected and therefore $R \equiv Alg_{ges}\big(P \cup \{p\}, P' \setminus \{p\}, nb', Msgs, add \setminus \{(p, q)\}\big)$ is an undirected correct configuration according to Lemma 9.

□

**Corollary 11: $UNTM = UG_{LIN}$ increasing**

Let $A \equiv Alg_{ges}\big(P, P', nb, Msgs, add\big)$ be an arbitrary configuration of the algorithm and $A$ is an undirected correct configuration i. e., $UNTM(A) = UG_{LIN}$, then it holds for every reachable configuration $R$ i. e., $A \Longmapsto R$ according to Corollary 9 and Lemma 13 that

$$UNTM(R) = UG_{LIN} \quad \wedge \quad UNT(A) \subseteq UNT(R)$$

According to definition 1 in order to show that the algorithm is self-stabilizing, we have to prove the properties *convergence* and *closure*. In order to be a linearization algorithm according to definition 4, the system is in a legal state if and only if the topology of the system is the linear graph on the nodes of the system. The unique linear Graph $G_{LIN}$ is defined in Definition 14 and describes a sorted doubly-linked-list according to the total order on the process ids. The topology of a configuration is the linear graph if and only if the configuration is a correct configuration as defined in 31. The correct configuration is unique up to structural congruence, the local state of the processes and the number of messages in the system as shown in Lemma 8. The *closure* property states whenever a correct configuration is reached, provided that no fault occurs, the system stays in a correct configuration. In order to prove the *closure* property of the algorithm, we have to show that whenever we have reached a correct configuration according to Definition 31 every possible step leads again to a correct configuration.

**Theorem 1: Closure for Correct Configurations**

Let $A$ be a correct configuration then it holds for every reachable configuration $C$ i. e., $A \Longmapsto C$, that $C$ is also a correct configuration.
*Proof*:
We prove this by induction over the step semantic i. e., every possible step leads again to a correct configuration. Let $A \equiv Alg_{ges}\left(P, P', nb, Msgs, add\right)$ be the process term of an arbitrary correct configuration i. e., $NTM(A) = G_{LIN} \wedge NT(A) = G_{LIN}$ according to Definition 31. According to Lemma 8 it holds that

$$\begin{aligned}
&P \cup P' = \mathscr{P} \quad \text{and} \quad P \cap P' = \emptyset,\\
&\forall p \in \mathscr{P}.nb(p) = \{succ(p), pred(p)\}_\perp,\\
&\forall (p,q) \in Msgs.(q = succ(p)) \vee (q = pred(p)) \ \text{ and}\\
&\forall p \in P'.(add(p) = succ(p)) \vee (add(p) = pred(p))
\end{aligned}$$

Since $\forall p \in \mathscr{P}.nb(p) = \{succ(p), pred(p)\}_\perp$ and the Definition of $select(findLin())$ it holds that $\forall p \in \mathscr{P}.select\left(findLin\left(p, nb(p)\right)\right) = \perp$. According to Corollary 2 there are (up to structural congruence) only the following steps possible:

- $\forall p \in \mathscr{P}.select\left(findLin\left(p, nb(p)\right)\right) = \perp \implies Alg_{ges}\left(P, P', nb, Msgs, add\right) \longmapsto A'$

  with $A' \equiv Alg_{ges}\left(P, P', nb, Msgs \cup \{(j,p) | j \in nb(p)\}, add\right)$

  Let $p \in \mathscr{P}$ be an arbitrary process with $select\left(findLin\left(p, nb(p)\right)\right) = \perp$. Since the only change in the configuration concerns the multiset of messages and $nb(p) = \{succ(p), pred(p)\}_\perp$, it still holds that

  $$\forall (p', q) \in Msgs \cup \{(j,p) | j \in nb(p)\}.(q = succ(p')) \vee (q = pred(p'))$$

  and therefore $A' \equiv Alg_{ges}\left(P, P', nb, Msgs \cup \{(j,p) | j \in nb(p)\}, add\right)$ is a correct configuration according to Lemma 8.

- $\forall p \in P.\exists q \in \mathscr{P}.(p,q) \in Msgs \implies Alg_{ges}\big(P, P', nb, Msgs, add\big) \longmapsto A'$

  $$\text{with } A' \equiv Alg_{ges}\big(P \setminus \{p\}, P' \cup \{p\}, nb, Msgs \setminus \{(p,q)\}, add \cup \{(p,q)\}\big)$$

  Let $p \in P$ be an arbitrary process and $q \in \mathscr{P}$ with $(p,q) \in Msgs$. According to the assumptions it holds that $q = succ(p) \vee q = pred(p)$ and therefore

  $$\forall p' \in P' \cup \{p\}.(add(p') = succ(p')) \vee (add(p') = pred(p'))$$

  Additionally with the assumptions it obviously holds that:

  $$\big(P \setminus \{p\}\big) \cup \big(P' \cup \{p\}\big) = \mathscr{P} \quad \text{and} \quad \big(P \setminus \{p\}\big) \cap \big(P' \cup \{p\}\big) = \emptyset \text{ and}$$
  $$\forall (p', q') \in Msgs \setminus \{(p,q)\}.(q' = succ(p')) \vee (q' = pred(p'))$$

  Therefore, $A' \equiv Alg_{ges}\big(P \setminus \{p\}, P' \cup \{p\}, nb, Msgs \setminus \{(p,q)\}, add \cup \{(p,q)\}\big)$ is a correct configuration according to Lemma 8.

- $\forall p \in P'.\exists q \in \mathscr{P}.(p,q) \in add \wedge Alg_{ges}\big(P, P', nb, Msgs, add\big) \longmapsto A'$

  $$\text{with } A' \equiv Alg_{ges}\big(P \cup \{p\}, P' \setminus \{p\}, nb_{A'}, Msgs, add \setminus \{(p,q)\}\big)$$
  $$\text{and } nb_{A'}(x) = \begin{cases} nb(x), & \text{if } x \neq p \\ nb(p) \cup \{q\}, & \text{if } x = p \end{cases}$$

  Let $p \in P'$ be an arbitrary process and $q \in \mathscr{P}$ with $(p,q) \in add$. According to the assumptions it holds that $q = succ(p) \vee q = pred(p)$ and therefore

  $$nb(p) \cup \{q\} = \{succ(p), pred(p)\}_\perp \cup \{q\} = \{succ(p), pred(p)\}_\perp$$

  Additionally with the assumptions it obviously holds that:

  $$\big(P \cup \{p\}\big) \cup \big(P' \setminus \{p\}\big) = \mathscr{P} \quad \text{and} \quad \big(P \cup \{p\}\big) \cap \big(P' \setminus \{p\}\big) = \emptyset \text{ and}$$
  $$\forall p' \in P' \setminus \{p\}.(p', succ(p')) \in add \setminus \{(p,q)\}) \vee ((p', pred(p')) \in add \setminus \{(p,q)\})$$

  and therefore $A' \equiv Alg_{ges}\big(P \cup \{p\}, P' \setminus \{p\}, nb_{A'}, Msgs, add \setminus \{(p,q)\}\big)$ is a correct configuration according to Lemma 8.

□

## 3.5 Potential Functions

In this section, we introduce three potential functions in order to show in Section 3.6 several convergence properties for our algorithm. The first function takes all connections between non-consecutive processes into account, the second only considers the closest neighbor to the left and the closest neighbor to the right, and the third is a combination of both.

The first potential function $\Psi$ sums up the distances between all non-consecutive processes multiplied by the number of connections between them. Therefore, it sumps up the length of all non-correct edges in the topology with messages but takes also the number of connections that cause such an edge into account. For example, if process $q \notin \{succ(p), pred(p)\}$ is in the neighborhood of $p$ and there are additionally three messages of the form $\overline{p}\langle q\rangle$, the length of edge $(p, q)$ in the topology, and therefore the distance between $p$ and $q$, is added four times in the potential function.

**Definition 33: Potential Function $\Psi_p$ for Processes**

Let $p \in \mathscr{P}$ be an arbitrary process and $A$ be a process term of the algorithm according to Lemma 1. Let $Rec\colon (\mathscr{T} \times \mathscr{P}) \to \mathbb{N}^{\mathscr{P}}$ with

$$Rec(A, p) = \{\!\{q \in \mathscr{P} \mid (p, q) \in Msgs_A \wedge q \notin \{succ(p), pred(p)\}\}\!\}$$

be the multiset of all process ids that are sent to $p$ but still in transit and not a desired neighbor and $adding\colon (\mathscr{T} \times \mathscr{P}) \to \mathbb{N}$ with

$$adding(A, p) = \begin{cases} dist(p, q), & \text{if } add_A(p) = q \in \mathscr{P} \wedge q \notin \{succ(p), pred(p)\} \\ 0, & \text{otherwise} \end{cases}$$

The *potential function* $\Psi_p : (\mathscr{T} \times \mathscr{P}) \to \mathbb{N}$ sums up the distances (with respect to the linear order $\leq$) of all outgoing connections of the process $p$ while ignoring desired connections and is defined as follows:

$$\Psi_p(A, p) = \sum_{q \in (nb_A(p) \setminus \{succ(p), pred(p)\})} dist(p, q) + \sum_{q \in Rec(A,p)} dist(p, q) + adding(A, p)$$

**Definition 34: Potential Function $\Psi$ for Configurations**

Let $A$ be a process term of the algorithm according to Lemma 1. The *potential function for configurations* $\Psi : \mathscr{T} \to \mathbb{N}$ sums up the distances (with respect to the linear order $\leq$) between all the connections of processes while ignoring desired connections and is defined as:

$$\Psi(A) = \sum_{p \in \mathscr{P}} \Psi_p(A, p)$$

The potential function $\Psi$ is therefore minimal if every connection in a configuration is between consecutive processes and there are no other connections between processes. This is only the case for a (weakly) connected topology if and only if the topology is the linear graph. Hence, the potential function is minimal for (undirected) correct configurations.

**Lemma 14: Minimal Potential $\Psi$ for (undirected) Correct Configurations**

The potential function $\Psi$ is minimal for a process term $A \equiv Alg_{ges}(P, P', nb, Msgs, add)$ of the algorithm with a (weakly) connected (loop free) network topology iff $A$ is an (undirected) correct configuration i. e.,

$$\Psi(A) = 0 \quad \text{iff} \quad NTM(A) \subseteq G_{LIN} \ \text{ respectively } \ UNTM(A) = UG_{LIN}$$

*Proof*:
According to the Definitions 11, 12, 15, 28, 30 14 and 34 it holds that $\Psi(A) = 0$ if $NTM(A) \subseteq G_{LIN}$ respectively $UNTM(A) = UG_{LIN}$. We have to show that for every process term $A'$ with a (weakly) connected (loop free) topology Graph with $NTM(A') \not\subseteq G_{LIN}$ resp. $UNTM(A') \neq UG_{LIN}$ it holds that $\Psi_U(A') > 0$. It holds that $|\mathcal{P}| = n$ and therefore $|E_{NTM(A')}| \geq n-1$ resp. $|E_{UNTM(A')}| \geq n-1$. For every $p, q \in \mathcal{P}$ with $p \neq q$ it holds according to Definition 11 and 15 that $dist(p,q) \geq 1$. Since $NTM(A') \not\subseteq G_{LIN}$ resp. $UNTM(A') \neq UG_{LIN}$, there have to be $p, q \in \mathcal{P}$ with $p \notin \{q, succ(p), pred(q)\}$ and $(p,q) \in E_{NTM(A')}$ resp. $\{p,q\} \in E_{UNTM(A')}$ and therefore according to Definitions 28 and 30 $q \in nb_{A'}(p) \vee (p,q) \in Msgs_{A'} \vee add_{A'}(p) = q$. According to Definitions 15, 33 and 34 it then follows that $\Psi(A') > 0$. □

The potential function $\Psi$ is monotonically decreasing with exception of sending *keep-a-live*-messages to undesired neighbors and decreases with every linearization step as the new introduced connection is always strictly shorter than the removed one. Furthermore, it also decreases if an addition of an already known process is processed as the number of connections is then decreased by one. Whereby the sending, receiving and adding of desired process ids does not change anything as the potential function $\Psi$ considers only the connections that still need to be removed.

**Lemma 15: Progression of $\Psi$**

Let $A$ be an arbitrary configuration. For $A \longmapsto A'$ it holds:

- $\Psi(A') < \Psi(A)$ iff
  - an arbitrary process $p \in \mathcal{P}$ finds a left linearization step i. e.,

$$select(findLin(p, nb_A(p))) = (j,k) \ \wedge \ (j < k \wedge k < p)$$

it then holds

$$A' \equiv Alg_{ges}\big(P_A, P'_A, nb_{A'}, Msgs_A \cup \{(j,k)\}, add_A\big)$$
$$\text{and } nb_{A'}(x) = \begin{cases} nb_A(x), & \text{if } x \neq p \\ nb_A(p) \setminus \{j\}, & \text{if } x = p \end{cases}$$

– an arbitrary process $p \in \mathscr{P}$ finds a right linearization step i. e.,

$$select\big(findLin\big(p, nb_A(p)\big)\big) = (j,k) \;\wedge\; (j < k \wedge p < j)$$

it then holds

$$A' \equiv Alg_{ges}\big(P_A, P'_A, nb_{A'}, Msgs_A \cup \{(k,j)\}, add_A\big)$$
$$\text{and } nb_{A'}(x) = \begin{cases} nb_A(x), & \text{if } x \neq p \\ nb_A(p) \setminus \{k\}, & \text{if } x = p \end{cases}$$

– or an arbitrary process $p \in P'_A$ adds an already known process that is not a desired neighbor to its neighborhood i. e.,

$$add_A(p) = q \;\wedge\; q \notin \{succ(p), pred(p)\} \;\wedge\; q \in nb_A(p)$$

it then holds

$$A' \equiv Alg_{ges}\big(P_A \cup \{p\}, P'_A \setminus \{p\}, nb_{A'}, Msgs_A, add_A \setminus \{(p,q)\}\big)$$
$$\text{with } nb_{A'}(x) = \begin{cases} nb_A(x), & \text{if } x \neq p \\ nb_A(p) \cup \{q\}, & \text{if } x = p \end{cases}$$
$$\text{and } \Psi(A') = \Psi(A) - dist(p,q)$$

- $\Psi(A') = \Psi(A)$ iff

  – an arbitrary process $p \in \mathscr{P}$ sends *keep-alive*-messages to (a subset of) its desired neighbors i. e.,

  $$select\big(findLin\big(p, nb_A(p)\big)\big) = \bot \;\wedge\; nb_A(p) \subseteq \{succ(p), pred(p)\}$$

  it then holds $A' \equiv Alg_{ges}\big(P_A, P'_A, nb_A, Msgs_A \cup \{(j,p) | j \in nb_A(p)\}, add_A\big)$,

  – an arbitrary process $p \in P_A$ receives a message i. e.,

  $$q \in \mathscr{P} \;\wedge\; (p,q) \in Msgs_A$$

  it then holds $A' \equiv Alg_{ges}\big(P_A \setminus \{p\}, P'_A \cup \{p\}, nb_A, Msgs_A \setminus \{(p,q)\}, add_A \cup \{(p,q)\}\big)$,

  – an arbitrary process $p \in P'_A$ adds a former unknown process that is not a desired

neighbor to its neighborhood i. e.,

$$add_A(p) = q \wedge q \notin \{succ(p), pred(p)\} \wedge q \notin nb_A(p)$$

it then holds

$$A' \equiv Alg_{ges}\left(P_A \cup \{p\}, P'_A \setminus \{p\}, nb_{A'}, Msgs_A, add_A \setminus \{(p,q)\}\right)$$

$$\text{with } nb_{A'}(x) = \begin{cases} nb_A(x), & \text{if } x \neq p \\ nb_A(p) \cup \{q\}, & \text{if } x = p \end{cases}$$

– an arbitrary process $p \in P'_A$ adds a desired neighbor to its neighborhood i. e.,

$$add_A(p) = q \wedge q \in \{succ(p), pred(p)\}_\perp$$

it then holds

$$A' \equiv Alg_{ges}\left(P_A \cup \{p\}, P'_A \setminus \{p\}, nb_{A'}, Msgs_A, add_A \setminus \{(p,q)\}\right)$$

$$\text{with } nb_{A'}(x) = \begin{cases} nb_A(x), & \text{if } x \neq p \\ nb_A(p) \cup \{q\}, & \text{if } x = p \end{cases}$$

– or an arbitrary process $p \in \mathscr{P}$ tries to find a linearization step and none of the other cases is true. It then holds $A' \equiv A$.

- $\Psi(A') > \Psi(A)$ iff

  – an arbitrary process $p \in \mathscr{P}$ sends *keep-alive*-messages not exclusively to its desired neighbors i. e.,

$$select\left(findLin\left(p, nb_A(p)\right)\right) = \perp \wedge q \in nb_A(p) \wedge q \notin \{succ(p), pred(p)\}$$

it then holds $A' \equiv Alg_{ges}\left(P_A, P'_A, nb_A, Msgs_A \cup \{(j,p) | j \in nb_A(p)\}, add_A\right)$ and

$$\Psi(A') = \Psi(A) + \sum_{q \in \left(nb_A(p) \setminus \{succ(p), pred(p)\}\right)} dist(q,p)$$

and therefore $\Psi$ is monotone with exception of sending *keep-alive*-messages to undesired neighbors.

*Proof.*

According to Corollary 2 there are the following cases possible:

- $\forall p \in \mathscr{P}.select\left(findLin\left(p, nb_A(p)\right)\right) = \perp \implies A \longmapsto A'$ with

$$A' \equiv Alg_{ges}\left(P_A, P'_A, nb_A, Msgs_A \cup \{(j,p) | j \in nb_A(p)\}, add_A\right)$$

Let $p \in \mathscr{P}$ be the executing process. The only change in the configuration concerns the multiset of messages and $Msgs_{A'} = Msgs_A \cup \{(j,p) | j \in nb_A(p)\}$. Now there are two cases. If $nb_A(p) \subseteq \{succ(p), pred(p)\}$, it holds according to Definition 33 that

$$\forall q \in nb_A(p).Rec(A',q) = Rec(A,q) \quad \text{and since} \quad nb_{A'} = nb_A \quad \text{and} \quad add_{A'} = add_A$$

it follows for all $q \in \mathscr{P}$ that $\Psi_p(A',q) = \Psi_p(A,q)$ and therefore according to Definition 34 it holds $\Psi(A') = \Psi(A)$.

Otherwise, there is at least one process $q \in \mathscr{P}$ with $q \notin \{succ(p), pred(p)\}$ and $q \in nb_A(p)$ and therefore according to Definition 33 $Rec(A,q) = Rec(A',q) \cup \{p\}$. Since $nb_{A'} = nb_A$ and $add_{A'} = add_A$, according to Definition 33:

$$\begin{aligned}
\Psi_p(A',q) &= \sum_{q' \in (nb_{A'}(q) \setminus \{succ(q),pred(q)\})} dist(q,q') + \sum_{q' \in Rec(A',q)} dist(q,q') \\
&\quad + adding(A',q) \\
&= \sum_{q' \in (nb_A(q) \setminus \{succ(q),pred(q)\})} dist(q,q') + \sum_{q' \in Rec(A,q) \cup \{p\}} dist(q,q') \\
&\quad + adding(A,q) \\
&= \sum_{q' \in (nb_A(q) \setminus \{succ(q),pred(q)\})} dist(q,q') + \sum_{q' \in Rec(A,q)} dist(q,q') + dist(q,p) \\
&\quad + adding(A,q) \\
&= \Psi_p(A,q) + dist(q,p)
\end{aligned}$$

Hence with Definition 34 $\Psi(A') = \Psi(A) + \sum_{q \in (nb_A(p) \setminus \{succ(p),pred(p)\})} dist(q,p) > \Psi(A)$.

- $\forall p \in \mathscr{P}.select\big(findLin\big(p, nb_A(p)\big)\big) = (j,k) \wedge (j < k \wedge k < p) \implies A \longmapsto A'$ with

$$A' \equiv Alg_{ges}\left(P_A, P'_A, nb_{A'}, Msgs_A \cup \{(j,k)\}, add_A\right)$$
$$\text{and } nb_{A'}(x) = \begin{cases} nb_A(x), & \text{if } x \neq p \\ nb_A(p) \setminus \{j\}, & \text{if } x = p \end{cases}$$

Let $p \in \mathscr{P}$ be the executing process with $j,k \in \mathscr{P}$, $select\big(findLin\big(p, nb_A(p)\big)\big) = (j,k)$ and $j < k \wedge k < p$. The change of $nb_{A'}$ influences $\Psi_p(A,p)$ whereas the change in $Msgs_{A'}$ influences $\Psi_p(A',j)$. According to Corollary 1 it holds that $j < k \leq pred(p)$ hence $pred(p) \neq j$ and $succ(p) > p$ hence also $succ(p) \neq j$ and therefore $j \notin \{succ(p), pred(p)\}$. With the definition of $select\big(findLin\big(p, nb_A(p)\big)\big)$ it follows

$$j \in \big(nb_A(p) \setminus \{succ(p), pred(p)\}\big) \quad \text{but} \quad j \notin \big(nb_{A'}(p) \setminus \{succ(p), pred(p)\}\big)$$

According to Definition 33 it holds additionally

$$adding(A',p) = adding(A,p) \quad \text{and} \quad Rec(A',p) = Rec(A,p)$$

and further with Definition 33

$$\begin{aligned}
\Psi_p(A',p) &= \sum_{q'\in(nb_{A'}(p)\setminus\{succ(p),pred(p)\})} dist(p,q') + \sum_{q'\in Rec(A',p)} dist(p,q') \\
&\quad + adding(A',p) \\
&= \sum_{q'\in((nb_A(p)\setminus\{j\})\setminus\{succ(p),pred(p)\})} dist(p,q') + \sum_{q'\in Rec(A,p)} dist(p,q') \\
&\quad + adding(A,p) \\
&= \sum_{q'\in(((nb_A(p)\setminus\{j\})\setminus\{succ(p),pred(p)\})\cup\{j\})} dist(p,q') - dist(p,j) \\
&\quad + \sum_{q'\in Rec(A,p)} dist(p,q') + adding(A,p) \\
&= \sum_{q'\in(nb_A(p)\setminus\{succ(p),pred(p)\})} dist(p,q') + \sum_{q'\in Rec(A,p)} dist(p,q') \\
&\quad + adding(A,p) - dist(p,j) \\
&= \Psi_p(A,p) - dist(p,j)
\end{aligned}$$

For $\Psi_p(A',j)$ there are two cases. If $k = succ(j)$, then it holds according with Definition 33 that

$$adding(A',j) = adding(A,j) \quad \text{and} \quad Rec(A',j) = Rec(A,j)$$

and therefore $\Psi_p(A',j) = \Psi_p(A,j)$. Since $p \neq j$, it holds according to Definition 15 that $dist(p,j) > 0$. Hence, with Definition 34 it follows that $\Psi(A') < \Psi(A)$.
If $k \neq succ(j)$, it holds that $Rec(A',j) = Rec(A,j) \cup \{k\}$ since $j < k$ and therefore also $k \neq pred(j)$ according to Corollary 1. Hence, according to Definition 33

$$\begin{aligned}
\Psi_p(A',j) &= \sum_{q'\in(nb_{A'}(j)\setminus\{succ(j),pred(j)\})} dist(j,q') + \sum_{q'\in Rec(A',j)} dist(j,q') \\
&\quad + adding(A',j) \\
&= \sum_{q'\in(nb_A(k)\setminus\{succ(j),pred(j)\})} dist(j,q') + \sum_{q'\in(Rec(A,j)\cup\{k\})} dist(j,q') \\
&\quad + adding(A,j) \\
&= \sum_{q'\in(nb_A(j)\setminus\{succ(j),pred(j)\})} dist(j,q') + \sum_{q'\in(Rec(A,j)\cup\{k\})\setminus\{k\}} dist(j,q') \\
&\quad + dist(j,k) + adding(A,j) \\
&= \sum_{q'\in(nb_A(j)\setminus\{succ(j),pred(j)\})} dist(j,q') + \sum_{q'\in Rec(A,j)} dist(j,q') \\
&\quad + adding(A,j) + dist(j,k) \\
&= \Psi_p(A,j) + dist(j,k)
\end{aligned}$$

Since $\leq$ is according to Definition 11 a total order and $j < k \wedge k < p$, it follows with Definition 15 that $dist(p,j) > dist(j,k)$ and therefore according to Definition 34 it holds that $\Psi(A') < \Psi(A)$.

- $\forall p \in \mathscr{P}.select\big(findLin\big(p, nb_A(p)\big)\big) = (j,k) \wedge (j < k \wedge p < j) \implies A \longmapsto A'$ with

$$A' \equiv Alg_{ges}\left(P_A, P'_A, nb_{A'}, Msgs_A \cup \{(k,j)\}, add_A\right)$$
$$\text{and } nb_{A'}(x) = \begin{cases} nb_A(x), & \text{if } x \neq p \\ nb_A(p) \setminus \{k\}, & \text{if } x = p \end{cases}$$

Let $p \in \mathscr{P}$ be the executing process with $j,k \in \mathscr{P}$, $select\big(findLin\big(p, nb_A(p)\big)\big) = (j,k)$ and $j < k \wedge p < j$. The only changes in the configuration that influence $\Psi$ are $Msgs_{A'} = Msgs_A \cup \{(k,j)\}$ and $nb_{A'}$. The change of $nb_{A'}$ influences $\Psi_p(A,p)$ whereas the change in $Msgs_{A'}$ influences $\Psi_p(A',k)$. According to Corollary 1 it holds that $succ(p) \leq j < k$ hence $succ(p) \neq k$ and $pred(p) < p$ hence also $pred(p) \neq k$ and therefore $k \notin \{succ(p), pred(p)\}$. With the definition of $select\big(findLin\big(p, nb_A(p)\big)\big)$ it follows

$$k \in \big(nb_A(p) \setminus \{succ(p), pred(p)\}\big) \quad \text{but} \quad k \notin \big(nb_{A'}(p) \setminus \{succ(p), pred(p)\}\big)$$

According to Definition 33 it holds additionally

$$adding(A',p) = adding(A,p) \quad \text{and} \quad Rec(A',p) = Rec(A,p)$$

and further with Definition 33

$$\begin{aligned}
\Psi_p(A',p) &= \sum_{q' \in (nb_{A'}(p) \setminus \{succ(p), pred(p)\})} dist(p,q') + \sum_{q' \in Rec(A',p)} dist(p,q') \\
&\quad + adding(A',p) \\
&= \sum_{q' \in ((nb_A(p) \setminus \{k\}) \setminus \{succ(p), pred(p)\})} dist(p,q') + \sum_{q' \in Rec(A,p)} dist(p,q') \\
&\quad + adding(A,p) \\
&= \sum_{q' \in (((nb_A(p) \setminus \{k\}) \setminus \{succ(p), pred(p)\}) \cup \{k\})} dist(p,q') - dist(p,k) \\
&\quad + \sum_{q' \in Rec(A,p)} dist(p,q') + adding(A,p) \\
&= \sum_{q' \in (nb_A(p) \setminus \{succ(p), pred(p)\})} dist(p,q') + \sum_{q' \in Rec(A,p)} dist(p,q') \\
&\quad + adding(A,p) - dist(p,k) \\
&= \Psi_p(A,p) - dist(p,k)
\end{aligned}$$

For $\Psi_p(A',k)$ there are two cases. If $j = pred(k)$, then it holds according with Definition 33 that

$$adding(A',k) = adding(A,k) \quad \text{and} \quad Rec(A',k) = Rec(A,k)$$

and therefore $\Psi_p(A',k) = \Psi_p(A,k)$. Since $p \neq k$, it holds according to Definition 15 that $dist(p,k) > 0$. Hence, with Definition 34 it follows that $\Psi(A') < \Psi(A)$.
If $j \neq pred(k)$, it holds that $Rec(A',k) = Rec(A,k) \cup \{j\}$ since $j < k$ and therefore also $j \neq succ(k)$ according to Corollary 1. Hence, according to Definition 33

$$\begin{aligned}
\Psi_p(A',k) &= \sum_{q' \in (nb_{A'}(k) \setminus \{succ(k),pred(k)\})} dist(k,q') + \sum_{q' \in Rec(A',k)} dist(k,q') \\
&\quad + adding(A',k) \\
&= \sum_{q' \in (nb_A(k) \setminus \{succ(k),pred(k)\})} dist(k,q') + \sum_{q' \in (Rec(A,k) \cup \{j\})} dist(k,q') \\
&\quad + adding(A,k) \\
&= \sum_{q' \in (nb_A(k) \setminus \{succ(k),pred(k)\})} dist(k,q') + \sum_{q' \in (Rec(A,k) \cup \{j\}) \setminus \{j\}} dist(k,q') \\
&\quad + dist(k,j) + adding(A,k) \\
&= \sum_{q' \in (nb_A(k) \setminus \{succ(k),pred(k)\})} dist(k,q') + \sum_{q' \in Rec(A,k)} dist(k,q') \\
&\quad + adding(A,k) + dist(k,j) \\
&= \Psi_p(A,k) + dist(k,j)
\end{aligned}$$

Since $\leq$ is according to Definition 11 a total order and $j < k \wedge p < j$, it follows with Definition 15 that $dist(p,k) > dist(k,j)$ and therefore according to Definition 34 it holds that $\Psi(A') < \Psi(A)$.

- $\forall p \in \mathscr{P}.$if $select\big(findLin\big(p, nb_A(p)\big)\big)$ is something else $\Longrightarrow A \longmapsto A'$ with

  $$A' \equiv Alg_{ges}\big(P_A, P'_A, nb_A, Msgs_A, add_A\big)$$

  Obviously $\Psi(A') = \Psi(A)$.

- $\forall p \in P_A.\exists q \in \mathscr{P}.(p,q) \in Msgs_A \Longrightarrow A \longmapsto A'$ with

  $$A' \equiv Alg_{ges}\big(P_A \setminus \{p\}, P'_A \cup \{p\}, nb_A, Msgs_A \setminus \{(p,q)\}, add_A \cup \{(p,q)\}\big)$$

  Let $p \in P_A$ be the executing process with $q \in Procs$ and $(p,q) \in Msgs_A$. The only changes in the configuration that influence $\Psi$ are $Msgs_{A'} = Msgs_A \setminus \{(p,q)\}$ and $add_{A'} = add_A \cup \{(p,q)\}$. According to Definition 33 it holds (as $p \in P_A$ and therefore $add_A(p)$ is not defined) that

  $$adding(A,p) = 0 \quad \text{and} \quad nb_A(p) \setminus \{succ(p), pred(p)\} = nb_{A'}(p) \setminus \{succ(p), pred(p)\}$$

  If $q \notin \{succ(p), pred(p)\}$, it holds further according to Definition 33 that

  $$adding(A',p) = dist(p,q), \quad q \in Rec(A,p) \quad \text{and} \quad Rec(A',p) = Rec(A,p) \setminus \{q\}$$

and hence

$$\begin{aligned}
\Psi_p(A',p) &= \sum_{q'\in(nb_{A'}(p)\setminus\{succ(p),pred(p)\})} dist(p,q') + \sum_{q'\in Rec(A',p)} dist(p,q') \\
&\quad + adding(A',p) \\
&= \sum_{q'\in(nb_A(p)\setminus\{succ(p),pred(p)\})} dist(p,q') + \sum_{q'\in(Rec(A,p)\setminus\{q\})} dist(p,q') \\
&\quad + dist(p,q) \\
&= \sum_{q'\in(nb_A(p)\setminus\{succ(p),pred(p)\})} dist(p,q') + \sum_{q'\in(Rec(A,p)\setminus\{q\})\cup\{q\}} dist(p,q') + 0 \\
&= \sum_{q'\in(nb_A(p)\setminus\{succ(p),pred(p)\})} dist(p,q') + \sum_{q'\in Rec(A,p)} dist(p,q') \\
&\quad + adding(A,p) \\
&= \Psi_p(A,p)
\end{aligned}$$

Otherwise $q \in \{succ(p), pred(p)\}_\perp$ and therefore according to Definition 33 also $adding(A',p) = 0$ and $q \notin Rec(A,p) = Rec(A',p)$. Hence $\Psi_p(A,p) = \Psi_p(A',p)$. In both cases, it follows according to Definition 34 $\Psi(A) = \Psi(A')$.

- $\forall p \in P'_A . \exists q \in \mathscr{P}.(p,q) \in add_A \wedge A \longmapsto A'$ with

$$A' \equiv Alg_{ges}\big(P_A \cup \{p\}, P'_A \setminus \{p\}, nb_{A'}, Msgs_A, add_A \setminus \{(p,q)\}\big)$$

$$\text{with } nb_{A'}(x) = \begin{cases} nb_A(x), & \text{if } x \neq p \\ nb_A(p) \cup \{q\}, & \text{if } x = p \end{cases}$$

Let $p \in P'_A$ be the executing process and $add_A(p) = q$. The only changes in the configuration that influence $\Psi$ are $add_{A'} = add_A \setminus \{(p,q)\}$ and $nb_{A'}$. According to Definition 33 it holds that

$$adding(A',p) = 0$$

(as $p \in P_{A'}$ and therefore $add_{A'}$ is not defined). If $q \notin \{succ(p), pred(p)\}$, it holds further according to Definition 33 that

$$adding(A,p) = dist(p,q) \quad and \quad Rec(A',p) = Rec(A,p)$$

and since $p \neq q$ according to Definition 15 $adding(A,p) = dist(p,q) > 0$. Now there are two cases, $q$ could be already in the neighborhood of $p$ or not. If $q \in nb_A(p)$, then with $q \notin \{succ(p), pred(p)\}$ it follows

$$\begin{aligned}
nb_{A'}(p) \setminus \{succ(p), pred(p)\} &= \big(nb_A(p) \cup \{q\}\big) \setminus \{succ(p), pred(p)\} \\
&= nb_A(p) \setminus \{succ(p), pred(p)\}
\end{aligned}$$

Hence with Definition 33

$$
\begin{aligned}
\Psi_p(A',p) &= \sum_{q'\in(nb_{A'}(p)\setminus\{succ(p),pred(p)\})} dist(p,q') + \sum_{q'\in Rec(A',p)} dist(p,q') \\
&\quad + adding(A',p) \\
&= \sum_{q'\in(nb_A(p)\setminus\{succ(p),pred(p)\})} dist(p,q') + \sum_{q'\in Rec(A,p)} dist(p,q') + 0 \\
&= \sum_{q'\in(nb_A(p)\setminus\{succ(p),pred(p)\})} dist(p,q') + \sum_{q'\in Rec(A,p)} dist(p,q') \\
&\quad + dist(p,q) - dist(p,q) \\
&= \sum_{q'\in(nb_A(p)\setminus\{succ(p),pred(p)\})} dist(p,q') + \sum_{q'\in Rec(A,p)} dist(p,q') \\
&\quad + adding(A,p) - dist(p,q) \\
&= \Psi_p(A,p) - dist(p,q)
\end{aligned}
$$

Therefore, according to Definition 34 $\Psi(A) > \Psi(A) - dist(p,q) = \Psi(A')$.
If $q \notin nb_A(p)$, it follows that $q \notin \big(nb_A(p) \setminus \{succ(p), pred(p)\}\big)$ but $q \in nb_{A'}(p) = nb_A(p) \cup \{q\}$ and hence $q \in \big(nb_{A'}(p) \setminus \{succ(p), pred(p)\}\big)$. It follows with Definition 33 that

$$
\begin{aligned}
\Psi_p(A,p) &= \sum_{q'\in(nb_{A'} p\setminus\{succ(p),pred(p)\})} dist(p,q') + \sum_{q'\in Rec(A',p)} dist(p,q') \\
&\quad + adding(A',p) \\
&= \sum_{q'\in((nb_A(p)\cup\{q\})\setminus\{succ(p),pred(p)\})} dist(p,q') + \sum_{q'\in Rec(A,p)} dist(p,q') + 0 \\
&= \sum_{q'\in(nb_A(p)\setminus\{succ(p),pred(p)\})} dist(p,q') + dist(p,q) + \sum_{q'\in Rec(A,p)} dist(p,q') \\
&= \sum_{q'\in(nb_A(p)\setminus\{succ(p),pred(p)\})} dist(p,q') + \sum_{q'\in Rec(A,p)} dist(p,q') \\
&\quad + adding(A,p) \\
&= \Psi_p(A',p)
\end{aligned}
$$

Hence according to Definition 34 $\Psi(A) = \Psi(A')$.
Otherwise $q \in \{succ(p), pred(p)\}_\perp$ and therefore according to Definition 33 also

$$
\begin{aligned}
&adding(A,p) = 0 \quad \text{and} \\
&q \notin \big(nb_A(p) \setminus \{succ(p), pred(p)\}\big) = \big((nb_A(p) \cup \{q\}) \setminus \{succ(p), pred(p)\}\big) \\
&\qquad\qquad = \big(nb_{A'}(p) \setminus \{succ(p), pred(p)\}\big)
\end{aligned}
$$

Hence $\Psi_p(A,p) = \Psi_p(A',p)$. Hence according to Definition 34 $\Psi(A) = \Psi(A')$.

$\square$

The second potential function $\Psi_E$ considers the nearest left and the nearest right neighbor of a process while taking the messages in transit into account. The nearest left neighbor is the process with the greatest id in the left-neighborhood with messages, i. e., all processes with smaller ids to them the process has an edge in the topology with messages. The nearest right neighbor is defined correspondingly. The potential function $\Psi_E$ sums up the distances to the nearest neighbors with messages for all processes in the system while adding a value that is greater than any possible value for every process that does not have such a neighbor on one of the sides (and is not the smallest respectively greatest process).

**Definition 35: Right- and Left-neighborhood with Messages**

The *left-neighborhood with messages* of a process $LeftNM : \mathcal{T} \times \mathcal{P} \to 2^{\mathcal{P}}$ with

$$LeftNM\big(A,p\big) = \{q \in \mathcal{P} | \big(q \in nb_A(p) \vee \big(p \in P'_A \wedge add_A(p) = q\big) \vee (p,q) \in Msgs_A\big) \wedge q < p\}$$

describes all left neighbors of a process in the directed topology with messages. Therefore, it contains all smaller processes to whom $p$ has an outgoing connection.
The *right-neighborhood with messages* of a process $RightNM : \mathcal{T} \times \mathcal{P} \to 2^{\mathcal{P}}$ with

$$RightNM\big(A,p\big) = \{q \in \mathcal{P} | \big(q \in nb_A(p) \vee \big(p \in P'_A \wedge add_A(p) = q\big) \vee (p,q) \in Msgs_A\big) \wedge p < q\}$$

describes all right neighbors of a process in the directed topology with messages. Therefore, it contains all greater processes to whom $p$ has an outgoing connection.

**Definition 36: Nearest Neighbors of Process with Messages**

The *nearest left neighbor* of a process $ShortestLeftN : \mathcal{T} \times \mathcal{P} \to \mathcal{P} \cup \{\bot\}$ with

$$ShortestLeftN(A,p) = \begin{cases} max\big(LeftNM\big(A,p\big)\big) & \text{if } LeftNM\big(A,p\big) \neq \emptyset \\ \bot & \text{if } LeftNM\big(A,p\big) = \emptyset \end{cases}$$

is the greatest process in the left-neighborhood with messages of a process provided the left-neighborhood is not empty. Therefore, it is the process on the end of the shortest outgoing edge on the left of $p$ in the directed topology with messages.
The *nearest right neighbor* of a process $ShortestRightN : \mathcal{T} \times \mathcal{P} \to \mathcal{P} \cup \{\bot\}$ with

$$ShortestRightN(A,p) = \begin{cases} min\big(RightNM\big(A,p\big)\big) & \text{if } RightNM\big(A,p\big) \neq \emptyset \\ \bot & \text{if } RightNM\big(A,p\big) = \emptyset \end{cases}$$

is the smallest process in the right-neighborhood with messages of a process provided the right-neighborhood is not empty. Therefore, it is the process on the end of the shortest outgoing edge on the right of $p$ in the directed topology with messages.

**Definition 37: Potential Function $\Psi_E$**

The *potential function of left neighbors* for a process $\Psi_{E(p)l} : \mathcal{T} \times \mathcal{P} \to \mathbb{N}$ with

$$\Psi_{E(p)l}(A,p) = \begin{cases} dist(p, ShortestLeftN(A,p)) & \text{, if } ShortestLeftN(A,p) \neq \bot \\ (maxdist + 1) & \text{, if } ShortestLeftN(A,p) = \bot \wedge p \neq min(\mathcal{P}) \\ 0 & \text{, if } p = min(\mathcal{P}) \end{cases}$$

returns the distance to the nearest left neighbor of a process while taking the messages that are still in transit into account. If the process does not have any smaller neighbors with messages (and is not the process with the minimum id) it returns a value that is greater than any possible value.
Correspondingly, the *potential function of right neighbors* for a process $\Psi_{E(p)r} : \mathcal{T} \times \mathcal{P} \to \mathbb{N}$ with

$$\Psi_{E(p)r}(A,p) = \begin{cases} dist(p, ShortestRightN(A,p)) & \text{, if } ShortestRightN(A,p) \neq \bot \\ (maxdist + 1) & \text{, if } ShortestRightN(A,p) = \bot \wedge p \neq max(\mathcal{P}) \\ 0 & \text{, if } p = max(\mathcal{P}) \end{cases}$$

returns the distance to the nearest right neighbor of a process while taking the messages that are still in transit into account. If the process does not have any greater neighbors with messages (and is not the process with the maximum id) it returns a value that is greater than any possible value.
Therefore, the *potential function of neighbors* for a process $\Psi_{E(p)} : \mathcal{T} \times \mathcal{P} \to \mathbb{N}$ with

$$\Psi_{E(p)}(A,p) = \Psi_{E(p)l}(A,p) + \Psi_{E(p)r}(A,p)$$

returns the sum of the distances to the nearest left and the nearest right neighbor of the process.
The *potential function of neighbors* for a configuration $\Psi_E : \mathcal{T} \to \mathbb{N}$ with

$$\Psi_E(A) = \sum_{p \in \mathcal{P}} \Psi_{E(p)}(A,p)$$

sums up all potential functions of neighbors for all processes in the configuration.

The potential function $\Psi_E$ is therefore minimal if every process, with exception of the smallest and greatest process, has a neighbor with messages on both sides and this neighbor is the nearest neighbor that is possible in the system. This is the case if it holds for every process that the predecessor is in the right-neighborhood with messages and the successor is in the left-neighborhood with messages. This holds if every correct edge is already established in the topology with messages and therefore the linear graph is a subgraph of the topology with messages.

**Lemma 16: Potential $\Psi_E$ Minimal for $G_{LIN} \subseteq NTM$**

Let $A$ be an arbitrary configuration of the algorithm. Then $\Psi_E(A)$ is minimal if the desired topology is contained in the directed topology with messages i. e.,

$$\Psi_E(A) = 2 \cdot (n-1) \quad \text{iff} \quad G_{LIN} \subseteq NTM(A)$$

*Proof:*
It holds with Definitions 15, 35, 36 and 37 that

$$\forall p \in \mathscr{P}. \big(p \neq max(\mathscr{P}) \implies \Psi_{E(p)r}(A,p) \geq 1\big) \wedge \\ \big(p \neq min(\mathscr{P}) \in \mathscr{P} \implies \Psi_{E(p)l}(A,p) \geq 1\big)$$

Therefore, it follows with $|\mathscr{P}| = n$ that $\Psi_E(A) \geq 2 \cdot (n-1)$ for an arbitrary configuration $A$.
If $G_{LIN} \subseteq NTM(A)$, it holds according to Definitions 14 and 28 that

$$\forall p \in \mathscr{P}. \big(succ(p) \in \mathscr{P} \implies succ(p) \in nb_A(p) \vee succ(p) \in Msgs_A \vee add_A(p) = succ(p)\big) \wedge \\ \big(pred(p) \in \mathscr{P} \implies pred(p) \in nb_A(p) \vee pred(p) \in Msgs_A \vee add_A(p) = pred(p)\big)$$

Hence with Definition 35

$$\forall p \in \mathscr{P}. \big(succ(p) \in \mathscr{P} \implies succ(p) \in RightNM\big(A,p\big)\big) \wedge \\ \big(pred(p) \in \mathscr{P} \implies pred(p) \in LeftNM\big(A,p\big)\big)$$

Therefore, with Definition 12, Corollary 1 and Definition 36

$$\forall p \in \mathscr{P}. \big(succ(p) \in \mathscr{P} \implies ShortestRightN(A,p) = succ(p)\big) \wedge \\ \big(pred(p) \in \mathscr{P} \implies ShortestLeftN(A,p) = pred(p)\big)$$

Hence with Definitions 15 and 37 it holds

$$\forall p \in \mathscr{P}. \Psi_{E(p)l}(A,p) = \begin{cases} 1, & \text{if } p \neq min(\mathscr{P}) \\ 0, & \text{if } p = min(\mathscr{P}) \end{cases} \wedge \Psi_{E(p)r}(A,p) = \begin{cases} 1, & \text{if } p \neq max(\mathscr{P}) \\ 0, & \text{if } p = max(\mathscr{P}) \end{cases}$$

and therefore with $|\mathscr{P}| = n$ it holds that $\Psi_E(A) = 2 \cdot (n-1)$.
If $G_{LIN} \not\subseteq NTM(A)$, it holds according to Definitions 14 and 28 that

$$\exists p \in \mathscr{P}. \big(succ(p) \in \mathscr{P} \wedge succ(p) \notin nb_A(p) \wedge succ(p) \notin Msgs_A \wedge add_A(p) \neq succ(p)\big) \vee \\ \big(pred(p) \in \mathscr{P} \wedge pred(p) \notin nb_A(p) \wedge pred(p) \notin Msgs_A \wedge add_A(p) \neq pred(p)\big)$$

Let $p \in \mathscr{P}$ with $succ(p) \in \mathscr{P} \wedge succ(p) \notin nb_A(p) \wedge succ(p) \notin Msgs_A \wedge add_A(p) \neq succ(p)$ (the case for $pred(p)$ is similar). Hence $p \neq max(\mathscr{P})$ and with Definition 35 $succ(p) \notin RightNM(A, p)$. Now there are two cases. If $RightNM(A, p) = \emptyset$, it holds according to Definition 36 that $ShortestRightN(A, p) = \perp$. Therefore, with Definition 37 $\Psi_{E(p)r}(A, p) = maxdist + 1 = n$. If $RightNM(A, p) \neq \emptyset$, then there is a process $q \in RightNM(A, p)$ with $min(RightNM(A, p)) = q$ and therefore with Definition 36 $ShortestRightN(A, p) = q$ and according to Definition 35 $q > p$. Since $q \neq succ(p)$, it holds according to Corollary 1 and Definition 15 that $dist(p, q) > 1$ and therefore with Definition 37 $\Psi_{E(p)r}(A, p) > 1$.
Hence $\Psi_E(A) \geq 2 \cdot n - 1 > 2 \cdot (n-1)$. □

The potential function $\Psi_E$ is monotonically decreasing. The only steps that remove edges are linearization steps and they always remove the process that is further away. Since additionally no messages get lost, the nearest neighbor with message gets never removed. The function decreases whenever a connection is established to a process that is closer than all already known processes. This can happen through sending of *keep-alive*-messages or linearization steps.

**Lemma 17: Progression $\Psi_E$**

Let $A$ be an arbitrary configuration. For $A \longmapsto A'$ holds:

- $\Psi_E(A') < \Psi_E(A)$ iff
  - an arbitrary process $p \in \mathscr{P}$ finds a left linearization step and the greater process is nearer to the smaller process than every yet known right neighbor with messages i. e.,

$$\begin{aligned} & select(findLin(p, nb_A(p))) = (j, k) \wedge (j < k \wedge k < p) \\ \wedge \quad & \forall q \in RightNM(A, j) \,.\, k < q \end{aligned}$$

    it then holds

$$A' \equiv Alg_{ges}(P_A, P'_A, nb_{A'}, Msgs_A \cup \{(j,k)\}, add_A)$$
$$\text{with } nb_{A'}(x) = \begin{cases} nb_A(x), & \text{if } x \neq p \\ nb_A(p) \setminus \{j\}, & \text{if } x = p, \end{cases}$$

  - an arbitrary process $p \in \mathscr{P}$ finds a right linearization step and the smaller process is nearer to the greater process than every yet known left neighbor with messages i. e.,

$$\begin{aligned} & select(findLin(p, nb_A(p))) = (j, k) \wedge (j < k \wedge p < j) \\ \wedge \quad & \forall q \in LeftNM(A, k) \,.\, q < j \end{aligned}$$

it then holds

$$A' \equiv Alg_{ges}\left(P_A, P'_A, nb_{A'}, Msgs_A \cup \{(k,j)\}, add_A\right)$$

$$\text{with } nb_{A'}(x) = \begin{cases} nb_A(x), & \text{if } x \neq p \\ nb_A(p) \setminus \{k\}, & \text{if } x = p, \end{cases}$$

– or if an arbitrary process $p \in \mathcal{P}$ sends *keep-alive*-messages and is a nearer neighbor for at least one of its (at most two) neighbors i. e.,

$$\begin{aligned} & select\big(findLin\big(p, nb_A(p)\big)\big) = \bot \\ \wedge \quad & \big(\exists q \in LeftN\big(nb_A(p), p\big).\forall q' \in RightNM\big(A, q\big).p < q' \\ \vee \quad & \exists q \in RightN\big(nb_A(p), p\big).\forall q' \in LeftNM\big(A, q\big).q' < p\big) \end{aligned}$$

it then holds $A' \equiv Alg_{ges}\left(P_A, P'_A, nb_A, Msgs_A \cup \{(j,p) | j \in nb_A(p)\}, add_A\right)$

- $\Psi_E(A') = \Psi_E(A)$ iff
  – an arbitrary process $p \in P_A$ receives a message i. e.,

$$q \in \mathcal{P} \quad \wedge \quad (p,q) \in Msgs_A$$

it then holds $A' \equiv Alg_{ges}\left(P_A \setminus \{p\}, P'_A \cup \{p\}, nb_A, Msgs_A \setminus \{(p,q)\}, add_A \cup \{(p,q)\}\right)$,

– an arbitrary process $p \in P'_A$ adds a process to its neighborhood i. e.,

$$q \in \mathcal{P} \quad \wedge \quad add_A(p) = q$$

it holds

$$A' \equiv Alg_{ges}\left(P_A \cup \{p\}, P'_A \setminus \{p\}, nb_{A'}, Msgs_A, add_A \setminus \{(p,q)\}\right)$$

$$\text{with } nb_{A'}(x) = \begin{cases} nb(x), & \text{if } x \neq p \\ nb(p) \cup \{q\}, & \text{if } x = p, \end{cases}$$

– an arbitrary process $p \in \mathcal{P}$ finds a left linearization step but the smaller process has already a nearer right neighbor with messages than the greater process i. e.,

$$\begin{aligned} & select\big(findLin\big(p, nb_A(p)\big)\big) = (j,k) \wedge (j < k \wedge k < p) \\ \wedge \quad & \exists q \in RightNM\big(A, j\big).q \leq k \end{aligned}$$

it then holds

$$A' \equiv Alg_{ges}\left(P_A, P'_A, nb_{A'}, Msgs_A \cup \{(j,k)\}, add_A\right)$$

$$\text{with } nb_{A'}(x) = \begin{cases} nb_A(x), & \text{if } x \neq p \\ nb_A(p) \setminus \{j\}, & \text{if } x = p, \end{cases}$$

- an arbitrary process $p \in \mathscr{P}$ finds a right linearization step but the greater process has already a nearer left neighbor with messages than the smaller process i. e.,

$$\begin{aligned} & select\big(findLin\big(p, nb_A(p)\big)\big) = (j,k) \wedge (j < k \wedge p < j) \\ \wedge \quad & \exists q \in LeftNM(A,k).j \le q \end{aligned}$$

  it then holds

$$A' \equiv Alg_{ges}\big(P_A, P'_A, nb_{A'}, Msgs_A \cup \{(k,j)\}, add_A\big)$$
$$\text{with } nb_{A'}(x) = \begin{cases} nb_A(x), & \text{if } x \neq p \\ nb_A(p) \setminus \{k\}, & \text{if } x = p, \end{cases}$$

- if an arbitrary process $p \in \mathscr{P}$ sends *keep-alive*-messages and every neighbor already knows with messages a nearer process i. e.,

$$\begin{aligned} & select\big(findLin\big(p, nb_A(p)\big)\big) = \bot \\ \wedge \quad & \forall q \in LeftN\big(nb_A(p), p\big).\exists q' \in RightNM\big(A,q\big).q' \le p \\ \wedge \quad & \forall q \in RightN\big(nb_A(p), p\big).\exists q' \in LeftNM\big(A,q\big).p \le q' \end{aligned}$$

  it then holds $A' \equiv Alg_{ges}\big(P_A, P'_A, nb_A, Msgs_A \cup \{(j,p) | j \in nb_A(p)\}, add_A\big)$,
- or an arbitrary process $p \in \mathscr{P}$ tries to find a linearization step and none of the other cases is true. It then holds $A' \equiv A$.

and therefore $\Psi_E$ is monotonically decreasing.
*Proof*:
According to Corollary 2 there are the following cases possible:

- If a process $p \in \mathscr{P}$ tries to find a linearization step but $select\big(findLin\big(p, nb_A(p)\big)\big) = \bot$, it holds that

$$A' \equiv Alg_{ges}\big(P_A, P'_A, nb_A, Msgs_A \cup \{(j,p) | j \in nb_A(p)\}, add_A\big)$$

  According to Lemma 6 it holds

$$|nb_A(p)| \le 2 \wedge \big(|nb_A(p)| = 2 \implies LeftN\big(nb_A(p), p\big) = 1 \wedge RightN\big(nb_A(p), p\big) = 1\big)$$

  - if $nb_A(p) = \emptyset$, it holds that $A' \equiv A$ and therefore obviously $\Psi_{E(p)}(A', p) = \Psi_{E(p)}(A, p)$ and $\Psi_E(A') = \Psi_E(A)$.
  - if $|LeftN\big(nb_A(p), p\big)| = 1 \wedge |RightN\big(nb_A(p), p\big)| = 0$:
    Let $j$ be the process with $LeftN\big(nb_A(p), p\big) = \{j\}$. According to the definition of *LeftN* it holds that $j < p$.
    * if $RightNM\big(A, j\big) = \emptyset$:

It holds $RightNM\big(A', j\big) = \{p\}$ and therefore obviously $min\big(RightNM\big(A', j\big)\big) = p$. With Definition 36 it holds

$$ShortestRightN(A, j) = \bot \quad \text{and} \quad ShortestRightN(A', j) = p$$

Hence with Definition 37

$$\begin{aligned}
\Psi_{E(p)}(A', j) &= \Psi_{E(p)l}(A', j) + \Psi_{E(p)r}(A', j) \\
&= \Psi_{E(p)l}(A', j) + dist(j, ShortestRightN(A', j)) \\
&= \Psi_{E(p)l}(A, j) + dist(j, ShortestRightN(A', j)) \\
&< \Psi_{E(p)l}(A, j) + maxdist + 1 \\
&= \Psi_{E(p)l}(A, j) + \Psi_{E(p)r}(A, j) \\
&= \Psi_{E(p)}(A, j)
\end{aligned}$$

Therefore, with Definition 37

$$\Psi_E(A') < \Psi_E(A)$$

* if $RightNM\big(A, j\big) \neq \emptyset \wedge \forall q \in RightNM\big(A, j\big).p < q$:
  if for all processes $q \in RightNM\big(A, j\big)$ it holds that $p < q$, then it follows with Definition 36

$$ShortestRightN(A', j) = p$$

Let $r$ be the process with $ShortestRightN(A, j) = r$. Since $j < p < r$, it holds according to Definition 15 $dist(j, p) < dist(j, r)$ and therefore with Definition 37

$$\begin{aligned}
\Psi_{E(p)}(A', j) &= \Psi_{E(p)l}(A', j) + \Psi_{E(p)r}(A', j) \\
&= \Psi_{E(p)l}(A', j) + dist(j, ShortestRightN(A', j)) \\
&= \Psi_{E(p)l}(A, j) + dist(j, p) \\
&< \Psi_{E(p)l}(A, j) + dist(j, r) \\
&= \Psi_{E(p)l}(A, j) + dist(j, ShortestRightN(A, j)) \\
&= \Psi_{E(p)l}(A, j) + \Psi_{E(p)r}(A, j) \\
&= \Psi_{E(p)}(A, j)
\end{aligned}$$

Therefore, with Definition 37

$$\Psi_E(A') < \Psi_E(A)$$

* if $\exists q \in RightNM\big(A, j\big).q \leq p$:
  if there is a process $q \in RightNM\big(A, j\big)$ with $q \leq p$, then it holds according to

Definition 36 that

$$\begin{aligned} ShortestRightN(A', j) &= min\big(RightNM\big(A', j\big)\big) \\ &= min\big(RightNM\big(A, j\big) \cup \{p\}\big) \\ &= min\big(RightNM\big(A, j\big)\big) \\ &= ShortestRightN(A, j) \end{aligned}$$

Therefore, with Definition 37 $\Psi_{E(p)}(A', j) = \Psi_{E(p)}(A, j)$ and hence

$$\Psi_E(A') = \Psi_E(A)$$

– if $|RightN\big(nb_A(p), p\big)| = 1 \wedge |LeftN\big(nb_A(p), p\big)| = 0$:
Let $k$ be the process with $RightN\big(nb_A(p), p\big) = \{k\}$. According to the definition of $RightN$ it holds that $p < k$.

  * if $LeftNM(A, k) = \emptyset$:
    It holds $LeftNM\big(A', k\big) = \{p\}$ and therefore it obviously $max\big(LeftNM\big(A', k\big)\big) = p$. With Definition 36 it holds

$$ShortestLeftN(A, k) = \perp \quad \text{and} \quad ShortestLeftN(A', k) = p$$

    Hence with Definition 37

$$\begin{aligned} \Psi_{E(p)}(A', k) &= \Psi_{E(p)l}(A', k) + \Psi_{E(p)r}(A', k) \\ &= dist(k, ShortestLeftN(A', k)) + \Psi_{E(p)r}(A', k) \\ &= dist(k, ShortestRightN(A', k)) + \Psi_{E(p)r}(A, k) \\ &< maxdist + 1 + \Psi_{E(p)r}(A, k) \\ &= \Psi_{E(p)l}(A, k) + \Psi_{E(p)r}(A, k) \\ &= \Psi_{E(p)}(A, k) \end{aligned}$$

    Therefore, with Definition 37

$$\Psi_E(A') < \Psi_E(A)$$

  * if $LeftNM(A, k) \neq \emptyset \wedge \forall q \in LeftNM(A, k).q < p$:
    if for all processes $q \in LeftNM(A, k)$ it holds that $q < p$, then it follows with Definition 36

$$ShortestLeftN(A', k) = j$$

    Let $r$ be the process with $ShortestLeftN(A, k) = r$. Since $r < p < k$, it holds according to Definition 15 $dist(k, p) < dist(k, r)$ and therefore with Definition

37

$$\begin{aligned}\Psi_{E(p)}(A',k) &= \Psi_{E(p)l}(A',k) + \Psi_{E(p)r}(A',k)\\ &= dist(k, ShortestLeftN(A',k)) + \Psi_{E(p)r}(A',k)\\ &= dist(k,p) + \Psi_{E(p)r}(A,k)\\ &< dist(k,r) + \Psi_{E(p)r}(A,k)\\ &= dist(k, ShortestRightN(A,k)) + \Psi_{E(p)r}(A,k)\\ &= \Psi_{E(p)l}(A,k) + \Psi_{E(p)r}(A,k)\\ &= \Psi_{E(p)}(A,k)\end{aligned}$$

Therefore, with Definition 37

$$\Psi_E(A') < \Psi_E(A)$$

* if $\exists q \in LeftNM(A,k).p \leq q$:
  if there is a process $q \in LeftNM(A,k)$ with $p \leq q$, then it holds according to Definition 36 that

$$\begin{aligned}ShortestLeftN(A',k) &= max\big(RightNM\big(A',k\big)\big)\\ &= max\big(LeftNM(A,k) \cup \{p\}\big)\\ &= max(LeftNM(A,k))\\ &= ShortestLeftN(A,k)\end{aligned}$$

  Therefore, with Definition 37 $\Psi_{E(p)}(A',k) = \Psi_{E(p)}(A,k)$ and hence

$$\Psi_E(A') = \Psi_E(A)$$

– if $|LeftN\big(nb_A(p),p\big)| = 1 \wedge |RightN\big(nb_A(p),p\big)| = 1$:
  Let $j$ be the process with $LeftN\big(nb_A(p),p\big) = \{j\}$ and let $k$ be the process with $RightN\big(nb_A(p),p\big) = \{k\}$.

* if $RightNM\big(A,j\big) = \emptyset \vee \big(RightNM\big(A,j\big) \neq \emptyset \wedge \forall q \in RightNM\big(A,j\big).p < q\big) \vee$ $LeftNM(A,k) = \emptyset \vee \big(LeftNM(A,k) \neq \emptyset \wedge \forall q \in LeftNM(A,k).q < p\big)$:
  It holds that either

$$\begin{aligned}&\Psi_{E(p)}(A',j) < \Psi_{E(p)}(A,j) \wedge \Psi_{E(p)}(A',k) = \Psi_{E(p)}(A,k),\\ &\Psi_{E(p)}(A',j) = \Psi_{E(p)}(A,j) \wedge \Psi_{E(p)}(A',k) < \Psi_{E(p)}(A,k) \text{ or}\\ &\Psi_{E(p)}(A',j) < \Psi_{E(p)}(A,j) \wedge \Psi_{E(p)}(A',k) < \Psi_{E(p)}(A,k)\end{aligned}$$

  Therefore

$$\Psi_E(A') < \Psi_E(A)$$

* if $\exists q \in RightNM(A, j).q \le p \wedge \exists q \in LeftNM(A, k).p \le q$: It holds that $\Psi_{E(p)}(A', j) = \Psi_{E(p)}(A, j)$ and $\Psi_{E(p)}(A', k) = \Psi_{E(p)}(A, k)$. Therefore

$$\Psi_E(A') = \Psi_E(A)$$

  The proof is similar to the proofs in the previous cases.

- If a process $p \in \mathscr{P}$ finds a left linearization step i. e., $select\big(findLin\big(p, nb_A(p)\big)\big) = (j, k) \wedge (j < k \wedge k < p)$, it holds that

$$A' \equiv Alg_{ges}\left(P_A, P'_A, nb_{A'}, Msgs_A \cup \{(j,k)\}, add_A\right)$$
$$\text{with } nb_{A'}(x) = \begin{cases} nb_A(x), & \text{if } x \ne p \\ nb_A(p) \setminus \{j\}, & \text{if } x = p \end{cases}$$

Since $j < p$ and $k < p$ and $j, k \in nb_{A'}(p)$, it holds according to the definition of *LeftN* that $j, k \in LeftN\big(nb_A(p), p\big)$ and with $j < k$ therefore $max\big(LeftN\big(nb_A(p), p\big)\big) \ne j$. Hence with Definition 35 also $max\big(LeftNM\big(A, p\big)\big) \ne j$. Since $j < p$, it holds according to Definition 35 that $j \notin RightNM\big(A, p\big)$. With $nb_{A'}(p) = nb_A(p) \setminus \{j\}$, $Msgs_{A'} = Msgs_A \cup \{(j,k)\}$ and $add_{A'} = add_A$ it holds according to Definition 35 $LeftNM\big(A', p\big) = LeftNM\big(A, p\big)$ and $RightNM\big(A', p\big) = RightNM\big(A, p\big)$. Therefore, with Definition 36 also

$$ShortestLeftN(A', p) = ShortestLeftN(A, p) \qquad \text{and}$$
$$ShortestRightN(A', p) = ShortestRightN(A, p)$$

Hence with Definition 37 $\Psi_{E(p)}(A', p) = \Psi_{E(p)}(A, p)$. Since $nb_{A'}(j) = nb_A(j)$, $add_{A'} = add_A$, $Msgs_{A'} = Msgs_A \cup \{(j,k)\}$ and $j < k$, it holds with Definition 35 that $LeftNM\big(A', j\big) = LeftNM\big(A, j\big)$ and $RightNM\big(A', j\big) = RightNM\big(A, j\big) \cup \{k\}$. With Definition 36 it holds

$$ShortestLeftN(A', j) = ShortestLeftN(A, j)$$

There are the following cases for $\Psi_{E(p)}(A', j)$ since with $j < k$ it holds $j \ne max(\mathscr{P})$:

  – if $RightNM\big(A, j\big) = \emptyset$: It holds $RightNM\big(A', j\big) = \{k\}$ and therefore it holds obviously $min\big(RightNM\big(A', j\big)\big) = k$. With Definition 36

$$ShortestRightN(A, j) = \bot \quad \text{and} \quad ShortestRightN(A', j) = k$$

  Hence with Definition 37

$$\begin{aligned}\Psi_{E(p)}(A', j) &= \Psi_{E(p)l}(A', j) + \Psi_{E(p)r}(A', j) \\ &= \Psi_{E(p)l}(A', j) + dist(j, ShortestRightN(A', j)) \\ &= \Psi_{E(p)l}(A, j) + dist(j, ShortestRightN(A', j)) \\ &< \Psi_{E(p)l}(A, j) + maxdist + 1\end{aligned}$$

$$= \Psi_{E(p)l}(A,j) + \Psi_{E(p)r}(A,j)$$
$$= \Psi_{E(p)}(A,j)$$

Therefore, with Definition 37

$$\Psi_E(A') < \Psi_E(A)$$

– if $RightNM(A,j) \neq \emptyset \wedge \forall q \in RightNM(A,j).k < q$: if for all processes $q \in RightNM(A,j)$ it holds that $k < q$, then it follows with Definition 36

$$ShortestRightN(A',j) = k$$

Let $r$ be the process with $ShortestRightN(A,j) = r$. Since $j < k < r$, it holds according to Definition 15 $dist(j,k) < dist(j,r)$ and therefore with Definition 37

$$\begin{aligned}
\Psi_{E(p)}(A',j) &= \Psi_{E(p)l}(A',j) + \Psi_{E(p)r}(A',j) \\
&= \Psi_{E(p)l}(A',j) + dist(j, ShortestRightN(A',j)) \\
&= \Psi_{E(p)l}(A,j) + dist(j,k) \\
&< \Psi_{E(p)l}(A,j) + dist(j,r) \\
&= \Psi_{E(p)l}(A,j) + dist(j, ShortestRightN(A,j)) \\
&= \Psi_{E(p)l}(A,j) + \Psi_{E(p)r}(A,j) \\
&= \Psi_{E(p)}(A,j)
\end{aligned}$$

Therefore, with Definition 37

$$\Psi_E(A') < \Psi_E(A)$$

– if $\exists q \in RightNM(A,j).q \leq k$: if there is a process $q \in RightNM(A,j)$ with $q \leq k$, then it holds according to Definition 36 that

$$\begin{aligned}
ShortestRightN(A',j) &= min(RightNM(A',j)) \\
&= min(RightNM(A,j) \cup \{k\}) \\
&= min(RightNM(A,j)) \\
&= ShortestRightN(A,j)
\end{aligned}$$

Therefore, with Definition 37 $\Psi_{E(p)}(A',j) = \Psi_{E(p)}(A,j)$ and hence

$$\Psi_E(A') = \Psi_E(A)$$

- If a process $p \in \mathscr{P}$ finds a right linearization step i. e., $select(findLin(p, nb_A(p))) =$

$(j,k) \wedge (j<k \wedge p<j)$ it holds that

$$A' \equiv Alg_{ges}\left(P_A, P'_A, nb_{A'}, Msgs_A \cup \{(k,j)\}, add_A\right)$$

$$\text{and } nb_{A'}(x) = \begin{cases} nb_A(x), & \text{if } x \neq p \\ nb_A(p) \setminus \{k\}, & \text{if } x = p \end{cases}$$

Since $j<k$ and $p<j$ and $j,k \in nb_{A'}(p)$, it holds according to the definition of $RightN$ that $j,k \in RightN\big(nb_A(p),p\big)$ and with $j<k$ therefore $min\big(RightN\big(nb_A(p),p\big)\big) \neq k$. Hence with Definition 35 also $min\big(RightNM\big(nb_A(p),p\big)\big) \neq k$. Since $p<k$, it holds according to Definition 35 that $k \notin LeftNM\big(A,p\big)$. With $nb_{A'}(p) = nb_A(p) \setminus \{k\}$, $Msgs_{A'} = Msgs_A \cup \{(k,j)\}$ and $add_{A'} = add_A$ it holds according to Definition 35 $LeftNM\big(A',p\big) = LeftNM\big(A,p\big)$ and $RightNM\big(A',p\big) = RightNM\big(A,p\big)$. Therefore, with Definition 36 also

$$ShortestLeftN(A',p) = ShortestLeftN(A,p) \qquad \text{and}$$
$$ShortestRightN(A',p) = ShortestRightN(A,p)$$

Hence with Definition 37 holds $\Psi_{E(p)}(A',p) = \Psi_{E(p)}(A,p)$. Since $nb_{A'}(k) = nb_A(k)$, $add_{A'} = add_A$, $Msgs_{A'} = Msgs_A \cup \{(k,j)\}$ and $j<k$, it follows with Definition 35 that $RightNM\big(A',k\big) = RightNM(A,k)$ and $LeftNM\big(A',k\big) = LeftNM(A,k) \cup \{j\}$. With Definition 36 therefore

$$ShortestRightN(A',j) = ShortestRightN(A,j)$$

There are the following cases for $\Psi_{E(p)}(A',k)$ since with $j<k$ it holds $k \neq min(\mathscr{P})$:

- if $LeftNM(A,k) = \emptyset$: It holds $LeftNM\big(A',k\big) = \{j\}$ and therefore it holds obviously $max\big(LeftNM\big(A',k\big)\big) = j$. With Definition 36

  $$ShortestLeftN(A,k) = \bot \quad \text{and} \quad ShortestLeftN(A',k) = j$$

  Hence with Definition 37

  $$\begin{aligned} \Psi_{E(p)}(A',k) &= \Psi_{E(p)l}(A',k) + \Psi_{E(p)r}(A',k) \\ &= dist(k, ShortestLeftN(A',k)) + \Psi_{E(p)r}(A',k) \\ &= dist(k, ShortestRightN(A',k)) + \Psi_{E(p)r}(A,k) \\ &< maxdist + 1 + \Psi_{E(p)r}(A,k) \\ &= \Psi_{E(p)l}(A,k) + \Psi_{E(p)r}(A,k) \\ &= \Psi_{E(p)}(A,k) \end{aligned}$$

  Therefore, with Definition 37

  $$\Psi_E(A') < \Psi_E(A)$$

– if $LeftNM(A,k) \neq \emptyset \wedge \forall q \in LeftNM(A,k).q < j$: if for all processes $q \in LeftNM(A,k)$ it holds that $q < j$, then it follows with Definition 36

$$ShortestLeftN(A',k) = j$$

Let $r$ be the process with $ShortestLeftN(A,k) = r$. Since $r < j < k$, it holds according to Definition 15 $dist(k,j) < dist(k,r)$ and therefore with Definition 37

$$\begin{aligned}
\Psi_{E(p)}(A',k) &= \Psi_{E(p)l}(A',k) + \Psi_{E(p)r}(A',k) \\
&= dist(k, ShortestLeftN(A',k)) + \Psi_{E(p)r}(A',k) \\
&= dist(k,j) + \Psi_{E(p)r}(A,k) \\
&< dist(k,r) + \Psi_{E(p)r}(A,k) \\
&= dist(k, ShortestRightN(A,k)) + \Psi_{E(p)r}(A,k) \\
&= \Psi_{E(p)l}(A,k) + \Psi_{E(p)r}(A,k) \\
&= \Psi_{E(p)}(A,k)
\end{aligned}$$

Therefore, with Definition 37

$$\Psi_E(A') < \Psi_E(A)$$

– if $\exists q \in LeftNM(A,k).j \leq q$: if there is a process $q \in LeftNM(A,k)$ with $j \leq q$, then it holds according to Definition 36 that

$$\begin{aligned}
ShortestLeftN(A',k) &= max\big(RightNM\big(A',k\big)\big) \\
&= max\big(LeftNM(A,k) \cup \{j\}\big) \\
&= max(LeftNM(A,k)) \\
&= ShortestLeftN(A,k)
\end{aligned}$$

Therefore, with Definition 37 $\Psi_{E(p)}(A',k) = \Psi_{E(p)}(A,k)$ and hence

$$\Psi_E(A') = \Psi_E(A)$$

- If for a process $p \in \mathscr{P}$ it could be that $select\big(findLin\big(p, nb_A(p)\big)\big)$ is something else, it holds that $A' \equiv A$. Therefore, obviously $\Psi_E(A') = \Psi_E(A)$

- If a process $p \in P_A$ receives a message with the id of $q \in \mathscr{P}$ i. e., $(p,q) \in Msgs_A$, it holds

$$A' \equiv Alg_{ges}\left(P_A \setminus \{p\}, P'_A \cup \{p\}, nb_A, Msgs_A \setminus \{(p,q)\}, add_A \cup \{(p,q)\}\right)$$

Since $nb_{A'} = nb_A$, $Msgs_{A'} = Msgs_A \cup \{(p,q)\}$ and $add_{A'}(p) = q$, it holds according to Definition 35 $LeftNM\big(A',p\big) = LeftNM\big(A,p\big)$ and $RightNM\big(A',p\big) = RightNM\big(A,p\big)$. There-fore, with Definition 36 also

$$ShortestLeftN(A', p) = ShortestLeftN(A, p) \qquad \text{and}$$
$$ShortestRightN(A', p) = ShortestRightN(A, p)$$

Hence with Definition 37 $\Psi_{E(p)}(A', p) = \Psi_{E(p)}(A, p)$ and

$$\Psi_E(A') = \Psi_E(A)$$

- If a process $p \in P'_A$ adds a process $q \in \mathscr{P}$ to its neighborhood i. e., $(p, q) \in add_A$ it holds

$$A' \equiv Alg_{ges}\big(P_A \cup \{p\}, P'_A \setminus \{p\}, nb_{A'}, Msgs_A, add_A \setminus \{(p, q)\}\big)$$
$$\text{with } nb_{A'}(x) = \begin{cases} nb(x), & \text{if } x \neq p \\ nb(p) \cup \{q\}, & \text{if } x = p \end{cases}$$

Since $nb_{A'}(p) = nb_A(p) \cup \{p\}$, $Msgs_{A'} = Msgs_A$ and $add_A(p) = q$, it holds according to Definition 35 $LeftNM\big(A', p\big) = LeftNM\big(A, p\big)$ and $RightNM\big(A', p\big) = RightNM\big(A, p\big)$. Therefore, with Definition 36 also

$$ShortestLeftN(A', p) = ShortestLeftN(A, p) \qquad \text{and}$$
$$ShortestRightN(A', p) = ShortestRightN(A, p)$$

Hence with Definition 37 $\Psi_{E(p)}(A', p) = \Psi_{E(p)}(A, p)$ and

$$\Psi_E(A') = \Psi_E(A)$$

□

**Corollary 12: Monotonicity of $\Psi_E$**

Let $A$ be an arbitrary configuration. It holds for every reachable configuration $R$ i. e., $A \Longmapsto R$ that $\Psi_E(R) \leq \Psi_E(A)$ and therefore $\Psi_E$ is monotonically decreasing.

The third potential function $\Psi_\Sigma$ is the weighted sum and therefore a combination of the other two functions. Therefore, the potential function $\Psi_\Sigma$ is minimal if both composite potential functions $\Psi$ and $\Psi_E$ are minimal. This is the case if the topology with messages is the linear graph. The functions are weighted in a way that $\Psi_\Sigma$ decreases with every linearization step and whenever a process gets to know a closer process than every yet known neighbor with messages on this side. The sending of *keep-alive*-messages to undesired neighbors gets compensated by establishing a connection to a nearer process. Therefore, whenever a process sends *keep-a-live*-messages it holds that if at least one of the at most two receiving processes gets to know a nearer process, the potential function decreases. The potential function is monotonically decreasing with exception of sending *keep-alive*-messages to undesired neighbors that already have nearer neighbors.

**Definition 38: Potential $\Psi_{\sum}$**

The *potential sumfunction* of a configuration $\Psi_{\sum} : \mathcal{T} \to \mathbb{N}$ with

$$\Psi_{\sum}(A) = \Psi(A) + n \cdot \Psi_E(A)$$

is the weighted sum of the potential function of neighbors $\Psi_E$ and the potential function for configurations $\Psi$.

**Lemma 18: Minimal Potential $\Psi_{\sum}$ for $\boldsymbol{NTM} = \boldsymbol{G_{LIN}}$**

Let $A$ be an arbitrary configuration. It holds $\Psi_{\sum}(A)$ is minimal iff the directed topology with messages is the linear graph i. e.,

$$\Psi_{\sum}(A) = 2 \cdot n \cdot (n-1) \quad \text{iff} \quad NTM(A) = G_{LIN}$$

*Proof:*
According of Definition 38 $\Psi_{\sum}(A)$ is minimal iff $\Psi(A)$ and $\Psi_E(A)$ are both minimal. $\Psi_E(A)$ is according to Lemma 16 minimal with $\Psi_E(A) = 2 \cdot (n-1)$ iff $G_{LIN} \subseteq NTM(A)$. $\Psi(A)$ is minimal with $\Psi(A) = 0$ according to Lemma 14 iff $NTM(A) \subseteq G_{LIN}$. Therefore, with Definition 38 $\Psi_{\sum}(A)$ is minimal with $\Psi_{\sum}(A) = 2 \cdot n \cdot (n-1)$ iff $NTM(A) = G_{LIN}$. □

**Lemma 19: Progression $\Psi_{\sum}$**

Let $A$ be an arbitrary configuration. For $A \longmapsto A'$ it holds:

- $\Psi_{\sum}(A') < \Psi_{\sum}(A)$ iff
  - an arbitrary process $p \in \mathcal{P}$ finds a left linearization step i. e.,

    $$select\big(findLin\big(p, nb_A(p)\big)\big) = (j,k) \wedge (j < k \wedge k < p)$$

    and it holds that

    $$A' \equiv Alg_{ges}\big(P_A, P'_A, nb_{A'}, Msgs_A \cup \{(j,k)\}, add_A\big)$$
    $$\text{with } nb_{A'}(x) = \begin{cases} nb_A(x), & \text{if } x \neq p \\ nb_A(p) \setminus \{j\}, & \text{if } x = p \end{cases}$$

  - an arbitrary process $p \in \mathcal{P}$ finds a right linearization step i. e.,

    $$select\big(findLin\big(p, nb_A(p)\big)\big) = (j,k) \wedge (j < k \wedge p < j)$$

and it holds that

$$A' \equiv Alg_{ges}\big(P_A, P'_A, nb_{A'}, Msgs_A \cup \{(k,j)\}, add_A\big)$$
$$\text{with } nb_{A'}(x) = \begin{cases} nb_A(x), & \text{if } x \neq p \\ nb_A(p) \setminus \{k\}, & \text{if } x = p \end{cases}$$

– or an arbitrary process $p \in \mathscr{P}$ sends *keep-alive*-messages and is a nearer neighbor for at least one of its (at most two) neighbors i. e.,

$$\begin{aligned} & select\big(findLin\big(p, nb_A(p)\big)\big) = \bot \\ \wedge\quad & \big(\exists q \in LeftN\big(nb_A(p), p\big).\forall q' \in RightNM\big(A, q\big).p < q' \\ \vee\quad & \exists q \in RightN\big(nb_A(p), p\big).\forall q' \in LeftNM\big(A, q\big).q' < p\big) \end{aligned}$$

and it holds that $A' \equiv Alg_{ges}\big(P_A, P'_A, nb_A, Msgs_A \cup \{(j,p) | j \in nb_A(p)\}, add_A\big)$

- $\Psi_\Sigma(A') = \Psi_\Sigma(A)$ iff

  – an arbitrary process $p \in P_A$ receives a message i. e.,

$$q \in \mathscr{P} \quad \wedge \quad (p,q) \in Msgs_A$$

and it holds that $A' \equiv Alg_{ges}\big(P_A \setminus \{p\}, P'_A \cup \{p\}, nb_A, Msgs_A \setminus \{(p,q)\}, add_A \cup \{(p,q)\}\big)$,

  – an arbitrary process $p \in P'_A$ adds a process to its neighborhood i. e.,

$$q \in \mathscr{P} \quad \wedge \quad add_A(p) = q$$

and it holds that

$$A' \equiv Alg_{ges}\big(P_A \cup \{p\}, P'_A \setminus \{p\}, nb_{A'}, Msgs_A, add_A \setminus \{(p,q)\}\big)$$
$$\text{with } nb_{A'}(x) = \begin{cases} nb(x), & \text{if } x \neq p \\ nb(p) \cup \{q\}, & \text{if } x = p \end{cases}$$

  – an arbitrary process $p \in \mathscr{P}$ sends *keep-alive*-messages exclusively to (a subset of) desired neighbors and is already with messages known by them i. e.,

$$\begin{aligned} & select\big(findLin\big(p, nb_A(p)\big)\big) = \bot \\ \wedge\quad & \forall q \in LeftN\big(nb_A(p), p\big).p = succ(q) \wedge p \in RightNM\big(A, q\big) \\ \wedge\quad & \forall q \in RightN\big(nb_A(p), p\big).p = pred(q) \wedge p \in LeftNM\big(A, q\big) \end{aligned}$$

and it holds that $A' \equiv Alg_{ges}\big(P_A, P'_A, nb_A, Msgs_A \cup \{(j,p) | j \in nb_A(p)\}, add_A\big)$,

  – or an arbitrary process $p \in \mathscr{P}$ tries to find a linearization step and none of the other cases is true. It then holds $A' \equiv A$.

- $\Psi_{\sum}(A') > \Psi_{\sum}(A)$ iff
  - an arbitrary process $p \in \mathscr{P}$ sends *keep-alive*-messages and every neighbor already knows with messages a nearer process i. e.,

$$\begin{aligned} & select\big(findLin\big(p, nb_A(p)\big)\big) = \bot \\ \wedge \quad & \forall q \in LeftN\big(nb_A(p), p\big).\exists q' \in RightNM\big(A, q\big).q' \leq p \\ \wedge \quad & \forall q \in RightN\big(nb_A(p), p\big).\exists q' \in LeftNM\big(A, q\big).p \leq q' \\ \wedge \quad & \big(\exists q \in LeftN\big(nb_A(p), p\big).p \neq succ(p) \\ & \vee\ \exists q \in RightN\big(nb_A(p), p\big).p \neq pred(q)\big) \end{aligned}$$

and it holds that $A' \equiv Alg_{ges}\big(P_A, P'_A, nb_A, Msgs_A \cup \{(j,p) | j \in nb_A(p)\}, add_A\big)$ then

$$\Psi_{\sum}(A') = \Psi_{\sum}(A) + \sum_{q \in (nb_A(p) \setminus \{succ(p), pred(p)\})} dist(q,p)$$

and therefore $\Psi_{\sum}$ is monotonically decreasing with exception of sending *keep-alive*-messages to undesired neighbors that already have nearer neighbors.
*Proof.*

- If a process $p \in \mathscr{P}$ tries to find a linearization step but $select\big(findLin\big(p, nb_A(p)\big)\big) = \bot$, it holds that

$$A' \equiv Alg_{ges}\big(P_A, P'_A, nb_A, Msgs_A \cup \{(j,p) | j \in nb_A(p)\}, add_A\big)$$

  According to Lemma 6 it holds that $p$ has at most two neighbors and

$$|nb_A(p)| \leq 2 \wedge \big(|nb_A(p)| = 2 \implies LeftN\big(nb_A(p), p\big) = 1 \wedge RightN\big(nb_A(p), p\big) = 1\big)$$

  - if $nb_A(p) = \emptyset$, it holds that $A' \equiv A$ and therefore obviously $\Psi_{\sum}(A') = \Psi_{\sum}(A)$.
  - if $|LeftN\big(nb_A(p), p\big)| = 1 \wedge |RightN\big(nb_A(p), p\big)| = 0$: Let $j$ be the process with $LeftN\big(nb_A(p), p\big) = \{j\}$ and therefore according to the definition of *LeftN* it holds $j < p$.
    * if $p$ will be a nearer right neighbor than all yet from $j$ known processes in the topology with messages i. e., it holds that $RightNM\big(A, j\big) = \emptyset \vee \big(RightNM\big(A, j\big) \neq \emptyset \wedge \forall q \in RightNM\big(A, j\big).p < q\big)$: According to Lemma 15 there are two cases. If $p = succ(j)$, it holds that $\Psi(A') = \Psi(A)$ otherwise $\Psi(A') = \Psi(A) + dist(j,p)$. In both cases, it holds that $\Psi(A') \leq \Psi(A) + maxdist$. According to Lemma 17 it holds that $\Psi(A') < \Psi(A)$. Therefore, with Definition 38

$$\Psi_{\sum}(A') = \Psi(A') + n \cdot \Psi_E(A')$$

$$
\begin{aligned}
&\leq \Psi(A) + maxdist + n \cdot \Psi_E(A') \\
&\leq \Psi(A) + maxdist + n \cdot (\Psi_E(A) - 1) \\
&= \Psi(A) + (n-1) + n \cdot \Psi_E(A) - n \\
&= \Psi(A) + n \cdot \Psi_E(A) - 1 \\
&< \Psi(A) + n \cdot \Psi_E(A) \\
&= \Psi_\Sum(A)
\end{aligned}
$$

  * if $p$ is a desired neighbor of $j$ but already known in the topology with messages i. e., $p = succ(j) \land p \in RightNM(A, j)$: According to Lemma 15 it holds that $\Psi(A') = \Psi(A)$ and according to Lemma 15 it holds that $\Psi_E(A') = \Psi_E(A)$. Therefore, with Definition 38 it holds that

$$\Psi_\Sum(A') = \Psi_\Sum(A)$$

  * if $j$ already has a closer right neighbor than $p$ in the topology with messages i. e., $p \neq succ(j) \land \exists q \in RightNM(A, j).q \leq p$: It holds according to Lemma 17 that $\Psi_E(A') = \Psi_E(A)$. According to Lemma 15 it holds with $p \neq succ(j)$ that $\Psi(A') = \Psi(A) + dist(j, p)$. Hence with Definition 38

$$\Psi_\Sum(A') = \Psi_\Sum(A) + dist(j, p)$$

- if $|LeftN(nb_A(p), p)| = 0 \land |RightN(nb_A(p), p)| = 1$: Let $k$ be the process with $RightN(nb_A(p), p) = \{k\}$ and therefore according to the definition of *RightN* it holds $p < k$.
  * if $p$ will be a nearer left neighbor than all yet from $k$ known processes in the topology with messages i. e., it holds that $LeftNM(A, k) = \emptyset \lor (LeftNM(A, k) \neq \emptyset \land \forall q \in LeftNM(A, k).q < p)$: It holds

$$\Psi_\Sum(A') < \Psi_\Sum(A)$$

  * if $p$ is a desired neighbor of $k$ but already known in the topology with messages i. e., $p = pred(k) \land p \in LeftNM(A, k)$: It holds that

$$\Psi_\Sum(A') = \Psi_\Sum(A)$$

  * if $k$ already has a closer right neighbor than $p$ in the topology with messages i. e., $p \neq pred(k) \land \exists q \in LeftNM(A, k).p \leq q$: It holds

$$\Psi_\Sum(A') = \Psi_\Sum(A) + dist(k, p)$$

  The proofs are similar to the previous case.

- if $|LeftN(nb_A(p), p)| = 1 \land |RightN(nb_A(p), p)| = 1$: Let $j$ be the process with $LeftN(nb_A(p), p) = \{j\}$ and let $k$ be the process with $RightN(nb_A(p), p) = \{k\}$.

* if at least one of the processes gets to know a nearer process i. e.,

$$RightNM(A,j) = \emptyset \vee \left(RightNM(A,j) \neq \emptyset \wedge \forall q \in RightNM(A,j).p < q\right) \vee$$
$$LeftNM(A,k) = \emptyset \vee \left(LeftNM(A,k) \neq \emptyset \wedge \forall q \in LeftNM(A,k).q < p\right)$$

Assume $\left(RightNM(A,j) \neq \emptyset \wedge \forall q \in RightNM(A,j).p < q\right) \vee RightNM(A,j) = \emptyset$ (The case for $LeftNM(A,k)$ is similar). According to Lemma 15 there are four cases. If $p = succ(j)$ and $p = pred(k)$ it holds that $\Psi(A') = \Psi(A)$, if $p \neq succ(j)$ and $p = pred(k)$ it holds that $\Psi(A') = \Psi(A) + dist(j,p)$, if $p = succ(j)$ and $p \neq pred(k)$ it holds that $\Psi(A') = \Psi(A) + dist(k,p)$, and if $p \neq succ(j)$ and $p \neq pred(k)$ it holds that $\Psi(A') = \Psi(A) + dist(j,p) + dist(k,p)$. According to the definitions of *LeftN* and *RightN* it holds that $j < p < k$ and therefore with Definition 15 it holds in all cases that $\Psi(A') \leq \Psi(A) + maxdist$. According to Lemma 17 it holds that $\Psi(A') < \Psi(A)$. Therefore, with Definition 38

$$\begin{aligned}
\Psi_{\sum}(A') &= \Psi(A') + n \cdot \Psi_E(A') \\
&\leq \Psi(A) + maxdist + n \cdot \Psi_E(A') \\
&\leq \Psi(A) + maxdist + n \cdot (\Psi_E(A) - 1) \\
&= \Psi(A) + (n-1) + n \cdot \Psi_E(A) - n \\
&= \Psi(A) + n \cdot \Psi_E(A) - 1 \\
&< \Psi(A) + n \cdot \Psi_E(A) \\
&= \Psi_{\sum}(A)
\end{aligned}$$

* if $p$ is a desired neighbors of both but already known i. e.,

$$p = succ(j) \wedge p \in RightNM(A,j) \wedge$$
$$p = pred(k) \wedge p \in LeftNM(A,k)$$

According to Lemma 15 it holds that $\Psi(A') = \Psi(A)$ and according to Lemma 15 it holds that $\Psi_E(A') = \Psi_E(A)$. Therefore, with Definition 38 it holds that

$$\Psi_{\sum}(A') = \Psi_{\sum}(A)$$

* if both processes already know a closer process i. e.,

$$\left(p \neq succ(j) \vee p \neq pred(k)\right) \wedge$$
$$\exists q \in RightNM(A,j).q \leq p \wedge \exists q \in LeftNM(A,k).p \leq q$$

· if $p \neq succ(j) \wedge p = pred(k)$: It holds according to Lemma 17 that $\Psi_E(A') = \Psi_E(A)$. According to Lemma 15 it holds with $p \neq succ(j)$ and $p = pred(k)$ that $\Psi(A') = \Psi(A) + dist(j,p)$. Hence with Definition 38

$$\Psi_{\sum}(A') = \Psi_{\sum}(A) + dist(j,p)$$

- if $p = succ(j) \wedge p \neq pred(k)$: It holds according to Lemma 17 that $\Psi_E(A') = \Psi_E(A)$. According to Lemma 15 it holds with $p = succ(j)$ and $p \neq pred(k)$ that $\Psi(A') = \Psi(A) + dist(k,p)$. Hence with Definition 38

$$\Psi_\sum(A') = \Psi_\sum(A) + dist(k,p)$$

- if $p \neq succ(j) \wedge p \neq pred(k)$: It holds according to Lemma 17 that $\Psi_E(A') = \Psi_E(A)$. According to Lemma 15 it holds with $p \neq succ(j)$ and $p \neq pred(k)$ that $\Psi(A') = \Psi(A) + dist(j,p) + dist(k,p)$. Hence with Definition 38

$$\Psi_\sum(A') = \Psi_\sum(A) + dist(j,p) + dist(k,p)$$

- If a process $p \in \mathscr{P}$ finds a left linearization step i. e., $select\big(findLin\big(p, nb_A(p)\big)\big) = (j,k) \wedge (j < k \wedge k < p)$: it holds that

$$A' \equiv Alg_{ges}\big(P_A, P'_A, nb_{A'}, Msgs_A \cup \{(j,k)\}, add_A\big)$$
$$\text{with } nb_{A'}(x) = \begin{cases} nb_A(x), & \text{if } x \neq p \\ nb_A(p) \setminus \{j\}, & \text{if } x = p \end{cases}$$

According to Lemma 15 it holds that $\Psi(A') < \Psi(A)$ and according to Lemma 17 $\Psi_E(A') \leq \Psi_E(A)$. Hence with Definition 38 it holds that

$$\Psi_\sum(A') < \Psi_\sum(A)$$

- If a process $p \in \mathscr{P}$ finds a right linearization step i. e., $select\big(findLin\big(p, nb_A(p)\big)\big) = (j,k) \wedge (j < k \wedge p < j)$:

$$A' \equiv Alg_{ges}\big(P_A, P'_A, nb_{A'}, Msgs_A \cup \{(k,j)\}, add_A\big)$$
$$\text{and } nb_{A'}(x) = \begin{cases} nb_A(x), & \text{if } x \neq p \\ nb_A(p) \setminus \{k\}, & \text{if } x = p \end{cases}$$

According to Lemma 15 it holds that $\Psi(A') < \Psi(A)$ and according to Lemma 17 $\Psi_E(A') \leq \Psi_E(A)$. Hence with Definition 38 it holds that

$$\Psi_\sum(A') < \Psi_\sum(A)$$

- If for a process $p \in \mathscr{P}$ it could be that $select\big(findLin\big(p, nb_A(p)\big)\big)$ is something else it holds that $A' \equiv A$. Obviously, it holds $\Psi_\sum(A') = \Psi_\sum(A)$
- If a process $p \in P_A$ receives a message with the id of $q \in \mathscr{P}$ i. e., $(p,q) \in Msgs_A$: it holds

$$A' \equiv Alg_{ges}\big(P_A \setminus \{p\}, P'_A \cup \{p\}, nb_A, Msgs_A \setminus \{(p,q)\}, add_A \cup \{(p,q)\}\big)$$

According to Lemma 15 it holds that $\Psi(A') = \Psi(A)$ and according to Lemma 17 $\Psi_E(A') =$

$\Psi_E(A)$. Hence with Definition 38 it holds that

$$\Psi_{\sum}(A') = \Psi_{\sum}(A)$$

- If a process $p \in P'_A$ adds a process $q \in \mathcal{P}$ to its neighborhood i. e., $(p, q) \in add_A$: it holds

$$A' \equiv Alg_{ges}\left(P_A \cup \{p\}, P'_A \setminus \{p\}, nb_{A'}, Msgs_A, add_A \setminus \{(p, q)\}\right)$$

$$\text{with } nb_{A'}(x) = \begin{cases} nb(x), & \text{if } x \neq p \\ nb(p) \cup \{q\}, & \text{if } x = p \end{cases}$$

According to Lemma 15 it holds that $\Psi(A') \leq \Psi(A)$ and according to Lemma 17 $\Psi_E(A') = \Psi_E(A)$. Hence with Definition 38 it holds that

$$\Psi_{\sum}(A') \leq \Psi_{\sum}(A)$$

□

## 3.6 Convergence

In this section, we prove strong convergence for two special cases and weak convergence, i. e., for every arbitrary initial configuration there exist executions that reach a correct configuration, in the general case. Strong convergence is proven in case that either there are possibly correct edges missing but no non-correct edges existent in the topology with messages i. e., the topology is a subgraph of the linear graph, or there are possibly still non-correct edges existent but at least all correct edges are contained in the topology with messages i. e., the linear graph is a subgraph of the topology.

In case that there are only correct edges missing, the system converges to a correct configuration as between every pair of consecutive processes there is at least one connection. Therefore, no more possible linearization steps exist and every process sends *keep-alive*-messages to the already known subset of desired neighbors. All these messages are received and processed eventually and therefore all missing correct edges are eventually established.

**Lemma 20: Convergence for $UNTM = UG_{LIN}$**

Let $A \equiv Alg_{ges}\left(P, P', nb, Msgs, add\right)$ be an arbitrary configuration of the algorithm according to Lemma 1. If the undirected network topology with messages is the desired undirected topology i. e., $UNTM(A) = UG_{LIN}$, then a correct configuration $C$ is reached after a finite number of steps i. e.,

$$A \Longmapsto C \quad \wedge \quad NTM(C) = G_{LIN} \wedge NT(C) = G_{LIN}$$

*Proof.*
Since $UNTM(A) = UG_{LIN}$, it holds according to Definition 32 that $A$ is an undirected correct configuration and therefore according to Lemma 9 that:

$$\begin{aligned}
& P \cup P' = \mathscr{P} \quad \text{and} \quad P \cap P' = \emptyset, \\
& \forall p \in \mathscr{P}.nb(p) \subseteq \{succ(p), pred(p)\}_{\perp}, \\
& \forall (p,q) \in Msgs.(q = succ(p)) \vee (q = pred(p)), \\
& \forall p \in P'.(add(p) = succ(p)) \vee (add(p) = pred(p)) \quad \text{and} \\
& \forall p,q \in \mathscr{P}. \quad \big(succ(p) = q \implies \big(q \in nb(p) \vee p \in nb(q) \vee (p,q) \in Msgs \vee (q,p) \in Msgs \\
& \qquad\qquad \vee (p \in P' \wedge add(p) = q) \vee (q \in P' \wedge add(q) = p)\big)\big).
\end{aligned}$$

Let $U : \mathscr{T} \to \mathscr{P}$ with $U(C) = \{p \in \mathscr{P} | nb_C(p) \subset \{succ(p), pred(p)\}_{\perp}\}$. If $U(A) = \emptyset$ i. e., $|U(A)| = 0$, then $\forall p \in \mathscr{P}.nb(p) = \{succ(p), pred(p)\}_{\perp}$ and it holds according to Lemma 8 that $A$ itself is a correct configuration and $NTM(A) = G_{LIN} \wedge NT(A) = G_{LIN}$.
Otherwise we can show that if $U(A) \neq \emptyset$, then $|U(A)| \leq \mathscr{P}$ will always eventually decrease. Therefore, a correct configuration will be reached after a finite number of steps since every process in $U$ adds the missing desired neighbor(s) after a finite number of steps.

Let $p \in U(A)$ be an arbitrary process with $succ(p) \in \mathscr{P}$ and $succ(p) \notin nb(p)$. We have to show that after a finite number of steps an undirected correct configuration $C'$ is reached with $succ(p) \in nb_{C'}(p)$. According to the assumption at least one of the following cases is true:

- $p \in P' \wedge add(p) = succ(p)$: Then it holds that the subprocess $Alg_{add}\big(p, succ(p)\big)$ is enabled and according to Lemma 3 it stays enabled until it executes a step. According to the fairness assumption 3

  $$Alg_{add}\big(p, succ(p)\big) = nb_p\big(y\big).\Big(\overline{nb_p}\langle y \cup \{succ(p)\}\rangle \mid Alg_{rec}\big(p\big)\Big)$$

  will execute a step after a finite number of steps. Let $R$ be the configuration before this step. Then it holds according to Corollary 2

  $$R \longmapsto C' \equiv Alg_{ges}\big(P_R \cup \{p\}, P'_R \setminus \{p\}, nb_{C'}, Msgs_R, add_R \setminus \{(p, succ(p))\}\big)$$
  $$\text{with } nb_{C'}(x) = \begin{cases} nb_R(x), & \text{if } x \neq p \\ nb_R(p) \cup \{succ(p)\}, & \text{if } x = p \end{cases}$$

  and therefore $succ(p) \in nb_{C'}(p)$. According to Corollary 11 it holds $C'$ is an undirected correct configuration i. e., $UNTM(C') = UG_{LIN}$ and with $succ(p) \notin nb(p)$ and $succ(p) \in nb_{C'}(p)$ that $UNT(A) \subset UNT(C')$.

- $(p, succ(p)) \in Msgs$: According to the Assumption 2 that there is no message loss, the message will be eventually (after a finite number of steps) received. According to Lemma 2, $p$ is the only process that can receive this message. Let $R$ be the configuration before this step. Then it holds according to Corollary 2 $R \longmapsto R'$

  $$\text{with } R' \equiv Alg_{ges}\big(P_R \setminus \{p\}, P'_R \cup \{p\}, nb_R, Msgs_R \setminus \{(p, succ(p))\}, add_R \cup \{(p, succ(p))\}\big)$$

  and therefore $p \in P'_{R'}$ and $add_{R'}(p) = succ(p)$. According to Corollary 11 it holds that $R'$ is an undirected correct configuration i. e., $UNTM(R') = UG_{LIN}$ and $UNT(A) \subseteq UNT(R')$. Therefore, we reduced the problem to the previous case.

- $p \in nb(succ(p))$: According to to Lemma 3 the subprocess $Alg_{match}\big(succ(p)\big)$ is enabled and stays enabled until it executes a step. According to the fairness assumption 3, $Alg_{match}\big(succ(p)\big)$ will execute a step after a finite number of steps. Let $R$ be the configuration before this step. Since $R$ has to be an undirected correct configuration according to Lemma 13, it holds that

  $$nb_R(succ(p)) \subseteq \{succ(succ(p)), pred(succ(p))\}_\perp = \{succ(succ(p)), p\}_\perp$$

  and therefore $select\big(findLin\big(succ(p), nb_R succ(p)\big)\big) = \perp$ and according to Corollary 2 $R \longmapsto R'$

  $$\text{with } R' \equiv Alg_{ges}\big(P_R, P'_R, nb_R, Msgs_R \cup \{(j, succ(p)) | j \in nb_R(succ(p))\}, add_R\big)$$

and therefore $(p, succ(p)) \in Msgs_{R'}$. According to Corollary 11 it holds that $R'$ is an undirected correct configuration i. e., $UNTM(R') = UG_{LIN}$ and $UNT(A) \subseteq UNT(R')$. Therefore, we reduced the problem to the previous case.

- $succ(p) \in P' \wedge add(succ(p)) = p$: Then it holds that the subprocess $Alg_{add}\big(succ(p), p\big)$ is enabled and according to Lemma 3 it stays enabled until it executes a step. According to the fairness assumption 3

$$Alg_{add}\big(succ(p), p\big) = nb_{succ(p)}\big(y\big).\Big(\overline{nb_{succ(p)}}\big\langle y \cup \{p\}\big\rangle \mid Alg_{rec}\big(succ(p)\big)\Big)$$

will execute a step after a finite number of steps. Let $R$ be the configuration before this step. Then it holds according to Corollary 2 $R \longmapsto R'$ with

$$R' \equiv Alg_{ges}\big(P_R \cup \{succ(p)\}, P'_R \setminus \{succ(p)\}, nb_{R'}, Msgs_R, add_R \setminus \{(succ(p), p)\}\big)$$

$$\text{and } \; nb_{R'}(x) = \begin{cases} nb_R(x), & \text{if } x \neq succ(p) \\ nb_R(p) \cup \{p\}, & \text{if } x = succ(p) \end{cases}$$

and therefore $p \in nb_{R'}(succ(p))$. According to Corollary 11 it holds that $R'$ is an undirected correct configuration i. e., $UNTM(R') = UG_{LIN}$ and $UNT(A) \subseteq UNT(R')$. Therefore, we reduced the problem to the previous case.

- $(succ(p), p) \in Msgs$: According to the Assumption 2 that there is no message loss, the message will be eventually (after a finite number of steps) received. According to Lemma 2, $succ(p)$ is the only process that can receive this message. Let $R$ be the configuration before this step. Then it holds according to Corollary 2 $R \longmapsto R'$ with $R' \equiv$

$$Alg_{ges}\big(P_R \setminus \{succ(p)\}, P'_R \cup \{succ(p)\}, nb_R, Msgs_R \setminus \{(succ(p), p)\}, add_R \cup \{(succ(p), p)\}\big)$$

and therefore $succ(p) \in P'_{R'}$ and $add_{R'}(succ(p)) = p$. According to Corollary 11 it holds that $R'$ is an undirected correct configuration i. e., $UNTM(R') = UG_{LIN}$ and $UNT(A) \subseteq UNT(R')$. Therefore, we reduced the problem to the previous case.

The case for $pred(p) \in \mathscr{P}$ and $pred(p) \notin nb(p)$ is similar. Therefore, we always reach an undirected correct configuration $C''$ with $|U(A)| < |U(C'')|$. □

In case that there are possibly too many edges in the topology but at least all correct edges are already established i. e., the linear graph is a subgraph of the topology, the system also converges to a correct configuration. In this case, the only processes that can send *keep-alive*-messages are processes that know exactly their desired neighbors. This holds as every process knows at least its desired neighbors and whenever a process has additionally neighbors, there is a possible linearization step that can be executed by the process. Therefore, with every linearization step the topology gets closer to the desired topology. This is shown via the potential function $\Psi$.

**Lemma 21: Convergence for $G_{LIN} \subseteq NTM \wedge G_{LIN} \subseteq NT$**

Let $A \equiv Alg_{ges}\big(P, P', nb, Msgs, add\big)$ be an arbitrary configuration of the algorithm according to Lemma 1. If $G_{LIN} \subseteq NTM(A) \wedge G_{LIN} \subseteq NT(A)$, then a correct configuration $C$ is reached after a finite number of steps i. e.,

$$A \Longmapsto C \quad \wedge \quad NTM(C) = G_{LIN} \wedge NT(C) = G_{LIN}$$

*Proof:*
If $\Psi(A) = 0$, then the potential is minimal and according to Lemma 14 $NTM(A) \subseteq G_{LIN}$ respectively $UNTM(A) = UG_{LIN}$. From the assumptions it holds that $G_{LIN} \subseteq NTM(A)$ and therefore $NTM(A) = G_{LIN}$. Since $G_{LIN} \subseteq NT(A)$ it holds from Definitions 14 and 27 that $\forall p \in \mathscr{P}.\{succ(p), pred(p)\}_\perp \subseteq nb(p)$. Additionally $A$ is an undirected correct configuration as $UNTM(A) = UG_{LIN}$ and with Lemma 9 it holds

$$P \cup P' = \mathscr{P} \quad \text{and} \quad P \cap P' = \emptyset,$$
$$\forall p \in \mathscr{P}.nb(p) \subseteq \{succ(p), pred(p)\}_\perp,$$
$$\forall (p,q) \in Msgs.(q = succ(p)) \vee (q = pred(p)) \text{ and}$$
$$\forall p \in P'.(add(p) = succ(p)) \vee (add(p) = pred(p))$$

Hence $\forall p \in \mathscr{P}.nb(p) = \{succ(p), pred(p)\}_\perp$ and with Lemma 8 $NTM(A) = G_{LIN} \wedge NT(A) = G_{LIN}$ and $A$ is a correct configuration itself.
Otherwise it holds that $\Psi(A) > 0$ but the potential function will decrease over the execution and eventually it holds that a configuration $C$ is reached i. e., $A \Longmapsto C$ with $\Psi(C) = 0$. This follows from the two observations that the potential function can not increase for a process term whose network topology graph contains the desired graph and if the term is not a correct configuration there will always eventually steps executed which decrease the potential function strictly.
Therefore, first, for every reachable configuration $R$ with $A \Longmapsto R'$ it holds that

$$\Psi(R') \leq \Psi(A)$$

Let $R$ be an arbitrary reachable configuration $A \Longmapsto R$ with $\Psi(R) \leq \Psi(A)$, we have to show that for every configuration $R'$ with $R \longmapsto R'$ it holds $\Psi(R') \leq \Psi(R)$. According to Lemma 15 the only case with $\Psi(R') > R$ would be if a process sends *keep-alive*-messages to processes that are not its desired neighbors i. e., there is $p, q \in \mathscr{P}$ with $select\big(findLin\big(p, nb_R(p)\big)\big) = \perp$, $q \notin \{succ(p), pred(p)\}$ and $q \in nb_R(p)$ and the executed step is $R \longmapsto R' \equiv Alg_{ges}\big(P_R, P'_R, nb_R, Msgs_R \cup \{(j,p) | j \in nb_R(p)\}, add_R\big)$. For every reachable configuration $R$ i. e., $A \Longmapsto R$ it holds according to Corollary 10 that $G_{LIN} \subseteq NT(R)$ and therefore with Definitions 14 and 27 that $\forall p \in \mathscr{P}.\{succ(p), pred(p)\}_\perp \subseteq nb_R(p)$. For every process $q \neq p$ with $q \notin \{succ(p), pred(p)\}$ follows according to Corollary 1 that $pred(p) < q \vee q > succ(p)$. Since $\{succ(p), pred(p)\}_\perp \subseteq nb_R(p)$ it would then hold that there is an enabled linearization step i. e., according to the definition of $select(findLin())$ either $select\big(findLin\big(p, nb_R(p)\big)\big) = (j,k) \wedge (j < k \wedge k < p)$ or $select\big(findLin\big(p, nb_R(p)\big)\big) = (j,k) \wedge (j < k \wedge p < j)$ and therefore this is a contradiction to $select\big(findLin\big(p, nb_R(p)\big)\big) = \bot$.
Second, for every reachable configuration $R$ with $A \Longmapsto R$ and $\Psi(R) > 0$ there is always eventually a configuration $R'$ reached i. e., $R \Longmapsto R'$ with

$$\Psi(R') < \Psi(R)$$

According to Definitions 33 and 34 and $\Psi(R) > 0$ it holds that there is are processes $p, q \in \mathscr{P}$ with $p \neq q$ and $q \notin \{succ(p), pred(p)\}$ and one of the following cases is true:

- $q \in nb_R(p))$: According to to Lemma 3 the subprocess $Alg_{match}\big(p\big)$ is enabled and stays enabled until it executes a step. According to the fairness assumption 3 $Alg_{match}\big(p\big)$ will execute a step after a finite number of steps. Let $H$ be the configuration before this step. It holds according to Corollary 10 that $G_{LIN} \subseteq NT(H)$ and therefore with Definitions 14 and 27 that $\{succ(p), pred(p)\}_\bot \subseteq nb_H(p)$. Since $p \neq q$ and $q \notin \{succ(p), pred(p)\}$, it holds according to Corollary 1 that $pred(p) < q \vee q > succ(p)$.
  Therefore, it follows with the definition of $select(findLin())$ that there has to be an enabled linearization step for $p$ and hence either $select\big(findLin\big(p, nb_H(p)\big)\big) = (j,k) \wedge (j < k \wedge k < p)$ or $select\big(findLin\big(p, nb_H(p)\big)\big) = (j,k) \wedge (j < k \wedge p < j)$ and with Lemma 15 for $H \longmapsto R'$ that $\Psi(R') < \Psi(H)$. From the monotonicity of the potential function it follows that $\Psi(H) \leq \Psi(R)$ and hence $\Psi(R') < \Psi(R)$.

- $p \in P'_R \wedge add_R(p) = q$: Then it holds that the subprocess $Alg_{add}\big(p,q\big)$ is enabled and according to Lemma 3 it stays enabled until it executes a step. According to the fairness assumption 3

  $$Alg_{add}\big(p,q\big) = nb_p\big(y\big).\Big(\overline{nb_p}\langle y \cup \{q\}\rangle \mid Alg_{rec}\big(p\big)\Big)$$

  will execute a step after a finite number of steps. Let $H$ be the configuration before this step. Then it holds according to Corollary 2

  $$H \longmapsto H' \equiv Alg_{ges}\big(P_H \cup \{p\}, P'_H \setminus \{p\}, nb_{H'}, Msgs_H, add_H \setminus \{(p,q)\}\big)$$
  $$\text{with } nb_{H'}(x) = \begin{cases} nb_H(x), & \text{if } x \neq p \\ nb_H(x) \cup \{q\}, & \text{if } x = p \end{cases}$$

  and hence $q \in nb_{H'}(x)$ and according to the monotonicity $\Psi(H') \leq \Psi(R)$. Therefore, we can reduce the problem to the previous case.

- $(p,q) \in Msgs_R$: According to the Assumption 2 that there is no message loss, the message will be eventually (after a finite number of steps) received. According to Lemma 2, $p$ is the only process that can receive this message. Let $H$ be the configuration before this step. Then it holds according to Corollary 2 $H \longmapsto H'$

$$\text{with } H' \equiv Alg_{ges}\big(P_H \setminus \{p\}, P'_H \cup \{p\}, nb_H, Msgs_H \setminus \{(p,q)\}, add_H \cup \{(p,q)\}\big)$$

and hence $p \in P'_H \wedge add_H(p) = q$ and according to the monotonicity $\Psi(H') \leq \Psi(R)$. Therefore, we can reduce the problem to the previous case.

Therefore, we will eventually reach a configuration $C$ i. e., $A \Longmapsto C$ with $\Psi(C) = 0$. According to Corollary 10 it holds that $G_{LIN} \subseteq NTM(C) \wedge G_{LIN} \subseteq NT(C)$ and therefore with $\Psi(C) = 0$ as shown in the base case $NTM(C) = G_{LIN} \wedge NT(C) = G_{LIN}$. □

Convergence also holds if the desired topology is a subgraph of the topology with messages. Since every message that is in transit will eventually be received and processed, we always reach a configuration for that holds that the linear graph is a subgraph of the topology without messages. Then according to the last lemma, we always reach a correct configuration.

**Lemma 22: Convergence for $G_{LIN} \subseteq NTM$**

Let $A$ be an arbitrary configuration of the algorithm according to Lemma 1. If $G_{LIN} \subseteq NTM(A)$, then a correct configuration $C$ is reached after a finite number of steps i. e.,

$$A \Longmapsto C \quad \wedge \quad NTM(C) = G_{LIN} \wedge NT(C) = G_{LIN}$$

*Proof:*
Let $U : \mathcal{T} \to \mathcal{P}$ with $U(C) = \{p \in \mathcal{P} | nb_C(p) \not\supseteq \{succ(p), pred(p)\}_\perp\}$ . If $U(A) = \emptyset$ i. e., $|U(A)| = 0$, then $\forall p \in \mathcal{P}.nb_A(p) \supseteq \{succ(p), pred(p)\}_\perp$ and it holds according to Definitions 14 and 28 that $G_{LIN} \subseteq NTM(R) \wedge G_{LIN} \subseteq NT(R)$. Hence with Lemma 21 it holds that

$$A \Longmapsto C \quad \wedge \quad NTM(C) = G_{LIN} \wedge NT(C) = G_{LIN}$$

Otherwise we can show that if $U(A) \neq \emptyset$, then $|U(A)| \leq \mathcal{P}$ will always eventually decrease. Let $p \in U(A)$ be an arbitrary process with $succ(p) \in \mathcal{P}$ and $succ(p) \notin nb_A(p)$. We have to show that after a finite number of steps configuration $C'$ is reached with $G_{LIN} \subseteq NTM(C')$ and $succ(p) \in nb_{C'}(p)$. According to the assumption that $G_{LIN} \subseteq NTM(A)$ at least one of the following cases is true according to Definition 28:

- $p \in P'_A \wedge add_A(p) = succ(p)$: According to Lemma 4 it is a configuration $C'$ reached with $succ(p) \in nb_{C'}(p)$. According to Lemma 12 it holds with Definitions 14 and 28 that $G_{LIN} \subseteq NTM(C')$ and with Lemma 11 and $succ(p) \notin nb_A(p)$ and $succ(p) \in nb_{C'}(p)$ that $U(A) \subset U(C')$.
- $(p, succ(p)) \in Msgs$: According to Corollary 4 it is a configuration $C'$ reached with $succ(p) \in nb_{C'}(p)$. According to Lemma 12 it holds with Definitions 14 and 28 that $G_{LIN} \subseteq NTM(C')$ and with Lemma 11 and $succ(p) \notin nb_A(p)$ and $succ(p) \in nb_{C'}(p)$ that $U(A) \subset U(C')$.

The case for $pred(p) \in \mathscr{P}$ and $pred(p) \notin nb_A(p)$ is similar. Therefore, we will always reach a configuration $C''$ with $G_{LIN} \subseteq NTM(C'')$ and $|U(A)| < |U(C'')|$. Hence eventually a configuration $R$ is reached with $|U(R)| = 0$ and therefore according to the base case $G_{LIN} \subseteq NTM(R) \wedge G_{LIN} \subseteq NT(R)$. Therefore, according to Lemma 21 it holds that

$$A \Longmapsto R \Longmapsto C \quad \wedge \quad NTM(C) = G_{LIN} \wedge NT(C) = G_{LIN}$$

□

Therefore, we proved strong convergence for the special cases that there are only correct edges missing, i. e., the topology is a subgraph of the linear graph, and that there are only too many edges i. e., the linear graph is a subgraph of the topology (with messages). The proofs for both cases use the fact that *keep-alive*-messages are only exchanged between desired neighbors. However, proving strong convergence in the general case is much more difficult as the sending of *keep-alive*-messages to undesired neighbors can cause the reestablishing of connections that were already removed through linearization steps and should stay removed. This is explained in more detail in the Chapter 4. Therefore, we show weak convergence in the general case i. e., for every initial configuration there are executions that converge to a correct configuration. In order to prove weak convergence for any arbitrary configuration, we define a perfect oracle. A perfect oracle is a global omniscient instance that suppresses the sending of all *keep-alive*-messages with exception of *keep-alive*-messages that are necessary to resolve potential deadlocks. More concrete, whenever the system would deadlock without the sending of *keep-alive*-messages, in the sense that there are possible linearization steps in the undirected topology with messages but not in the directed topology with messages, the perfect oracle chooses processes that are allowed to send *keep-alive*-messages once, in order to resolve the deadlock. Example situations are depicted in Figure 3.3. A perfect oracle cannot be implemented in a distributed system and should therefore only be seen as a restriction on the set of executions. The sending is executed as recently as all corresponding neighborhood relations are established. Therefore, in the first line of Figure 3.3 *keep-alive*-messages can be send, but in the second one only after the system converged to the first line (i. e., after all messages are received and processed eventually).

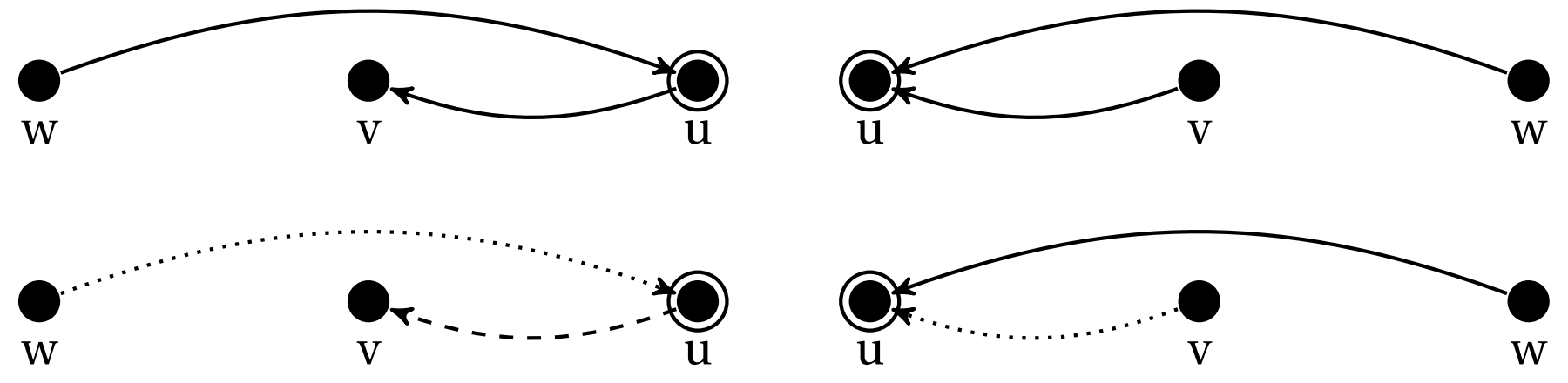

Figure 3.3: Linearization steps in the undirected topology with messages but not in the directed

**Definition 39: Perfect Oracle $O$**

A *perfect oracle* is a global omniscient instance that only let the processes send *keep-alive*-messages to resolve deadlocks and otherwise suppresses all *keep-alive*-messages . Therefore, whenever there is a linearization step in the directed topology with messages in a configuration $A$ i. e.,

$$\exists p,q,r \in \mathscr{P}.((p<q \wedge q<r) \vee (r<q \wedge q<p)) \wedge ((p,q) \in E_{NTM(A)} \wedge (p,r) \in E_{NTM(A)})$$

then no process is allowed to send *keep-alive*-messages i. e., for every process $p'$ with $select\big(findLin\big(p, nb_R(p)\big)\big) = \bot$ the subprocess $Alg_{match}\big(p'\big)$ is not allowed by $O$ to execute a step.
If there is a possible linearization step in the undirected topology with messages i. e.,

$$\exists p,q,r \in \mathscr{P}.\{p,q\},\{p,r\} \in E_{UNTM(A)} \wedge ((p<q \wedge q<r) \vee (r<q \wedge q<p))$$

but there is no possible linearization step in the directed topology with messages i. e.,

$$\forall p,q,r \in \mathscr{P}.((p<q \wedge p<r) \vee (q<p \wedge r<p)) \implies ((p,q) \notin E_{NTM(A)} \vee (p,r) \notin E_{NTM(A)})$$

the perfect oracle $O$ will choose one respectively two processes and allow them to send *keep-alive*-messages once in order to resolve the potential deadlock. If it is sufficient that one process sends *keep-alive*-messages i. e.,

$$\exists p,q,r \in \mathscr{P}.\big((p<q \wedge q<r) \vee (r<q \wedge q<p)\big) \wedge \\ \big(\big((p,q) \in E_{NTM(A)} \wedge (r,p) \in E_{NTM(A)}\big) \vee \big((p,r) \in E_{NTM(A)} \wedge (q,p) \in E_{NTM(A)}\big)\big)$$

then $O$ chooses one process $q \in \mathscr{P}$ such that there are $p,r \in \mathscr{P}$ and $(p<q \wedge q<r) \vee (r<q \wedge q<p) \vee (p<r \wedge r<q) \vee (q<r \wedge r<p)$ and $(p,r) \in E_{NTM(A)} \wedge (q,p) \in E_{NTM(A)}$, and allows $q$ to send *keep-alive*-messages once as soon as a configuration $C$ is reached with $r \in nb_C(p) \wedge p \in nb_C(q)$. Otherwise it is necessary that two processes send *keep-alive*-messages in order to resolve the potential deadlock i. e.,

$$\exists p,q,r \in \mathscr{P}.\big((p<q \wedge q<r) \vee (r<q \wedge q<p)\big) \wedge \big((q,p) \in E_{NTM(A)} \wedge (r,p) \in E_{NTM(A)}\big)$$

$O$ chooses two processes $q,r \in \mathscr{P}$ such that there is $p \in \mathscr{P}$ and $(p<q \wedge q<r) \vee (r<q \wedge q<p)$ and $(q,p) \in E_{NTM(A)} \wedge (r,p) \in E_{NTM(A)}$. Then $O$ allows $q$ and $r$ to send *keep-alive*-messages once as soon as a configuration $C$ is reached with $p \in nb_C(q) \wedge p \in nb_C(r)$.
If there is no more possible linearization step in the undirected topology with messages i. e.,

$$\forall p,q,r \in \mathscr{P}.((p<q \wedge p<r) \vee (q<p \wedge r<p)) \implies (\{p,q\} \notin E_{UNTM(D)} \vee \{p,r\} \notin E_{UNTM(D)})$$

and therefore $UNTM(A) \subseteq UG_{LIN}$, then all processes are allowed to send *keep-alive*-messages .

*Remark*: A perfect oracle $O$ is not contradictory to the fairness assumption 3 if starting from any initial configuration in every execution after a finite number of steps a configuration is reached where $O$ does not suppress the sending of *keep-alive*-messages anymore. In this case, it is possible that every continuously enabled subprocess can execute a step after a finite number of steps. We show in Lemma 23 that starting from an arbitrary initial configuration a correct configuration is reached after a finite number of steps. A perfect oracle $O$ does not suppress the sending of *keep-alive*-messages in a correct configuration. Once the system is in a correct configuration, it stays in a correct configuration according to the closure property as shown in Theorem 1. Therefore, a perfect oracle is not contradictory to the fairness assumption.

We show that every execution that is admissible under the restriction of a perfect oracle converges to a correct configuration. Since for every configuration this set of executions is non-empty, we prove weak convergence for the general case. We have introduced three potential functions and shown that whenever at least one of them is minimal, the system reaches a correct configuration in a finite number of steps. For every configuration there is exactly one of the three cases true: there is no more possible linearization step in the undirected topology with messages, there is a possible linearization step in the directed topology with messages, or there is a possible linearization step in the undirected but not in the directed topology with messages. In the first case, it follows that $\Psi$ is minimal and therefore strong convergence is ensured. In the second case, the perfect oracle precludes the sending of *keep-alive*-messages and therefore all three potential functions behave monotonically decreasing. Furthermore, with every executed linearization step $\Psi_{\sum}$ and $\Psi$ are strictly decreasing. In the third case, the perfect oracle allows chosen processes to send *keep-alive*-messages in order to resolve the potential deadlock and this leads to an increasing of $\Psi$ and possibly also $\Psi_{\sum}$. Nevertheless, $\Psi_E$ is still monotonically decreasing and we always reach a configuration with a strictly smaller value of $\Psi_E$ and $\Psi_{\sum}$ such that there is a possible linearization step in the directed topology with messages. Hence, the third case can not occur more often than the initial value of $\Psi_E$ without $\Psi_E$ getting minimal, as $\Psi_E$ is monotonically in every case and strictly decreases with every occurrence of the third case. Therefore, neither we can stay infinite long in the second case nor we can infinitely often alternate between the second and the third case without reaching a configuration in which at least one of the three potential functions is minimal and therefore convergence ensured.

**Lemma 23: Convergence with Perfect Oracle**

Let $I \equiv Alg_{ges}\left(P, P', nb, Msgs, add\right)$ be an arbitrary connected, i. e., $UNTM(I)$ is connected, initial configuration of the algorithm according to Definition 23 and $A$ an arbitrary reachable configuration $I \Longmapsto A$ according to Lemma 1. Assume there is a perfect oracle $O$. Then a correct configuration $C$ is reached after a finite number of steps i. e.,

$$A \Longmapsto C \quad \wedge \quad NTM(C) = G_{LIN} \wedge NT(C) = G_{LIN}$$

*Proof:*
If $\Psi_{\sum}(A)$ is minimal (i. e., $\Psi_{\sum}(A) = 2 \cdot n \cdot (n-1)$), it holds according to Lemma 18 that $NTM(A) = G_{LIN}$. Hence with Lemma 22 it holds that a correct configuration $C$ is reached after a finite number of steps i. e.,

$$A \Longmapsto C \quad \wedge \quad NTM(C) = G_{LIN} \wedge NT(C) = G_{LIN}$$

If $\Psi_E(A)$ is minimal (i. e., $\Psi_E(A) = 2 \cdot (n-1)$), it holds according to Lemma 16 that $NTM(A) \supseteq G_{LIN}$. Hence with Lemma 22 it holds that a correct configuration $C$ is reached after a finite number of steps i. e.,

$$A \Longmapsto C \quad \wedge \quad NTM(C) = G_{LIN} \wedge NT(C) = G_{LIN}$$

If $\Psi(A) = 0$, then the potential is minimal and according to Lemma 14 $NTM(A) \subseteq G_{LIN}$ respectively $UNTM(A) = UG_{LIN}$. Since $O$ does allow sending *keep-alive*-messages if $UNTM(A) = UG_{LIN}$, it holds with Lemma 20 that a correct configuration $C$ is reached after a finite number of steps i. e.,

$$A \Longmapsto C \quad \wedge \quad NTM(C) = G_{LIN} \wedge NT(C) = G_{LIN}$$

Otherwise it holds that $\Psi_{\sum}(A) > 2 \cdot n \cdot (n-1)$ but the potential functions will decrease over the execution and eventually it holds that a configuration $C'$ is reached i. e., $A \Longmapsto C'$ where $\Psi_{\sum}(C') = 2 \cdot n \cdot (n-1)$, $\Psi_E(C') = 2 \cdot (n-1)$, or $\Psi(C') = 0$.
For every reachable configuration $R$ with $A \Longmapsto R$, $\Psi_{\sum}(R) \leq \Psi_{\sum}(A) \wedge \Psi_E(R) \leq \Psi_E(A) \wedge \Psi(R) \leq \Psi(A)$ and $\Psi_{\sum}(R) > 2 \cdot n \cdot (n-1)$ where there is still a possible linearization step in the directed topology with messages i. e., $\exists p, q, r \in \mathscr{P}.((p < q \wedge q < r) \vee (r < q \wedge q < p)) \wedge ((p, q) \in E_{NTM(R)} \wedge (p, r) \in E_{NTM(R)})$ it holds that $\forall R'.R \longmapsto R' \implies \Psi_{\sum}(R') \leq \Psi_{\sum}(R) \wedge \Psi_E(R') \leq \Psi_E(R) \wedge \Psi(R') \leq \Psi(R)$ and there is always eventually a configuration $S$ reached i. e., $R \Longmapsto S$ with

$$\Psi_{\sum}(S) < \Psi_{\sum}(R) \wedge \Psi(S) < \Psi(R) \wedge \Psi_E(S) \leq \Psi_E(R)$$

Since there is a perfect oracle and there is a possible linearization step in the directed topology with messages i. e., $\exists p, q, r \in \mathscr{P}.((p < q \wedge q < r) \vee (r < q \wedge q < p)) \wedge ((p, q) \in E_{NTM(R)} \wedge (p, r) \in E_{NTM(R)})$, it holds that no process can send *keep-alive*-messages according to 39. Therefore, it is not possible that a process $p$ executes a step with $select\big(findLin\big(p, nb_R(p)\big)\big) = \bot$ and $R \longmapsto R' \equiv Alg_{ges}\big(P_R, P'_R, nb_R, Msgs_R \cup \{(j, p) | j \in nb_R(p)\}, add_R\big)$. According to Lemmata 19, 17 and 15 it holds therefore that $\forall R'.R \longmapsto R' \implies \Psi_{\sum}(R') \leq \Psi_{\sum}(R) \wedge \Psi_E(R') \leq \Psi_E(R) \wedge \Psi(R') \leq \Psi(R)$.
Let $p, q, r \in \mathscr{P}$ with $((p < q \wedge q < r) \vee (r < q \wedge q < p)) \wedge ((p, q) \in E_{NTM(R)} \wedge (p, r) \in E_{NTM(R)})$. Assume $p < q \wedge q < r$ (the proof for $r < q \wedge q < p$ is similar). Furthermore, assume that in every of the below reached configurations $R'$ i. e., $R \Longmapsto R'$ the neighborhood of $p$ has only grown i. e., $nb_R(p) \subseteq nb_{R'}(p)$, otherwise according to Lemma 10 there has to be intermediate configuration $S$ such that the subprocess $Alg_{match}(p)$ executes a step and $select\big(findLin\big(p, nb_S(p)\big)\big) = (j, k) \wedge (j < k \wedge k < p)$ or $select\big(findLin\big(p, nb_S(p)\big)\big) = (j, k) \wedge (j < k \wedge p' < j)$, but then $p$ exe-cutes a linearization step in $S$ and according to Lemmata 15, 19 and 17 $\Psi_{\sum}(S) < \Psi_{\sum}(R) \wedge \Psi(S) < \Psi(R) \wedge \Psi_E(S) \le \Psi_E(R)$. According to Definition 28 there are the following cases:

- Both processes are in the neighborhood of p:
  it holds $q, r \in nb_R(p)$. According to Lemma 3 the subprocess $Alg_{match}(q)$ is enabled in $R$ and stays enabled until it executes a step. According to the fairness assumption 3 $Alg_{match}(q)$ will execute a step after a finite number of steps. Let $R'$ be the configuration before this step. Since up to this point there was at least one possible linearization step in the directed topology with messages, it holds that $\Psi_{\sum}(R') \le \Psi_{\sum}(R) \wedge \Psi_E(R') \le \Psi_E(R) \wedge \Psi(R') \le \Psi(R)$. Since $Alg_{match}(q)$ does not executed a step it holds according to Lemma 10 that $nb_{R'}(p) \supseteq nb_R(p)$ and therefore $q, r \in nb_{R'}(p)$. Hence with $p < q \wedge p < r$ and the definition of $RightN$ it holds that $|nb_{R'}(p)| \ge 2 \wedge |RightN(nb_{R'}(p), p)| \ge 2$. Therefore, with Lemma 6 $select(findLin(p, nb_{R'}(p))) = (j, k) \wedge j < k \wedge ((k < p) \vee (p < j))$ and with Lemmata 19, 15 and 17 for $R' \longmapsto S$ that $\Psi_{\sum}(S) < \Psi_{\sum}(R') \wedge \Psi(S) < \Psi(R') \wedge \Psi_E(S) \le \Psi_E(R')$. Hence with $\Psi_{\sum}(R') \le \Psi_{\sum}(R) \wedge \Psi(R') \le \Psi(R) \wedge \Psi_E(R') \le \Psi_E(R)$ it holds that $\Psi_{\sum}(S) < \Psi_{\sum}(R) \wedge \Psi(S) < \Psi(R) \wedge \Psi_E(S) \le \Psi_E(R)$.

- There is just an add-step necessary to enable the linearization step:
  assume $q \in nb_R(p) \wedge (p \in P'_R \wedge add_R(p) = r)$. According to Lemma 4 there is a configuration $R'$ reached with $r \in nb_{R'}(p)$. According to the assumptions it holds that $q \in nb_{R'}(p)$ and $p < q \wedge q < r$. Therefore, we reduced the problem to the previous case.
  The case $(p \in P'_R \wedge add_R(p) = q) \wedge r \in nb_R(p)$ is similar.

- There is just one message left to receive:
  assume $q \in nb_R(p) \wedge (p, r) \in Msgs_R$. According to Corollary 4 we reach a configuration $R'$ with $r \in nb_{R'}(p)$. According to the assumptions it holds that $q \in nb_{R'}(p)$ and $p < q < r$. Therefore, we reduced the problem to the first case.
  The case $(p, q) \in Msgs_R \wedge r \in nb_R(p)$ is similar.

- There is an add-step necessary and a message left to receive:
  assume $(p \in P'_R \wedge add_R(p) = q) \wedge (p, r) \in Msgs_R$. According to Lemma 4 there are configurations $C \longmapsto C'$ reached with

  $$Alg_{ges}(P_C \cup \{p\}, P'_C \setminus \{p\}, nb_{C'}, Msgs_C, add_C \setminus \{(p, q)\})$$

  $$\text{with } nb_{C'}(x) = \begin{cases} nb_C(x), & \text{if } x \neq p \\ nb_C(p) \cup \{q\}, & \text{if } x = p \end{cases}$$

  and $q \in nb_{C'}(p)$. Up to this point $p$ cannot receive any message as the subprocess $Alg_{rec}(p)$ is not enabled as long as $Alg_{add}(p, q)$ is enabled and according to 2 messages cannot get lost and according to Lemma 2 $p$ is the only process that can receive this message. Therefore, it holds that $(p, r) \in Msgs_{C'}$. According to Corollary 4 a configuration $R'$ is reached with $r \in nb_{R'}(p)$. According to the assumptions it holds that $q \in nb_{R'}(p) \supseteq nb_{C'}(p)$ and $p < q < r$. Therefore, we reduced the problem to the first case.
The case for $(p,q) \in Msgs_R \wedge \big(p \in P'_R \wedge add_R(p) = r\big)$ is similar.

- There are two messages left to receive:
$(p,q) \in Msgs_R \wedge (p,r) \in Msgs_R$. According to Corollary 4 a configuration $C$ is reached with $q \in nb_C(p)$ and a configuration $C'$ is reached with $r \in nb_{C'}(p)$ and either $R \Longmapsto C \Longmapsto C'$ or $R \Longmapsto C' \Longmapsto C$. Assume $R \Longmapsto C \Longmapsto C'$ (the second case is similar). Then according to the assumptions $nb_C(p) \subseteq nb_{C'}(p)$ and therefore $q, r \in nb_{C'}(p)$ and $p < q < r$. Therefore, we reduced the problem to the first case.

Assume first, there is never a configuration $D$ reached i. e., $A \Longmapsto D$ that has a deadlock in the sense that there is a possible linearization step in the undirected network topology with message i. e., $\exists p,q,r \in \mathscr{P}.\{p,q\},\{p,r\} \in E_{UNTM(D)} \wedge ((p < q \wedge q < r) \vee (r < q \wedge q < p))$ but no possible linearization step in the directed topology with messages i. e., $\forall p,q,r \in \mathscr{P}.((p < q \wedge p < r) \vee (q < p \wedge r < p)) \implies ((p,q) \notin E_{NTM(D)} \vee (p,r) \notin E_{NTM(D)})$. Since there is a perfect oracle, it holds that the potential function can not increase and if the configuration is not an (undirected) correct configuration, there are always eventually steps executed which decrease the potential functions strictly as shown before. Therefore, we eventually reach a configuration $C'$ i. e., $A \Longmapsto C'$ with $\Psi_\Sigma(C') = 2 \cdot n \cdot (n-1)$, $\Psi(C') = 0$ or $\Psi_E(C') = 2 \cdot (n-1)$. As shown in the base case it holds then for $C'$ that

$$A \Longmapsto C' \Longmapsto C \quad \wedge \quad NTM(C) = G_{LIN} \wedge NT(C) = G_{LIN}$$

Assume a configuration $D$ is reached where there is no possible linearization step in the directed topology with messages i. e., $\forall p,q,r \in \mathscr{P}.((p < q \wedge p < r) \vee (q < p \wedge r < p)) \implies ((p,q) \notin E_{NTM(D)} \vee (p,r) \notin E_{NTM(D)})$ but still a possible linearization step in the undirected topology with messages i. e., $\exists p,q,r \in \mathscr{P}.\{p,q\},\{p,r\} \in E_{UNTM(D)} \wedge ((p < q \wedge q < r) \vee (r < q \wedge q < p))$. If there would be no possible linearization step in the undirected topology with messages i. e., $\forall p,q,r \in \mathscr{P}.((p < q \wedge p < r) \vee (q < p \wedge r < p)) \implies (\{p,q\} \notin E_{UNTM(D)} \vee \{p,r\} \notin E_{UNTM(D)})$, then it holds that $UNTM(D) = UG_{LIN}$ according to Lemma 7 and $I \Longmapsto D$ whereby $I$ is a connect initial configuration. Therefore, according to Lemma 14 it holds that $\Psi(D) = 0$ and as shown in the base case it holds that $D$ is an undirected correct configuration and

$$A \Longmapsto D \Longmapsto C \quad \wedge \quad NTM(C) = G_{LIN} \wedge NT(C) = G_{LIN}$$

Hence $\Psi(D) > 0$ and therefore $\Psi_\Sigma(D) > 2 \cdot n \cdot (n-1)$. We show that a configuration $D''$ is reached where again at least one linearization step in the directed topology with messages is possible and

$$\Psi_\Sigma(D'') < \Psi_\Sigma(D) \wedge \Psi_E(D'') < \Psi_E(D)$$

Now the perfect oracle $O$ allows to send *keep-alive*-messages. There are two cases. It is either sufficient that one process is sending *keep-alive*-messages or there are two process that have to send *keep-alive*-messages to resolve the deadlock.
For the first case, it holds

$$\exists p,q,r \in \mathscr{P}.\{p,q\},\{p,r\} \in E_{UNTM(D)} \land ((p<q \land q<r) \lor (r<q \land q<p)) \land \\ \big(((p,q) \in E_{NTM(D)} \land (r,p) \in E_{NTM(D)}\big) \lor \big((p,r) \in E_{NTM(D)} \land (q,p) \in E_{NTM(D)})\big)$$

Let $p,q,r \in \mathscr{P}$ with $\{p,q\},\{p,r\} \in E_{UNTM(D)}$ and $((p<q \land q<r) \lor (r<q \land q<p))$ and $\big((p,q) \in E_{NTM(D)} \land (r,p) \in E_{NTM(D)}\big) \lor \big((p,r) \in E_{NTM(D)} \land (q,p) \in E_{NTM(D)}\big)$ and $r$ respectively $q$ is the process that is allowed to send *keep-alive*-messages by $O$ as soon as it knows $p$ and $p$ knows the other process. Assume $p<q \land q<r$ (the proof for $r<q \land q<p$ is similar). Furthermore, assume that in every of the below reached configurations $R$ i. e., $D \Longmapsto R$ the neighborhood of $p' \in \{p,q,r\}$ has only grown i. e., $nb_D(p') \subseteq nb_R(p')$, otherwise according to Lemma 10 there has to be an intermediate configuration $S$ in which the subprocess $Alg_{match}(p')$ executes a step and $select\big(findLin(p',nb_S(p'))\big) = (j,k) \land (j<k \land k<p)$ or $select\big(findLin(p',nb_S(p'))\big) = (j,k) \land (j<k \land p'<j)$, but then $p'$ executes a linearization step in $S$ which is a contradiction to the assumption that there is no possible linearization step in the directed topology with messages and this can only be resolved by sending *keep-alive*-messages as all possible steps are receiving messages that are already in transit and the addition of previously received process ids to the neighborhood. According to Definition 28 there are the following cases:

- $p$ has one process in its neighborhood and the other process has $p$ in its neighborhood: There are two cases $q \in nb_D(p) \land p \in nb_D(r)$ or $p \in nb_D(q) \land r \in nb_D(p)$.
  If $q \in nb_D(p) \land p \in nb_D(r)$ and $r$ is the process that is allowed to send *keep-alive*-messages by $O$, it holds according to Lemma 3 the subprocess $Alg_{match}(r)$ is enabled in $D$ and stays enabled until it executes a step. According to the fairness assumption 3 $Alg_{match}(r)$ will execute a step after a finite number of steps and $O$ allows $r$ to send *keep-alive*-messages according to the assumptions. Let $D'$ be the configuration before this step. It holds according to Lemmata 19, 15 and 17 that $\Psi_\sum(D') \le \Psi_\sum(D) \land \Psi(D') \le \Psi(D) \land \Psi_E(D') \le \Psi_E(D)$ since up to $D'$ no process was allowed to send *keep-alive*-messages. It holds $select(findLin(r,nb_{D'}(r))) = \perp$ (otherwise there would be a possible linearization step for $r$ which is contradictory to the deadlock assumption). According to Corollary 2 it holds

  $$D' \longmapsto E \equiv Alg_{ges}\big(P_{D'},P'_{D'},nb_{D'},Msgs_{D'} \cup \{(j,r) \mid j \in nb_{D'}(r)\},add_{D'}\big)$$

  According to the assumptions $nb_D(r) \subseteq nb_{D'}(r)$. Therefore, it holds $(p,r) \in Msgs_E$. With $p<q<r$ and $q \in nb_E(p) \supseteq nb_D(p)$ it holds with Definition 35 that $RightNM\big(E,p\big) = \{q,r\}$. Therefore, there is a possible linearization step in the directed topology with messages of $E$. From now on $O$ prevents sending *keep-alive*-messages as long as another deadlocked configuration is reached. Now with Lemma 6 there are two cases $|nb_{D'}(r)| = 1$ or $|nb_{D'}(r)| = 2 \land RightN(nb_{D'}(r),r) = 1 \land LeftN(nb_{D'}(r),r) = 1$.
  If $|nb_{D'}(r)| = 1$, it holds with Lemmata 19 and 17 $\Psi_\sum(E) = \Psi_\sum(D') + dist(r,p) \land \Psi_E(E) =$

$\Psi_E(D')$ since $RightNM(D',p) = \{q\}$, $RightNM(E,p) = \{q,r\}$ and $q < r$. Since $(p,r) \in Msgs_E$, we will according to Corollary 4 reach a configuration $G$ with $r \in nb_G(p)$. According to the assumption it holds $q \in nb_G(p) \supseteq nb_D(p)$. Since no other process was allowed to send *keep-alive*-messages by $O$ and until now there was no enabled linearization step it holds $nb_G(p) = \{q,r\}$ and this remains unchanged until $p$ executes the linearization step. Therefore, according to Lemma 3 the subprocess $Alg_{match}(p)$ is enabled in $G$ and stays enabled until it executes a step. According to the fairness assumption 3 $Alg_{match}(p)$ will execute a step after a finite number of steps. Let $G'$ be the configuration before this step. Since up to this point there was one possible linearization step in the directed topology with messages, it holds that $\Psi_\Sigma(G') \leq \Psi_\Sigma(E) \wedge \Psi_E(G') \leq \Psi_E(E) \wedge \Psi(G') \leq \Psi(E)$. With $\Psi_\Sigma(D') \leq \Psi_\Sigma(D) \wedge \Psi_E(D') \leq \Psi_E(D)$ and $\Psi_\Sigma(E) = \Psi_\Sigma(D') + dist(r,p) \wedge \Psi_E(E) = \Psi_E(D')$ it holds that $\Psi_\Sigma(G') \leq \Psi_\Sigma(D) + dist(r,p) \wedge \Psi_E(G') \leq \Psi_E(D)$. Since $nb_{G'}(p) = \{q,r\}$, it holds with the definition of $select(findLin())$ that $select(findLin(p, nb_{G'}(p))) = (q,r)$. Since $q < r$ and $p < q$, it holds with Corollary 2 that

$$G' \longmapsto D'' \equiv Alg_{ges}\left(P_{G'}, P'_{G'}, nb_{G'}, Msgs_{G'} \cup \{(r,q)\}, add_{G'}\right)$$
$$\text{and } \; nb_{D''}(x) = \begin{cases} nb_{G'}(x), & \text{if } x \neq p \\ nb_{G''}(p) \setminus \{r\}, & \text{if } x = p \end{cases}$$

Hence with Lemma 15 it holds $\Psi(D'') < \Psi(G')$. It holds with the assumptions that $p \in nb_D(r) \supseteq nb_{G'}(r)$ and with $p < r$ and Definition 35 therefore that $LeftNM(G',r) = \{p\}$ (if $|LeftNM(G',r)| > 1$ another process must have send $r$ a message which is contradictory to the deadlock assumption and $O$). Hence with $(r,q) \in Msgs_{D''}$ and Definition 35 $LeftNM(D'',r) = \{p,q\}$. Therefore, with $p < q$ and Lemma 17 it holds $\Psi_E(D'') < \Psi_E(G')$. With $\Psi(D'') < \Psi(G')$ and Definition 38 it holds

$$\begin{aligned}
\Psi_\Sigma(D'') &= \Psi(D'') + n \cdot \Psi_E(D'') \\
&< \Psi(G') + n \cdot \Psi_E(D'') \\
&\leq \Psi(G') + n \cdot \left(\Psi_E(G') - 1\right) \\
&= \Psi(G') + n \cdot \Psi_E(G') - n \\
&< \Psi(G') + n \cdot \Psi_E(G') - (n-1) \\
&= \Psi(G') + n \cdot \Psi_E(G') - maxdist \\
&= \Psi_\Sigma(G') - maxdist
\end{aligned}$$

Hence with $\Psi_\Sigma(G') \leq \Psi_\Sigma(D) + dist(r,p) \wedge \Psi_E(G') \leq \Psi_E(D)$ and $dist(r,p) \leq maxdist$ it holds that $\Psi_\Sigma(D'') < \Psi_\Sigma(G') - maxdist \leq \Psi_\Sigma(D) + dist(r,p) - maxdist \leq \Psi_\Sigma(D)$. Therefore, with $LeftNM(D'',r) = \{p,q\}$ it holds that there is a possible linearization step in the directed topology with messages of $D''$ and

$$\Psi_\Sigma(D'') < \Psi_\Sigma(D) \wedge \Psi_E(D'') < \Psi_E(D)$$

If $|nb_{D'}(r)| = 2 \wedge RightN(nb_{D'}(r), r) = 1 \wedge LeftN(nb_{D'}(r), r) = 1$, it holds that there is a process $s \in \mathscr{P}$ with $RightN(D'r, r) = \{s\}$ and hence $(s, r) \in Msgs_E$. There are two cases. If $LeftNM(D', s) = \emptyset \vee (LeftNM(D', s) = \{q'\} \wedge q' < r)$ (if $|LeftNM(D', s)| > 1$, another process has sent $s$ a message which is contradictory to the deadlock assumption and $O$), it holds with Definition 36 $ShortestLeftN(D', s) = \perp$ or $ShortestLeftN(D', s) = q'$ with $q' < r$. With $(s, r) \in Msgs_E$ it holds therefore with Definition 35 that $r \in LeftNM(E, s) = LeftNM(D', s) \cup \{r\}$ and since $q' < r$ with Definition 36 $ShortestLeftN(E, s) = r$. Hence with Lemmata 17 and 19 it holds $\Psi_{\sum}(E) < \Psi_{\sum}(D') \wedge \Psi_E(E) < \Psi_E(D')$. Therefore, with $\Psi_{\sum}(D') \leq \Psi_{\sum}(D) \wedge \Psi_E(D') \leq \Psi_E(D)$ it follows $\Psi_{\sum}(E) < \Psi_{\sum}(D) \wedge \Psi_E(E) < \Psi_E(D)$. With $RightNM(E, p) = \{q, r\}$ it holds that there is a possible linearization step in the directed topology with messages of $E$ and

$$\Psi_{\sum}(E) < \Psi_{\sum}(D) \wedge \Psi_E(E) < \Psi_E(D)$$

and hence $E$ is the searched configuration $D''$ itself.

If $LeftNM(D', s) = \{q'\} \wedge r \leq q'$ (if $|LeftNM(D', s)| > 1$, another process has sent $s$ a message in contradiction to the deadlock assumption and $O$), it holds with Lemmata 17 and 19 that $\Psi_{\sum}(E) = \Psi_{\sum}(D') + dist(r, p) + dist(r, s) \wedge \Psi_E(E) = \Psi_E(D')$. With Definition 35 it holds $p < r < s$. Therefore, with Definition 15 it follows $dist(r, p) + dist(r, s) = dist(p, s) \leq maxdist$. Hence $\Psi_{\sum}(E) \leq \Psi_{\sum}(D') + maxdist \wedge \Psi_E(E) = \Psi_E(D')$. Since $(p, r) \in Msgs_E$, we reach according to Corollary 4 a configuration $G$ with $r \in nb_G(p)$. With $(s, r) \in Msgs_E$ and $LeftNM(D', s) = \{q'\}$ we reach with Definition 35, Lemma 4 and Corollary 4 a configuration $G'$ with $\{r, q'\} \subseteq nb_{G'}(s)$. Therefore, according to Lemma 3 the subprocess $Alg_{match}(p)$ is enabled in $G$ and the subprocess $Alg_{match}(s)$ is enabled in $G'$ and they stay enabled until they each execute a step. According to the fairness assumption 3, $Alg_{match}(p)$ executes a step after a finite number of steps. Let $H$ be the configuration before this step. Additionally, according to the fairness assumption 3, $Alg_{match}(s)$ executes a step after a finite number of steps. Let $H'$ be the configuration before this step. It holds either $D' \Longmapsto H \Longmapsto H'$ or $D' \Longmapsto H' \Longmapsto H$. Assume $E \Longmapsto H \Longmapsto H'$ (the proof for $E \Longmapsto H' \Longmapsto H$ is similar). Therefore, with the assumptions it holds that $q \in nb_H(p) \supseteq nb_D(p)$. Since no other process was allowed to send *keep-alive*-messages by $O$ and until now there was no other enabled linearization step that included $p$ it holds $nb_H(p) = \{q, r\}$. Since up to this point there was one possible linearization step in the directed topology with messages it holds that $\Psi_{\sum}(H) \leq \Psi_{\sum}(E) \wedge \Psi_E(H) \leq \Psi_E(E) \wedge \Psi(H) \leq \Psi(E)$. With $\Psi_{\sum}(D') \leq \Psi_{\sum}(D) \wedge \Psi_E(D') \leq \Psi_E(D)$ and $\Psi_{\sum}(E) \leq \Psi_{\sum}(D') + maxdist \wedge \Psi_E(E) = \Psi_E(D')$ it holds that $\Psi_{\sum}(H) \leq \Psi_{\sum}(D) + maxdist \wedge \Psi_E(H) \leq \Psi_E(D)$. Since $nb_H(p) = \{q, r\}$ and $p < q < r$, it holds with the definition of $select(findLin())$ that $select(findLin(p, nb_H(p))) = (q, r)$. Since $q < r$ and $p < q$, it holds with Corollary 2 that

$$H \longmapsto D'' \equiv Alg_{ges}(P_H, P'_H, nb_H, Msgs_H \cup \{(r, q)\}, add_H)$$

$$\text{and } nb_{D''}(x) = \begin{cases} nb_H(x), & \text{if } x \neq p \\ nb_H(p) \setminus \{r\}, & \text{if } x = p \end{cases}$$

Hence with Lemma 15 it holds $\Psi(D'') < \Psi(H)$. It holds with the assumptions that $p \in nb_D(r) \supseteq nb_H(r)$ and with $p < r$ and Definition 35 therefore that $LeftNM(H, r) = \{p\}$ (if $|LeftNM(H, r)| > 1$ another process must have send $r$ a message which is contradictory to $O$ and that it is the first executed linearization step after the deadlock resolving). Hence with $(r, q) \in Msgs_{D''}$ and Definition 35 $LeftNM\big(D'', r\big) = \{p, q\}$. Therefore, with $p < q$ and Lemma 17 it holds $\Psi_E(D'') < \Psi_E(H)$. With $\Psi(D'') < \Psi(H)$ and Definition 38 it holds

$$\begin{aligned}
\Psi_\Sigma(D'') &= \Psi(D'') + n \cdot \Psi_E(D'') \\
&< \Psi(H) + n \cdot \Psi_E(D'') \\
&\leq \Psi(H) + n \cdot (\Psi_E(H) - 1) \\
&= \Psi(H) + n \cdot \Psi_E(H) - n \\
&< \Psi(H) + n \cdot \Psi_E(H) - (n-1) \\
&= \Psi(H) + n \cdot \Psi_E(H) - maxdist \\
&= \Psi_\Sigma(H) - maxdist
\end{aligned}$$

Hence with $\Psi_\Sigma(H) \leq \Psi_\Sigma(D) + maxdist \wedge \Psi_E(H) \leq \Psi_E(D)$ it holds that $\Psi_\Sigma(D'') < \Psi_\Sigma(H) - maxdist \leq \Psi_\Sigma(D) + maxdist - maxdist = \Psi_\Sigma(D)$. Therefore, with $LeftNM\big(D'', r\big) = \{p, q\}$ it holds that there is a possible linearization step in the directed topology with messages of $D''$ and

$$\Psi_\Sigma(D'') < \Psi_\Sigma(D) \wedge \Psi_E(D'') < \Psi_E(D)$$

If $p \in nb_D(q) \wedge r \in nb_D(p)$ and $q$ is the process that is allowed to send *keep-alive*-messages by $O$ it holds according to Lemma 3 the subprocess $Alg_{match}\big(q\big)$ is enabled in $D$ and stays enabled until it executes a step. According to the fairness assumption 3, $Alg_{match}\big(q\big)$ executes a step after a finite number of steps and $O$ allows $q$ to send *keep-alive*-messages according to the assumptions. Let $D'$ be the configuration before this step. It holds according to Lemmata 19, 15 and 17 that $\Psi_\Sigma(D') \leq \Psi_\Sigma(D) \wedge \Psi(D') \leq \Psi(D) \wedge \Psi_E(D') \leq \Psi_E(D)$ since up to $D'$ no process was allowed to send *keep-alive*-messages . It holds $select\big(findLin\big(q, nb_{D'}(q)\big)\big) = \perp$ (otherwise there would be a possible linearization step for $q$ which is contradictory to the deadlock assumption). According to Corollary 2, it holds

$$D' \longmapsto D'' \equiv Alg_{ges}\big(P_{D'}, P'_{D'}, nb_{D'}, Msgs_{D'} \cup \{(j, q) | j \in nb_{D'}(q)\}, add_{D'}\big)$$

According to the assumptions $nb_D(q) \subseteq nb_{D'}(q)$. Therefore, it holds $(p, q) \in Msgs_{D''}$. Additionally also $nb_D(p) \subseteq nb_{D'}(p) \subseteq nb_{D''}(p)$ and therefore $r \in nb_{D'}(p)$ and $r \in nb_{D''}(p)$. With the assumptions and Definition 35 it holds that $RightNM\big(D', p\big) = \{r\}$ and with $(p, q) \in Msgs_{D''}$ it holds $RightNM\big(D'', p\big) = \{r, q\}$. Therefore, with $q < r$ and Lemmata 19 and 17 it holds $\Psi_\Sigma(D'') < \Psi_\Sigma(D') \wedge \Psi_E(D'') < \Psi_E(D')$. Hence with $\Psi_\Sigma(D') \leq \Psi_\Sigma(D) \wedge \Psi_E(D') \leq \Psi_E(D)$ it follows $\Psi_\Sigma(D'') < \Psi_\Sigma(D)$ and $\Psi_E(D'') < \Psi_E(D)$. Therefore, with $RightNM\big(D'', p\big) = \{r, q\}$ there is a possible linearization step in the directed topology with messages of $D''$ and

$$\Psi_{\sum}(D'') < \Psi_{\sum}(D) \wedge \Psi_E(D'') < \Psi_E(D)$$

- $p$ has one process in its neighborhood and for the other process is an add-step necessary:
  assume $q \in nb_D(p) \wedge \big(r \in P'_D \wedge add_D(r) = p\big)$. According to Lemma 4 there is a configuration $E$ reached with $p \in nb_E(r)$. According to the assumptions it holds that $q \in nb_E(p) \supseteq nb_D(p)$. Hence, we reduced the problem to the previous case.
  The proof for $\big(q \in P'_D \wedge add_D(q) = p\big) \wedge r \in nb_D(p)$ is similar.

- $p$ has one process in its neighborhood and the other process has a message left to receive:
  assume $q \in nb_D(p) \wedge (r, p) \in Msgs_D$. According to Corollary 4 we reach a configuration $E$ with $p \in nb_E(r)$. According to the assumptions it holds that $q \in nb_E(p) \supseteq nb_D(p)$. Hence, we reduced the problem to the first case.
  The proof for $(q, p) \in Msgs_D \wedge r \in nb_D(p)$ is similar.

- One process has $p$ in its neighborhood and for $p$ is an add-step necessary:
  assume $\big(p \in P'_D \wedge add_D(p) = q\big) \wedge p \in nb_D(r)$. According to Lemma 4 there is a configuration $E$ reached with $q \in nb_E(p)$. According to the assumptions it holds that $p \in nb_E(r) \supseteq nb_D(r)$. Hence, we reduced the problem to the first case.
  The proof for $p \in nb_D(q) \wedge \big(p \in P'_D \wedge add_D(p) = r\big)$ is similar.

- One process has $p$ in its neighborhood and for $p$ is a message left to receive: assume $(p, q) \in Msgs_D \wedge p \in nb_D(r)$. According to Corollary 4 we reach a configuration $E$ with $q \in nb_E(p)$. According to the assumptions it holds that $p \in nb_E(r) \supseteq nb_D(r)$. Hence, we reduced the problem to the first case.
  The proof for $p \in nb_D(q) \wedge (p, r) \in Msgs_D$is similar.

- There are two add-steps necessary:
  assume $\big(p \in P'_D \wedge add_D(p) = q\big) \wedge \big(r \in P'_D \wedge add_D(r) = p\big)$. According to Lemma 4 there is a configuration $E$ reached with $q \in nb_E(p)$ and a configuration $E'$ is reached with $p \in nb_{E'}(r)$ and either $D \Longmapsto E \Longmapsto E'$ or $D \Longmapsto E' \Longmapsto E$. Assume $D \Longmapsto E \Longmapsto E'$ (the second case is similar). Then according to the assumptions $nb_E(p) \subseteq nb_{E'}(p)$ and therefore $q \in nb_{E'}(p)$. Hence, we reduced the problem to the first case.
  The proof for $\big(q \in P'_D \wedge add_D(q) = p\big) \wedge \big(p \in P'_D \wedge add_D(p) = r\big)$ is similar.

- There is an add-step necessary and a message left to receive:
  assume $\big(p \in P'_D \wedge add_D(p) = q\big) \wedge (r, p) \in Msgs_D$. According to Lemma 4 there is a configuration $E$ reached with $q \in nb_E(p)$ and according to Corollary 4 a configuration $E'$ with $p \in nb_{E'}(r)$. Either $D \Longmapsto E \Longmapsto E'$ or $D \Longmapsto E' \Longmapsto E$. Assume $D \Longmapsto E \Longmapsto E'$ (the second case is similar). Then according to the assumptions $nb_E(p) \subseteq nb_{E'}(p)$ and therefore $q \in nb_{E'}(p)$. Hence, we reduced the problem to the first case.
The proof for $(q,p) \in Msgs_D \wedge \big(p \in P'_D \wedge add_D(p) = r\big)$ is similar and the proofs for $(p,q) \in Msgs_D \wedge \big(r \in P'_D \wedge add_D(r) = p\big)$ and $\big(q \in P'_D \wedge add_D(q) = p\big) \wedge (p,r) \in Msgs_D$ are similar with switched roles.

- There are two messages left to receive:
  assume $(p,q) \in Msgs_D \wedge (r,p) \in Msgs_D$. According to Corollary 4 there is a configuration $E$ reached with $q \in nb_E(p)$ and a configuration $E'$ is reached with $p \in nb_{E'}(r)$ and either $D \Longmapsto E \Longmapsto E'$ or $D \Longmapsto E' \Longmapsto E$. Assume $D \Longmapsto E \Longmapsto E'$ (the second case is similar). Then according to the assumptions $nb_E(p) \subseteq nb_{E'}(p)$ and therefore $q \in nb_{E'}(p)$. Hence, we reduced the problem to the first case.
  The proof for $(q,p) \in Msgs_D \wedge (p,r) \in Msgs_D$ is similar.

For the second case holds

$$\exists p,q,r \in \mathscr{P}.\{p,q\},\{p,r\} \in E_{UNTM(D)} \wedge ((p < q \wedge q < r) \vee (r < q \wedge q < p)) \quad \text{and}$$
$$\forall p,q,r \in \mathscr{P}.((p < q \wedge p < r) \vee (q < p \wedge r < p)) \wedge \{p,q\},\{p,r\} \in E_{UNTM(D)}$$
$$\Longrightarrow ((p,q) \notin E_{NTM(D)} \wedge (p,r) \notin E_{NTM(D)})$$

Let $p,q,r \in \mathscr{P}$ with $\{p,q\},\{p,r\} \in E_{UNTM(D)}$, $((p < q \wedge q < r) \vee (r < q \wedge q < p))$ and $r$ and $q$ are the processes that are allowed to send *keep-alive*-messages by $O$ as soon as they both know $p$. Furthermore, assume that in every of the below reached configurations $R$ i. e., $D \Longmapsto R$ the neighborhood of $p' \in \{p,q,r\}$ has only grown i. e., $nb_D(p') \subseteq nb_R(p')$, otherwise according to Lemma 10 there has to be intermediate configuration $S$ in which the subprocess $Alg_{match}\big(p'\big)$ executes a step and $select\big(findLin\big(p', nb_S(p')\big)\big) = (j,k) \wedge (j < k \wedge k < p)$ or $select\big(findLin\big(p', nb_S(p')\big)\big) = (j,k) \wedge (j < k \wedge p' < j)$, but then $p'$ executes a linearization step in $S$ which is a contradiction to the assumption that there is no possible linearization step in the directed topology with messages and this can only be resolved by sending *keep-alive*-messages as all possible steps are receiving messages that are already in transit and the addition of previously received process ids to the neighborhood. Assume $p < q < r$ (the proof for $r < q < p$ is similar). According to Definitions 28 and 30 there are the following cases:

- $p$ is in the neighborhood of both processes:
  $p \in nb_D(q) \wedge p \in nb_D(r)$. According to Lemma 3 the subprocesses $Alg_{match}\big(q\big)$ and $Alg_{match}(r)$ are enabled in $D$ and stay enabled until they each execute a step. According to the assumption regarding $O$, $Alg_{match}\big(q\big)$ executes a step after a finite number of steps. Let $E$ be the configuration before this step. It holds $select\big(findLin\big(q, nb_E(q)\big)\big) = \bot$ (otherwise there would be a possible linearization step for $q$). According to the assumption regarding $O$, $Alg_{match}(r)$ executes a step after a finite number of steps. Let $E'$ be the configuration before this step. It holds $select(findLin(r, nb_{E'}(r))) = \bot$ (otherwise there would be a possible linearization step for $q$). Now there are two cases regarding the order of this steps.

If $D \Longmapsto E' \Longmapsto E$, it holds according to Lemmata 19, 15 and 17 that $\Psi_{\sum}(E') \leq \Psi_{\sum}(D) \wedge \Psi(E') \leq \Psi(D) \wedge \Psi_E(E') \leq \Psi_E(D)$ since up to $E'$ no process was allowed to send *keep-alive*-messages . According to Corollary 2 it holds

$$E' \longmapsto F' \equiv Alg_{ges}\left(P_{E'}, P'_{E'}, nb_{E'}, Msgs_{E'} \cup \{(j,r) | j \in nb_{E'}(r)\}, add_{E'}\right)$$

and according to the assumptions $nb_A(r) \subseteq nb_{E'}(r)$. Therefore, it holds $(p,r) \in Msgs_{F'}$. It holds according to the assumptions and Definition 35 that $RightNM(E',p) = \emptyset$ and $RightNM(F',p) = \{r\}$. Therefore, with Lemmata 19 and 17 it holds $\Psi_{\sum}(F') < \Psi_{\sum}(E') \wedge \Psi_E(F') < \Psi_E(E')$. Since no other processes are allowed to send *keep-alive*-messages, it holds according to Lemmata 19, 15 and 17 that $\Psi_{\sum}(E) \leq \Psi_{\sum}(F') \wedge \Psi(E) \leq \Psi(F') \wedge \Psi_E(E) \leq \Psi_E(F')$. According to Corollary 2 it holds

$$E \longmapsto D'' \equiv Alg_{ges}\left(P_E, P'_E, nb_E, Msgs_E \cup \{(j,r) | j \in nb_E(q)\}, add_E\right)$$

and according to the assumptions $nb_A(q) \subseteq nb_E(q)$. Therefore, it holds $(p,q) \in Msgs_{D''}$. Since $(p,r) \in Msgs_{F'}$, it follows with Assumption 2 and Lemma 2 that $p$ is the only process that can receive this message and therefore $(p,r) \in Msgs_{D''} \vee \left(p \in P'_{D''} \wedge add_{D''}(p) = r\right) \vee r \in nb_{D''}(p)$. With the assumptions and Definition 35 it holds that $RightNM(E,p) = \{r\}$ and $RightNM(D'',p) = \{r,q\}$. Therefore, with $q < r$ and Lemmata 19 and 17 it holds $\Psi_{\sum}(D'') < \Psi_{\sum}(E) \wedge \Psi_E(D'') < \Psi_E(E)$. Hence with $\Psi_{\sum}(E') \leq \Psi_{\sum}(D) \wedge \Psi_E(E') \leq \Psi_E(D)$, $\Psi_{\sum}(F') < \Psi_{\sum}(E') \wedge \Psi_E(F') < \Psi_E(E')$ and $\Psi_{\sum}(E) \leq \Psi_{\sum}(F') \wedge \Psi_E(E) \leq \Psi_E(F')$ it follows $\Psi_{\sum}(D'') < \Psi_{\sum}(D)$. Therefore, with $RightNM(D'',p) = \{r,q\}$ there is a possible linearization step in the directed topology with messages of $D''$ and

$$\Psi_{\sum}(D'') < \Psi_{\sum}(D) \wedge \Psi_E(D'') < \Psi_E(D)$$

If $D \Longmapsto E \Longmapsto E'$, it holds according to Lemmata 19, 15 and 17 that $\Psi_{\sum}(E) \leq \Psi_{\sum}(D) \wedge \Psi(E) \leq \Psi(D) \wedge \Psi_E(E) \leq \Psi_E(D)$ since up to $E$ no process was allowed to send *keep-alive*-messages. According to Corollary 2 it holds

$$E \longmapsto F \equiv Alg_{ges}\left(P_E, P'_E, nb_E, Msgs_E \cup \{(j,r) | j \in nb_E(q)\}, add_E\right)$$

and according to the assumptions $nb_A(q) \subseteq nb_E(q)$. Therefore, it holds $(p,q) \in Msgs_F$. It holds according to the assumptions and Definition 35 that $RightNM(E,p) = \emptyset$ and $RightNM(F,p) = \{q\}$. Therefore, with Lemmata 17 and 19 it holds $\Psi_{\sum}(F) < \Psi_{\sum}(E) \wedge \Psi_E(F) < \Psi_E(E)$. Since no other processes are allowed to send *keep-alive*-messages, it holds according to Lemmata 19, 15 and 17 that $\Psi_{\sum}(E') \leq \Psi_{\sum}(F) \wedge \Psi(E') \leq \Psi(F) \wedge \Psi_E(E') \leq \Psi_E(F)$. According to Corollary 2 it holds

$$E' \longmapsto D' \equiv Alg_{ges}\left(P_{E'}, P'_{E'}, nb_{E'}, Msgs_{E'} \cup \{(j,r) | j \in nb_{E'}(q)\}, add_{E'}\right)$$

and according to the assumptions $nb_A(q) \subseteq nb_{E'}(q)$. Therefore, it holds $(p,r) \in Msgs_{D'}$.

Since $(p,q) \in Msgs_F$, it follows with Assumption 2 and Lemma 2 that $p$ is the only process that can receive this message and therefore $(p,q) \in Msgs_{D''} \vee \left(p \in P'_{D'} \wedge add_{D'}(p) = q\right) \vee q \in nb_{D''}(p)$. Therefore, with $(p,r) \in Msgs_{D'}$ there is a possible linearization step in the directed topology with messages of $D''$. And with $\Psi_\sum(E') \leq \Psi_\sum(F) \wedge \Psi_E(E') \leq \Psi_E(F)$, $\Psi_\sum(F) < \Psi_\sum(E) \wedge \Psi_E(F) < \Psi_E(E)$ and $\Psi_\sum(E) \leq \Psi_\sum(D) \wedge \Psi_E(E) \leq \Psi_E(D)$ it holds $\Psi_\sum(E') < \Psi_\sum(D) \wedge \Psi_E(E') < \Psi_E(D)$. Now with Lemma 6 there are two cases $|nb_{E'}(r)| = 1$ or $|nb_{E'}(r)| = 2 \wedge RightN(nb_{E'}(r), r) = 1 \wedge LeftN(nb_{E'}(r), r) = 1$.
If $|nb_{E'}(r)| = 2 \wedge RightN(nb_{E'}(r), r) = 1 \wedge LeftN(nb_{E'}(r), r) = 1$, it holds that there is a process $s \in \mathscr{P}$ with $RightN\left(E'r, r\right) = \{s\}$ and therefore $(s,r) \in Msgs_{D'}$. In holds with Definitions 35, 30 and 28 the assumptions that $LeftNM\left(E', s\right) = \emptyset \vee LeftNM\left(E', s\right) = \{q\}$ (otherwise it would have been sufficient if only $r$ sends *keep-alive*-messages and we would have been in the other case). Therefore, with Definition 36 $ShortestLeftN(E', s) = \perp \vee ShortestLeftN(E', s) = q$. With $(s,r) \in Msgs_{D'}$ it holds therefore $r \in LeftNM\left(D', s\right) = LeftNM\left(E', s\right) \cup \{r\}$ and since $q < r$ with Definition 36 $ShortestLeftN(D', s) = r$. Hence with Lemmata 17 and 19 it holds $\Psi_\sum(D') < \Psi_\sum(E') \wedge \Psi_E(D') < \Psi_E(E')$. With $\Psi_\sum(E') < \Psi_\sum(D) \wedge \Psi_E(E') < \Psi_E(D)$ it holds that there is a possible linearization step in the directed topology with messages of $D'$ and

$$\Psi_\sum(D') < \Psi_\sum(D) \wedge \Psi_E(D') < \Psi_E(D)$$

and hence $D'$ is the searched configuration $D''$ itself.

If $|nb_{E'}(r)| = 1$, it holds with Lemmata 19 and 17 $\Psi_\sum(D') = \Psi_\sum(E') + dist(r,p) \wedge \Psi_E(D') = \Psi_E(E')$ since $RightNM\left(E', p\right) = \{q\}$, $RightNM\left(D', p\right) = \{q, r\}$ and $q < r$. From now on $O$ prevents sending *keep-alive*-messages as long as another deadlocked configuration is reached. Since $(p,q) \in Msgs_{D'} \vee \left(p \in P'_{D'} \wedge add_{D'}(p) = q\right) \vee q \in nb_{D'}(p)$ and $(p,r) \in Msgs_{D'}$, we reach according to Lemma 4 and Corollary 4 configurations $G$ and $G'$ with $q \in nb_G(p)$ and $r \in nb_{G'}(p)$. Assume $D' \Longmapsto G \Longmapsto G'$ (the proof for $D' \Longmapsto G' \Longmapsto G$ is similar). According to the assumption it holds $q \in nb_{G'}(p) \supseteq nb_G(p)$. Since no other process was allowed to send *keep-alive*-messages by $O$ and until now there was no enabled linearization step, it holds $nb_{G'}(p) = \{q, r\}$ and this remains unchanged until $p$ executes the linearization step. Therefore, according to Lemma 3 the subprocess $Alg_{match}(p)$ is enabled in $G'$ and stays enabled until it executes a step. According to the fairness assumption 3, $Alg_{match}(p)$ executes a step after a finite number of steps. Let $G''$ be the configuration before this step. Since up to this point there was one possible linearization step in the directed topology with messages, it holds that $\Psi_\sum(G'') \leq \Psi_\sum(D') \wedge \Psi_E(G'') \leq \Psi_E(D') \wedge \Psi(G'') \leq \Psi(D')$. With $\Psi_\sum(E') < \Psi_\sum(D) \wedge \Psi_E(E') < \Psi_E(D)$ and $\Psi_\sum(D') = \Psi_\sum(E') + dist(r,p) \wedge \Psi_E(D') = \Psi_E(E')$ it holds that $\Psi_\sum(G'') < \Psi_\sum(D) + dist(r,p) \wedge \Psi_E(G'') < \Psi_E(D)$. Since $nb_{G''}(p) = \{q, r\}$ and $p < q < r$, it holds with the definition of $select(findLin())$ that $select\left(findLin\left(p, nb_{G''}(p)\right)\right) = (q, r)$. Since $q < r$ and $p < q$, it holds with Corollary 2 that

$$G'' \longmapsto D'' \equiv Alg_{ges}\left(P_{G''}, P'_{G''}, nb_{G''}, Msgs_{G''} \cup \{(r,q)\}, add_{G''}\right)$$

$$\text{and } nb_{D''}(x) = \begin{cases} nb_{G''}(x), & \text{if } x \neq p \\ nb_{G''}(p) \setminus \{r\}, & \text{if } x = p \end{cases}$$

Hence with Lemma 15 it holds $\Psi(D'') < \Psi(G'')$. It holds with the assumptions that $p \in nb_D(r) \supseteq nb_{G''}(r)$ and with $p < r$ and Definition 35 therefore that $\mathit{LeftNM}(G'', r) = \{p\}$ (if $|\mathit{LeftNM}(G'', r)| > 1$ another process must have send $r$ a message which is contradictory to the deadlock assumption and $O$). Hence with $(r, q) \in \mathit{Msgs}_{D''}$ and Definition 35 $\mathit{LeftNM}(D'', r) = \{p, q\}$. Therefore, with $p < q$ and Lemma 17 it holds $\Psi_E(D'') < \Psi_E(G'')$. With $\Psi(D'') < \Psi(G'')$ and Definition 38 it holds

$$\begin{aligned}
\Psi_\Sigma(D'') &= \Psi(D'') + n \cdot \Psi_E(D'') \\
&< \Psi(G'') + n \cdot \Psi_E(D'') \\
&\leq \Psi(G'') + n \cdot \big(\Psi_E(G'') - 1\big) \\
&= \Psi(G'') + n \cdot \Psi_E(G'') - n \\
&< \Psi(G'') + n \cdot \Psi_E(G'') - (n-1) \\
&= \Psi(G'') + n \cdot \Psi_E(G'') - \mathit{maxdist} \\
&= \Psi_\Sigma(G'') - \mathit{maxdist}
\end{aligned}$$

Hence with $\Psi_\Sigma(G'') < \Psi_\Sigma(D) + \mathit{dist}(r, p) \wedge \Psi_E(G'') < \Psi_E(D)$ and $\mathit{dist}(r, p) \leq \mathit{maxdist}$ it holds that $\Psi_\Sigma(D'') < \Psi_\Sigma(G'') - \mathit{maxdist} < \Psi_\Sigma(D) + \mathit{dist}(r, p) - \mathit{maxdist} \leq \Psi_\Sigma(D)$. Therefore, with $\mathit{LeftNM}(D'', r) = \{p, q\}$ it holds that there is a possible linearization step in the directed topology with messages of $D''$ and

$$\Psi_\Sigma(D'') < \Psi_\Sigma(D) \wedge \Psi_E(D'') < \Psi_E(D)$$

- $p$ is in the neighborhood of one process and for the other one is an add-step necessary: assume $p \in nb_D(q) \wedge \big(r \in P'_D \wedge \mathit{add}_D(r) = p\big)$. According to Lemma 4 there will be a configuration $E$ reached with $p \in nb_E(r)$. According to the assumptions it holds that $p \in nb_E(q) \supseteq nb_D(q)$. Hence, we reduced the problem to the previous case.
  The proof for $\big(q \in P'_D \wedge \mathit{add}_D(q) = p\big) \wedge p \in nb_D(r)$ is similar.

- $p$ is in the neighborhood of one process and the other one has a message left to receive: $p \in nb_D(q) \wedge (r, p) \in \mathit{Msgs}_D$. According to Corollary 4 we reach a configuration $E$ with $p \in nb_E(r)$. According to the assumptions it holds that $p \in nb_E(q) \supseteq nb_D(q)$. Hence, we reduced the problem to the first case.
  The proof for $(q, p) \in \mathit{Msgs}_D \wedge p \in nb_D(r)$ is similar.

- For both processes is an add-step necessary:
  $\big(q \in P'_D \wedge \mathit{add}_D(q) = p\big) \wedge \big(r \in P'_D \wedge \mathit{add}_D(r) = p\big)$. According to Lemma 4 there is a configuration $E$ reached with $p \in nb_E(q)$ and a configuration $E'$ is reached with $p \in nb_{E'}(r)$ and either $D \Longmapsto E \Longmapsto E'$ or $D \Longmapsto E' \Longmapsto E$. Assume $D \Longmapsto E \Longmapsto E'$ (the second case is similar). Then according to the assumptions $nb_E(q) \subseteq nb_{E'}(q)$ and therefore $p \in nb_{E'}(q)$.

Hence, we reduced the problem to the first case.

- For one process is an add-step necessary and the other one has a message left to receive: assume $\left(q \in P'_D \wedge add_D(q) = p\right) \wedge (r,p) \in Msgs_D$. According to Lemma 4 there is a configuration $E$ reached with $p \in nb_E(q)$ and according to Corollary 4 a configuration $E'$ with $p \in nb_{E'}(r)$. Either $D \Longmapsto E \Longmapsto E'$ or $D \Longmapsto C' \Longmapsto C$. Assume $D \Longmapsto E \Longmapsto E'$ (the second case is similar). Then according to the assumptions $nb_E(q) \subseteq nb_{E'}(q)$ and therefore $p \in nb_{E'}(q)$. Hence, we reduced the problem to the first case.
The proof for $(q,p) \in Msgs_D \wedge \left(r \in P'_D \wedge add_D(r) = p\right)$ is similar.

- Both processes have a message left to receive:
$(q,p) \in Msgs_D \wedge (r,p) \in Msgs_D$ . According to Corollary 4 there is a configuration $E$ reached with $p \in nb_E(q)$ and a configuration $E'$ is reached with $p \in nb_{E'}(r)$ and either $D \Longmapsto E \Longmapsto E'$ or $D \Longmapsto E' \Longmapsto E$. Assume $D \Longmapsto E \Longmapsto E'$ (the second case is similar). Then according to the assumptions $nb_E(q) \subseteq nb_{E'}(q)$ and therefore $p \in nb_{E'}(q)$. Hence, we reduced the problem to the first case.

Therefore, for every reachable configuration $R$ i. e., $A \Longmapsto R$ it holds for that either there is no more possible linearization step in the undirected topology with message or there is at least one.
In the first case, it holds as shown that

$$\Psi(R) = 0$$

If there is still a possible linearization step in the undirected topology with messages, then there are two cases. Either there is still a possible linearization step in the directed topology with messages or not.
In the first case, we reach a configuration $S$ i. e., $A \Longmapsto R \Longmapsto S$ with

$$\Psi_\sum(S) < \Psi_\sum(R) \wedge \Psi(S) < \Psi(R) \wedge \Psi_E(S) \leq \Psi_E(R)$$

In the second case, we reach a configuration $S'$ i. e., $A \Longmapsto R \Longmapsto S'$ such that there is again a possible linearization step in the directed topology with messages of $S'$ with

$$\Psi_\sum(S') < \Psi_\sum(R) \wedge \Psi_E(S') < \Psi_E(R)$$

Therefore, by induction we eventually reach a configuration $C'$ i. e., $A \Longmapsto C'$ with $\Psi_\sum(C') = 2 \cdot n \cdot (n-1)$, $\Psi(C') = 0$ or $\Psi_E(C') = 2 \cdot (n-1)$. As shown in the base case, it holds for $C'$ that

$$A \Longmapsto C' \Longmapsto C \quad \wedge \quad NTM(C) = G_{LIN} \wedge NT(C) = G_{LIN}$$

□

# 4 Strong Convergence

In this chapter, first of all, we describe the problem in proving strong convergence. However, we are convinced that the presented algorithm not only converges weakly, but also strongly. Therefore, we give strong arguments for strong convergence in the general case. We will introduce further already proven lemmata and informally further open lemmata with proof sketches, that could be helpful in a formal proof. Afterwards, we discuss several approaches for a proof of the strong convergence property in the general case.

## 4.1 Problem Description

Although *keep-alive*-messages are necessary to prevent deadlocks and termination of the algorithm, they also introduce problems in proving strong convergence. This is mainly based on the fact that such a message can reestablish an edge that was already deleted through a linearization step. Consider, for example, a configuration with a topology without messages as depicted in Figure 4.1. Process $w$ can send *keep-alive*-messages to process $u$. Assume process $u$ executes its possible linearization step and deletes $w$ from its neighborhood, but afterwards there are still messages of the form $\overline{u}\langle w\rangle$ in transit. With reception of such a message, $u$ reestablishes its edge to $w$. Therefore, $u$ potentially has to execute multiple linearization steps in order to remove the edge to $w$ permanently.

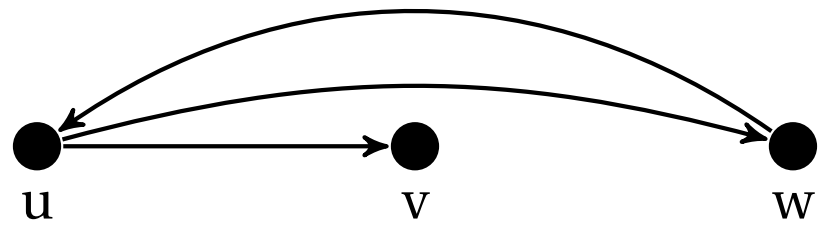

Figure 4.1: Topology with sending *keep-alive*-messages and linearization step

We proved strong convergence holds whenever either only too many edges exist or only desired edges are missing. Unfortunately, for arbitrary configurations where desired edges are still missing and undesired edges existing, the system does not converge always to the same of these situations first. Therefore, it holds neither in the general case that first all undesired edges are deleted nor that first all desired edges are established. In fact, it is possible in a configuration in which both not hold, that with the execution of the last linearization step the desired topology, up to receiving the sent message, is reached directly. To make matters worse, for the same initial configuration there can be depending on the order of steps executions where first all undesired edges are deleted, first all desired edges are established, and the desired topology is reached directly. This holds for example in a configuration with a topology as depicted in Figure 4.2. If first $v$ and $x$ are sending *keep-alive*-messages and the messages are received by $w$ before $u$ executed its linearization step, a configuration is reached where only too many edges exist. If $u$ executes its linearization step (and no more $\overline{u}\langle w\rangle$ messages are in transit) and $w$ receives the sent message before $x$ sends *keep-alive*-messages, a configuration that only lacks desired edges is reached. If $x$ sends *keep-alive*-messages and $w$ receives such a message before $u$ executes its linearization step (and $v$ does not send *keep-alive*-messages in the meantime), the desired topology, up to message reception, is with execution of the linearization step reached directly. Therefore, it is not possible to show for the general case that always a weaker property, like every process has a neighbor on both sides, holds, before the system reaches a correct configuration.

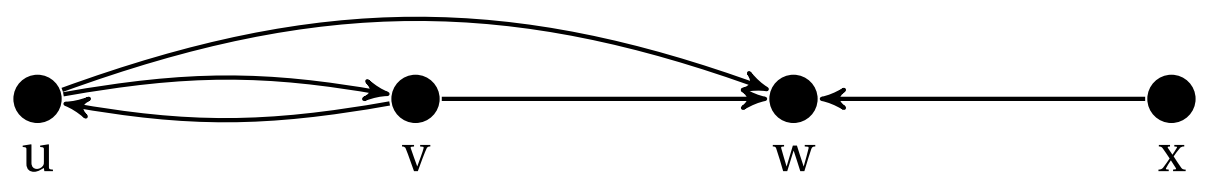

Figure 4.2: Topology where all three cases are possible

We can show that, whenever undesired edges still exist, eventually another linearization step is executed. However, we have no control or knowledge which linearization step will get executed and how the system evolves in the meantime. This is actually a problem for most of the liveness properties. For example, we proved that messages always are received and processed. Therefore, the carried process id will be at some point in the neighborhood of the receiving process. However, we do not know which steps were executed in the meantime and therefore how the rest of the configuration looks like.

A standard way to prove strong convergence is to define a monotonically decreasing potential function that is minimal for a correct configuration. In order to show progress, it is necessary that one can show that this function always eventually decreases strictly as long as a correct configuration is not yet reached. With the property that whenever the system is not stabilized there will be eventually a linearization step executed, this would be achieved most easily if the potential function strictly decreases with every executed linearization step. Unfortunately, this is not readily possible. If we want to design a potential function that decreases with every executed linearization step, we have to take the number of connections (i. e., neighborhood, adding and messages in transit) into account. We have to take the messages in transit and adding of processes into account, because otherwise the potential function would increase whenever a process adds a process id it does not already has in its neighborhood. In order to have a strict decrease with every linearization step, we also have to take the number of connec-tions into account, as some linearization steps do not change the topology with messages, but only change the number of connections. This is depicted in Figure 4.3, let the shown graph be the topology with and without messages of a configuration. Assume that additionally to the depicted neighborhoods there are messages of the form $\overline{u}\langle w\rangle$ and $\overline{w}\langle u\rangle$ in transit. Then, the execution of both possible linearization steps does not change the topology with messages. The only change is that there is one connection less of the form $(u, w)$ (respectively $(w, u)$) and one more of the form $(w, v)$ (respectively $(u, v)$). Nevertheless, this leads always to a function that at least increases whenever a process is sending *keep-alive*-messages to processes that already know it while taking messages into account. These steps lead to no further change in the topology with messages besides to increase the number of such messages and therefore the number of connections between these processes. Therefore, a potential function that decreases strictly with every linearization step, cannot be monotonically decreasing in every case.

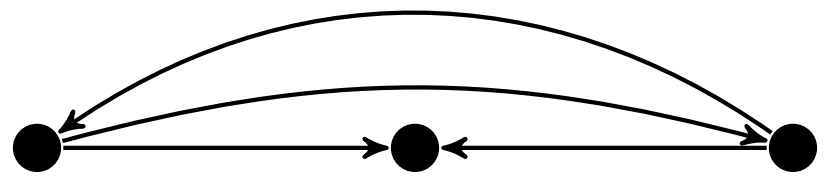

Figure 4.3: Topology that potentially not change with linearization step

Therefore, the best that can be achieve with a monotonically decreasing potential functions is, to design a function that strictly decreases for (linearization) steps that have certain properties. Such properties are, for example, removing the last longest connection or acquaint a process with a nearer process than all already known one. However, we do not only have to show that such steps exist and the potential function strictly decreases whenever such a step is executed, but also that such steps are always reached if the system is not stabilized yet. The problem in this is that such decreasing steps are often triggered by further steps that other processes have to execute previously.

It is crucial to be very careful with simply assuming that certain steps will be executed in a subsystem and therefore lead to a certain configuration for the whole system. The assumption that the system reaches a configuration with certain properties presupposes in general a form of livelock freedom in a subsystem or the overall system. Strong convergence itself shows a special form of livelock freedom in the system whenever the system is not yet stabilized. Assumptions based on the livelock freedom in a certain subsystem are therefore very prone to introduce circular reasoning. Without careful consideration of such assumptions, it could easily happen that one only shows that strong convergence holds if strong convergence holds and therefore nothing at all. Consider a configuration with a topology without messages as depicted in Figure 4.4. It is not possible without further ado to assume that $v$ will eventually send *keep-alive*-messages to $u$ and therefore, with reception of such a message, to enable $u$ to execute a linearization step. In order to ensure that $v$ will eventually send *keep-alive*-messages it is necessary that $v$ previously removes $w$ and $x$ permanently from its neighborhood. This is only the case if there is no livelock in the subsystem containing the processes $v$, $w$ and $x$.

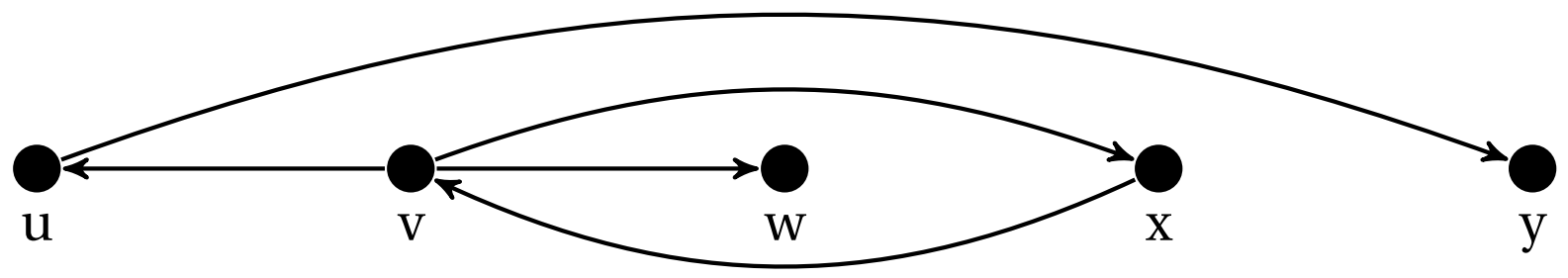

Figure 4.4: Topology with potential livelock

## 4.2 Arguments for Strong Convergence

New edges are only established by receiving a corresponding message. A message to a process $p$ can only be send from a neighboring process $q$ that sends *keep-alive*-messages or executes a linearization step. While sending *keep-alive*-messages two new edges are established at most, as a process can only send such messages if it has at most two neighbors and they are not on the same side. The established edges have the same length as the edges that are used to send the messages. Furthermore, if a new edge is established in a linearization step, it is always strictly shorter than the potentially removed one. A process $p$ only executes a linearization step if it has two neighbors that have either both smaller or both greater ids than the executing process and are therefore on the same side. The process sends the further away process $r$ the id of the nearer one $q$ and deletes the further away one from its neighborhood. This means that the number of connections always stays the same in a linearization step. The number of connections from $r$ to $q$ is increased by one while the connections from $p$ to $r$ decrease by one. The distance between the executing process $p$ and the more distant process $r$ is strictly greater than the distance between $r$ and the process in the middle $q$. Hence, the removed connection is strictly longer than the new established one. Therefore, in every linearization step a longer connection is exchanged through a strictly shorter one. Altogether, it can never be an edge established that is longer than the longest existing edge in the topology with messages.

Only a process itself can remove another process from its own neighborhood. The only steps that lead to such a removal are linearization steps. An enabled linearization step only exists if the process has at least two neighbors on the same side and while executing a linearization always only one process is removed. Therefore, whenever a process knows a process on one of its sides, it always has a neighbor on this side. Additionally, as always the further away process is removed, a process never removes a desired neighbor from its neighborhood. Furthermore, it follows that as soon as the neighborhood of a process contains all desired neighbors it will never send undesired *keep-alive*-messages again.

Messages can be send through *keep-alive*-messages or linearization steps. All these messages are eventually received and processed. If a process $p$ receives a message there are four cases, depending on the neighborhood of $p$ before the addition of the received id $q$. If the neighborhood of the receiving process does not contain any process on the same side like $q$, then after the addition of this id, there will always be a process on this side in the neighborhood of $p$. This holds, as processes never remove the current closest ids in the neighborhood to both sides. If the neighborhood of $p$ already contains another process on the same side like $q$, independent whether this process is closer or more distant to $p$ than $q$, then the addition of $q$ will enable $p$ to execute a linearization step. This holds, since there are now at least two neighbors on this side. If the distance to $q$ is smaller than to any other previously known process on this side, the nearest neighbor on this side is permanently improved. Furthermore, $p$ will never again send *keep-alive*-messages on this side to a process that is further away than $q$. This holds as $p$ can only remove $q$ again from its neighborhood after the reception of an even closer id on this side and processes can only send *keep-alive*-messages if they only have at most one process in the neighborhood on each side. If a process receives a message with an id that is already in its neighborhood, the number of this connection is decreased by one during addition. Therefore, the only steps that do not yield to any progress in the system are the sending of *keep-alive*-messages to processes that already know the sending process in the topology with messages. These steps only increase the number of corresponding connections.

Whenever an already deleted edge is reestablished through a message sent during a linearization step, an even longer connection was removed for this. Additionally, it holds that edges that were already deleted during linearization steps in the topology without messages, can be reestablished through *keep-alive*-messages that are already in transit. However, when executing a linearization step it is ensured that the more distant process eventually forever stops to send *keep-alive*-messages to the executing process. The executing process $p$ sends the id of the intermediate process $q$ to the further away process $r$. This message will eventually be received and processed by $r$. After the addition of $q$, at the latest, $r$ has a process in its neighborhood that is on the same side then $p$, but closer. As mentioned before $r$ can never send *keep-alive*-messages to $p$ again. From this point on, the number of connections from $p$ to $r$ can only be increased if another process sends a $\overline{p}\langle r\rangle$ message during a linearization step. A linearization step that leads to the sending of such a message can again not be disturbed by further *keep-alive*-messages. For the executing process $s$ of such a linearization step, $p$ itself has to be the further away process. It holds that $p$ already cannot send *keep-alive*-messages to $s$ ever again, as it will always have a closer neighbor than $s$ on this side, currently $r$. This is depicted in Figure 4.5, whereby there could be further intermediate not depicted processes that would not change anything on this situation. Therefore, there cannot be a livelock through *keep-alive*-messages in this sense between two overlapping linearization steps.

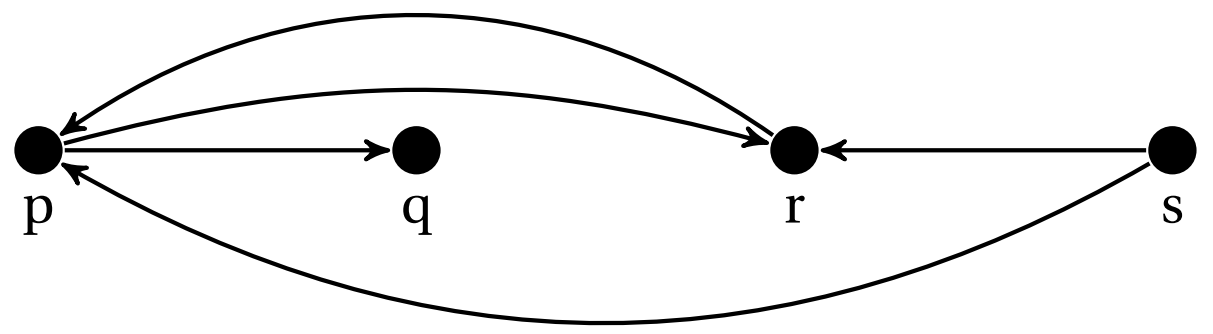

Figure 4.5: No two processes with *keep-alive*-messages

Furthermore, the current longest edges in the system cannot be reestablished through linearization steps, as the established connection is always strictly shorter than the removed one. Therefore, whenever a process $p$ removes a longest edge during a linearization step where the process $r$ on the other end already cannot send *keep-alive*-messages and there are no further messages in transit that can reestablish this edge, it is permanently removed. As mentioned before, the sending of such *keep-alive*-messages stops at the latest with the reception of the message of the first linearization attempt. From this on, the number of connections $m$ for this longest edge $(p,r)$ can never increase again. If there are still messages in transit that reestablish this edge, all of them will eventually be received and processed. Every processing of such a message that occurs while $r$ is currently in the neighborhood of $p$ decreases the number of connections by one. Furthermore, whenever $r$ is in the neighborhood of $p$, there is always a possible enabled linearization step involving $r$, as $p$ will always have a closer neighbor on the same side. Otherwise the linearization step would not have been possible at the first time. Therefore, the edge $(p,r)$ is permanently removed, whenever $p$ executes at most $m$ more linearization steps that involve $r$. Thereby, the intermediate process can, but not necessarily has to be the same process.

Every process that is only known in the topology with messages by (a subset of) its desired neighbors will never receive any other messages than *keep-alive*-messages from these desired neighbors, unless it sends its own id through *keep-alive*-messages to other processes or the desired neighbor sends its id during a linearization step to another process. Further, such a process cannot receive a message from a desired neighbor during a linearization step as the messages are always sent to the further away process. Every process that only knows (a subset of) its desired neighbors will never send any other messages than *keep-alive*-messages to these desired neighbors, unless it receives the id of another process. Therefore, a process that in the topology with messages only knows (a subset of) its desired neighbors and is only known by (a subset of) its desired neighbors is permanent partially stabilized. This means, that it only can send and receive *keep-alive*-messages to and from desired neighbors, until one of them sends its id to another process during a linearization step. Furthermore, if a process is contained in the neighborhood of all its desired neighbors and its neighborhood contains them, this property remains invariant, as the desired neighbors never remove the process during a linearization step and the process never removes them.

If the minimum or maximum process of the system is in the topology with messages only in the neighborhood of desired neighbors and its neighborhood only contains desired neighbors, it is completely stabilized permanently. As mentioned, the process itself can only send *keep-alive*-messages to desired neighbors, that does not change anything for the topology with messages. Furthermore, a desired neighbor cannot know another process on the same side and will never send the process id during a linearization step to another process. Therefore, the process will never receive another process id anymore and stays stabilized. This property is then transferred to the next process inwards the system, i. e., for the minimum process the next greater process and for the maximum process the next smaller, making this process the next border between the stabilized part of the system and the unstabilized one. Figure 4.6 depicts the topology with messages of a configuration with the border process $u$ and $y$. Therefore, starting with the minimum and maximum process of the system, every border process that is stabilized once, remains stabilized forever.

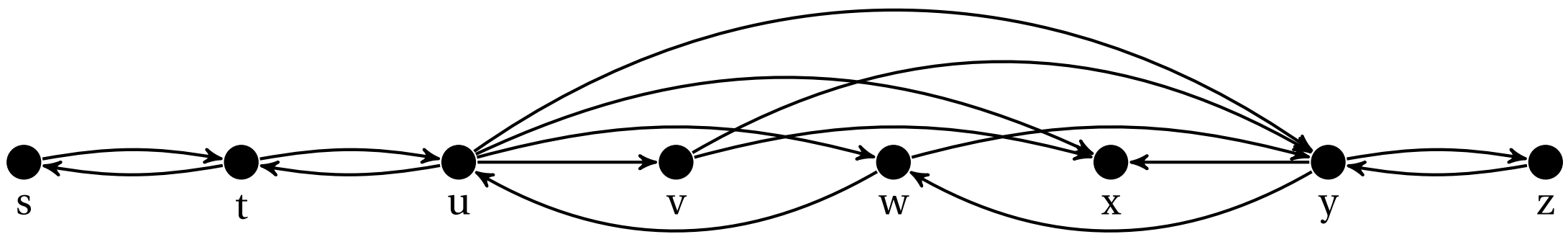

Figure 4.6: System with $u$ and $y$ as border processes

### 4.2.1 Further Proven Lemmata

We show that the system does not deadlock i. e., whenever there are still possible linearization steps in the undirected topology with messages, there is always a configuration reached where again a linearization step is possible in the directed topology without messages. This holds since every process that cannot find a linearization step can send *keep-alive*-messages to its current neighbors and all messages are received and processed eventually.

**Lemma 24: No Deadlock**

Let $A$ be a configuration of the algorithm according to Lemma 1. If there is a possible linearization step in the undirected topology with messages i. e., $\exists p,q,r \in \mathscr{P}.\{p,q\},\{p,r\} \in E_{UNTM(A)} \wedge ((p < q \wedge q < r) \vee (r < q \wedge q < p))$ but there is no possible linearization step in the directed topology i. e., $\forall p,q,r \in \mathscr{P}.((p < q \wedge q < r) \vee (r < q \wedge q < p)) \implies ((p,q) \notin E_{NT(A)} \vee (p,r) \notin E_{NT(A)})$ then eventually a configuration will be reached where a linearization step is possible in the directed network topology.

*Proof*:

Since there is no possible linearization step in the directed network topology it holds according to Lemma 6 that $\forall p' \in \mathscr{P}.nb_A(p') \leq 2$. Since there is a possible linearization step in the undirected topology with messages but not in the directed topology there have to be $p,q,r \in \mathscr{P}$ with $\{p,q\},\{p,r\} \in E_{UNTM(A)}$ and $p < q < r$ (possible left linearization step) or $r < q < p$ (possible right linearization step) . Assume $p < q < r$, the proof for $r < q < p$ is similar. Furthermore, we assume that in every reached configurations $C, C'$ i. e., $A \Longmapsto C \Longmapsto C'$ the neighborhood of $p' \in \{p,q,r\}$ only grow i. e., $nb_A(p') \subseteq nb_C(p') \subseteq nb_{C'}(p')$, otherwise according to Lemma 10 there had to be intermediate configuration $R$ where the subprocess $Alg_{match}(p')$ executed a step and $select(findLin(p', nb_R(p'))) = (j,k) \wedge (j < k \wedge k < p')$ or $select(findLin(p', nb_R(p'))) = (j,k) \wedge (j < k \wedge p' < j)$ but then there was an linearization step possible (and executed) in $R$. According to Definition 30 one of the following cases is true: The cases when $p$ knows both other processes in the directed topology with messages are:

- There is just an add-step necessary to enable a linearization step:

assume $q \in nb_A(p) \wedge \left(p \in P'_A \wedge add_A(p) = r\right)$. According to Lemma 4 there will be a configuration $C$ reached with $r \in nb_C(p)$. According to the assumptions it holds that $q \in nb_C(p) \supseteq nb_A(p)$ and $p < q < r$. Therefore, in $C$ is a linearization step possible.
The case $\left(p \in P'_A \wedge add_A(p) = q\right) \wedge r \in nb_A(p)$ is similar.

- There is just one message left to receive:
  assume $q \in nb_A(p) \wedge (p, r) \in Msgs_A$. According to Corollary 4 we reach a configuration $C$ with $r \in nb_C(p)$. According to the assumptions it holds that $q \in nb_C(p) \supseteq nb_A(p)$ and $p < q < r$ therefore in $C$ is a linearization step possible.
  The case $(p, q) \in Msgs_A \wedge r \in nb_A(p)$ is similar.

- There is an add-step necessary and a message left to receive:
  assume $\left(p \in P'_A \wedge add_A(p) = q\right) \wedge (p, r) \in Msgs_A$. According to Lemma 4 there will be configurations $C \longmapsto C'$ reached with

  $$Alg_{ges}\left(P_C \cup \{p\}, P'_C \setminus \{p\}, nb_{C'}, Msgs_C, add_C \setminus \{(p, q)\}\right)$$
  $$\text{with } nb_{C'}(x) = \begin{cases} nb_C(x), & \text{if } x \neq p \\ nb_C(p) \cup \{q\}, & \text{if } x = p \end{cases}$$

  and $q \in nb_{C'}(p)$. Up to this point $p$ cannot receive any message as the subprocess $Alg_{rec}\left(p\right)$ is not enabled as long as $Alg_{add}\left(p, q\right)$ is enabled and according to 2 messages cannot get lost and according to Lemma 2 $p$ is the only process that can receive this message. Therefore it holds that $(p, r) \in Msgs_{C'}$. According to Corollary 4 a configuration $C''$ is reached with $r \in nb_{C''}(p)$. According to the assumptions it holds that $q \in nb_{C''}(p) \supseteq nb_{C'}(p)$ and $p < q < r$. Therefore, in $C''$ is a linearization step possible.
  The case for $(p, q) \in Msgs_A \wedge \left(p \in P'_A \wedge add_A(p) = r\right)$ is similar.

- There are two messages left to receive:
  $(p, q) \in Msgs_A \wedge (p, r) \in Msgs_A$. According to Corollary 4 a configuration $C$ is reached with $q \in nb_C(p)$ and a configuration $C'$ is reached with $r \in nb_{C'}(p)$ and either $A \Longmapsto C \Longmapsto C'$ or $A \Longmapsto C' \Longmapsto C$. Assume $A \Longmapsto C \Longmapsto C'$ (the second case is similar). Then according to the assumptions $nb_C(p) \subseteq nb_{C'}(p)$ and therefore $q, r \in nb_{C'}(p)$ and $p < q < r$. Therefore, in $C'$ is a linearization step possible.

The cases when $p$ knows one process and the other process knows $p$ in the directed topology with messages are:

- $p$ has one process in its neighborhood and the other process has $p$ in its neighborhood:
  assume $q \in nb_A(p) \wedge p \in nb_A(r)$ . According to Lemma 3 the subprocess $Alg_{match}(r)$ is enabled in $A$ and stays enabled until it executes a step. According to the fairness assumption 3 $Alg_{match}(r)$ will execute a step after a finite number of steps. Let $C$ be the configuration before this step. Assume $select(findLin(r, nb_{A'}(r))) = \bot$ (otherwise there would be a possible linearization step for $r$ and nothing left to show). According

to Corollary 2 it holds

$$C \longmapsto C' \equiv Alg_{ges}\left(P_C, P'_C, nb_C, Msgs_C \cup \{(j,r) | j \in nb_C(r)\}, add_C\right)$$

According to the assumptions $nb_A(r) \subseteq nb_C(r)$. Therefore it holds $(p,r) \in Msgs_{C'}$. Additionally also $nb_A(p) \subseteq nb_{C'}(p)$ and therefore $q \in nb_{C'}(p)$. Hence we reduced the problem to the second case.
The case for $p \in nb_A(q) \wedge r \in nb_A(p)$ is similar.

- $p$ has one process in its neighborhood and for the other process is an add-step necessary: assume $q \in nb_A(p) \wedge \left(r \in P'_A \wedge add_A(r) = p\right)$ . According to Lemma 4 there will be a configuration $C$ reached with $p \in nb_C(r)$. According to the assumptions it holds that $q \in nb_C(p) \supseteq nb_A(p)$. Hence, we reduced the problem to the previous case.
The case for $\left(q \in P'_A \wedge add_A(q) = p\right) \wedge r \in nb_A(p)$ is similar.

- $p$ has one process in its neighborhood and the other process has a message left to receive:
assume $q \in nb_A(p) \wedge (r,p) \in Msgs_A$ . According to Corollary 4 we reach a configuration $C$ with $p \in nb_C(r)$. According to the assumptions it holds that $q \in nb_C(p) \supseteq nb_A(p)$. Hence, we reduced the problem to the fifth case.
The case for $(q,p) \in Msgs_A \wedge r \in nb_A(p)$ is similar.

- One process has $p$ in its neighborhood and for $p$ is an add-step necessary:
assume $\left(p \in P'_A \wedge add_A(p) = q\right) \wedge p \in nb_A(r)$. According to Lemma 4 there will be a configuration $C$ reached with $q \in nb_C(p)$. According to the assumptions it holds that $p \in nb_C(r) \supseteq nb_A(r)$. Hence, we reduced the problem to the fifth case.
The case for $p \in nb_A(q) \wedge \left(p \in P'_A \wedge add_A(p) = r\right)$ is similar.

- One process has $p$ in its neighborhood and for $p$ is a message left to receive:
assume $(p,q) \in Msgs_A \wedge p \in nb_A(r)$. According to Corollary 4 we reach a configuration $C$ with $q \in nb_C(p)$. According to the assumptions it holds that $p \in nb_C(r) \supseteq nb_A(r)$. Hence, we reduced the problem to the fifth case.
The case for $p \in nb_A(q) \wedge (p,r) \in Msgs_A$is similar.

- There are two add-steps necessary:
assume $\left(p \in P'_A \wedge add_A(p) = q\right) \wedge \left(r \in P'_A \wedge add_A(r) = p\right)$. According to Lemma 4 there will be a configuration $C$ reached with $q \in nb_C(p)$ and a configuration $C'$ is reached with $p \in nb_{C'}(r)$ and either $A \Longmapsto C \Longmapsto C'$ or $A \Longmapsto C' \Longmapsto C$. Assume $A \Longmapsto C \Longmapsto C'$ (the second case is similar). Then according to the assumptions $nb_C(p) \subseteq nb_{C'}(p)$ and therefore $q \in nb_{C'}(p)$. Hence, we reduced the problem to the fifth case.
The case for $\left(q \in P'_A \wedge add_A(q) = p\right) \wedge \left(p \in P'_A \wedge add_A(p) = r\right)$ is similar.

- There is an add-step necessary and a message left to receive:
assume $\left(p \in P'_A \wedge add_A(p) = q\right) \wedge (r,p) \in Msgs_A$ . According to Lemma 4 there will be a configuration $C$ reached with $q \in nb_C(p)$ and according to Corollary 4 a configuration $C'$ with $p \in nb_{C'}(r)$. Either $A \Longmapsto C \Longmapsto C'$ or $A \Longmapsto C' \Longmapsto C$. Assume $A \Longmapsto C \Longmapsto C'$ (the second case is similar). Then according to the assumptions $nb_C(p) \subseteq nb_{C'}(p)$ and therefore $q \in nb_{C'}(p)$. Hence, we reduced the problem to the fifth case.
The case for $(q,p) \in Msgs_A \wedge \big(p \in P'_A \wedge add_A(p) = r\big)$ is similar and the cases for $(p,q) \in Msgs_A \wedge \big(r \in P'_A \wedge add_A(r) = p\big)$ and $\big(q \in P'_A \wedge add_A(q) = p\big) \wedge (p,r) \in Msgs_A$ are similar with switched roles.

- There are two messages left to receive:
  assume $(p,q) \in Msgs_A \wedge (r,p) \in Msgs_A$. According to Corollary 4 there will be a configuration $C$ reached with $q \in nb_C(p)$ and a configuration $C'$ is reached with $p \in nb_{C'}(r)$ and either $A \Longmapsto C \Longmapsto C'$ or $A \Longmapsto C' \Longmapsto C$. Assume $A \Longmapsto C \Longmapsto C'$ (the second case is similar). Then according to the assumptions $nb_C(p) \subseteq nb_{C'}(p)$ and therefore $q \in nb_{C'}(p)$. Hence, we reduced the problem to the fifth case.
  The case for $(q,p) \in Msgs_A \wedge (p,r) \in Msgs_A$ is similar.

The cases when both other processes know $p$ in the undirected topology with messages are:

- $p$ is in the neighborhood of both processes:
  $p \in nb_A(q) \wedge p \in nb_A(r)$ . According to Lemma 3 the subprocess $Alg_{match}\big(q\big)$ is enabled in $A$ and stays enabled until it executes a step. According to the fairness assumption 3 $Alg_{match}\big(q\big)$ will execute a step after a finite number of steps. Let $C$ be the configuration before this step. Assume $select\big(findLin\big(q, nb_C(q)\big)\big) = \perp$ (otherwise there would be a possible linearization step for $q$ and nothing left to show). According to Corollary 2 it holds

  $$C \longmapsto C' \equiv Alg_{ges}\big(P_C, P'_C, nb_C, Msgs_C \cup \{(j,r) | j \in nb_C(q)\}, add_C\big)$$

  and according to the assumptions $nb_A(q) \subseteq nb_C(q)$. Therefore it holds $(p,q) \in Msgs_{C'}$. Additionally also $nb_A(r) \subseteq nb_{C'}(r)$ and therefore $p \in nb_{C'}(r)$. Hence, we reduced the problem to the ninth case.

- $p$ is in the neighborhood of one process and for the other one is an add-step necessary:
  assume $p \in nb_A(q) \wedge \big(r \in P'_A \wedge add_A(r) = p\big)$. According to Lemma 4 there will be a configuration $C$ reached with $p \in nb_C(r)$. According to the assumptions it holds that $p \in nb_C(q) \supseteq nb_A(q)$. Hence, we reduced the problem to the previous case.
  The case for $\big(q \in P'_A \wedge add_A(q) = p\big) \wedge p \in nb_A(r)$ is similar.

- $p$ is in the neighborhood of one process and the other one has a message left to receive:
  $p \in nb_A(q) \wedge (r,p) \in Msgs_A$. According to Corollary 4 we reach a configuration $C$ with $p \in nb_C(r)$. According to the assumptions it holds that $p \in nb_C(q) \supseteq nb_A(q)$. Hence, we reduced the problem to the thirteenth case.
  The case for $(q,p) \in Msgs_A \wedge p \in nb_A(r)$ is similar.

- For both processes is an add-step necessary:
  $\big(q \in P'_A \wedge add_A(q) = p\big) \wedge \big(r \in P'_A \wedge add_A(r) = p\big)$. According to Lemma 4 there will be

a configuration $C$ reached with $p \in nb_C(q)$ and a configuration $C'$ is reached with $p \in nb_{C'}(r)$ and either $A \Longmapsto C \Longmapsto C'$ or $A \Longmapsto C' \Longmapsto C$. Assume $A \Longmapsto C \Longmapsto C'$ (the second case is similar). Then according to the assumptions $nb_C(q) \subseteq nb_{C'}(q)$ and therefore $p \in nb_{C'}(q)$ . Hence, we reduced the problem to the thirteenth case.

- For one process is an add-step necessary and the other one has a message left to receive: assume $\left(q \in P'_A \wedge add_A(q) = p\right) \wedge (r, p) \in Msgs_A$. According to Lemma 4 there will be a configuration $C$ reached with $p \in nb_C(q)$ and according to Corollary 4 a configuration $C'$ with $p \in nb_{C'}(r)$. Either $A \Longmapsto C \Longmapsto C'$ or $A \Longmapsto C' \Longmapsto C$. Assume $A \Longmapsto C \Longmapsto C'$ (the second case is similar). Then according to the assumptions $nb_C(q) \subseteq nb_{C'}(q)$ and therefore $p \in nb_{C'}(q)$. Hence, we reduced the problem to the thirteenth case.
  The case for $(q, p) \in Msgs_A \wedge \left(r \in P'_A \wedge add_A(r) = p\right)$ is similar.

- Both processes have a message left to receive:
  $(q, p) \in Msgs_A \wedge (r, p) \in Msgs_A$ . According to Corollary 4 there will be a configuration $C$ reached with $p \in nb_C(q)$ and a configuration $C'$ is reached with $p \in nb_{C'}(r)$ and either $A \Longmapsto C \Longmapsto C'$ or $A \Longmapsto C' \Longmapsto C$. Assume $A \Longmapsto C \Longmapsto C'$ (the second case is similar). Then according to the assumptions $nb_C(q) \subseteq nb_{C'}(q)$ and therefore $p \in nb_{C'}(q)$. Hence, we reduced the problem to the thirteenth case.

□

Therefore, through fairness it holds that whenever the system is not stabilized yet, there will always be linearization steps executed. Nevertheless, it is not possible to show without further ado that a certain linearization step will get enabled or executed.

A crucial fact for convergence is that it is not possible to establish longer edges than the current existing edges. Therefore, whenever $m$ is the length of the longest edges in the system and all edges with length $m$ are removed, there will never be edges with length $m$ again.

**Definition 40: Longest Edge(s)**

The *length set* $Lengths : \mathcal{T} \rightarrow 2^{\mathbb{N}}$ with

$$Lengths(A) = \{len(e) | e \in E_{NTM(A)}\} = \{len(e) | e \in E_{UNTM(A)}\}$$

is the set of the lengths of all edges in the (undirected) topology with messages of a configuration.
The *length of the longest edge(s)* of a configuration is then defined as $lenmaxEdge : \mathcal{T} \rightarrow \mathbb{N}$ with

$$lenmaxEdge(A) = max\big(Lengths(A)\big)$$

whereby $max()$ returns the maximum value. Since there are always edges in the (undirected) network topology with messages (according to Lemma 7) and only a finite set of processes according to Definition 11 (and therefore finitely many edges), the maximum is always defined. The *set of longest edges* in the *undirected* network topology with messages of a configuration is defined as $maxUMEdge : \mathscr{T} \to 2^{\mathscr{P}\times\mathscr{P}}$ with

$$maxUMEdge(A) = \{e | e \in E_{UNTM(A)} \wedge len(e) = lenmaxEdge(A))\}$$

and the *set of longest edges* in the network topology with messages of a configuration as $maxEdge : \mathscr{T} \to 2^{\mathscr{P}\times\mathscr{P}}$ with

$$maxEdge(A) = \{e | e \in E_{NTM(A)} \wedge len(e) = lenmaxEdge(A))\}$$

New edges in the topology with messages can only be established through the sending of *keep-alive*-messages or linearization steps. Every linearization step only introduces (if not already existent) a shorter edge than the deleted (but eventual through messages still existent) one, every sending of *keep-alive*-messages establishes at most two equal long edges than the edges used to send them. Therefore, no edge can be established that is strictly longer than the already existing edges.

**Lemma 25: Only shorter Edges established**

Let $A$ be an arbitrary configuration of the algorithm according to Lemma 1. Let $A'$ be an arbitrary configuration with $A \longmapsto A'$. It holds two more edges are established at most and every new edge is shorter or equal to another already existing edge. Therefore, it holds

$$\begin{aligned}
&|E_{NT(A')}| \in \{|E_{NT(A)}|, |E_{NT(A)}| - 1, |E_{NT(A)}| + 1\},\\
&\forall e \in E_{NT(A')}.\exists e' \in E_{NTM(A)}.len(e) \le len(e'),\\
&|E_{NTM(A')}| \in \{|E_{NTM(A)}|, |E_{NTM(A)}| - 1, |E_{NTM(A)}| + 1, |E_{NTM(A)}| + 2\} \quad \text{and}\\
&\forall e \in E_{NTM(A')}.\exists e' \in E_{NTM(A)}.len(e) \le len(e')
\end{aligned}$$

*Proof*:
According to Corollary 2 there are only the following cases:

- If a process $p \in \mathscr{P}$ tries to find a linearization step, but $select\big(findLin\big(p, nb_A(p)\big)\big) = \perp$ it holds that

  $$A' \equiv Alg_{ges}\big(P_A, P'_A, nb_A, Msgs_A \cup \{(j,p) | j \in nb_A(p)\}, add_A\big)$$

  The (undirected) network topology does not change because the neighborhood of all processes stays the same. Since $nb_{A'} = nb_A$ it holds according to Definitions 27 and 29

  $$E_{NT(A')} = E_{NT(A)} \quad \text{and} \quad E_{UNT(A')} = E_{UNT(A)}$$

  The undirected network topology with messages does not change since it is only pos-sible to send *keep-alive*-messages to neighbors. From $select\big(findLin\big(p, nb_A(p)\big)\big) = \bot$ it follows with Lemma 6 that $|nb_A(p)| \leq 2$ and $|nb_A(p)| = 2 \implies |LeftN\big(A,p\big)| = 1 \wedge |RightN\big(A,p\big)| = 1$. For every new message $m \in Msgs_{A'} \setminus Msgs_A$ it holds that $m = (j,p)$ with $j \in nb_A(p)$. Let $j \in \mathscr{P}$ be an arbitrary process with $j \in nb_A(p)$. According to Definition 27 it holds $(p,j) \in E_{NT(A)}$. Therefore, with $nb_{A'} = nb_A$, $add_{A'} = add_A$ and Definition 30 it holds that

$$E_{UNTM(A')} = E_{UNTM(A)}$$

The network topology with messages changes depending on whether the (at most) two neighbors of $p$ already have $p$ in their neighborhood or not. There are two cases. If $(j,p) \in E_{NTM(A)}$, then also $(j,p) \in E_{NTM(A')}$. If $(j,p) \notin E_{NTM(A)}$, then with $(j,p) \in Msgs_{A'}$ according to Definition 28 $(j,p) \in E_{NTM(A')}$ and with Definitions 16 and 15 $len((j,p)) = len((p,j))$. Therefore, it holds

- $E_{NTM(A')} = E_{NTM(A)}$ iff

$$nb_A(p) = \emptyset \quad \vee \quad \big(nb_A(p) = \{q\} \wedge (q,p) \in E_{NTM(A)}\big) \quad \vee$$
$$\big(nb_A(p) = \{q,r\} \wedge q \neq r \wedge (q,p) \in E_{NTM(A)} \wedge (r,p) \in E_{NTM(A)}\big)$$

- $E_{NTM(A')} = E_{NTM(A)} \cup \{(q,p)\} \wedge |E_{NTM(A')}| = |E_{NTM(A)}| + 1 \wedge (p,q) \in E_{NT(A)} \wedge len((q,p)) = len((p,q))$ iff

$$\big(nb_A(p) = \{q\} \wedge (q,p) \notin E_{NTM(A)}\big) \quad \vee$$
$$\big(nb_A(p) = \{q,r\} \wedge q \neq r \wedge (q,p) \notin E_{NTM(A)} \wedge (r,p) \in E_{NTM(A)}\big)$$

- $E_{NTM(A')} = E_{NTM(A)} \cup \{(q,p),(r,p)\} \wedge |E_{NTM(A')}| = |E_{NTM(A)}| + 2 \wedge (p,q) \in E_{NT(A)} \wedge len((q,p)) = len((p,q)) \wedge (p,r) \in E_{NT(A)} \wedge len((r,p)) = len((p,r))$ iff

$$nb_A(p) = \{q,r\} \wedge q \neq r \wedge (q,p) \notin E_{NTM(A)} \wedge (r,p) \notin E_{NTM(A)}$$

- If a process $p \in \mathscr{P}$ finds a left linearization step i. e., $select\big(findLin\big(p, nb_A(p)\big)\big) = (j,k) \wedge (j < k \wedge k < p)$, it holds that

$$A' \equiv Alg_{ges}\big(P_A, P'_A, nb_{A'}, Msgs_A \cup \{(j,k)\}, add_A\big)$$
$$\text{with } nb_{A'}(x) = \begin{cases} nb_A(x), & \text{if } x \neq p \\ nb_A(p) \setminus \{j\}, & \text{if } x = p \end{cases}$$

In every linearization step $p$ removes the further away process from its neighborhood, hence the network topology without messages is reduced by this edge. Therefore, with $j \in nb_A(p)$ according to the definition of $select\big(findLin\big(p, nb_A(p)\big)\big)$ and Definition 27 it holds that

$$E_{NT(A')} = E_{NT(A)} \setminus \{(p,j)\} \wedge |E_{NT(A')}| = |E_{NT(A)}| - 1$$

This reduction holds also for the undirected network topology but only if $p$ is not in the neighborhood of $q$. It holds $(j,p) \in E_{NT(A)}$ iff $(j,p) \in E_{NT(A')}$ and therefore with Definition 29 that

$$E_{UNT(A')} = E_{UNT(A)} \text{ iff } (j,p) \in E_{NT(A)}$$
$$E_{UNT(A')} = E_{UNT(A)} \setminus \{\{p,j\}\} \land |E_{UNT(A')}| = |E_{UNT(A)}| - 1 \text{ iff } (j,p) \notin E_{NT(A)}$$

The network topology with messages changes depending on whether there are still messages in transit those will lead to an adding of $q$ to the neighborhood of $p$ again and whether the deleted process $j$ already has $k$ in its neighborhood (or it is yet a corresponding message in transit). According to Definition 27 it holds $(p,j) \in E_{NT(A)}$. Since $j < k < p$ it holds according to Definitions 11, 15 and 16 that $len((j,k)) < len((p,j))$. Therefore, it holds with $add_{A'} = add_A$ according to Definition 28:

- $E_{NTM(A')} = E_{NTM(A)}$ iff

$$\big((p,j) \in Msgs_A \quad \lor \quad \big(p \in P'_A \land add_A(p) = j\big)\big) \quad \land \quad (j,k) \in E_{NTM(A)}$$

- $E_{NTM(A')} = E_{NTM(A)} \setminus \{(p,j)\} \land |E_{NT(A')}| = |E_{NT(A)}| - 1$ iff

$$\big((p,j) \notin Msgs_A \quad \land \quad \big(p \notin P'_A \lor add_A(p) \neq j\big)\big) \quad \land \quad (j,k) \in E_{NTM(A)}$$

- $E_{NTM(A')} = E_{NTM(A)} \cup \{(j,k)\} \land |E_{NT(A')}| = |E_{NT(A)}| + 1 \land (p,j) \in E_{NT(A)} \land len((j,k)) < len((p,j))$ iff

$$\big((p,j) \in Msgs_A \quad \lor \quad \big(p \in P'_A \land add_A(p) = j\big)\big) \quad \land \quad (j,k) \notin E_{NTM(A)}$$

- $E_{NTM(A')} = \big(E_{NTM(A)} \setminus \{(p,j)\}\big) \cup \{(j,k)\} \land |E_{NT(A')}| = |E_{NT(A)}| \land E_{NTM(A')} \neq E_{NTM(A)} \land (p,j) \in E_{NT(A)} \land len((j,k)) < len((p,j))$ iff

$$\big((p,j) \notin Msgs_A \quad \land \quad \big(p \notin P'_A \lor add_A(p) \neq j\big)\big) \quad \land \quad (j,k) \notin E_{NTM(A)}$$

The undirected topology with messages changes similarly, but also takes the (potential) back edges into account. Since $(k,j) \in E_{NTM(A')}$ iff $(k,j) \in E_{NTM(A)}$ and $(j,p) \in E_{NTM(A')}$ iff $(j,p) \in E_{NTM(A)}$ it holds with Definition 30:

- $E_{UNTM(A')} = E_{UNTM(A)}$ iff

$$\big((p,j) \in Msgs_A \quad \lor \quad \big(p \in P'_A \land add_A(p) = j\big) \quad \lor \quad (j,p) \in E_{NTM(A)}\big) \quad \land$$
$$\big((j,k) \in E_{NTM(A)} \quad \lor \quad (k,j) \in E_{NTM(A)}\big)$$

- $E_{UNTM(A')} = E_{UNTM(A)} \setminus \{\{p,j\}\} \land |E_{NT(A')}| = |E_{NT(A)}| - 1$ iff

$$\big((p,j) \notin Msgs_A \quad \land \quad \big(p \notin P'_A \lor add_A(p) \neq j\big) \quad \land \quad (j,p) \notin E_{NTM(A)}\big) \quad \land$$
$$\big((j,k) \in E_{NTM(A)} \quad \lor \quad (k,j) \in E_{NTM(A)}\big)$$

- $E_{NTM(A')} = E_{NTM(A)} \cup \{\{j,k\}\} \wedge |E_{NT(A')}| = |E_{NT(A)}|+1 \wedge (p,j) \in E_{NT(A)} \wedge len(\{j,k\}) < len(\{p,j\})$ iff

$$\big((p,j) \in Msgs_A \quad \vee \quad \big(p \in P'_A \wedge add_A(p) = j\big) \quad \vee \quad (j,p) \in E_{NTM(A)}\big) \quad \wedge$$
$$\big((j,k) \notin E_{NTM(A)} \quad \wedge \quad (k,j) \notin E_{NTM(A)}\big)$$

- $E_{NTM(A')} = \big(E_{NTM(A)} \setminus \{\{p,j\}\}\big) \cup \{\{j,k\}\} \wedge |E_{NT(A')}| = |E_{NT(A)}| \wedge E_{NTM(A')} \neq E_{NTM(A)} \wedge (p,j) \in E_{NT(A)} \wedge len(\{j,k\}) < len(\{p,j\})$ iff

$$\big((p,j) \notin Msgs_A \quad \wedge \quad \big(p \notin P'_A \vee add_A(p) \neq j\big) \quad \wedge \quad (j,p) \notin E_{NTM(A)}\big) \quad \wedge$$
$$\big((j,k) \notin E_{NTM(A)} \quad \wedge \quad (k,j) \notin E_{NTM(A)}\big)$$

- If a process $p \in \mathscr{P}$ finds a right linearization step i.e., $select\big(findLin\big(p, nb_A(p)\big)\big) = (j,k) \wedge (j < k \wedge p < j)$, it holds that

$$A' \equiv Alg_{ges}\big(P_A, P'_A, nb_{A'}, Msgs_A \cup \{(k,j)\}, add_A\big)$$
$$\text{and } nb_{A'}(x) = \begin{cases} nb_A(x), & \text{if } x \neq p \\ nb_A(p) \setminus \{k\}, & \text{if } x = p \end{cases}$$

In every linearization step $p$ removes the further away process from its neighborhood, hence the network topology without messages is reduced by this edge. Therefore, with $k \in nb_A(p)$ according to the definition of $select\big(findLin\big(p, nb_A(p)\big)\big)$ and Definition 27 it holds that

$$E_{NT(A')} = E_{NT(A)} \setminus \{(p,k)\} \wedge |E_{NT(A')}| = |E_{NT(A)}| - 1$$

This reduction holds also for the undirected network topology but only if $p$ is not in the neighborhood of $q$. It holds $(k,p) \in E_{NT(A)}$ iff $(k,p) \in E_{NT(A')}$ and therefore with Definition 29 that

$$E_{UNT(A')} = E_{UNT(A)} \text{ iff } (k,p) \in E_{NT(A)}$$
$$E_{UNT(A')} = E_{UNT(A)} \setminus \{\{p,k\}\} \wedge |E_{UNT(A')}| = |E_{UNT(A)}| - 1 \text{ iff } (k,p) \notin E_{NT(A)}$$

The network topology with messages changes depending on whether there are still messages in transit those will lead to an adding of $q$ to the neighborhood of $p$ again and whether the deleted process $k$ already has $j$ in its neighborhood (or it is yet a corresponding message in transit). According to Definition 27 it holds $(p,k) \in E_{NT(A)}$. Since $p < j < k$ it holds according to Definitions 11, 15 and 16 that $len((k,j)) < len((p,k))$. Therefore, it holds with $add_{A'} = add_A$ according to Definition 28:

- $E_{NTM(A')} = E_{NTM(A)}$ iff

$$\big((p,k) \in Msgs_A \quad \vee \quad \big(p \in P'_A \wedge add_A(p) = k\big)\big) \quad \wedge \quad (k,j) \in E_{NTM(A)}$$

– $E_{NTM(A')} = E_{NTM(A)} \setminus \{(p,k)\} \ \wedge \ |E_{NT(A')}| = |E_{NT(A)}| - 1$ iff

$$\big((p,k) \notin Msgs_A \quad \wedge \quad \big(p \notin P'_A \vee add_A(p) \neq k\big)\big) \quad \wedge \quad (k,j) \in E_{NTM(A)}$$

– $E_{NTM(A')} = E_{NTM(A)} \cup \{(k,j)\} \wedge |E_{NT(A')}| = |E_{NT(A)}| + 1 \wedge (p,k) \in E_{NT(A)} \wedge len((k,j)) < len((p,k))$ iff

$$\big((p,k) \in Msgs_A \quad \vee \quad \big(p \in P'_A \wedge add_A(p) = k\big)\big) \quad \wedge \quad (k,j) \notin E_{NTM(A)}$$

– $E_{NTM(A')} = \big(E_{NTM(A)} \setminus \{(p,k)\}\big) \cup \{(k,j)\} \wedge |E_{NT(A')}| = |E_{NT(A)}| \wedge E_{NTM(A')} \neq E_{NTM(A)} \wedge (p,k) \in E_{NT(A)} \ \wedge \ len((k.j)) < len((p,k))$ iff

$$\big((p,k) \notin Msgs_A \quad \wedge \quad \big(p \notin P'_A \vee add_A(p) \neq k\big)\big) \quad \wedge \quad (k,j) \notin E_{NTM(A)}$$

The undirected topology with messages changes similarly, but also takes the (potential) back edges into account. Since $(k,j) \in E_{NTM(A')}$ iff $(k,j) \in E_{NTM(A)}$ and $(k,p) \in E_{NTM(A')}$ iff $(k,p) \in E_{NTM(A)}$ it holds with Definition 30:

– $E_{UNTM(A')} = E_{UNTM(A)}$ iff

$$\big((p,k) \in Msgs_A \quad \vee \quad \big(p \in P'_A \wedge add_A(p) = k\big) \quad \vee \quad (k,p) \in E_{NTM(A)}\big) \quad \wedge$$
$$\big((k,j) \in E_{NTM(A)} \quad \vee \quad (j,k) \in E_{NTM(A)}\big)$$

– $E_{UNTM(A')} = E_{UNTM(A)} \setminus \{\{p,k\}\} \ \wedge \ |E_{NT(A')}| = |E_{NT(A)}| - 1$ iff

$$\big((p,k) \notin Msgs_A \quad \wedge \quad \big(p \notin P'_A \vee add_A(p) \neq k\big) \quad \wedge \quad (k,p) \notin E_{NTM(A)}\big) \quad \wedge$$
$$\big((k,j) \in E_{NTM(A)} \quad \vee \quad (j,k) \in E_{NTM(A)}\big)$$

– $E_{NTM(A')} = E_{NTM(A)} \cup \{\{k,j\}\} \wedge |E_{NT(A')}| = |E_{NT(A)}| + 1 \wedge (p,k) \in E_{NT(A)} \wedge len(\{k,j\}) < len(\{p,k\})$ iff

$$\big((p,k) \in Msgs_A \quad \vee \quad \big(p \in P'_A \wedge add_A(p) = k\big) \quad \vee \quad (k,p) \in E_{NTM(A)}\big) \quad \wedge$$
$$\big((k,j) \notin E_{NTM(A)} \quad \wedge \quad (j,k) \notin E_{NTM(A)}\big)$$

– $E_{NTM(A')} = \big(E_{NTM(A)} \setminus \{\{p,k\}\}\big) \cup \{\{k,j\}\} \wedge |E_{NT(A')}| = |E_{NT(A)}| \wedge E_{NTM(A')} \neq E_{NTM(A)} \wedge (p,k) \in E_{NT(A)} \ \wedge \ len(\{k,j\}) < len(\{p,k\})$ iff

$$\big((p,k) \notin Msgs_A \quad \wedge \quad \big(p \notin P'_A \vee add_A(p) \neq k\big) \quad \wedge \quad (k,p) \notin E_{NTM(A)}\big) \quad \wedge$$
$$\big((k,j) \notin E_{NTM(A)} \quad \wedge \quad (j,k) \notin E_{NTM(A)}\big)$$

- If for a process $p \in \mathscr{P}$ it could be that $select\big(findLin\big(p, nb_A(p)\big)\big)$ is something else, it holds that $A' \equiv A$. Therefore, obviously

$$E_{NT(A')} = E_{NT(A)}, \qquad E_{NTM(A')} = E_{NTM(A)},$$
$$E_{UNT(A')} = E_{UNT(A)} \qquad \text{and} \qquad E_{UNTM(A')} = E_{UNTM(A)}$$

- If a process $p \in P_A$ receives a message with the id of $q \in \mathscr{P}$ i. e., $(p,q) \in Msgs_A$, it holds

  $$A' \equiv Alg_{ges}\left(P_A \setminus \{p\}, P'_A \cup \{p\}, nb_A, Msgs_A \setminus \{(p,q)\}, add_A \cup \{(p,q)\}\right)$$

  The (undirected) network topology does not change since $q$ is not yet added to the neighborhood of $p$. Since $nb_{A'} = nb_A$ it holds according to Definitions 27 and 29

  $$E_{NT(A')} = E_{NT(A)} \quad \text{and} \quad E_{UNT(A')} = E_{UNT(A)}$$

  The (undirected) network topology with messages does not change since the edge is already established through the message in transit. Since $(p,q) \in Msgs_A$ it holds according to Definitions 28 and 30 that $(p,q) \in E_{NTM(A)}$ and $\{p,q\} \in E_{UNTM(A)}$. With $p \in P'_{A'}$ and $add_{A'}(p) = q$ it holds further that also $(p,q) \in E_{NTM(A')}$ and $\{p,q\} \in E_{UNTM(A')}$. Therefore

  $$E_{NTM(A')} = E_{NTM(A)} \quad \text{and} \quad E_{UNTM(A')} = E_{UNTM(A)}$$

- If a process $p \in P'_A$ adds a process $q \in \mathscr{P}$ to its neighborhood i. e., $(p,q) \in add_A$ it holds

  $$A' \equiv Alg_{ges}\left(P_A \cup \{p\}, P'_A \setminus \{p\}, nb_{A'}, Msgs_A, add_A \setminus \{(p,q)\}\right)$$
  $$\text{with } nb_{A'}(x) = \begin{cases} nb(x), & \text{if } x \neq p \\ nb(p) \cup \{q\}, & \text{if } x = p \end{cases}$$

  If $q$ is already in the neighborhood of $p$ then the topology does not change, otherwise a new edge is established in the network topology without message.

  – $E_{NT(A')} = E_{NT(A)}$ iff

  $$(p,q) \in E_{NT(A)}$$

  – $E_{NT(A')} = E_{NT(A)} \cup \{(p,q)\} \wedge |E_{NT(A')}| = |E_{NT(A)}| + 1 \wedge (p,q) \in E_{NTM(A)}$ iff

  $$(p,q) \notin E_{NT(A)}$$

  The undirected network topology does not change if either $q$ is already in the neighborhood of $p$ or vice versa.

  – $E_{UNT(A')} = E_{UNT(A)}$ iff

  $$\{p,q\} \in E_{UNT(A)}$$

  – $E_{UNT(A')} = E_{UNT(A)} \cup \{\{p,q\}\} \wedge |E_{UNT(A')}| = |E_{UNT(A)}| + 1 \wedge (p,q) \in E_{NTM(A)}$ iff

  $$\{p,q\} \notin E_{UNT(A)}$$

The (undirected) network topology with messages does not change as the edge is already established. Since $p \in P'_A$ and $add_A(p) = q$ it holds according to Definitions 28 and 30 that $(p,q) \in E_{NTM(A)}$ and $\{p,q\} \in E_{UNTM(A)}$. With $q \in nb_{A'}(p)$ it holds further that also $(p,q) \in E_{NTM(A')}$ and $\{p,q\} \in E_{UNTM(A')}$. Therefore

$$E_{NTM(A')} = E_{NTM(A)} \quad \text{and} \quad E_{UNTM(A')} = E_{UNTM(A)}$$

□

**Corollary 13: Never longer Edges**

Let $A$ be an arbitrary configuration of the algorithm. Let $l$ be the length of the longest edge(s) in the network topology with messages $NTM(A)$, i. e., $l = lenmaxEdge(A)$. Then for all reachable configurations it holds that the longest edge is shorter or similar to $l$ i. e.,

$$\forall C. A \Longmapsto C \implies lenmaxEdge(C) \leq l$$

It holds that as soon as a process knows a process on one of its sides i. e., a smaller respectively greater process, its neighborhood on this side is never empty again. Furthermore, the nearest process on this side can never get worse in the sense that it is a process that is further away than the current nearest process.

**Lemma 26: Neighborhoods never get empty again and Nearest Neighbors only get better**

Let $A$ be an arbitrary configuration of the algorithm. Whenever a process $p \in \mathcal{P}$ knows a smaller resp. greater process it will always know a smaller resp. greater process. Furthermore, the greatest process from its smaller neighbors can only get greater and the smallest process from its greater neighbors can only get smaller. Let $R$ be an arbitrary reachable configuration i. e., $A \Longmapsto R$, then it holds that

$$\begin{aligned}
\forall p \in \mathcal{P}. (\, LeftN\big(nb_A(p), p\big) \neq \emptyset \implies \big(&LeftN\big(nb_R(p), p\big) \neq \emptyset \ \wedge \\
&max\big(LeftN\big(nb_R(p), p\big)\big) \geq max\big(LeftN\big(nb_A(p), p\big)\big)\big)\big) \quad \wedge \\
(\, RightN\big(nb_A(p), p\big) \neq \emptyset \implies \big(&RightN\big(nb_R(p), p\big) \neq \emptyset \ \wedge \\
&min\big(RightN\big(nb_R(p), p\big)\big) \leq min\big(RightN\big(nb_A(p), p\big)\big)\big)\big)\big)
\end{aligned}$$

*Proof*:
We prove this by induction over the step semantics.
Let $p \in \mathcal{P}$ be an arbitrary process with $LeftN\big(nb_A(p), p\big) \neq \emptyset$ since $\mathcal{P}$ is finite and $\leq$ is a total order according to Definition 11 $max\big(LeftN\big(nb_A(p), p\big)\big)$ is defined. Obviously it holds

$$\begin{aligned}
&LeftN\big(nb_A(p), p\big) = LeftN\big(nb_A(p), p\big) \quad \text{and} \\
&max\big(LeftN\big(nb_A(p), p\big)\big) \geq max\big(LeftN\big(nb_A(p), p\big)\big)
\end{aligned}$$

Let $A'$ be an arbitrary configuration with $A \Longmapsto A' \longmapsto R$ and

$$LeftN\big(nb_{A'}(p), p\big) \neq \emptyset \ \wedge \ max\big(LeftN\big(nb_{A'}(p), p\big)\big) \geq max\big(LeftN\big(nb_A(p), p\big)\big)$$

We have to show that then it holds that

$$LeftN\big(nb_R(p), p\big) \neq \emptyset \ \wedge \ max\big(LeftN\big(nb_R(p), p\big)\big) \geq max\big(LeftN\big(nb_A(p), p\big)\big)$$

According to Lemma 10 it holds that if the executed step $A' \longmapsto R$ is not performed by the subprocess $Alg_{match}(p)$ or it holds that $select\big(findLin\big(p, nb_{A'}(p)\big)\big) \neq (j,k) \wedge (j < k \wedge k < p)$ and $select\big(findLin\big(p, nb_{A'}(p)\big)\big) \neq (j,k) \wedge (j < k \wedge p < j)$ it holds that $nb_{A'}(p) \subseteq nb_R(p)$ and therefore with $LeftN\big(nb_{A'}(p), p\big) \neq \emptyset$ it follows

$$LeftN\big(nb_R(p), p\big) \neq \emptyset$$

according to the definition of $LeftN$ and $LeftN\big(nb_{A'}(p), p\big) \subseteq LeftN\big(nb_R(p), p\big)$ and hence $max\big(LeftN\big(nb_{A'}(p), p\big)\big) \in LeftN\big(nb_R(p), p\big)$.
Now there are two cases. It could be possible that all new left neighbors are smaller than the previous greatest left neighbor or $p$ has a new greatest left neighbor. If $max\big(LeftN\big(nb_R(p), p\big)\big) = max\big(LeftN\big(nb_{A'}(p), p\big)\big)$ it follows with $max\big(LeftN\big(nb_{A'}(p), p\big)\big) \geq max\big(LeftN\big(nb_A(p), p\big)\big)$ that also

$$max\big(LeftN\big(nb_R(p), p\big)\big) \geq max\big(LeftN\big(nb_A(p), p\big)\big)$$

If $max\big(LeftN\big(nb_R(p), p\big)\big) \neq max\big(LeftN\big(nb_{A'}(p), p\big)\big)$ it follows according to the definition of $LeftN$ and $max\big(LeftN\big(nb_{A'}(p), p\big)\big) \in LeftN\big(nb_R(p), p\big)$ that $max\big(LeftN\big(nb_R(p), p\big)\big) > max\big(LeftN\big(nb_{A'}(p), p\big)\big)$ and therefore with $max\big(LeftN\big(nb_{A'}(p), p\big)\big) \geq max\big(LeftN\big(nb_A(p), p\big)\big)$ also

$$max\big(LeftN\big(nb_R(p), p\big)\big) \geq max\big(LeftN\big(nb_A(p), p\big)\big)$$

If the subprocess $Alg_{match}(p)$ executed the step, there are only the following cases left:

- $select\big(findLin\big(p, nb_{A'}(p)\big)\big) = (j,k) \wedge (j < k \wedge k < p)$:

  $$Alg_{ges}\big(P_{A'}, P'_{A'}, nb_{A'}, Msgs_{A'}, add_{A'}\big) \longmapsto R$$
  with $R \equiv Alg_{ges}\big(P_{A'}, P'_{A'}, nb_R, Msgs_{A'} \cup \{(j,k)\}, add_{A'}\big)$
  and $nb_R(x) = \begin{cases} nb_{A'}(x), & \text{if } x \neq p \\ nb_{A'}(p) \setminus \{j\}, & \text{if } x = p \end{cases}$

  Since $j < p$ and $k < p$ and $j, k \in nb_{A'}(p)$ it holds according to the definition of $LeftN$ that $j, k \in LeftN\big(nb_{A'}(p), p\big)$ and with $j < k$ therefore $max\big(LeftN\big(nb_{A'}(p), p\big)\big) \neq j$. Hence, $max\big(LeftN\big(nb_{A'}(p), p\big)\big) \in nb_R(p) = nb_{A'}(p) \setminus \{j\}$. Since additionally no process is added to the neighborhood it holds that $max\big(LeftN\big(nb_R(p), p\big)\big) = max\big(LeftN\big(nb_{A'}(p), p\big)\big)$ and therefore with $max\big(LeftN\big(nb_{A'}(p), p\big)\big) \geq max\big(LeftN\big(nb_A(p), p\big)\big)$ that also

$$max\big(LeftN\big(nb_R(p), p\big)\big) \geq max\big(LeftN\big(nb_A(p), p\big)\big)$$

- $select\big(findLin\big(p, nb_{A'}(p)\big)\big) = (j, k) \wedge (j < k \wedge p < j)$:

$$\begin{aligned}
&Alg_{ges}\left(P_{A'}, P'_{A'}, nb_{A'}, Msgs_{A'}, add_{A'}\right) \longmapsto R \\
&\text{with } R \equiv Alg_{ges}\left(P_{A'}, P'_{A'}, nb_R, Msgs_{A'} \cup \{(k, j)\}, add_{A'}\right) \\
&\text{and } nb_R(x) = \begin{cases} nb_{A'}(x), & \text{if } x \neq p \\ nb_{A'}(p) \setminus \{k\}, & \text{if } x = p \end{cases}
\end{aligned}$$

Since $p > k$ it holds according to the definition of $LeftN$ that $k \notin LeftN\big(nb_{A'}(p), p\big)$ and therefore $max\big(LeftN\big(nb_{A'}(p), p\big)\big) \neq k$. Since additionally no process is added to the neighborhood it holds that $LeftN\big(nb_R(p), p\big) = LeftN\big(nb_{A'}(p), p\big)$ and hence $max\big(LeftN\big(nb_R(p), p\big)\big) = max\big(LeftN\big(nb_{A'}(p), p\big)\big)$. It holds $max\big(LeftN\big(nb_{A'}(p), p\big)\big) \geq max\big(LeftN\big(nb_A(p), p\big)\big)$ and therefore it follows that

$$max\big(LeftN\big(nb_R(p), p\big)\big) \geq max\big(LeftN\big(nb_A(p), p\big)\big)$$

The proof for $RightN\big(nb_R(p), p\big)$ is similar. □

This also induces that the shortest edges of each process never gets longer.

**Corollary 14: Shortest Edge to each Side only gets shorter**

Let $A$ be an arbitrary configuration of the algorithm. Let $R$ be an arbitrary reachable configuration i. e., $A \Longmapsto R$. It holds that the shortest outgoing edge of any process in $R$ in each direction is shorter or equal to the shortest outgoing edge of this process in $A$ in the same direction. This follows directly with Definitions 15, 27 and 16.

It follows moreover that whenever a process knows a process on each side, its neighborhood always contains at least two (or in the case of the smallest and greatest process one) other processes.

**Corollary 15: Neighborhood stays ≥ 2 (resp. 1 for Min and Max)**

Let $A$ be an arbitrary configuration of the algorithm. Let $R$ be an arbitrary reachable configuration i. e., $A \Longmapsto R$. It holds

$$\begin{aligned}
&\forall p \in \mathcal{P} \setminus \{min(\mathcal{P}), max(\mathcal{P})\}.\ |LeftN\big(nb_A(p), p\big)| \geq 1 \wedge |RightN\big(nb_A(p), p\big)| \geq 1 \implies \\
&\qquad |LeftN\big(nb_R(p), p\big)| \geq 1 \wedge |RightN\big(nb_R(p), p\big)| \geq 1 \wedge |nb_R(q)| \geq 2 \\
&\forall q \in \{min(\mathcal{P}), max(\mathcal{P})\}. |nb_A(q)| \geq 1 \implies |nb_R(q)| \geq 1
\end{aligned}$$

Whenever this is the case, the reception of every unknown process id leads to an enabled linearization step and stops the process from sending *keep-alive*-messages for now (if this was possible previously).

### 4.2.2 Open Lemmata

The sending of *keep-alive*-messages can reestablish an edge in the topology without message that was already removed through a linearization step. However, if a process executes a linearization step, it prevent the further away process eventually from ever sending *keep-alive*-messages to it again.

**Lemma 27: Stop sending *Keep-alive*-messages during Linearization Step**

If a process $p$ executes a linearization step, then the further away process $r$ eventually can never send *keep-alive*-messages to $p$ again.
*Proofsketch*:
Let $C$ be an arbitrary configuration and an arbitrary process $p$ executes a linearization step in $C$, i. e., $(select(findLin(p, nb_C(p))) = (q,r) \wedge (q < r \wedge p < q)) \vee (select(findLin(p, nb_C(p))) = (r,q) \wedge (r < q \wedge q < p))$ and $C \longmapsto C'$ with

$$C' \equiv Alg_{ges}\left(P_C, P'_C, nb_{C'}, Msgs_C \cup \{(r,q)\}, add_C\right)$$
$$\text{and } nb_{C'}(x) = \begin{cases} nb_C(x), & \text{if } x \neq p \\ nb_C(p) \setminus \{r\}, & \text{if } x = p \end{cases}$$

Assume $select(findLin(p, nb_C(p))) = (q,r) \wedge (q < r \wedge p < q)$ (the other case is similar). Since $(r,q) \in Msgs_{C'}$ it holds according to Corollary 4 that a configuration $C''$ is reached with $q \in nb_{C''}(r)$. For every reachable configuration $R$ i. e., $C'' \Longmapsto R$ it holds according to Lemma 26 with $q < r \wedge p < q$ that there is a process $s \in \mathscr{P}$ with $s \in LeftN(nb_R(r), r)$ and $s > p$. Since $p < r$ it holds if $p \in nb_R(r)$, then $p \in LeftN(nb_R(r), r)$ and $|nb_R(r)| \geq 2$. Therefore, there is an enabled linearization step for $r$ in $R$ and $r$ cannot send *keep-alive*-messages according to Lemma 6. Hence, it holds for every from $C''$ reachable configuration $R$, if $r$ can send *keep-alive*-messages then $p \notin nb_R(r)$ and therefore $r$ can never send *keep-alive*-messages to $p$ again.

If there are only three processes in the system, then an enabled linearization step is successful after a finite number of steps i. e., the longer edge is removed and will never be reestablished. With a similar argumentation it holds that an edge which cannot be reestablished through other linearization steps is removed permanently under the assumption that a corresponding linearization step is chosen often enough.

**Lemma 28: Linearization Step for Three Processes**

Let $|Procs| = 3$ and $\mathscr{P} = p, q, r$ with $p < q < r$. Let $A$ be an arbitrary configuration with $nb_A(p) = \{q, r\}$ (respectively $nb_A(r) = \{p, q\}$). After a finite number of steps there is a configuration $C$ reached with $(p, r) \notin E_{NTM(C)}$ (respectively $(r, p) \notin E_{NTM(C)}$). It holds then for every reachable configuration $R$ i. e., $C \Longmapsto R$ that $(p,r) \notin E_{NTM(R)}$ (respectively $(r,p) \notin E_{NTM(R)}$).
*Proofsketch:*
According to Lemma 3 the subprocess $Alg_{match}(p)$ is enabled and stays enabled until it executes a step. Let $A'$ be the configuration before this step. According to Lemma 10 it holds that $nb_{A'}(p) = \{q,r\}$. Since $p < q < r$ and $\mathcal{P} = p,q,r$ it holds with the definition of *select*(*findLin*$(\cdot)$) that $select(findLin(p, nb_A(p))) = (q,r)$. Hence, according to Corollary 2 $A' \longmapsto C'$ with

$$C' \equiv Alg_{ges}\left(P_{A'}, P'_{A'}, nb_{C'}, Msgs_{A'} \cup \{(r,q)\}, add_{A'}\right)$$
$$\text{and } nb_{C'}(x) = \begin{cases} nb_{A'}(x), & \text{if } x \neq p \\ nb_{A'}(p) \setminus \{r\}, & \text{if } x = p \end{cases}$$

According to Lemma 27 a configuration $C''$ is reached with $q \in nb_{C''}(r)$ and $r$ can never send *keep-alive*-messages again to $p$. Let $m$ be the number of connections from $p$ to $r$ in $C''$, i. e., $m = m_1 + m_2 + m_3$ whereby $m_1$ is the number of $\overline{p}\langle r\rangle$ messages in the system that are still in transit in $C''$ $m_1 = |\{\!\{x \in Msgs_{C''} | x = (p,r)\}\!\}|$ , $m_2$ adds one if $p$ wants to add $r$ to its neighborhood i. e., $m_2 = \begin{cases} 1 & \text{if } add_{C''}(p) = r \\ 0 & \text{else} \end{cases}$ and $m_3$ adds one if $r$ is already in the neighborhood of $p$ i. e., $m_3 = \begin{cases} 1 & \text{if } r \in nb_{C''}(p) \\ 0 & \text{else} \end{cases}$. It holds by induction that a configuration $C$ is reached with $(p,r) \notin E_{NTM(C)}$ and for every reachable configuration $R$ i. e., $C \Longmapsto R$ it holds further that $(p,r) \notin E_{NTM(R)}$. If $m = 0$, then it holds $(p,r) \notin E_{NTM(C'')}$ according to Definition 28. Since there is no connection from $p$ to $r$ , $r$ can never again send *keep-alive*-messages to $p$ and there is never a message $\overline{p}\langle r\rangle$ sent in a linearization step, it holds for every reachable configuration $R$ i. e., $C'' \Longmapsto R$ that $(p,r) \notin E_{NTM(R)}$. Assume $m > 0$ connections and the claim holds for all $m' < m$. Since $r$ can never again send *keep-alive*-messages to $p$ and there is never a message $\overline{p}r$ sent in a linearization step, it holds for every reachable configuration that the number of connections cannot be increasing. Since $m > 0$ it holds according to 28 that $\overline{p}\langle r\rangle \in Msgs_{C''}$, $add_{C''}(p) = r$ or $r \in nb_{C''}(p)$. With Lemma 4 and Corollary 4 in all cases a configuration $L$ is reached with $r \in nb_L(p)$. According to Lemma 3 the subprocess $Alg_{match}(p)$ is enabled and stays enabled until it executes a step. Let $L'$ be the configuration before this step. According to Lemma 10 it holds that $r \in nb_{L'}(p)$. Since $p < q < r$ and $\mathcal{P} = p,q,r$ it holds that $q = succ(p)$ and therefore with Lemma 11 follows $nb_{L'}(p) = \{q,r\}$. It holds with the definition of *select*(*findLin*$(\cdot)$) that $select(findLin(p, nb_{L'}(p))) = (q,r)$. Hence, according to Corollary 2 $L' \longmapsto L''$ with

$$L'' \equiv Alg_{ges}\left(P_{L'}, P'_{L'}, nb_{L''}, Msgs_{L'} \cup \{(r,q)\}, add_{L'}\right)$$
$$\text{and } nb_{L''}(x) = \begin{cases} nb_{L'}(x), & \text{if } x \neq p \\ nb_{L'}(p) \setminus \{r\}, & \text{if } x = p \end{cases}$$

Hence, there are with Definition 28 in $L''$ now $m'' < m'$ connections from $p$ to $r$ and the claim holds with the induction hypothesis.

If there are only three processes in the system, then strong convergence holds for any arbitrary initial configuration. If there are only edges missing, then the configuration is an undirected correct configuration and strong convergence is ensured. Otherwise the possible linearization step(s) will be successful finished after a finite number of steps and therefore a undirected configuration reached.

**Lemma 29: Convergence for Three Processes**

Let $|Procs| = 3$ and $\mathscr{P} = p, q, r$ with $p < q < r$. Let $I$ be an arbitrary connected initial configuration i. e., the undirected topology with messages of $I$ is connected. A correct configuration $C$ is reached after a finite number of steps i. e.,

$$I \Longmapsto C \quad \wedge \quad NTM(C) = G_{LIN} \wedge NT(C) = G_{LIN}$$

*Proofsketch:*
If $C$ is a correct configuration itself, then the claim follows directly. Otherwise there are three cases. If there are only too many edges in the topology with messages, it holds that $NTM(I) \supseteq G_{LIN}$. Hence, with Lemma 22 it holds that a correct configuration $C$ is reached after a finite number of steps i. e., $I \Longmapsto C$ and $NTM(C) = G_{LIN} \wedge NT(C) = G_{LIN}$. If there is no more possible linearization step in the undirected topology with message, then its holds with the assumption that $I$ is connected that $NTM(I) \subseteq G_{LIN}$ respectively $UNTM(I) = UG_{LIN}$. Hence, with Lemma 20 it holds that a correct configuration $C$ is reached after a finite number of steps i. e., $I \Longmapsto C$ and$NTM(C) = G_{LIN} \wedge NT(C) = G_{LIN}$. If there is still a possible linearization step in the undirected topology with messages, than there are two cases. Either there is a possible linearization step in the directed topology with messages or not. In both cases with Lemma 24 it holds a configuration $L$ is reached in which there is an enabled linearization step i. e., it either holds $nb_L(p) = \{q, r\}$ or $nb_L(r) = \{p, q\}$ or both. If both and therefore all possible linearization steps are enabled, it holds with Lemma 28 that there are configurations $L'$ with $(p, r) \notin E_{NTM(L')}$ and $L''$ with $(r, p) \notin E_{NTM(L'')}$ reached. It holds either $I \Longmapsto L' \Longmapsto L''$ or $I \Longmapsto L'' \Longmapsto L'$. Assume $I \Longmapsto L' \Longmapsto L''$ (the proof for the second case is similar). With $L' \Longmapsto L''$ and $(p, r) \notin E_{NTM(L')}$ it holds that $(p, r) \notin E_{NTM(L'')}$. Since the undirected topology with messages of $I$ is connected, it holds with $I \Longmapsto L''$ and Lemma 7 that also $L''$ is connected. Since $L''$ is connected and $\mathscr{P} = p, q, r$, but $(p, r), (r, p) \notin E_{NTM(L'')}$, it holds with Definitions 28 and 32 that $L''$ is an undirected correct configuration. Therefore, with Lemma 20 it holds that a correct configuration $C$ is reached after a finite number of steps i. e., $I \Longmapsto L' \Longmapsto L'' \Longmapsto C$ and $NTM(C) = G_{LIN} \wedge NT(C) = G_{LIN}$. For the case that only one linearization step is enabled, assume $nb_L(p) = \{q, r\}$ (the proof for $nb_L(r) = \{p, r\}$ is similar). The enabled linearization step will be successful after a finite number of steps according to Lemma 28 and therefore a configuration $L'$ with $(p, r) \notin E_{NTM(L')}$ is reached. If furthermore $(r, p) \notin E_{NTM(L')}$, then $L'$ is according to Definitions 28 and 32 an undirected correct configuration, since $L'$ is connected with Lemma 7 and $\mathscr{P} = p, q, r$, but $(p, r), (r, p) \notin E_{NTM(L')}$. It then follows with Lemma 20 that a correct configuration $C$ is reached after a finite number of steps i. e., $I \Longmapsto L \Longmapsto L' \Longmapsto C$ and $NTM(C) = G_{LIN} \wedge NT(C) = G_{LIN}$. Otherwise it holds that $(r, p) \in E_{NTM(L')}$. With Definition 28 it holds therefore $p \in nb_{L'}(r)$, $add_{L'}(r) = p$ or $(r,p) \in Msgs_{L'}$. Since the other linearization step is permanently finished it holds additionally $q \in nb_{L'}(r)$. Therefore, there is a possible linearization step in the directed topology with messages and according to Lemma 4 and Corollary 4 a configuration $L''$ is reached with $p \in nb_{L''}(r)$. Since $\mathcal{P} = p,q,r$ with $p < q < r$ it holds $q = pred(r)$ and therefore with Lemma 11 that $q \in nb_{L''}(r)$. Therefore for $r$ is a linearization step enabled in $L''$ and with Lemma 28 a configuration $L'''$ is reached with $(r,p) \notin E_{NTM(L''')}$. Since $L'''$ is connected according to Lemma 7 and $\mathcal{P} = p,q,r$, but $(p,r),(r,p) \notin E_{NTM(L''')}$ it holds with Definitions 28 and 32 that $L'''$ is an undirected correct configuration. Therefore, with Lemma 20 it holds that a correct configuration $C$ is reached after a finite number of steps i. e., $I \Longmapsto L \Longmapsto L' \Longmapsto L'' \Longmapsto L''' \Longmapsto C$ and $NTM(C) = G_{LIN} \wedge NT(C) = G_{LIN}$.

If a process sends *keep-alive*-messages at several times in an execution, then the processes it sends these to, get never further away over the execution. Therefore, the distances only decrease and the sending of *keep-alive*-messages can only get less disadvantageous.

**Lemma 30: Ids in *Keep-alive*-messages get only Closer**

Let $p \in \mathcal{P}$ be an arbitrary process. Let $H$ and $H'$ be arbitrary configurations in which $p$ sends *keep-alive*-messages with $H \Longmapsto H'$. It holds that

$$\forall q \in LeftN\big(p, nb_H(p)\big).\forall q' \in LeftN\big(p, nb_{H'}(p)\big).dist(p,q) \geq dist(p,q') \text{ and}$$
$$\forall q \in RightN\big(p, nb_H(p)\big).\forall q' \in RightN\big(p, nb_{H'}(p)\big).dist(p,q) \geq dist(p,q')$$

*Proofsketch:*
According to Lemma 6 it holds that $|LeftN(p, nb_H(p))| \leq 1$, $|RightN(p, nb_H(p))| \leq 1$, $|LeftN(p, nb_{H'}(p))| \leq 1$ and $|RightN(p, nb_{H'}(p))| \leq 1$ since $p$ sends *keep-alive*-messages in $H$ and $H'$. If $|LeftN(p, nb_H(p))| = 0$ or $|RightN(p, nb_H(p))| \leq 0$ the corresponding claim holds trivially. For $|LeftN(p, nb_H(p))| = 1$ respectively $|RightN(p, nb_H(p))| = 1$ it holds with Lemma 26 that $|LeftN(p, nb_{H'}(p))| = 1$ and $max\big(LeftN(nb_{H'}(p), p)\big) \geq max\big(LeftN(nb_H(p), p)\big)$ respectively $|RightN(p, nb_{H'}(p))| = 1$ and $min\big(RightN(nb_{H'}(p), p)\big) \geq min\big(RightN(nb_H(p), p)\big)$. Let be $l = max\big(LeftN(nb_H(p), p)\big)$, $l' = max\big(LeftN(nb_{H'}(p), p)\big)$, $r = min\big(RightN(nb_H(p), p)\big)$ and $r' = min\big(RightN(nb_{H'}(p), p)\big)$. With the definitions of *LeftN* and *RightN* and Definition 15 it follows $dist(p,l) \geq dist(p,l')$ and $dist(p,r) \geq dist(p,r')$. Every execution of a linearization step means progress for the stabilization of the system.

**Lemma 31: Progress Linearization step**

Let $A$ be an arbitrary configuration and $p$ executes a linearization step meanwhile it sends a message $\overline{q}\langle r \rangle$. If $q$ does not know any process on this side, it will eventually get to know a process and then always knows one on this side. If $q$ only knows further away processes than $r$ on this side, then $q$ will improve its nearest neighbor on this side at the latest with reception of the sent message and will be able to execute another linearization step itself, whereby the next time $q$ can send *keep-alive*-messages its to a nearer process on this side than before and therefore it never sends such messages to $p$ or another further away process than $r$ again. If $q$ already knows with or without messages a closer process than $r$, then $q$ also will eventually be able to execute a linearization step itself, meaning again more progress for the stabilization. In every case, a connection between two more distant processes is exchanged through a shorter connection.
*Proofsketch:*
The message $\overline{q}\langle r\rangle$ will according to Corollary 4 eventually received and processed and therefore a configuration is reached in which $q$ has a process in its left or right neighborhood (dependent on whether $r$ is smaller or greater than $q$). According to Lemma 26 $q$ will then always have a process in the neighborhood on this side. If $q$ already had another neighbor on this side before the processing of this message, then according to Lemma 6 it will enable $q$ to execute a linearization step itself. If $r$ is nearer than every neighbor that $q$ had before on this side then according to Lemma 26 the closest neighbor on this side is improved permanently. Since processes can only send *keep-alive*-messages if they only have at most one process on this side and they never remove their closest neighbor on a side, it holds additionally that $q$ will never send such messages again to a further away process than $r$ on this side (especially not $p$). It holds for the executed linearization step that either $p < r < q$ or $q < r < p$ and therefore with definition 15 that $dist(q, r) < dist(p, q)$. Since one connection from $p$ to $q$ is removed and one from $q$ to $r$ is added, it holds that a connection between two more distant processes is exchanged through a shorter connection.

As soon as a process knows all of its desired neighbors it never sends undesired *keep-alive*-messages again and therefore no further step of this process increases $\Psi$ and $\Psi_{\Sigma}$. If every process knows all desired neighbors, then the desired topology is a subgraph of the topology with and without messages and strong convergence is ensured through Lemma 21.

**Lemma 32: Stop sending undesired *Keep-alive*-messages**

Let $C$ be an arbitrary configuration and $p \in \mathscr{P}$ an arbitrary process with $\{succ(p), pred(p)\}_{\perp} \subseteq nb_C(p)$. It holds for every reachable configuration $R$ with $C \Longmapsto R$ that $p$ cannot send *keep-alive*-messages in $R$ to a process $q \notin \{succ(p), pred(p)\}_{\perp}$.
*Proofsketch*:
For every reachable configuration $R$ holds with Lemma 11 that $\{succ(p), pred(p)\}_{\perp} \subseteq nb_R(p)$. A process sends *keep-alive*-messages always to its complete neighborhood. Therefore, if $p$ could send *keep-alive*-messages in $R$ to $q \notin \{succ(p), pred(p)\}_{\perp}$, it holds $q \in nb_R(p)$. But since all desired neighbors are in the neighborhood of $p$ i. e., $\{succ(p), pred(p)\}_{\perp} \subseteq nb_R(p)$, it holds that there is an enabled linearization step for $p$ and therefore according to 6 $p$ cannot send *keep-alive*-messages at all.

## 4.3 Approaches

The first approach is a structural induction over the topology with messages. If a border process, starting with the process with minimum or the process with the maximum id, has in the topology with messages only edges to and from its desired neighbors, then every step that it executes in every execution can only be the sending to or receiving of *keep-alive*-messages from its desired neighbors. It can can only send *keep-alive*-messages as it does not have two neighbors on the same side. Since the only process that could have it in its neighborhood is a desired neighbor, it will never receive another message than a *keep-alive*-messages anymore, as messages in linearization steps are always sent to the further away process. The process closest to the current border process inwards the system is then the new border process with the same properties. Therefore, from the outside inwards the system stabilizes. Nevertheless, it is not simply possible to prove that the current border processes eventual permanently remove all their connections to undesired neighbors. Even harder to prove is that all other processes remove their connections to them. This is based on the fact that the border process and the inner system influence each other. There are, for example, configurations as depicted in Figure 4.4 where first the inner system has to execute certain steps to enable a border process to remove its connections to undesired neighbors. This requires some form of livelock freedom in the inner system without introducing circular reasoning. In the depicted situation the induction hypothesis could help as the border process can not influence the inner system as long as the inner system does not communicate with a border process. However, as long as there are connections from the border processes to or from undesired neighbors it is possible that we alter between situations where the border processes have to execute certain steps to enable the inner system to execute certain steps and vice versa. In such situations the induction hypothesis can not always be used. Consider a configuration with a topology with messages like depicted in Figure 4.7. The border process $z$ has to execute its linearization step in order to enable the inner system containing $v$, $w$, $x$ and $y$ to execute a linearization step. The border processes $u$ and $z$ can do nothing in the meantime except sending *keep-alive*-messages, which does not enable further linearization steps in the inner system. If the inner process $v$ does not send *keep-alive*-messages before the reception of the message from $z$, it has to execute its linearization step in order to be again able to send *keep-alive*-messages to $u$. This is only ensured to happen if $v$ can permanently remove its edge to $x$, which potentially requires some form of livelock freedom in the inner system although the induction hypothesis can not be used in this case. Only by sending these *keep-alive*-messages the border process $u$ can execute a linearization step after reception.

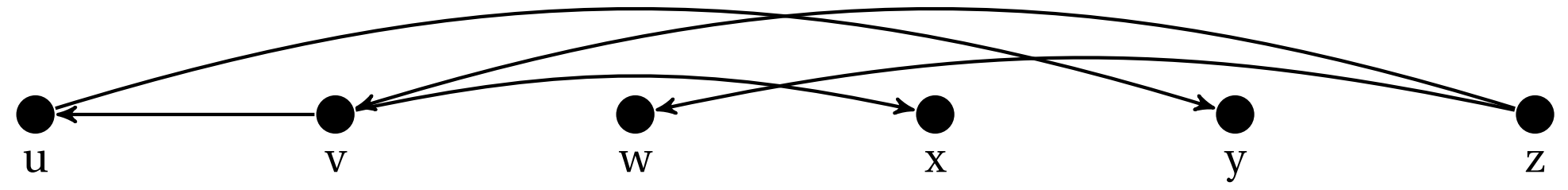

Figure 4.7: Topology where the border processes and the inner system alternately influence each other

Neither *keep-alive*-messages nor linearization steps establishing edges that are longer as the current longest edge in the topology with messages. The potentially new edge that is established through a linearization step is even strictly shorter than the at least temporarily removed one. If one could show that always the current longest edges are removed permanently, strong convergence would hold. This could be shown through induction. The induction variable could be either the length of the longest edge, or one could go for a nested induction over this length and the number of nodes. Another similar possibility is a proof through convergence stairs. The maximum length of an edge in a system with $n$ processes is $n-1$. Show that for a configuration with a maximum edge length of $k > 1$ in the topology with messages, eventually in every execution a configuration is reached with a maximum length of $k-1$. A configuration with a maximum edge length of one in the topology with messages is an undirected correct configuration and therefore strong convergence holds. It could be helpful for this approach to change the algorithm to the variant $LIN_{max}$, which could be simply done by adapting the *select* functions of the processes. In this variant we have more knowledge about which linearization steps are executed because the processes always select linearization steps that, at least temporarily, remove their longest edge. Nevertheless, the problem remains to show that the linearization steps regarding the longest edges in the topology with messages always get enabled without introducing circular reasoning.

The potential function $\Psi_E$ is monotonically decreasing with every step in every case. Therefore, if we can show that whenever the system is not stabilized yet, that always eventually a step is executed that leads to a strict decrease of $\Psi_E$ strong convergence would be proven. Since $\Psi_E$ is minimal if and only if the topology with messages contains only too many edges, and therefore with Lemma 22 convergence is ensured. Unfortunately, this is not easy to prove. The function $\Psi_E$ is the sum of functions for every process. The decrease of the function from a process $p$ is always caused by the sending of messages of another process $q$. This makes it harder to identify situations that lead to a decrease and reason about them. To make matters worse, there are configurations in which first certain linearization steps have to be executed before a situation is reached where such a strictly decreasing step is enabled (an example is depicted in Figure 4.8). It is very difficult to prove that such a situation is always reached, especially without further knowledge about how the rest of the system evolves in the meantime and no further properties about livelock freedom.

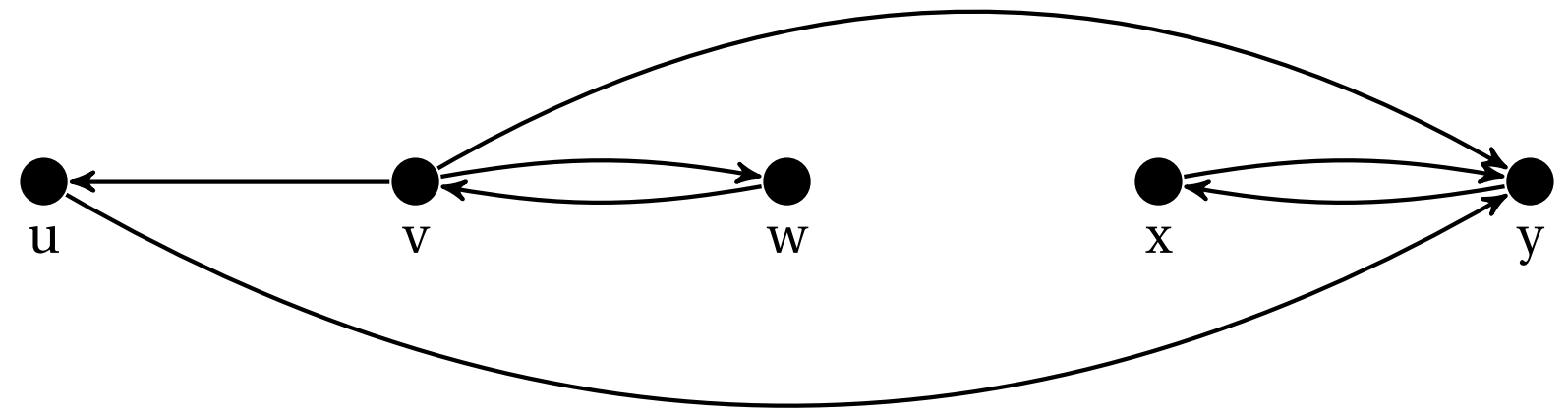

Figure 4.8: Topology with messages without a strict decreasing step

In general, perhaps a more global potential function could help. Indeed, it is really challenging to capture certain global properties formally. Also a monotonically decreasing function only leads to a relatively easy proof of convergence if it is possible to show for every configuration that, if the system is not stabilized yet, the function strictly decreases eventually. Finding a monotonically function, like $\Psi_{\sum}$ or simply the length of the longest edges in the system, is therefore not the hardest part of the convergence proof. But without proving that there are always eventually steps executed in which such a function strictly decreases, it does not prove convergence and all our previous attempts failed for this reason.

A lot of issues in proving strong convergence arise from the fact that almost every lemma regarding liveness properties can just state that eventually a certain property for the system or a certain process holds. Without further invariants, that leads to the problem that we have no knowledge about how the rest of the system evolves in the meantime.

Therefore, it could be more promising to try another approach in proving strong convergence and concentrate even more on properties that could be helpful for a proof through convergence stairs. Here it is especially challenging, for the same reasons as for strong convergence itself, to prove that the next stair i. e., a configuration that satisfies the next predicate, eventually is reached in every execution. Therefore, it is crucial to identify further properties that are weaker than convergence itself, that prevent the system from being captured in a livelock. This would be helpful in all previously stated approaches.

# 5 Conclusion

## 5.1 Conclusion

We redesigned the algorithm for shared memory from Gall et al. (2014) in such a way that our algorithm works in an asynchronous message-passing system. The algorithm of Gall et al. (2014) works in a system where potentially all processes must have access to the whole memory and is therefore very restricted. In the redesigned algorithm, processes communicate via message-passing. Furthermore, we do not require any assumptions about the time a process needs to execute a step or for message delivery. The algorithm works therefore in a completely asynchronous distributed message-passing system. This is a significantly weaker requirement and corresponds more to real life system properties. Therefore, the redesigned algorithm is applicable in more systems than the original one.

In order to model the algorithm for a message-passing system, we extended the localized $\pi$-calculus. This provides us through its precise defined syntax and semantics, with the possibility to model the algorithm in an unambiguous way. Furthermore, this is the basis for proving properties about the algorithm in a formal way. The usage of a kind of standardform of a configuration helps us to simplify the proof by identifying every possible reachable process term with a structural equivalent representative and therefore reducing the cases to be analyzed. Further, this enables us to explicitly keep track of the global state of the system in a convenient way. This allows us to execute our proofs in a state-based fashion, which is more traditional for distributed algorithms (Lynch, 1996), than in an action-based style, that is more natural for process calculi.

An algorithm is self-stabilizing, if it satisfies the properties closure and convergence. We formally proved the closure property, i. e., if the system reaches a correct configuration, then it stays in a correct configuration and therefore every step that can be executed leads again to a correct configuration. The convergence property postulates that from any arbitrary initial configuration a correct configuration is reached. There are two forms of convergence. Starting with an arbitrary initial configuration, strong convergence requires that in every execution a correct configuration is reached, whereas weak convergence only claims the existence of such an execution. We proved strong convergence for two special cases. First, whenever the topology with messages of a connected initial configuration only lacks desired edges but no undesired ones exist, strong convergence holds. This means more concrete that every message to a process in the system only carries the id of the successor or predecessor, the adding in progress involves only consecutive processes, the neighborhood of each process contains only desired neighbors, and between each pair of consecutive processes is at least one kind of these connections in one direction. Second, strong convergence also holds, whenever in the topology with messages of an initial configuration there are just too many edges, but no desired ones are missing. This means, every process knows its desired neighbors, while also taking the messages in transit into account as well as the adding in progress. For the general case, i. e., an arbitrary connected initial configuration in which there are in the topology with messages desired edges missing as well as still undesired edges exist, we proved weak convergence. For this proof, we introduced a global omniscient entity, called a perfect oracle. A perfect oracle is not implementable in reality, but also can, and should, be seen as a restriction of the set of executions for an arbitrary initial configuration. Then, we showed that every execution that is admissible under the assumption of a perfect oracle ensures strong convergence. Since for every initial configuration this is a non-empty set of executions, weak convergence holds in the general case.

Furthermore, we described the problems and challenges for a strong convergence proof in the general case. These problems arise mainly from undesired *keep-alive*-messages which can reestablish already deleted edges. These *keep-alive*-messages are necessary to prevent the system from deadlocks and the algorithm from termination, which would be contradictory for a distributed message-passing self-stabilizing algorithm. In the proof for weak convergence, unnecessary undesired *keep-alive*-messages are suppressed by the perfect oracle. This is not implementable by the nodes themselves or in a real distributed system, as it requires global knowledge about the system. Additionally, without further weaker livelock freedom properties than convergence itself, there is the threat to introduce circular reasoning. Nevertheless, we gave strong arguments for strong convergence for every arbitrary initial configuration and introduced further already proven, and open lemmata, including for example strong convergence in case of three processes, with proof sketches. These lemmata identify further properties that can support a number of different variants of a formal proof of strong convergence. Conclusively, we discussed several approaches for a proof of strong convergence in the general case.

Our proof is, using the precisely defined calculus as basis, much more formal than the original proof. Therefore, the chance of failures in the proof is significantly lower. Furthermore, it has a higher level of detail. This again reduces the risk of failures and leads to a better insight in the properties of the algorithm and overall to a better understanding. In comparison to the original algorithm, our algorithm is not only applicable in a wider field of system properties, but also reduces the threat of partitioning the system in case of faults. In the original algorithm, for example, even the single loss of a connection in a correct configuration leads to a loss of connectivity in the whole system, that has to be reestablished through an external mechanism.

However, the single loss of a connection cannot lead to a disconnection of the system in a correct configuration in our algorithm. A partitioning of the system can only occur, if all connections in both directions between two processes are getting lost through faults. In a correct configuration, there are at least two connections between each pair of consecutive processes.

Self-stabilizing algorithms are very difficult to design. For a distributed message-passing algorithm we need to ensure that the communication between neighboring processes continues in a correct configuration and therefore the algorithm does not terminate. In order to realize this, we introduced the sending of *keep-alive*-messages in case a process does not find a linearization step. Such messages can lead to a reestablishing of already deleted edges. This cannot be simply prevented by the use of additional variables. It is, for example, not possible to introduce a variable in which a process stores which edges it already has deleted through linearization steps and suppress the reestablishing in case of receiving a corresponding id. Every variable in a self-stabilizing algorithm has to be initializable with an arbitrary value. Therefore, in an initial configuration such a variable could contain a desired neighbor. This would prevent the reachability of a correct configuration, as this desired neighbor would never be added to the neighborhood of the process. For each process the reception of messages and searching for linearization step has to be executable concurrently and cannot be executed alternating as this could lead to deadlocks and prevent the stabilization of the system. This implies that a lot of problems in proving strong convergence are also existent in a weaker form, even with synchronous communication between the processes. Therefore, we decided to model the algorithm directly for asynchronous message-passing to lower the system assumptions as much as possible. Consequently, there are always many different sequences of events possible. Progress properties are mostly only eventually statements. Therefore, certain things will eventually happen, but we have no certainty when they happen. To make matters worse, we have no control or knowledge which steps are executed by the processes in the meantime. Therefore, without further invariants we do not know how the rest of the system evolves. Altogether, this leads for each initial configuration to an enormous number of executions that have to be considered. Additionally, every possible configuration has to be regarded as an initial configuration. Since strong convergence is a property that has to consider all executions of all initial configurations, the sheer number of executions by itself, makes a formal proof a challenging task.

## 5.2 Related Work

Self-stabilization in the context of distributed computing was first introduced by Dijkstra (1974). The fundamentals of topological self-stabilization and linearization for this work are originated in Gall et al. (2014). This paper is also the foundation for the algorithm and the main idea of the proofs. The basics in designing a self-stabilizing algorithm and main proof techniques, as well as a general introduction and overview can be found in Dolev (2000).

We used an extended localized $\pi$-calculus to model our algorithm in an unambiguous way and as a formal basis for proofs. The basic localized $\pi$-calculus is based on Merro and Sangiorgi (1998) and extended in a way similar to Wagner and Nestmann (2014). Furthermore, we introduce a kind of standardform based on ideas similar to Wagner and Nestmann (2014) which enables us to explicitly keep track of the global state and therefore the topology of the system. The basis of this idea was, in a very simple form, already introduced by Milner (1999).

## 5.3 Future Work

Regarding the future work, the next step is to prove strong convergence in the general case. The primary goal is therefore a convergence proof in the general case without any oracle necessary. On the way, it could be interesting to lower the restrictions on the set of executions by an oracle and therefore consider a weaker oracle to acquire further insight in properties that could be helpful for a proof without an oracle. Thereby, the weaker the oracle the more executions are proven to ensure convergence.

There are several approaches in order to prove strong convergence in the general case without any oracle. The first task is to identify further properties that ensure livelock freedom in the system that are weaker than strong convergence itself. Such properties, for example, can be used to prove that either the monotone decreasing function $\Psi_E$ always decreases strictly eventually whenever the system is not stabilized yet or to show that linearization steps involving the longest edges in the system are eventually enabled and executed. With one of the two properties a convergence proof for the general case can be achieved relatively easy. Further livelock freedom properties would also be helpful if considering a potential function that utilizes more global knowledge about the system, as well as a proof using no potential function at all, but works with convergence stairs.

Another future line of work is the consideration of confluence properties of the system. Further, it would be interesting to investigate the possibility of the parallel execution of steps. This would build the basis to analyze the complexity of the algorithm. Also the scalability of the algorithm is of further interest in this context. The algorithm could then be compared to other algorithms or variants of the algorithm itself. For example, the usage of counter variables that only allow to trigger *keep-alive*-messages every $n-$th time, could decrease the sending of undesired unnecessary *keep-alive*-messages significantly.

Furthermore, it would be interesting to consider how far the algorithm could be adapted in order to establish more complex overlay topologies, for example a hypercube. A line is the basic element for a lot of shapes, therefore it should be possible to simulate an algorithm that establishes a more complex topology, for example by running several instances of our algorithm. To analyze this possibility in comparison to other algorithms is an interesting future task. The investigation of algorithms for more complex topologies is in general a promising further field of work.

# List of Figures

# List of Tables

# List of Definitions

# List of Lemmata

# List of Corollaries

Hagit Attiya and Jennifer Welch. *Distributed Computing: Fundamentals, Simulations and Advanced Topics*. Wiley, 2. edition, 2004.

Paul-David Brodmann. Distributability of Asynchronous Process Calculi. Master's thesis, Technische Universität Berlin, Germany, October 2014.

Edmund M. Clarke and Jeannette M. Wing. Formal Methods: State of the Art and Future Directions. *ACM Comput. Surv.*, 28(4):626–643, December 1996.

Edsger W. Dijkstra. Self-stabilizing Systems in Spite of Distributed Control. *Communications of the ACM*, 17(11):643–644, November 1974.

Shlomi Dolev. *Self-Stabilization*. The MIT Press, Cambridge, Massachusetts, 2000.

Dominik Gall, Riko Jacob, Andrea Richa, Christian Scheideler, Stefan Schmid, and Hanjo Täubig. A Note on the Parallel Runtime of Self-Stabilizing Graph Linearization. *Theory of Computing Systems*, 55(1):110–135, 2014.

Felix C. Gärtner. Fundamentals of Fault-tolerant Distributed Computing in Asynchronous Environments. *ACM Comput. Surv.*, 31(1):1–26, March 1999.

Mohamed G. Gouda. The Triumph and Tribulation of System Stabilization. In *Proceedings of the 9th International Workshop on Distributed Algorithms*, WDAG '95, pages 1–18, London, UK, UK, 1995. Springer-Verlag.

Rachid Guerraoui and Luís Rodrigues. *Introduction to Reliable Distributed Programming*. Springer, 2006.

Leslie Lamport. Proving the Correctness of Multiprocess Programs. *IEEE Trans. Softw. Eng.*, 3 (2):125–143, March 1977.

Nancy Gail Leveson and Clark Savage Turner. An Investigation of the Therac-25 Accidents. *IEEE Computer*, 26(7):18–41, Juli 1993.

Nancy A. Lynch. *Distributed Algorithms*. Morgan Kaufmann Publishers Inc., San Francisco, CA, USA, 1996.

Mars Climate Orbiter Mishap Investigation Board, Jet Propulsion Laboratory (U.S.), and United States National Aeronautics and Space Administration (NASA). *Mars Climate Orbiter Mishap Investigation Board: Phase I Report.* Jet Propulsion Laboratory, 1999.

Massimo Merro and Davide Sangiorgi. On Asynchrony in Name-Passing Calculi. In KimG. Larsen, Sven Skyum, and Glynn Winskel, editors, *Automata, Languages and Programming*, volume 1443 of *Lecture Notes in Computer Science*, pages 856–867. Springer Berlin Heidelberg, 1998.

Robin Milner. *communicating and mobile systems: the $pi$-calculus.* Cambridge university press, 1999.

Robin Milner, Joachim Parrow, and David Walker. A Calculus of Mobile Processes, I. *Inf. Comput.*, 100(1):1–40, September 1992.

Kirstin Peters, Uwe Nestmann, and Ursula Goltz. On Distributability in Process Calculi. In Matthias Felleisen and Philippa Gardner, editors, *Programming Languages and Systems*, volume 7792 of *Lecture Notes in Computer Science*, pages 310–329. Springer Berlin Heidelberg, 2013.

Davide Sangiorgi and David Walker. *The pi-calculus: a Theory of Mobile Processes.* Cambridge university press, 2003.

Marco Schneider. Self-stabilization. *ACM Comput. Surv.*, 25(1):45–67, March 1993.

Andrew S. Tanenbaum and Maarten van Steen. *Verteilte Systeme: Prinzipien und Paradigmen.* Pearson Studium, 2. edition, 2008.

Rob van Glabbeek, Ursula Goltz, and Jens-Wolfhard Schicke. On Synchronous and Asynchronous Interaction in Distributed Systems. In Edward Ochmański and Jerzy Tyszkiewicz, editors, *Mathematical Foundations of Computer Science 2008*, volume 5162 of *Lecture Notes in Computer Science*, pages 16–35. Springer Berlin Heidelberg, 2008.

Christoph Wagner and Uwe Nestmann. States in Process Calculi. In *Proceedings of EXPRESS/-SOS*, volume 160 of *EPTCS*, pages 48–62, 2014.